%% file: buta.tex
\documentclass[cup6b]{cupbook}

\usepackage{graphicx} 
\usepackage{color}
\usepackage{morefloats}
\title[ XXIII Canary Islands Winter School of Astrophysics]{Secular evolution of galaxies}
\author{J. Falc\'on-Barroso and J.~H.~Knapen} 
\date{20th March 2012}

\begin{document}

\maketitle
\cleardoublepage
\input{chapter_buta}

\end{document}

%% file: chapter_buta.tex
\def\thechapter{}

\author[Ronald J. Buta]{Ronald J. Buta\\                              
        Department of Physics and Astronomy, University of Alabama\\
        Box 870324, Tuscaloosa, AL 35487, USA\\
        buta@sarah.astr.ua.edu}                                                                       
                                                                                                    
\chapter{Galaxy morphology}                                                                         
                                                                                                  
\abstract{Galaxy morphology has many structures that are suggestive of various\linebreak
processes or stages of secular evolution.  Internal perturbations such as bars
can drive secular evolution through gravity torques that move gas into the
central regions and build up a flattened, disk-like central bulge, or which may
convert an open spiral pseudoring into a more closed ring. Interaction between
individual components of a galaxy, such as between a bar and a dark halo, a bar
and a central mass concentration, or between a perturbation and the basic state
of a stellar disk, can also drive secular transformations. In this series of
lectures, I review many aspects of galaxy morphology with a view to delineating
some of the possible evolutionary pathways between different galaxy types.}

\def\thechapter{2}

%
%

\section{Introductory remarks}
A principal goal of extragalactic studies has been to understand what drives the
morphology of galaxies. It is important to determine the dynamical and
evolutionary mechanisms that underlie the bewildering array of forms that define
the various galaxy classification schemes used today (e.g., Sandage \& Bedke
1994; Buta \textit{et al.} 2007), because this will allow us to establish the
relationships, if any, between different galaxy types. Physical interpretations
of galaxy morphology have revolved around two different domains: (1) formative
evolution, where rapid, violent processes, such as hierarchical clustering and
merging, led to formation of major galactic components, such as bulges, disks,
haloes, and presumably, the Hubble sequence (e.g., White \& Rees 1978; Firmani \&
Avila-Reese 2003); and (2) secular evolution, where disk material is slowly
rearranged through the collective interaction of instabilities, such as bars,
ovals, spirals, and triaxial dark matter haloes\linebreak (Kormendy \& Kennicutt 2004,
hereafter KK04). KK04 argued that the Universe is in a state of transition,
where secular evolution is becoming the dominant process of morphological
change.

Galaxy morphology is a classical subject with a rich history in astronomy.
Associated with many famous names, such as the Herschels, Lord Rosse, Curtis,
Shapley, Hubble, Lundmark, Reynolds, Sandage, Morgan, and de Vaucouleurs,
morphology and classification are the first step in the study of galaxies as
fundamental units of matter in space. As elegantly noted by Peng \textit{et al.}
(2002), clues to galaxy formation and evolution `are hidden in the fine details
of galaxy structure.'

With this series of lectures, I not only have the great privilege of laying out
the fine details of galaxy morphology in the tradition of these earlier
observers, but also I must do so in the context of secular evolution. This is
very challenging, not just because the subject is so broad, but because we are
only beginning to understand how secular processes operate in galaxies.
Fortunately, my colleagues at this School bring a considerable expertise on this
subject, and I feel like I can cover galaxy morphology at the level of detail
that I think is needed.

Galaxy morphology may be a classical subject in astronomy, but it has a
surprising freshness that has defied predictions of its impending irrelevance to
true understanding of galaxies. Classical galaxy morphology and classification
have survived into the modern era for several reasons: the \textit{Hubble Space
Telescope} (\textit{HST}), the Sloan Digital Sky Survey (SDSS), and the improved
understanding of the meaning of different morphological types through extensive
theoretical and observational studies. There can be little doubt that morphology
holds the key to recognising the processes of galactic evolution, and that it
will continue to provide insight as the relationships between different types of
galaxies become better established.

My goals with this series of lectures on galaxy morphology are to:\linebreak (1) provide
a historical overview of galaxy morphology and classification;\linebreak (2) illustrate
phenomenology and highlight notation; (3) introduce non-optical galaxy
classification; (4) describe interpretative galaxy classification; (5) describe
environmental impacts on galaxy morphology; (6) describe the important
quantitative tools used for modern morphological studies and the use of large
surveys to explore morphology on an unprecedented scale; and finally (7) 
highlight the importance and relevance of morphology to the\linebreak evidence for secular
evolution in galaxies.

In describing morphology, I will draw heavily on two recent sources:
the de Vaucouleurs Atlas of Galaxies (dVA, Buta \textit{et al.} 2007) and the
major review on galaxy morphology by Buta (2013, hereafter B13) for the new
series: Planets, Stars, and Stellar Systems.

%
%

\section{How is morphology relevant to secular evolution?}

Morphology is the key to understanding secular processes. Galaxies are
susceptible to internal instabilities or component interactions that can\linebreak affect
a galaxy's basic structure and slowly change it over time. Even subtle
interactions with the environment can produce long-term secular evolution
(Kormendy \& Bender 2012).  Relevant questions are like this: when you see a
galaxy of type SA(s)b or SB(r)c, etc., has the galaxy always had this type, or
have these types evolved from other types? Is there a specific direction of
evolution, e.g., from late to early type, from barred to nonbarred, from
pseudoringed to ringed, from spiral to nonspiral? What guidance can we get from
theory regarding these questions?

\begin{figure}
\centerline{
\includegraphics[width=\textwidth]{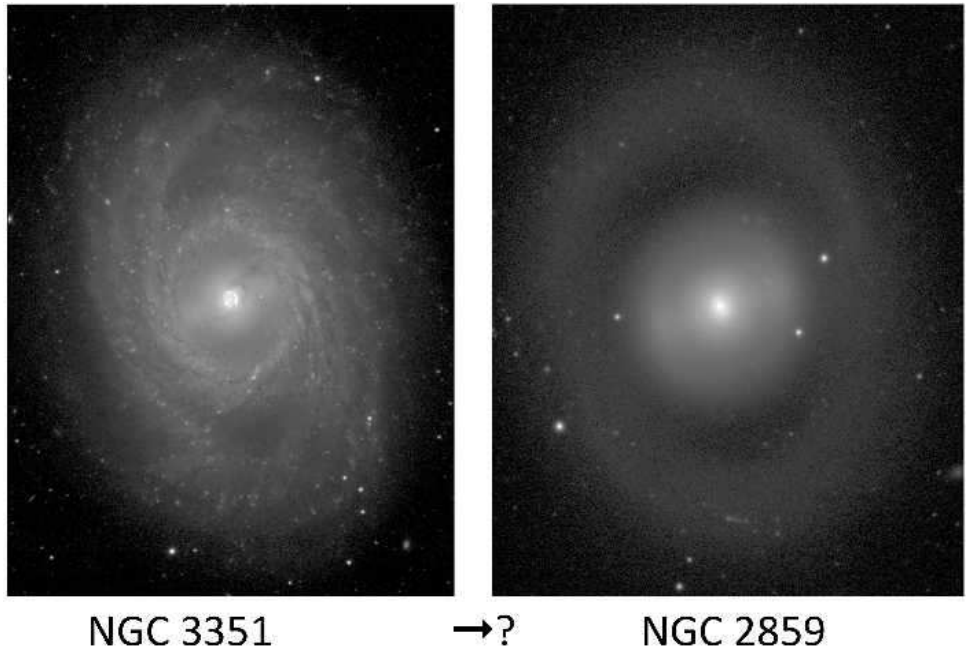}}
\caption{Does the similarity between these two galaxies, one an
intermediate-type spiral and the other a late S0, imply an evolutionary
connection?}
\label{ngc3351-2859} 
\end{figure}

It is not hard to find speculative examples of possible morphological evolution.
For example, Fig.~\ref{ngc3351-2859} shows two morphologically similar galaxies
that differ in a few ways: one (NGC\,3351) is a clear intermediate-type spiral
while the other (NGC\,2859) is a `late' S0; the spiral has very little bulge
while the S0 has a more prominent bulge; the bar in the S0 looks weaker than
that in the spiral, while the rings in the spiral are better described as
pseudorings compared to what is seen in the S0. Looking at these two galaxies,
one might wonder if a galaxy like NGC\,2859 might be a possible end-product of
some long-term evolutionary process in a galaxy like NGC\,3351. Indeed, it was
from examining such possible relationships that Kormendy (1979) first proposed
the idea that secular evolution takes place in barred galaxies: he noticed a
special relationship between bars and features called `lenses' that suggested
to him that bars may dissolve over time into a more axisymmetric state, the
engine of dissolution being an interaction between the spheroidal component and
the bar. This idea was not far off the mark: Bournaud \& Combes (2002) examined
bar dissolution and rejuvenation in models with and without external gas
accretion. In the models without accretion, the bar evolves to a lens-like
structure.

In examining the impact of secular evolution on galaxy morphology, we should be
mindful of the intrinsic and extrinsic factors that have an impact on
morphology. Here is a brief summary of these factors:

\paragraph{Random orientations of symmetry axes.}
Inclination of the symmetry plane to the line of sight, and the accompanying
projection effects and enhanced influence of dust obscuration, is probably the
most important\linebreak extrinsic factor affecting the morphology of nearby galaxies. As
a disk-shaped galaxy is viewed from a face-on orientation to an edge-on
orientation, the appearance of familiar morphological features can change. For
example, a bar may become so foreshortened that it is not recognisable. If a bar
has significant three-dimensional structure, its face-on shape can be lost while
its edge-on shape becomes its distinguishing characteristic.  Rings and spiral
patterns can be lost or less recognisable, although, as shown in B13, these
features may still be evident even at inclinations as high as 81$^{\circ}$.
\vspace{-0.25cm}

\begin{figure}
\centerline{
\includegraphics[width=\textwidth]{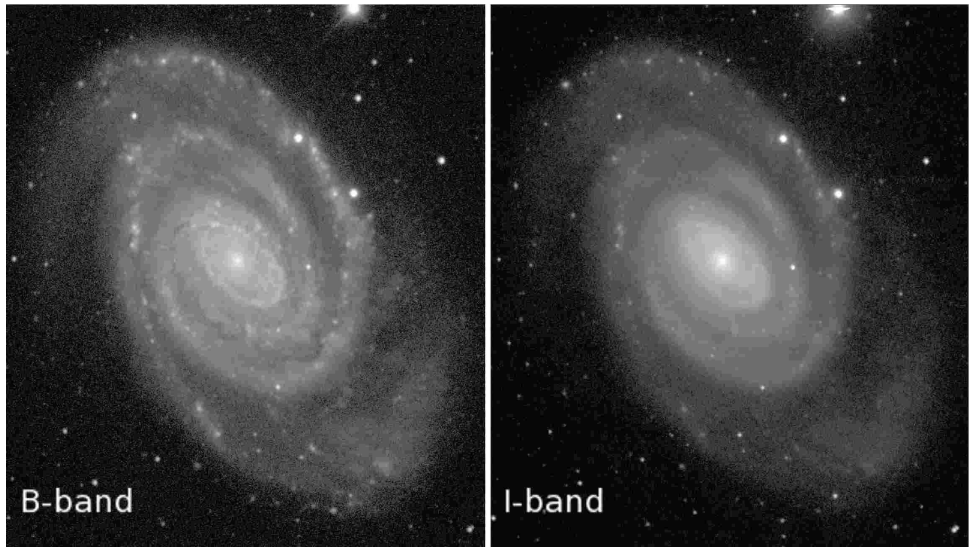}}
\caption{NGC\,5364 in two passbands. The $B$-band emphasises dust and star
formation, while the $I$-band is less sensitive to dust and emphasises an older,
more smoothly distributed stellar population.}
\label{ngc5364b-and-i}
\end{figure}

\paragraph{Wavelength of observation.}
The influence of dust and star formation on spiral galaxy morphology has a
strong wavelength dependence (Fig.~\ref{ngc5364b-and-i}). The blue ($B$) band,
the historical waveband of galaxy classification studies, is sensitive to
reddening and extinction by dust, and to the hot blue stars associated with
star-forming regions. As wavelength increases from $B$ to the near-infrared
(IR), the dust becomes more transparent, reddening and extinction are reduced,
and the influence of star-forming regions diminishes, giving spiral galaxies a
smoother appearance in the red and near-IR. However, a curious thing happens in
the mid-IR. In this wavelength domain, the ultraviolet energy absorbed by dust
grains in star-forming regions is re-emitted strongly, and is already evident at
3.6$\mu$m by the return of the prominence of star-forming regions even as the
extinction diminishes to only 5\% of that in the $V$-band. As is shown in
Lecture~3 (Section~\ref{sec:lecture3}), 3.6$\mu$m galaxy morphology is astonishingly
similar to $B$-band morphology, absent the effects of extinction and reddening.

\begin{figure}
\centerline{
\includegraphics[width=\textwidth]{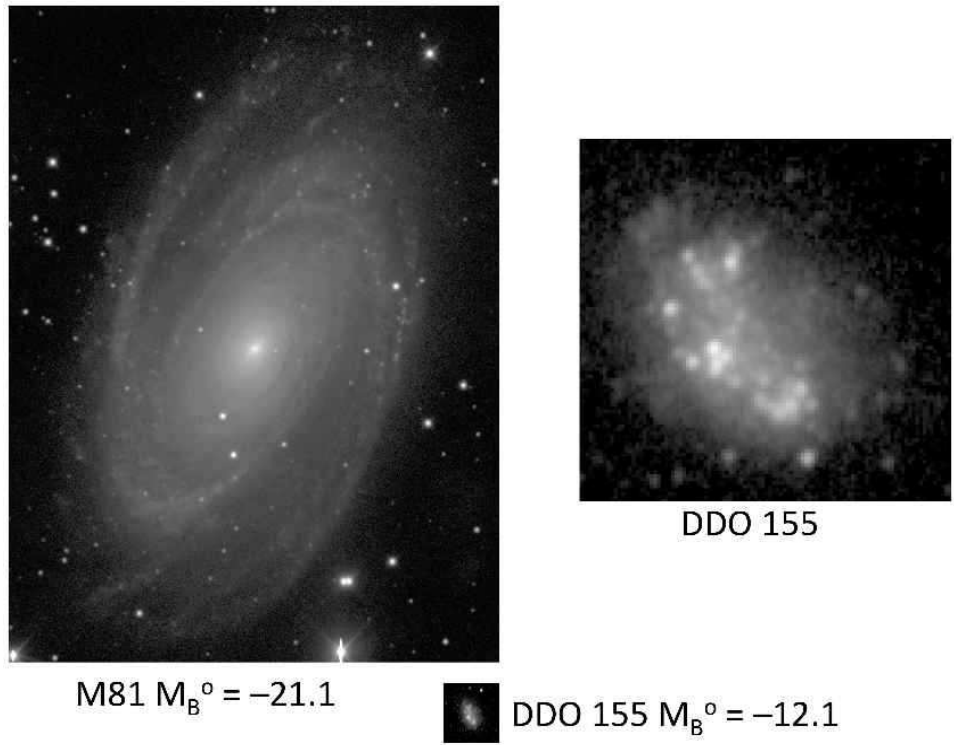}}
\caption{The effect of luminosity on $B$-band morphology, comparing the
giant spiral M\,81 with the nearby dwarf irregular DDO\,155 (shown on the same 
scale in the lower-left panel).}
\label{m81andddo155}
\end{figure}

\paragraph{Total mass and luminosity.}
Figure~\ref{m81andddo155} shows the strong dependence of galaxy morphology on
total mass and luminosity. M\,81 is a giant spiral having a $B$-band absolute
magnitude of $-21.1$ and shows extremely organised and well-developed high-surface 
brightness structure. DDO\,155, a dwarf having $M_B^{\rm o}$\,=\,$-12.1$, is in
contrast a very small, low surface brightness, irregular-shaped galaxy.  Van den
Bergh (1960a,b) and Sandage \& Tammann (1981) effectively used such differences
to define galactic {\it luminosity classes} (van den Bergh 1998; dVA; B13).

\begin{figure}
\centerline{
\includegraphics[width=\textwidth]{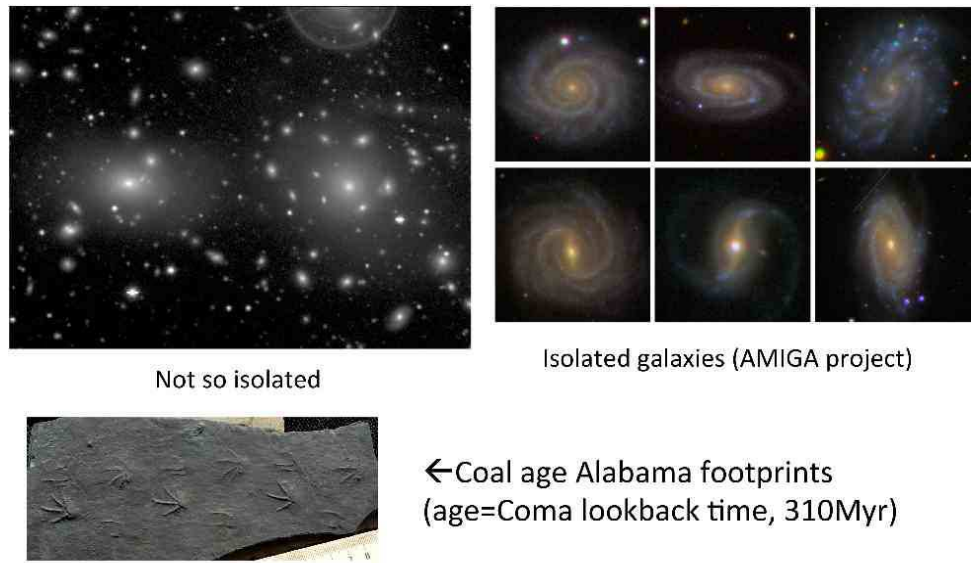}}
\caption{A high-density environment like the central region of the Coma
cluster is rich in early-type galaxies, while spirals are more
prevalent in less dense environments. The Analysis of the Interstellar
Medium in Isolated Galaxies (AMIGA) project is described by
Verdes-Montenegro \textit{et al.} (2005).}
\label{env-density}
\end{figure}

\paragraph{Environmental density, interaction, and merger history.}
The strong correlation between environmental density and galaxy morphology was
first described by Hubble \& Humason (1935) and studied in greater detail by
Dressler (1980). The morphology-density relation, as it is called, is such that
denser environments like rich galaxy clusters have a preponderance of early-type
(E, S0) galaxies compared to lower-density environments
(Fig.~\ref{env-density}). Even in lower-density environments where spirals are
abundant, such as the Virgo cluster, morphology can show evidence of an
interaction with the intra-cluster medium (e.g., Koopmann \& Kenney 2004; see
Lecture~4, Section~\ref{sec:lecture4}).

\paragraph{Star formation history.}
Galaxies that formed all of their stars many Gyr ago tend to look very different
from those that did not. Galaxies which are not currently forming any stars are
redder, smoother, more centrally concentrated, and more symmetric than those
which are. The bluest normal galaxies are Magellanic spirals and irregulars.

\paragraph{Lookback time.}
Morphology can be significantly affected by the lookback time to a galaxy. When
the redshift is high, the lookback time can be so great that we see an early
phase of morphological evolution. Galaxies tend to have more irregular shapes
and relatively small linear sizes for $z\geq1$ (see B13 and Lecture~4,
Section~\ref{sec:lecture4}).

%
%
\section{Historical overview}
\label{sec:lecture1_historical}

Galaxy morphology became evident when large and effective telescopes
began to be used to observe the sky. In the late 18th century, English
astronomer William Herschel built 18.7\,in speculum metal reflectors
which he used to `sweep' the sky for anything out of the ordinary, such
as `nebulae'. Herschel and those who followed him saw different kinds
of nebulae: those that appeared to lie within the band of light called
the Milky Way (`galactic nebulae'), and those which were found mainly
away from the Milky Way (`non-galactic nebulae'). For a while, all
types of nebulae were thought to be distant but unresolved stellar
systems, a popular but largely speculative idea at the time. Prior to
the 1870s, all illustrations of nebulae were visual sketches based on
what was seen through an eyepiece. One of the most detailed early
sketches was that of M\,51 made by Herschel's son John in the 1820s
(Hoskin 1982).

The non-galactic nebulae seen by Herschel had a variety of interesting
shapes, ranging from round to highly elongated, and showed varying
degrees of central brightness. Herschel invented a simple descriptive
classification of these objects based on brightness, size, shape, and
central concentration.\linebreak This approach was also used by John Herschel to
describe all nebulous objects compiled to the 1860s (Herschel 1864).

\begin{figure}
\centerline{
\includegraphics[width=\textwidth]{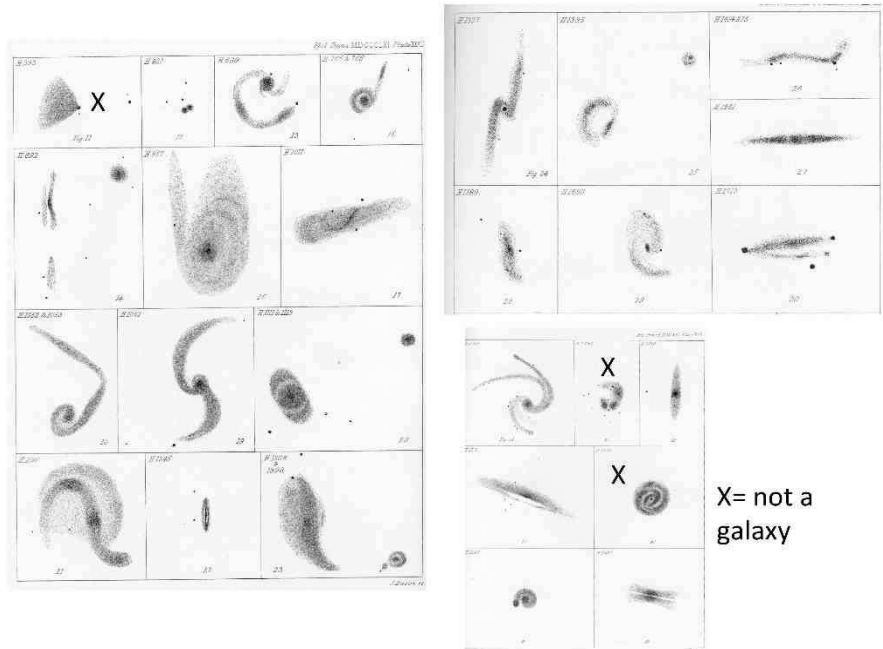}}
\caption{The first galaxy morphology atlas, based on visual sketches
made with the Birr Castle `Leviathan' (Parsons 1880).}
\label{parsons-atlas}
\end{figure}

Even though the Herschel telescopes could reveal thousands of
non-galactic nebulae, the finer details of galaxy morphology were
largely elusive.  Galaxy morphology `came alive' in 1845 when William
Parsons, 3rd Earl of Rosse, observed nebulae with a much larger
telescope, the 1.8\,m `Leviathan of Parsonstown'. This telescope was
constructed on the grounds of Birr Castle in central Ireland, and was
the largest telescope in the world for nearly 75~years. The Leviathan
observations are replete with visually seen details of galaxy
morphology, and from the sketches that were made, one can tell that the
observers had seen spiral arms, bars, rings, dust lanes, star-forming
regions, tidal features, even Magellanic barred spirals. In the
extensive set of notes published by the 4th Earl (Parsons 1880),
one can determine that spiral structure was seen in 75 `nebulae' later
found to be galaxies. Figure~\ref{parsons-atlas} shows what could be
viewed as the first galaxy morphology {\it atlas}: a compilation of
Birr Castle Leviathan sketches made by a variety of observers.

Although the Leviathan of Parsonstown was the most powerful telescope
of its day, it was not capable of guided photography and therefore
could not get long exposure images of things like spiral nebulae.
Nebular photography became possible in the 1880s with the availability
of silver bromide dry emulsion plates and telecopes designed for
accurate guiding. Isaac Roberts (1829--1904) obtained in 1888 the first
long-exposure photograph of the Andromeda Nebula, which first revealed
the spiral structure in the faint outer parts of the nebula (Roberts
1893). As the number of plates accumulated, the first classification
systems emerged. Max Wolf (1863-1932) published a simple system of
letters to describe 17 different types of non-galactic nebulae, ranging
from amorphous inclined types to patchy, well-developed spirals (Wolf
1908). This system was used over a period of 30 years by a number of
well-known nebular researchers, including Hubble who thought it was a
useful temporary system until accumulation of more data allowed
something better to come along.

\begin{figure}
\centerline{
\includegraphics[width=\textwidth]{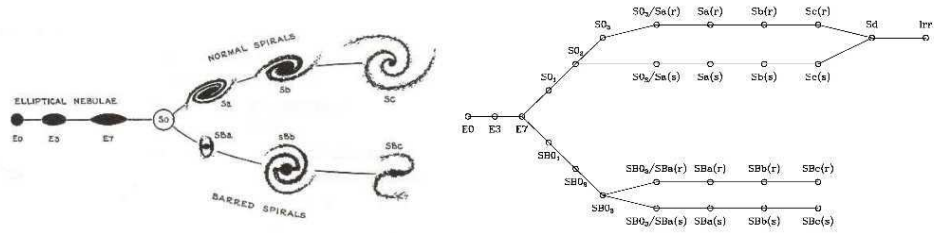}}
\caption{The original Hubble (1936) `tuning fork' classification
(left), and a schematic of the revised tuning fork outlined in the
Hubble Atlas (Sandage 1961; Buta \textit{et al.} 2007).}
\label{tuning-forks}
\end{figure}

A big photographic survey described by Curtis (1918) helped set the stage for
the Hubble classification system.  Photographs of hundreds of nebulae, galactic
and non-galactic, were taken over a nearly 20-year period with the the 36\,inch
Lick Crossley reflector by, in addition to Curtis, well-known photographers
James E. Keeler and Edward E. Barnard.  The main specific galaxy morphology
Curtis recognised was `$\phi$-type spirals', later renamed barred spirals by
Hubble (1926). Curtis believed all non-galactic nebulae were spirals and that
any that didn't look spiral would eventually be found to be such. Hubble (1922)
disputed this conclusion; he noted that genuine bright but definitely non-spiral
non-galactic nebulae existed.  After obtaining and inspecting many available
plates, Hubble (1926) published a new classification system to replace the Wolf
(1908) system. This system placed galaxies on a sequence ranging from amorphous
elliptical-shaped objects to well-developed, patchy-armed spirals. The spiral
part of the sequence was split between non-barred (`normal') and barred
spirals. However, Hubble believed that his 1926 system was flawed because the
transition from the flattest-looking E galaxies to Sa spirals looked too sharp
to be real. He hypothesised that there had to be armless but highly flattened
disk-shaped galaxies in the transition from types E to Sa. This was shown in his
famous `tuning fork' illustration in his book, The Realm of the Nebulae,
published in 1936 (Fig.~\ref{tuning-forks}, left). It is thought that Hubble was
inspired to illustrate his classification this way because this is how Sir James
Jeans illustrated it in his 1928 book, {\it Astronomy and Cosmogony} (Block et
al. 2004). Sandage (2005) has also noted that the arrangement of galaxies on a
sequence ranging from amorphous to highly structured spirals was outlined
independently by Reynolds (1920), but never referenced by Hubble.

Hubble had planned to revise his classification system further but died before
completing it. Allan Sandage used Hubble's notes to prepare the monumental {\it
Hubble Atlas of Galaxies} (Sandage 1961). This firmly cemented Hubble's ideas
into astronomy. Inclusion of S0 galaxies, as well as splitting each tuning fork
prong into ringed and non-ringed varieties, made the classification more
complicated. Van den Bergh (1976) commented that the addition of S0s destroyed
the `simple beauty' of the original 1926 classification. The revised Hubble-Sandage
(RHS) system was later expanded and further revised in the {\it Carnegie Atlas
of Galaxies} (Sandage \& Bedke 1994).

\begin{figure}[t]
\centerline{
\includegraphics[width=0.45\textwidth]{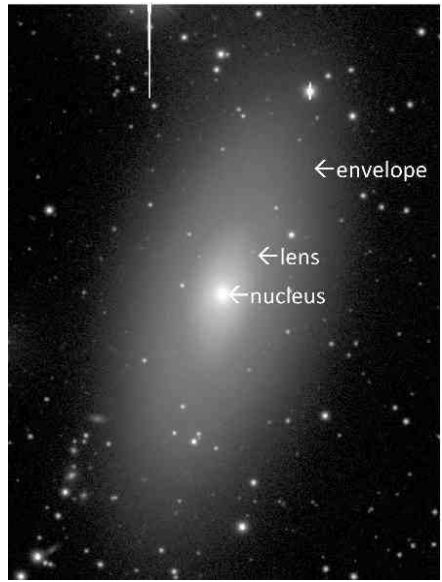}}
\caption{NGC\,2784 displays the main elements -- nucleus, lens, and
envelope, that define an S0 galaxy (Sandage 1961).}
\label{ngc2784}
\end{figure}

\begin{figure}[h!]
\centerline{
\includegraphics[width=\textwidth]{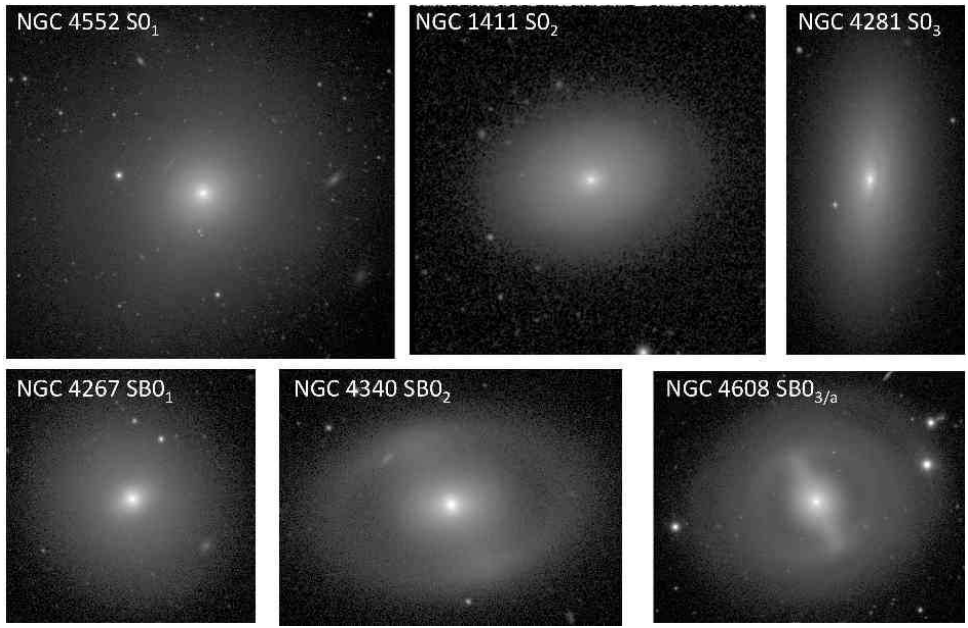}}
\caption{These galaxies show the main S0 galaxy categories descibed in
the Hubble and Carnegie atlases of galaxies.}
\label{s0gals-carnegie}
\end{figure}

S0 galaxies, probably the most enigmatic type in the whole Hubble sequence, are
thoroughly described in the Hubble Atlas and Carnegie Atlas. An excellent
example of an S0 is NGC\,2784, shown in Fig.~\ref{ngc2784}, while the RHS
classification of S0s is summarised in Fig.~\ref{s0gals-carnegie}. The three
main components of an S0 are the {\it nucleus}, the {\it lens}, and the {\it
envelope}. A lens is a distinct feature that appears as a well-defined region
having a shallow brightness gradient interior to a sharp edge. The enhancement
can be very slight as in NGC\,2784, or more distinct as in NGC\,1411
(Fig.~\ref{s0gals-carnegie}, top middle). The RHS subclassification of nonbarred
S0s, S0$_1$, S0$_2$, and S0$_3$, depends on structure differentiation, while
that for barred S0s:  SB0$_1$, SB0$_2$, and SB0$_3$, is based on the development
of the bar.  The differences between nonbarred and barred S0s are important. For
example, dust rings are a common feature of type S0$_3$, while they may not
factor in at all in type SB0$_3$.


Around the same time as the Hubble Atlas was being prepared, W.W. Morgan 
(1906-1994) proposed a classification system that combined galaxy form with 
stellar population defined by (Morgan 1958):

\begin{enumerate}[(a)]\listsize
\renewcommand{\theenumi}{(\alph{enumi})}

\item {\it population group}: a, af, f, fg, g, gk, k for dominant spectral types
A, AF, F, FG, G, GK, and K. This is estimated solely from {\it central
concentration};

\item {\it form family}: S (spirals), B (barred spirals), E (ellipticals), I 
(irregulars), D (like S0s), plus others;

\item inclination class flattening index.

\end{enumerate}

\begin{figure}
\centering
\begin{minipage}[t]{\linewidth}
\centering
\vspace{1cm}
\includegraphics[width=\textwidth]{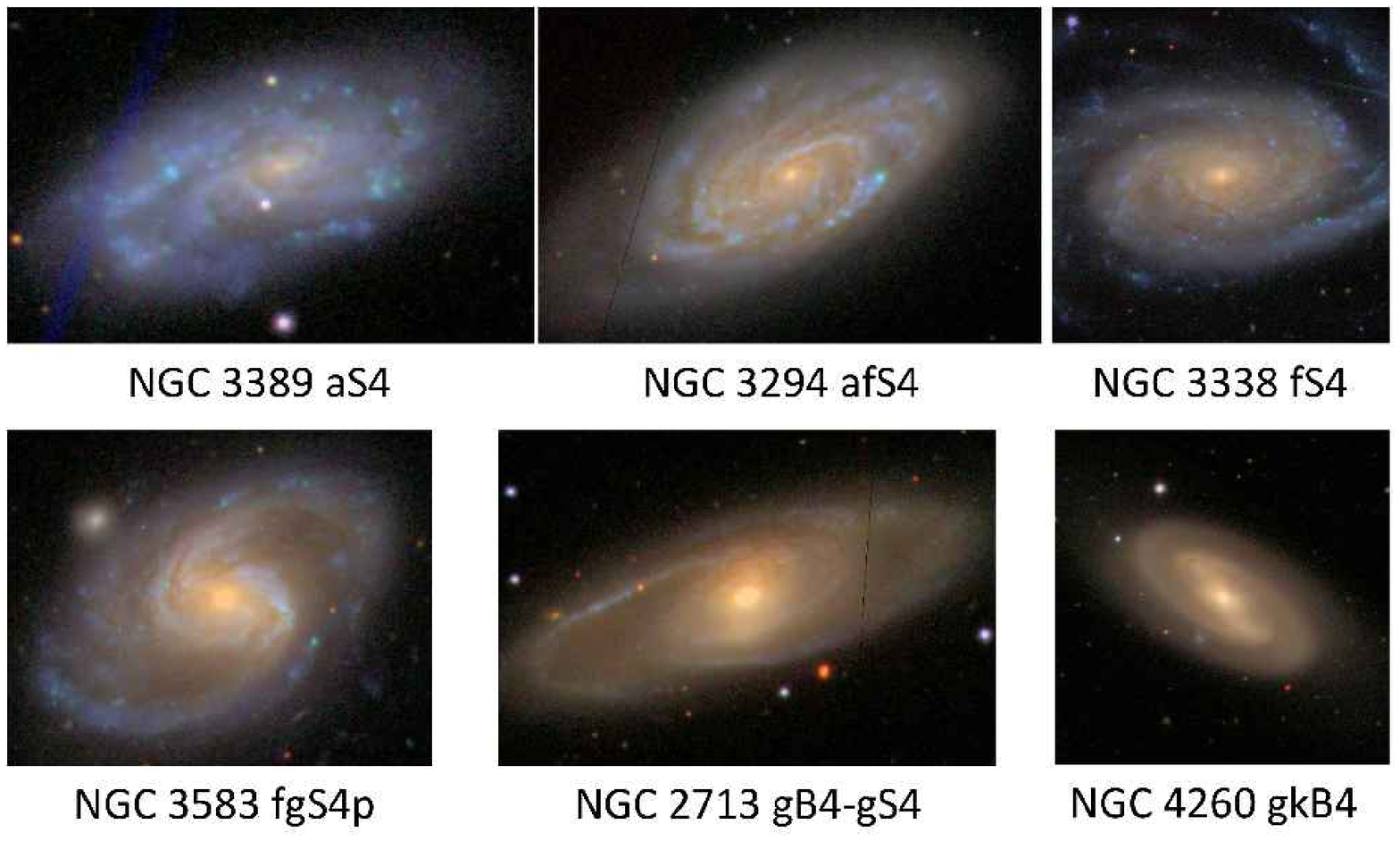}
\caption{Examples of the stellar-population/form class classification
of Morgan (1958).}
\label{morgan-type}
\end{minipage}%
\hspace{1cm}%
\begin{minipage}[b]{\linewidth}
\vspace{1cm}
\centering
\includegraphics[width=\textwidth]{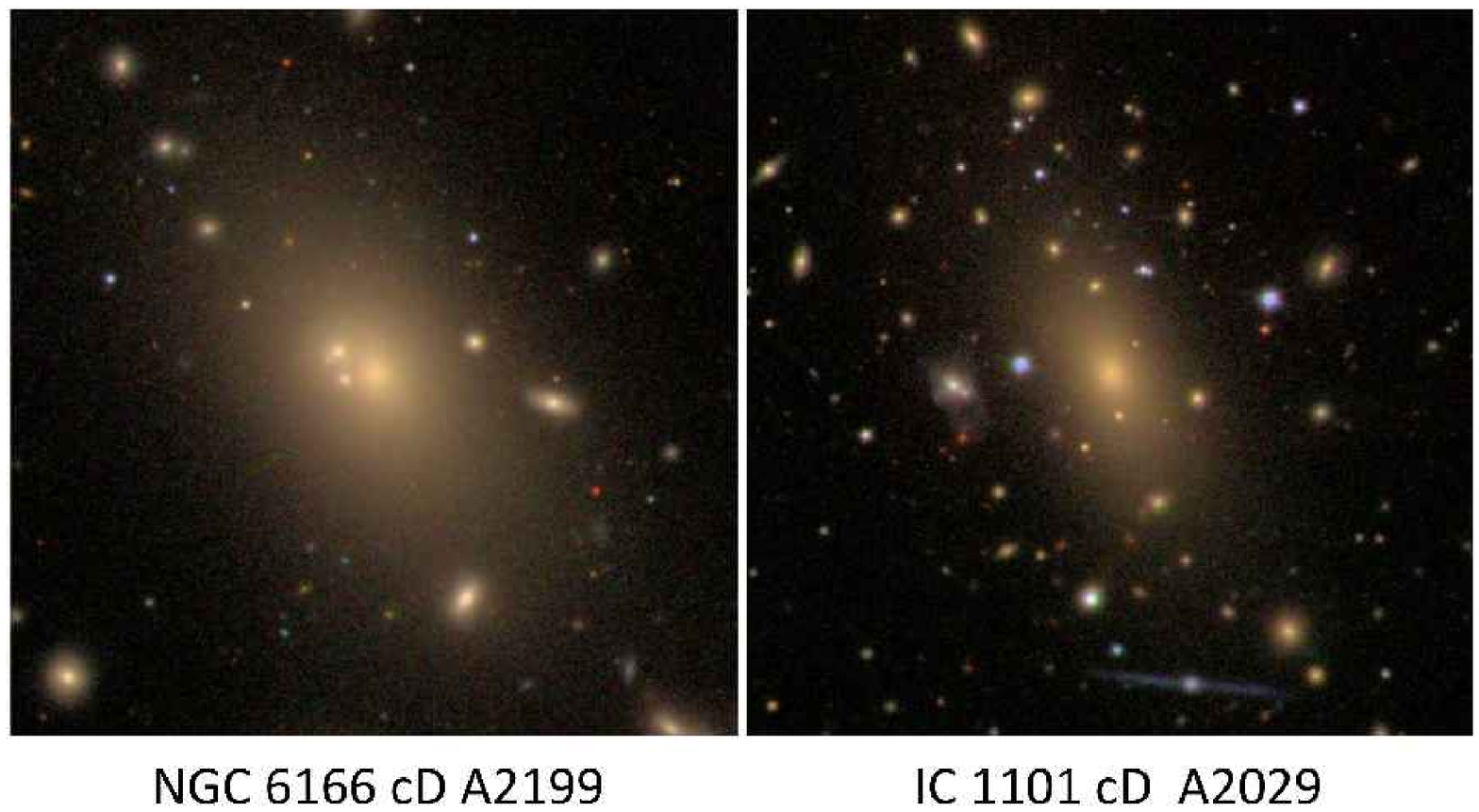}
\caption{The Morgan cD class of supergiant galaxies.}
\label{morgan-cDs}
\end{minipage}
\end{figure}

In Fig.~\ref{morgan-type}, a Morgan sequence of population groups, based on
classifications from Morgan (1958) and illustrated using SDSS colour images,
shows the effectiveness of his approach. The sequence a through k is a colour
sequence from bluish to yellow-orange. The system did not have the impact that
Hubble's did, perhaps because the population groups were closely analogous to
Hubble types Sa, Sb, and Sc. For example, in Fig.~\ref{morgan-type}, NGC\,3389
(type aS4) is Hubble type Sc, NGC\,3583 (type fgS4p) is Hubble type Sb, and
NGC\,4260 (type gkB4) is Hubble type Sa.

The most recognisable and important of Morgan's form classes is the cD
galaxy, a {\it supergiant} version of the D form family found in the
centres of rich clusters (Fig.~\ref{morgan-cDs}). These objects, also
known as `brightest cluster members', were extensively studied by
Schombert (1986, 1987, 1988).

\begin{figure}
\centering
\begin{minipage}[t]{\linewidth}
\centering
\vspace{2cm}
\includegraphics[width=\textwidth]{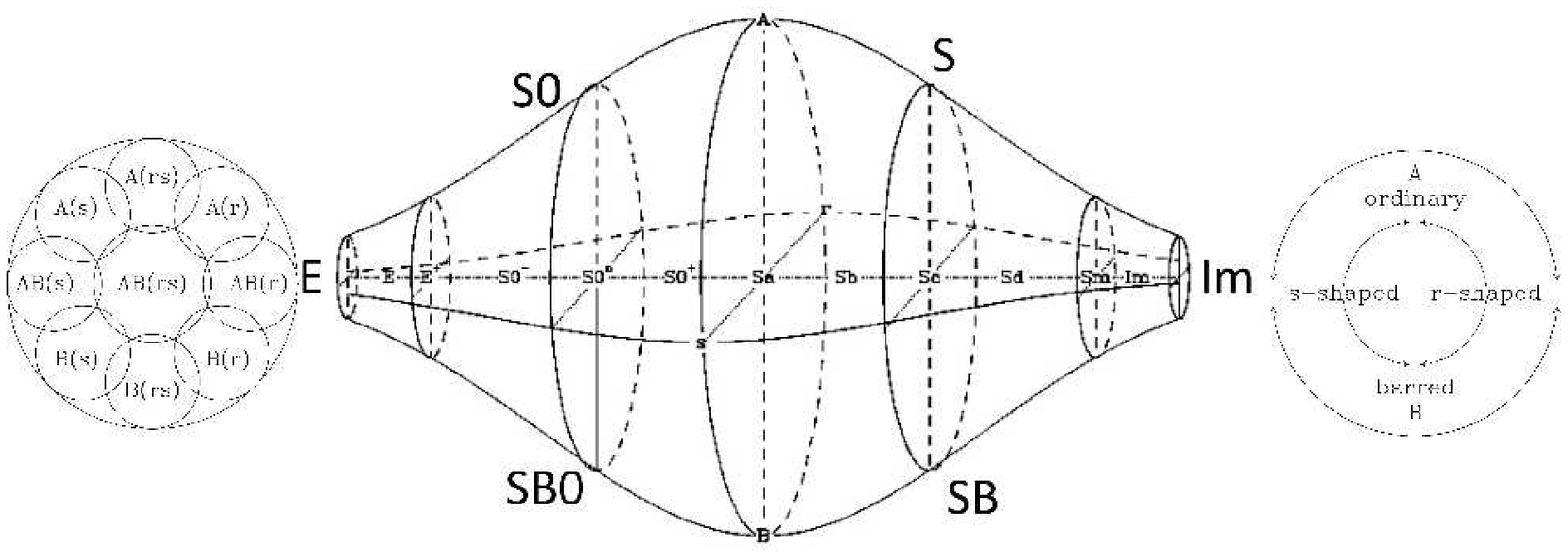}
\caption{The de Vaucouleurs (1959) revised Hubble-Sandage classification system.}
\label{deV}
\end{minipage}%
\hspace{1cm}%
\vfill
\begin{minipage}[b]{\linewidth}
\vspace{2.5cm}
\centering
\includegraphics[width=\textwidth]{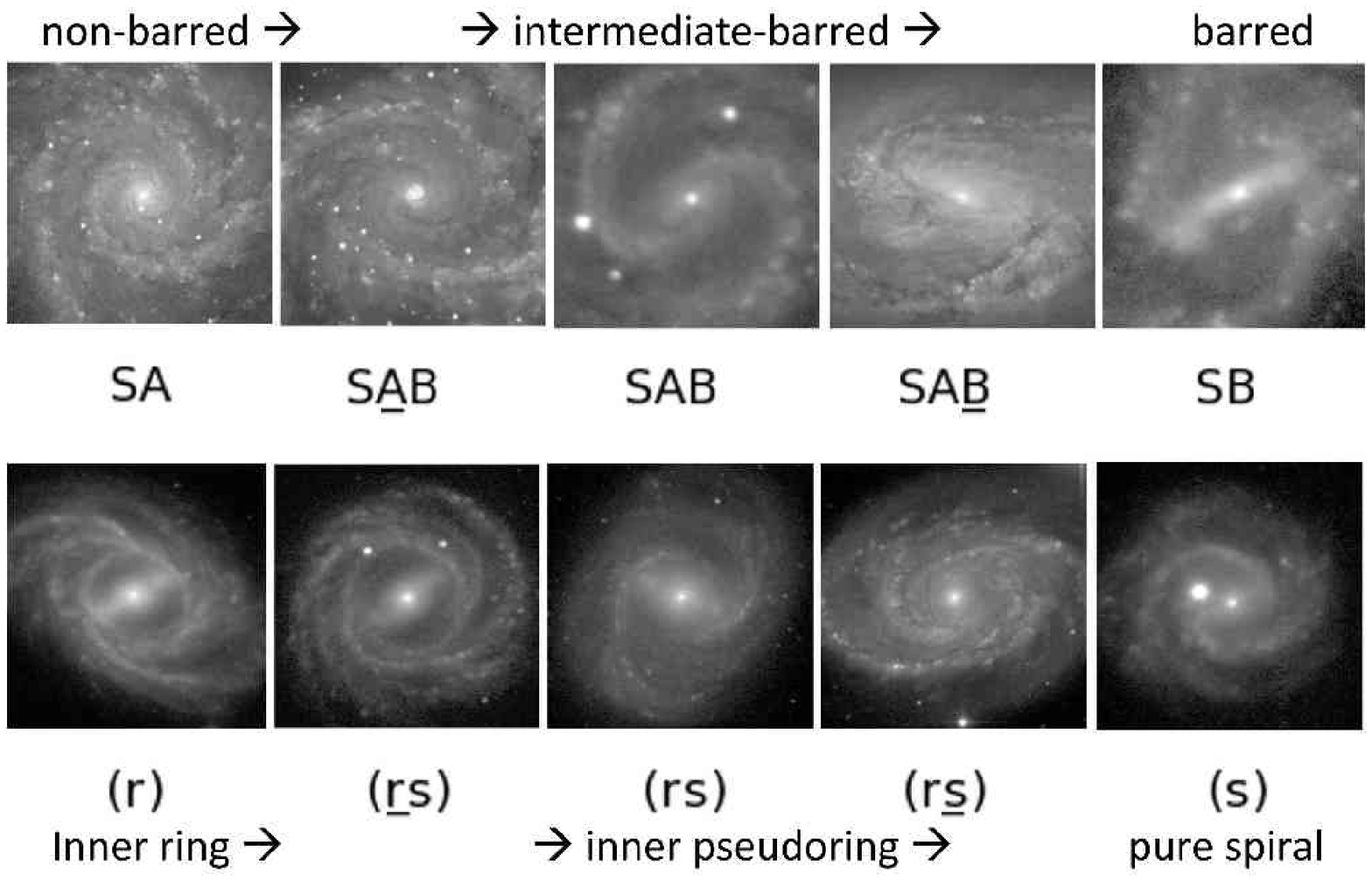}
\caption{Family and variety in the VRHS as continuous characteristics (from B13).}
\label{fam-var}
\end{minipage}
\end{figure}

In 1953, non-Palomar firebrand Gerard de Vaucouleurs (1918-1995)
carried galaxy morphology into the southern hemisphere, developing his
own interpretation of the Hubble-Sandage classifications on the way.
Figure~\ref{deV} shows de Vaucouleurs's (1959) `classification volume'
(the VRHS, or `three-pronged swirling two-handled tuning fork').  The
long axis defines the `stages' E, E$^+$, S0$^-$, S0$^{\rm o}$, S0$^+$, S0/a,
Sa, Sab, Sb, Sbc, Sc, Scd, Sd, Sdm, Sm, Im.

De Vaucouleurs viewed galaxy morphology as a continuous sequence of forms. He
was also artistically talented and made a sketch (see Kormendy's contribution,
this volume) of a cross-section of his classification volume in ~1962. The
sketch shows the arrangement of families (apparent bar strength) and varieties
(presence or absence of an inner ring) near stage Sb. Families and varieties can
be thought of as continuous secondary traits (de Vaucouleurs 1963);
Fig.~\ref{fam-var} shows the use of the underline notation for these
characteristics.

%

\begin{figure}
\centerline{\includegraphics[width=\textwidth]{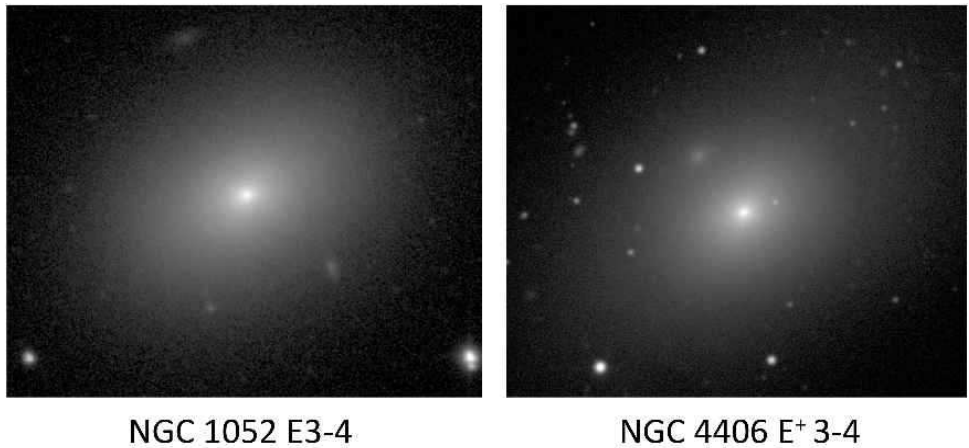}}
\caption{An elliptical galaxy (E) and a `late elliptical' galaxy (E$^+$).} 
\label{egals} 
\end{figure}

The {\it stage} is the primary dimension of the VRHS.  Elliptical galaxies are
amorphous systems with a smoothly declining brightness gradient
(Fig.~\ref{egals}). They are not disk-shaped. Elliptical galaxies have two VRHS
stages: E and E$^+$. The number after the letter E, as in E2, is the flattening
index $n = 10(1-b/a)$, a numerical specification of how elliptical the galaxy
is. The E$n$ sequence is not, however, physically significant. Type E$^+$ is a
transition stage to the S0 class. These are E-like galaxies showing slight
traces of differentiated structure, usually subtle evidence of lenses or faint
outer envelopes. Type E$^+$ has also been used by de Vaucouleurs as a `home'
for Morgan cD galaxies.

In the VRHS, S0 galaxies have three stages: S0$^-$, S0$^{\rm o}$, and S0$^+$, in
a\linebreak sequence of increasing structure. Type S0$^-$ generally has barely
differentiated structure, often in the form of subtle lenses. Type S0$^{\rm o}$
tends to have stronger lenses and is more obvious as an S0. Type S0$^+$ often
has well-defined ring structures, both inner and outer, but can also have subtle
spiral structure.  Types such as SB(r)0$^{\rm o}$ and SB(s)0$^+$ are possible,
as are SAB types. Note that this sequence is based on development of structure
and NOT on bulge-to-total luminosity ratio.

\begin{figure}
\centerline{\includegraphics[width=\textwidth]{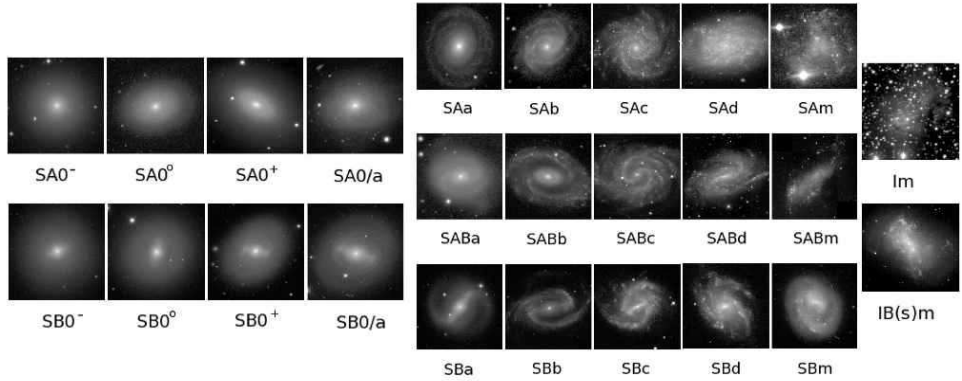}}
\caption{Examples of the S0 and spiral stages along the VRHS sequence.}
\label{dev-types} 
\end{figure}

In the VRHS, spirals have 9 stages:  S0/a, Sa, Sab, Sb, Sbc, Sc, Scd, Sd, Sdm,
and Sm. These are still recognised using Hubble's three criteria: the relative
size of the bulge, the degree of openness of the spiral arms, and the degree of
resolution of the arms into knots. Small-bulge early-type galaxies, especially
barred spirals, are the biggest violators of these rules. The VRHS introduces
{\it extreme late-type spirals}: Sd, Sdm, and Sm, to the Hubble sequence.
Figure~\ref{dev-types} shows how well these types fit into the sequence with
clear, easily distinguishable characteristics. The intermediate spiral types
(like Sab, Sbc, etc.) are almost as common as the main types. The most common
spirals in magnitude-limited samples are of types \hbox{Sb--Sc}. A special
hallmark of the VRHS is the recognition of the Magellanic Clouds as extreme
late-type barred spirals of the type SB(s)m that show a characteristic one-armed
asymmetry and offset bar (de Vaucouleurs \& Freeman 1972). Magellanic irregulars
form the last major stage along the VRHS, and are often barred (i.e., classified
as type IBm or IB(s)m, implying a subtle spiral variety). Examples highlighting
the S0, spiral, and irregular sequences are shown in Fig.~\ref{dev-types}.

\begin{figure}
\centerline{
\includegraphics[height=7.5cm]{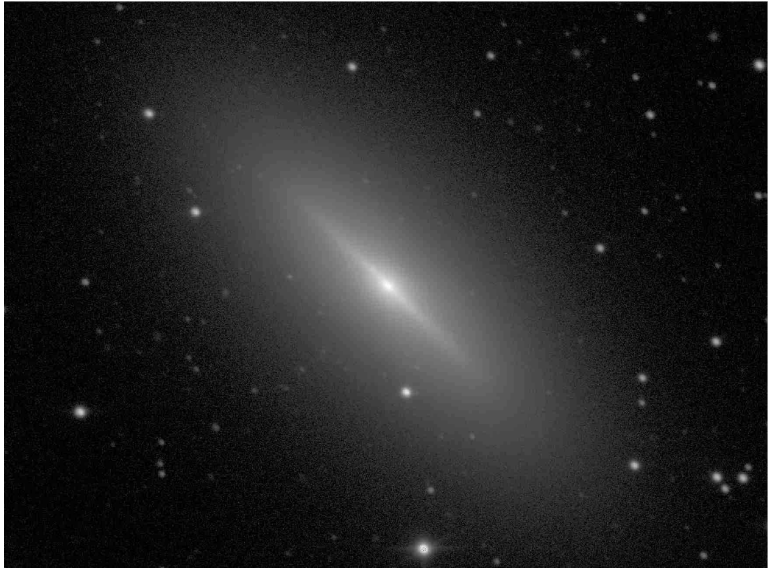}}
\caption{Edge-on S0 galaxy NGC\,3115 ($V$-band).}
\label{ngc3115}
\end{figure}

Spiral and S0 stages are still recognisable in edge-on galaxies, but
Fig.~\ref{ngc3115} shows one case that is distinctive. NGC\,3115 is an
original Hubble E7 type that was reclassified as type S0$_1$ in the
Hubble Atlas. It is now recognised as an S0 with a `thick' disk,
although it looks more like an E galaxy with an {\it embedded} disk 
(Fig.~\ref{ngc3115}).

The outer ring classification is the final original dimension of the
VRHS. Closed outer rings are symbolised by (R) preceding the other type
symbols. The hallmark of VRHS classifications is the recognition of
outer pseudorings (R$^{\prime}$) made of variable pitch angle outer
spiral arms. Examples are illustrated and described in Lecture~2 
(Section~\ref{sec:lecture2}).

\begin{figure}
\centerline{
\includegraphics[height=8cm]{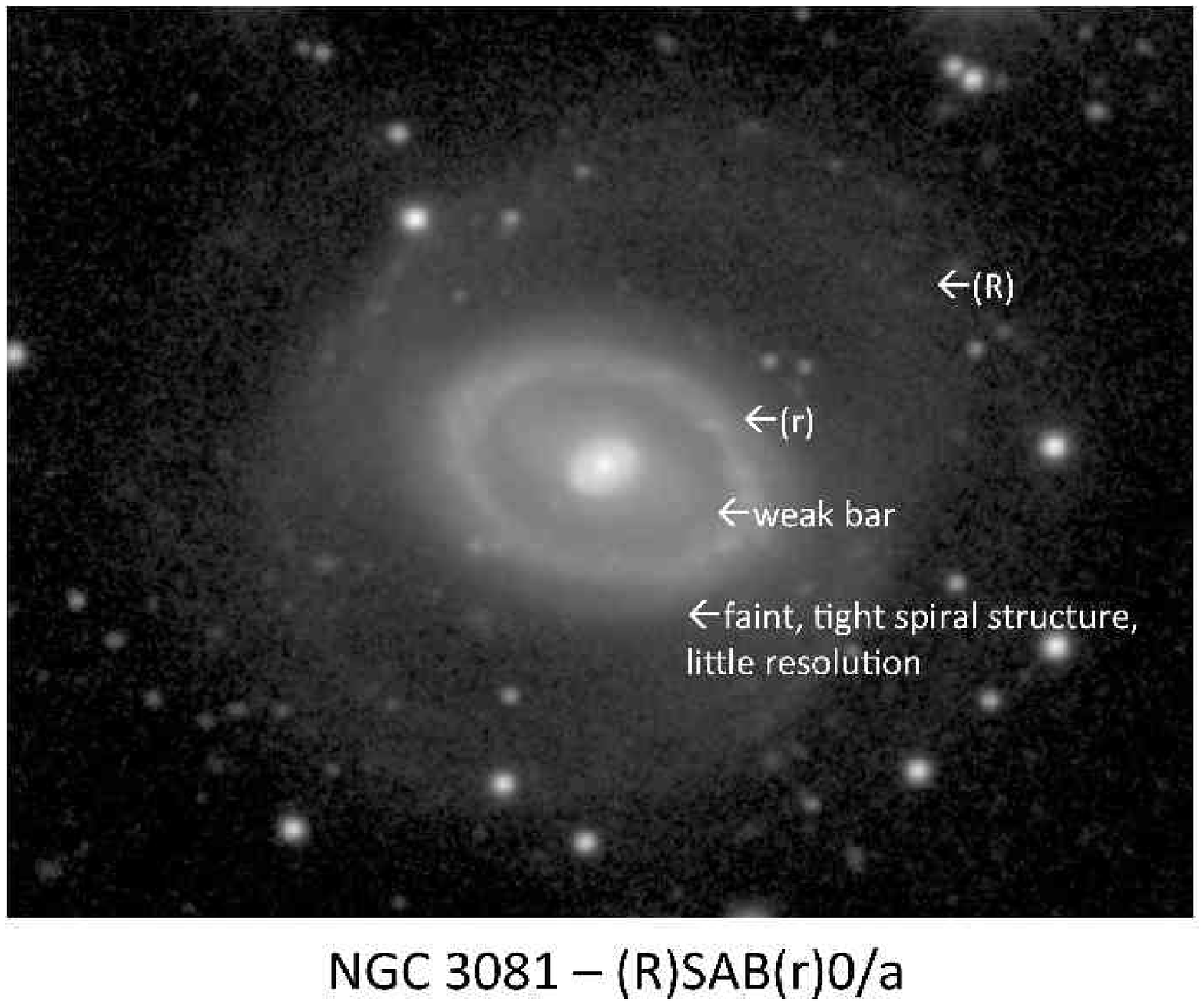}}
\caption{Classification of NGC\,3081.}
\label{ngc3081}
\end{figure}

The genesis of a galaxy classification is shown in Fig.~\ref{ngc3081}. The main
features of the galaxy are labelled and its basic type is (R)SAB(r)0/a. However,
its bulge is inconsistent with the type S0/a as is often the case for early-type
barred galaxies.

\begin{figure}
\centerline{\includegraphics[width=0.9\textwidth]{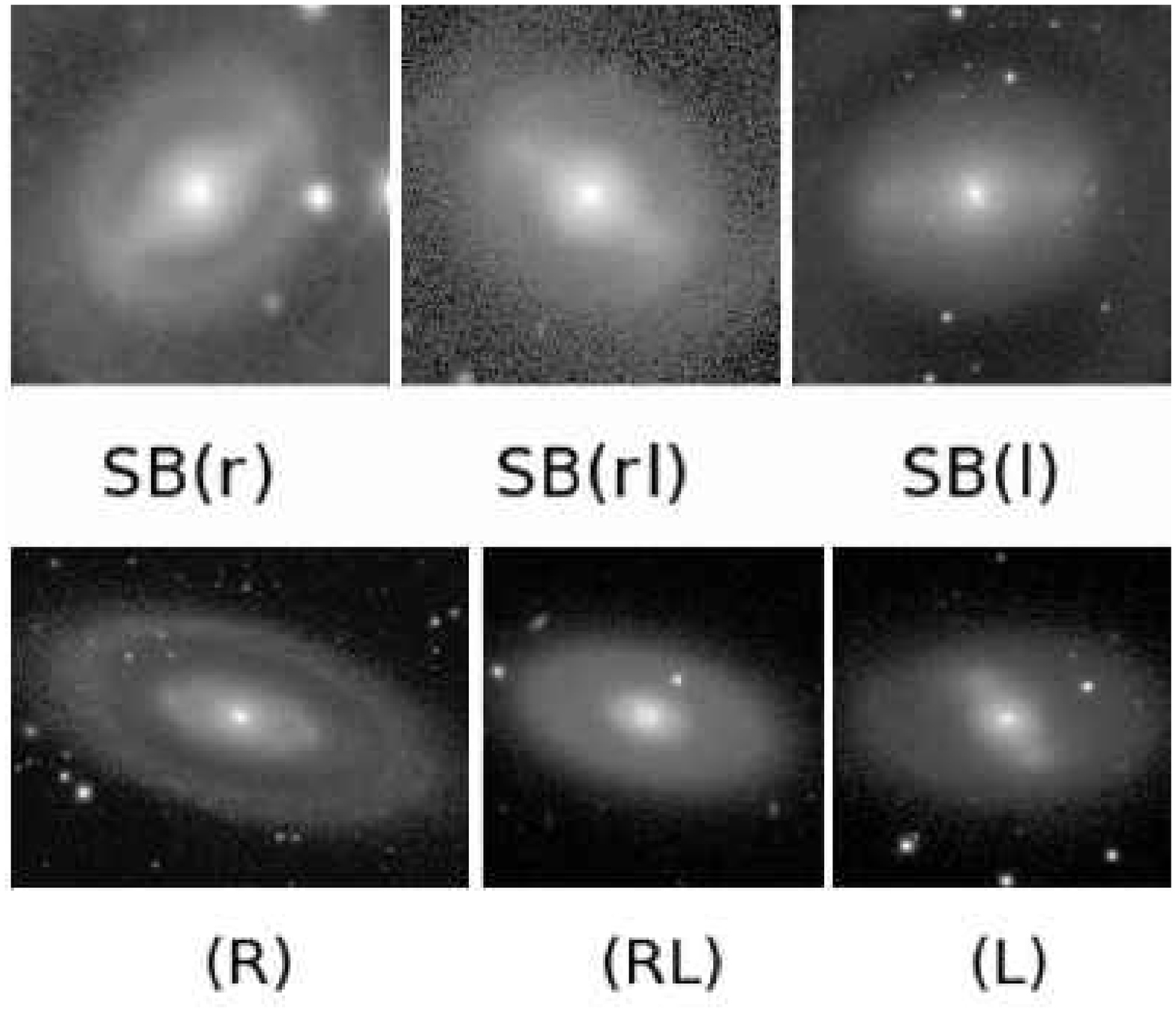}}
\caption{Examples showing the continuity of ring and lens morphologies. Other
examples may be found in B13.}
\label{lenses}
\end{figure}

The VRHS has had some additions and revisions in recent years. For example,
inner and outer lenses were added to the classification by Buta (1995) using
notation suggested by Kormendy (1979). Inner lenses are found roughly in the
same location as inner rings, while outer lenses are found where outer rings
often are seen. Figure~\ref{lenses} shows that there is a morphological
continuity between rings and lenses, and one possible interpretation is that
some lenses are highly evolved former star-forming rings.\looseness-1


The effect of total mass and luminosity on galaxy morphology was characterised
by van den Bergh (1960a,b) in terms of {\it luminosity classes}. (An
illustration of the luminosity class standards, van den Bergh 1998, is given in
B13). The classes are applicable only to intermediate to late-type spirals and
are analogous to those used in stellar spectral classification: luminosity class
I: supergiant spirals; II: bright giant spirals; III: giant spirals; IV:
subgiant spirals; and V: dwarf spirals and Magellanic irregulars. The main idea
is that more luminous spiral galaxies have the most well-organised structure
(Fig.~\ref{m81andddo155}). In massive, luminous spirals, the structure is more
symmetric and more ordered. In lower-mass spirals, the structure is more
chaotic; in fact, in dwarfs the spiral structure can be so weak or absent that
surface brightness is used as the criterion of luminosity classification.

\begin{figure}
\centerline{\includegraphics[width=0.75\textwidth]{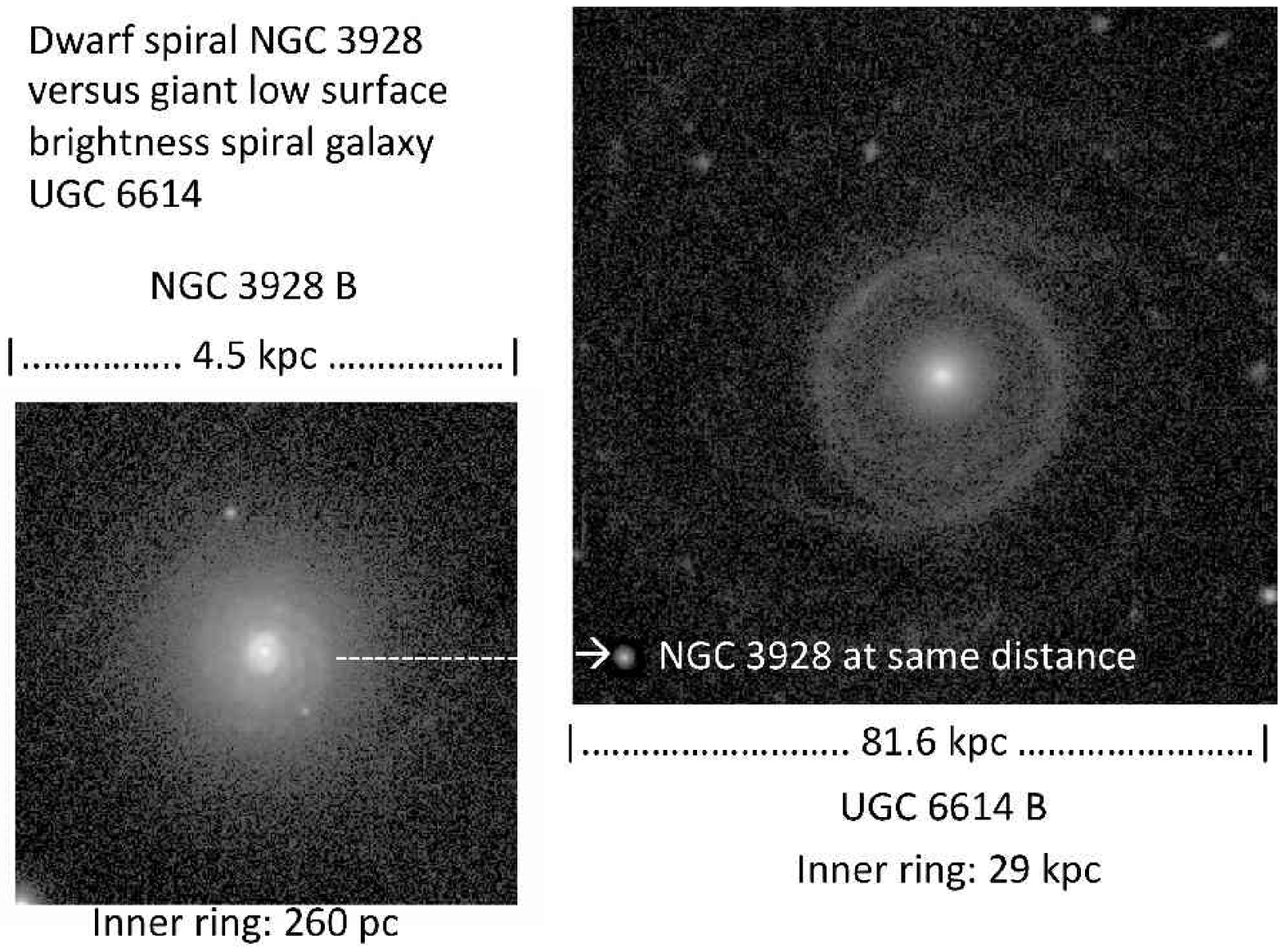}}
\caption{Dwarf spiral NGC\,3928 as compared to the supergiant spiral UGC\,6614.}
\label{ngc3928}
\end{figure}

Dwarf galaxies in the Virgo cluster were the subject of an exceptional study by
Binggeli {\it et al.} (1985). The main types of dwarfs classified by these
authors are: cE - compact ellipticals; dE - dwarf ellipticals; dE,N - nucleated
dwarf ellipticals; dS0 - dwarf S0s; dS0,N - nucleated dwarf S0; BCD - blue
compact dwarf; and `large dE' or large dwarf ellipticals (all illustrated in
B13). If the dEs and dS0s were actually dwarf versions of normal E and S0
galaxies, their existence would widen the parameter space at the left end of the
VRHS. However, recent studies have indicated that dEs and dS0s are likely not
connected to actual E or S0 galaxies but to Magellanic irregulars. Kormendy \&
Bender (2012) have suggested that dEs and dS0s are environmentally modified
Magellanic irregulars, and that the true dwarf members of the E galaxy class are
objects like the compact Es in Fig.~\ref{ces}. Kormendy \& Bender have
proposed renaming all dE and dS0 galaxies `spheroidals'. This connects these
objects directly to galaxies referred to as `dwarf spheroidals' and dwarf
irregulars in the Local Group.  Detailed studies of Local Group dwarf
spheroidals and irregulars reveal systems with a complex star formation history
(Mateo 1998).

Dwarf spirals are a controversial subject; only a few genuine cases are known.
IC\,3328, a dE, was shown to be a true dwarf spiral by Jerjen {\it et al.}
(2000).  NGC\,3928 was recognised as a dwarf spiral by van den Bergh (1980). It
is shown relative to the supergiant spiral UGC\,6614 in Fig.~\ref{ngc3928}.

%
%

\section{Tuning fork controversy}
\label{sec:lecture1_fork}

The RHS and VRHS classifications have always had a problem with early-type 
galaxies:

\begin{enumerate}[(a)]\listsize
\renewcommand{\theenumi}{(\alph{enumi})}

\item the E$n$ classification has no physical significance, unlike the stage 
classification for spirals;

\item the properties of S0s do not support the idea of them being a transition 
class between ellipticals and Sa, SBa galaxies (van den Bergh 1976, 1998).

\end{enumerate}

\begin{figure}
\centerline{\includegraphics[width=\textwidth]{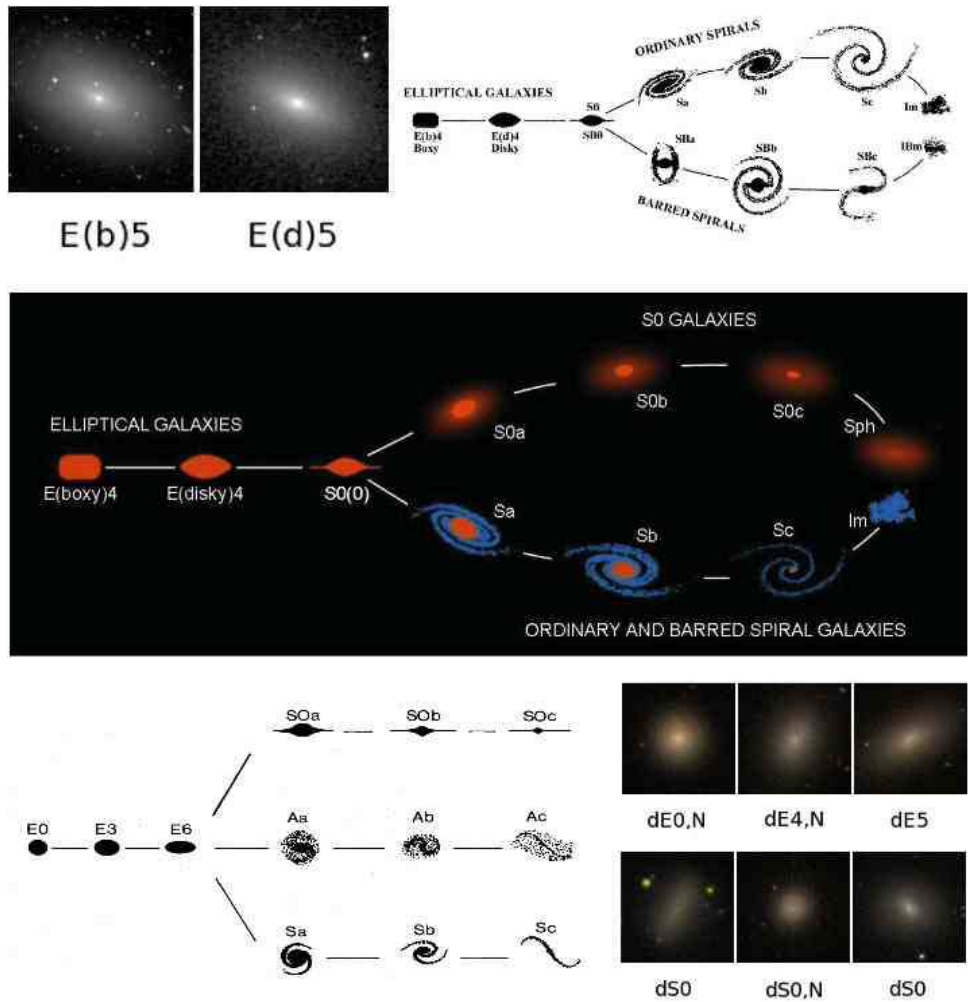}}
\caption{Two revisions of the Hubble tuning fork by Kormendy \& Bender (1996,
2012). The van den Bergh (1976) parallel sequence classification schematic is
shown at lower left, and several Sph galaxies are shown at lower right. All
three modified tuning forks are reproduced with permission of the authors}
\label{kb2012}
\end{figure}

Van den Bergh (1976) challenged the tuning fork by following the suggestion of
Spitzer \& Baade (1951; also Baade 1963) that S0s probably form a sequence
parallel to spirals, rather than being in the juncture of the tuning fork
between ellipticals and spirals. In a revised classification, van den Bergh
proposed that S0s be classified as a sequence S0a -- S0b -- S0c parallel to the
regular Hubble spiral sequence classification Sa -- Sb -- Sc (Fig.~\ref{kb2012},
lower left). In between these two sequences is a sequence of `anemic spirals':
Aa -- Ab -- Ac, meaning spirals whose lower than average star formation rate and
dust content implied a deficiency of H{\sc i} gas (discussed further in
Lecture~4, Section~\ref{sec:lecture4}). This view of the S0s is called `parallel sequence
classification'. Although a reasonable point of view, the van den Bergh
sequence did not gain much traction at least in part because of the absence,
until recently, of any S0s than might be identified as type S0c.

Several recent studies have provided strong support for the parallel sequence 
idea:

\begin{enumerate}[(a)]\listsize
\renewcommand{\theenumi}{(\alph{enumi})}

\item Kormendy \& Bender (2012): Sph galaxies \& E galaxy dichotomies; edge-on 
disks embedded in environmentally modified Sph galaxies; S0c galaxy identified;

\item Cappellari {\it et al.} (2011): ATLAS$^{3{\rm D}}$ kinematic classifications;

\item Laurikainen {\it et al.} (2011):  Near-IR S0 Survey (revised view of S0 bulges 
and identification of some of the first van den Bergh S0c galaxies).

\end{enumerate}

The problem with the E$n$ classification was at least partly solved by Kormendy
\& Bender (1996) who modified the Hubble E galaxy sequence to distinguish boxy
versus disky Es (Fig.~\ref{kb2012}, upper right). Boxy Es show isophotes that
have a negative value of the relative Fourier radius parameter $a_4/a$
(Jedrezjewski 1987), are more luminous, have less rotation, and more velocity
dispersion anisotropy than disky Es, which have a positive value of $a_4/a$.
Figure~\ref{kb2012} (upper left) shows two extreme examples of boxy and disky Es
where the character is evident by eye. The E(b)5 is NGC\,7029 while the E(d)5 is
NGC\,4697.

Further studies of E galaxies revealed that galaxies classified as dwarf
ellipticals in the Virgo cluster are NOT the low-luminosity versions of regular
ellipticals (Kormendy 1985; Kormendy {\it et al.} 2009). Instead, as we noted in
the previous section, dE and dS0 galaxies are the progeny of late-type galaxies.
Kormendy \& Bender's (2012) reclassification of these objects as Sph galaxies,
and their suggested location in parallel sequence classification is shown in
Fig.~\ref{kb2012}, middle. Several examples are shown in Fig.~\ref{kb2012},
lower right.

Kormendy {\it et al.} (2009) carried out a very detailed study of high- and
low-luminosity E galaxies and Sph galaxies. An E galaxy dichotomy exists in the
sense that luminous Es have a `core' (an inner radial zone of what seems like
`missing light', like a cone cut off at the top), while lower-luminosity Es
appear to lack this missing light core, instead showing `extra light' in the
manner of a power-law excess. B13 shows a comparison between three nucleated Sph
galaxies and three genuine E galaxies (one of the core type, and two of the
power-law type). Although the subtle differences are evident directly, the
distinctions are best seen in the profiles and parameter correlations described
by Kormendy {\it et al.} (2009).

\begin{figure}
\centerline{\includegraphics[width=\textwidth]{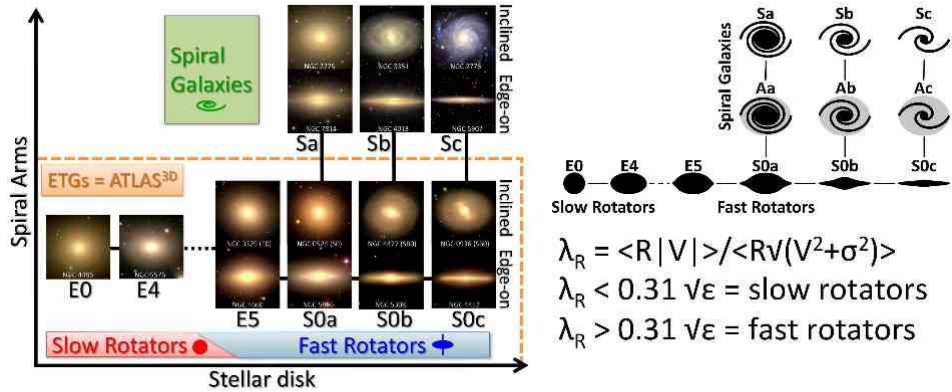}}
\caption{Atlas$^{3{\rm D}}$ kinematic basis for parallel sequence classification, 
from Cappellari {\it et al.} (2011, reproduced with permission).}
\label{atlas3d}
\end{figure}

ATLAS$^{3{\rm D}}$ was a massive study of the detailed kinematics of 260
early-type galaxies (Emsellem {\it et al.} 2007, 2011; Cappellari {\it et al.}
2011). Based on a kinematic parameter, $\lambda_{\rm R}$, Cappellari {\it et al.} 
(2011) independently came to the same conclusion as Kormendy \& Bender (2012): the
correct placement of S0s is parallel to spirals. Cappellari {\it et al.} (2011)
showed that the best way to order early-type galaxies is kinematically, not
morphologically. Their proposed revision to the Hubble tuning fork is shown in
Fig.~\ref{atlas3d}.

The contribution of Laurikainen {\it et al.} (2011) to the parallel sequence
classification idea involves the use of sophisticated two-dimensional
photometric decomposition that led to both a revised view of the
significance of S0 bulges (bulge-to-total luminosity ratio) and the
discovery of a few genuine S0c galaxies. This is discussed further in
Lecture~3 (Section~\ref{sec:lecture3}).

Note that none of this necessarily completely negates the value of tuning fork
classifications, because types such as E$n$, S0$_1$, S0$_2$, S0$_3$, or E$^+n$,
S0$^-$, S0$^{\circ}$, and S0$^+$ are still morphologically valid visual
categories. A galaxy may be classified as type SA(rl)0$^+$ and S0b, and the two
classifications are more complimentary than contradictory. The same is true for
the dE and dS0 categories. The revised tuning fork clarifies relationships
between types and, most importantly, finds a proper `home' for the enigmatic
Sph/dE/dS0 galaxies in the Hubble sequence.

The revisions to the tuning fork represent the coming to fruition of {\it
quantitative} and {\it interpretive} galaxy classification. Quantitative
morphology relies on derived photometric or kinematic parameters to determine
relations between galaxy types such as, for example, the relation between
spheroidal galaxies and Magellanic irregular galaxies, normal ellipticals and
compact ellipticals, disky ellipticals and disk galaxies, and between spirals and
S0s. Interpretive galaxy classification means `with a particular idea in
mind', such as the nature versus nurture origin of S0 galaxies, or, for
example, different subtypes of outer pseudorings (see Lecture~2, 
Section~\ref{sec:lecture2}).

\subsection{Does the continuity of galaxy morphology imply that secular
evolution must be occurring?}

To some extent it probably does, but apparent continuity can be misleading.  For
example, the VRHS classification of S0s from featureless systems with only a
trace of a lens to systems with obvious rings or traces of spiral structure is
clear continuity in galaxy morphology, but it does not necessarily mean that the
placement of S0s between ellipticals and spirals is correct. The type `S0/a'
automatically suggests a correct placement. Even so, some aspects of the VRHS
could imply secular evolution, such as:

\begin{enumerate}[(a)]\listsize
\renewcommand{\theenumi}{(\alph{enumi})}

\item The smooth variation in bar strength (with numerous examples filling the 
continuum of forms from SA to SB). Does this imply evolution from SB to SA or 
vice-versa, or both? The simulations of Bournaud \& Combes (2002) suggest that 
bar destruction by increasing central mass concentration followed by bar 
rejuvenation by external gas accretion could keep `apparent bar strength' as 
a continuum of forms just as recognised in the VRHS.

\item The seemingly smooth connection between rings and lenses. These features
are found in the same general locations relative to bars, and in particular the
radial profile of a lens can be very much like a ring of lower contrast on a
steeply declining background [e.g., the lens of NGC\,1553 (Kormendy 1984); as
compared to the bright stellar inner ring of NGC\,7702 (Buta 1991; see Fig.~20
of Buta \& Combes 1996)]. Do lenses simply represent the dissolution of bars or
bar-like features, or could many lenses be diffused (highly evolved) former
rings? The latter could follow from the mere existence of inner, outer, and
nuclear lenses in the same manner as inner, outer, and nuclear rings. 
Nevertheless, the virtually one-to-one connection between the sizes of bars and
the diameters of inner lenses still suggests, as noted by Kormendy (1979), that
such lenses formed by bar dissolution. A study of lens colours would aid greatly
in further establishing these connections.

\item The smooth variations in ring morphologies, ranging from completely closed
features to no ring at all, with a continuum of `pseudorings' in between. Does
this mean that pseudorings evolve secularly into closed rings? Does it also mean
that a galaxy with no ring could eventually become ringed? Even the simplest
numerical simulations, such as those of Schwarz (1981, 1984a), Simkin {\it et
al.} (1980), and Byrd {\it et al.} (1994), suggest that it might be possible for
gravity torques acting on spiral segments to evolve a pseudoring pattern and
eventually close it. It is not clear from such simulations, however, that pure
(s)-variety spirals could evolve into pure (r)-variety spirals. Such evolution
could depend on the evolution of the bar pattern speed as well.

\item The smooth variations in morphology along the spiral type sequence.  Could
a spiral evolve along this sequence?  Could it do this without also evolving
along the family and variety dimensions of the VRHS?  Stage evolution is perhaps
the most important question we could ask about secular galaxy evolution, because
it brings us into the realm of bulge formation and evolution. If many bulges are
built up by secular movement of disk material (as opposed to multiple mergers),
then secular evolution from late-to-early would indeed be possible (KK04). A big
question is, how many `steps' in stage could a galaxy evolve in this way
during a Hubble time? I did not discuss `pseudobulges' during my actual
lectures since these were covered in some detail by Kormendy (this volume), but
in the next set of notes I will describe the morphologies of such
bulges.

\end{enumerate}

%
%

\section{Lecture 2: Barred and spiral galaxies}
\label{sec:lecture2}

In my second lecture I would like to introduce the important morphological and
evolutionary issues connected with bars and spirals in galaxies. Although
related morphologically, bars are often considered as major\linebreak dynamical components
of galaxies, while spirals are seen as features possibly driven by bars (e.g.,
Kormendy 1979; Kormendy \& Norman 1979). Even so, a significant fraction of
normal galaxies are unbarred and still spiral. The role of bars and spirals in
secular evolution of galaxies is clearly important.

\begin{figure}
\centerline{\includegraphics[width=\textwidth]{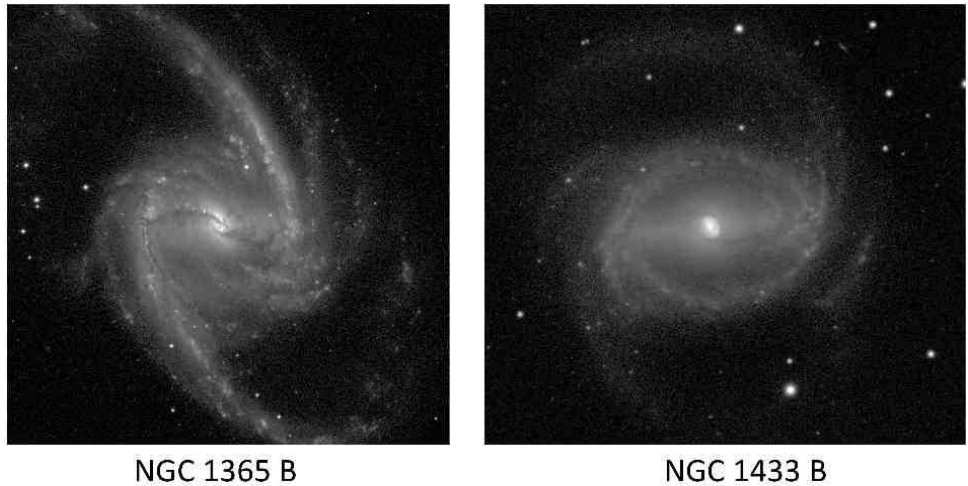}}
\caption{The bars in these galaxies are stronger than average and each
shows characteristics, such as the presence of a `pseudobulge', that
suggest secular evolution has occurred.}
\label{barred-gals}
\end{figure}

Barred galaxies can be considered `the ultimate' in galaxy morphology, by
which I mean they have some of the most organised structures known. This is
shown by the examples in Fig.~\ref{barred-gals}. In these galaxies the bar is
obviously a major perturbation, and one of the first questions we may ask is,
how many galaxies are barred? A large number of studies have examined this
question, often focussing on infrared observations since these penetrate dust
and make some obscured bars more visible.  One such study by Eskridge \textit{et
al.} (2000) found the following: from $H$-band (1.65\,$\mu$m) images of 186
bright nearby galaxies, 56\% were visually classified as `strongly barred',
16\% were classified as `weakly barred', and 27\% were `unbarred'. Although
it seems that strong bars are more common than weak ones, this is an illusion.
An optically classified SA galaxy may show a weak bar in the IR and get a new
classification of SAB, and an optically classified SAB galaxy may show a more
prominent bar in the IR and get classified as SB. However, even though the bar
of an optically classified SB galaxy may also look stronger in the IR, it has no
new classification bin to be placed in. Thus, IR imaging does not really change
the {\it rankings} of bars.  All bars in spirals at least look stronger in the
IR, and the rankings (what is actually strong and what is actually weak) remain
about the same.  From many studies (see Buta \textit{et al.} 2010b for a recent
summary), and including both SAB and SB types, the bar fraction ranges from
50-70\%.

\begin{figure}
\centerline{\includegraphics[width=\textwidth]{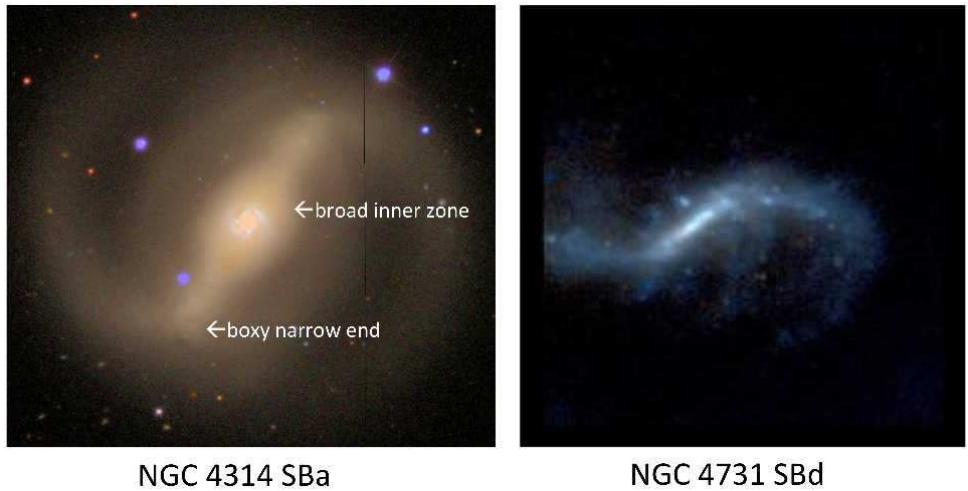}}
\caption{NGC\,4314 shows the characteristic features of an early-type
galaxy bar, including the broad inner zone and the boxy narrow ends,
while NGC\,4731 shows a typical late-type galaxy bar which lacks a broad
zone and has a much younger stellar population.}
\label{ngc4314-4731}
\end{figure}

Bars are characterised by their different morphologies and non-elliptical
shapes. A typical normal bar in an early-type galaxy has two sections: a broad
inner zone, and narrower ends (e.g., NGC\,4314 in Fig.~\ref{ngc4314-4731}). The
inner zone can be round or elliptical and can be mistaken for a large bulge. 
The broad section is not seen in late-type galaxies generally (e.g., NGC\,4731
in Fig.~\ref{ngc4314-4731}). Athanassoula \textit{et al.} (1990) fitted
generalised ellipses to the isophotal shapes of early-type galaxy bars and
demonstrated the typical boxy character of these features. In
early-to-intermediate-type galaxies, bars are made of an old stellar population,
while in very late-type galaxies, bars can include a much younger stellar
population.  These characteristics are well shown by the examples in
Fig.~\ref{ngc4314-4731}.

\begin{figure}
\centerline{\includegraphics[width=\textwidth]{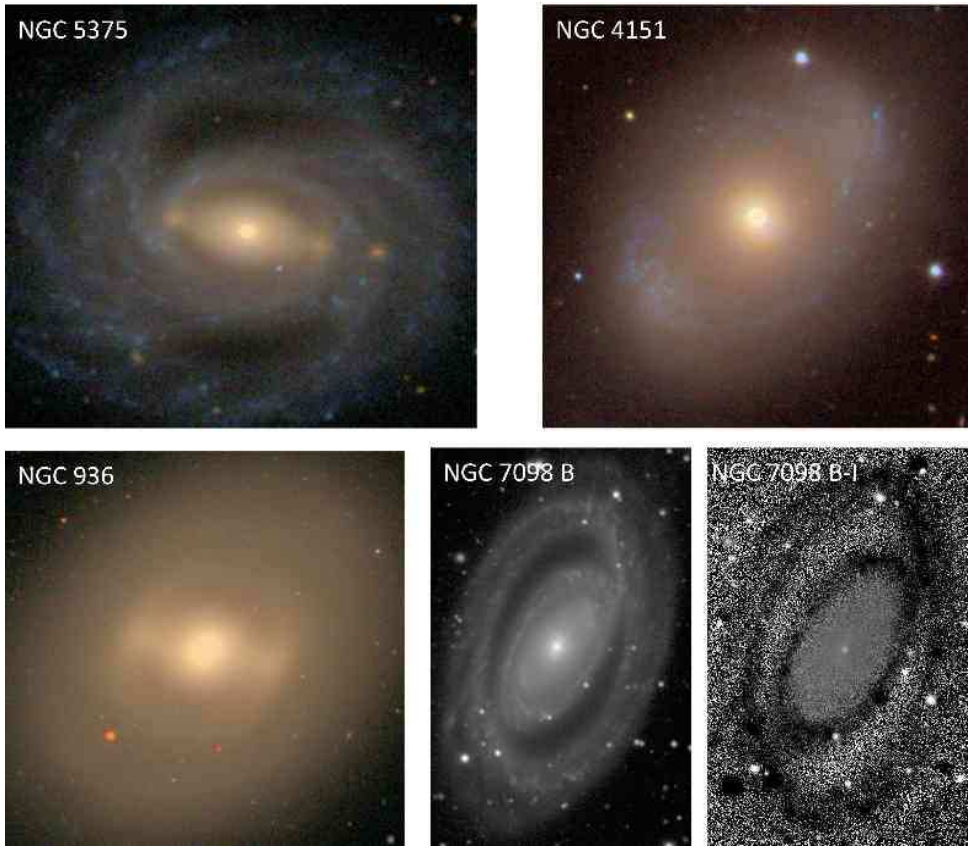}}
\caption{Colour SDSS images of three galaxies with ansae-type bars, and
a colour index map of one, that shows how ansae are often made of an old
stellar population but can include young stars, as in NGC\,4151.  The
$B-I$ colour index map of NGC\,7098 is coded such that bluer features are
dark and redder features are light.} 
\label{ansae} 
\end{figure}

In an ansae-type bar, the broad inner zone and the bar ends may be separated or
at least more distinct (Fig.~\ref{ansae}). Ansae bars are found in 40\% of
early-type barred galaxies (Mart\'inez-Valpuesta \textit{et al}. 2007) and are
very rare in later types. Even so, exceptional examples of ansae in
intermediate-type spirals are known, such as the SBb galaxy NGC\,5375 shown in
the upper left frame of Fig.~\ref{ansae}. Particularly intriguing, and not
really well understood, is the variety of ansae morphologies seen in early-type
barred galaxies. Some are roundish spots, others are relatively linear
enhancements, while still others are narrow arcs that appear to blend into an
inner ring or lens. In colour SDSS images or colour index maps like that shown
for NGC\,7098 in Fig.~\ref{ansae}, ansae are seen to be made of old stars,
although NGC\,4151 provides a counter example where the ansae are blue and
irregular due to star formation.

\begin{figure}
\centerline{\includegraphics[width=\textwidth]{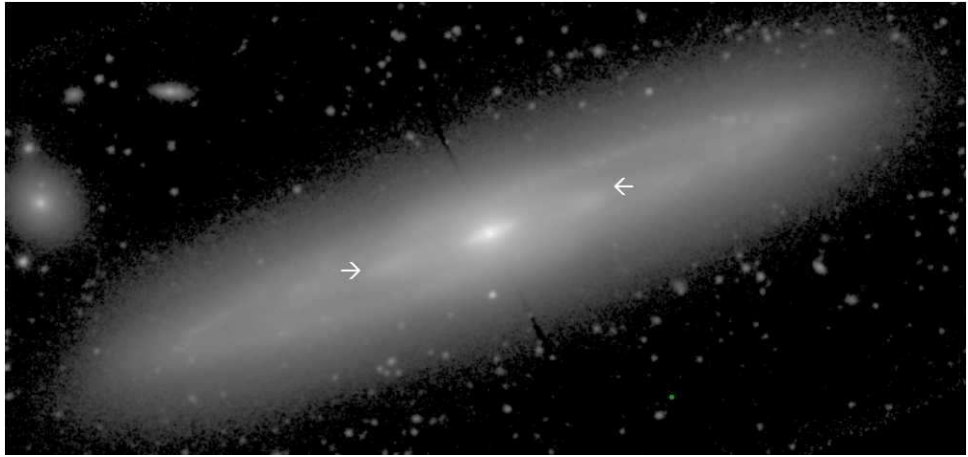}}
\caption{In this 3.6\,$\mu$m image, the nearly edge-on Sb spiral NGC\,4216
shows likely ansae (arrows) flanking an inner boxy zone.}
\label{ngc4216-3.6} 
\end{figure}

What are the three-dimensional shapes of ansae? The galaxy NGC\,4216
(Fig.~\ref{ngc4216-3.6}) shows a boxy bulge with bright enhancements that look
like ansae in a 3.6\,$\mu$m image. If this is what the enhancements are, then it
appears that ansae are flatter than the inner sections of bars.

\begin{figure}
\centerline{\includegraphics[width=\textwidth]{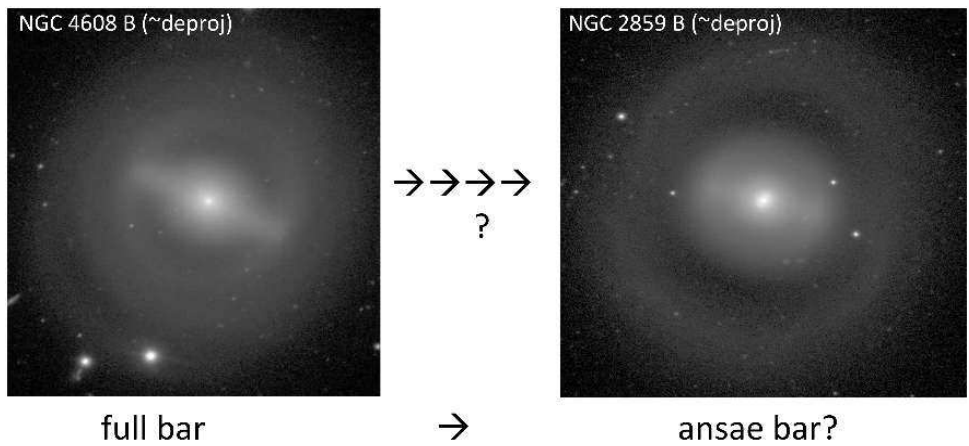}}
\caption{Possible scenario of secular evolution where a full bar
evolves to where the broad inner zone and the ends separate.}
\label{ansae-sec-ev} 
\end{figure}

The appearance of ansae, and the existence of many bars which do not have them,
suggests a possible interpretation of the features in terms of secular evolution
(Fig.~\ref{ansae-sec-ev}). We can ask: are ansae bars merely a distinct type of
bar, related to some aspect of bar formation, or is there a process that changes
a normal bar (e.g., as in NGC\,4608) into an ansae bar (e.g., as in NGC\,2859)?
Can this process separate the two parts of a normal bar and eventually stretch
the ansae into arcs?

\begin{figure}
\centerline{\includegraphics[width=\textwidth]{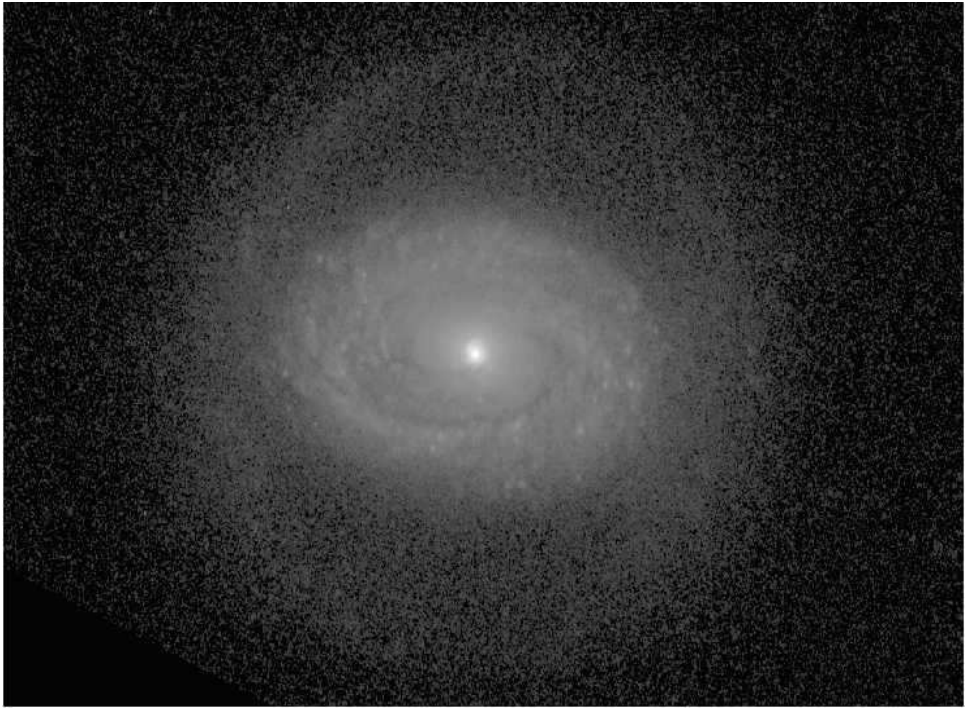}}
\caption{Deprojected blue light image of NGC\,4941, an outer-ringed galaxy with 
a prominent inner oval disk.} 
\label{ngc4941} 
\end{figure}

Ovals are broad, bar-like features having little Fourier amplitude above $m$=2,
as shown by the example in Fig.~\ref{ngc4941}. Other examples are illustrated by KK04 and B13. The effect
of an oval on galaxy morphology can be similar to that of a conventional bar
(KK04). The similarity between some oval galaxies and SB galaxies with outer
rings suggests that the oval is driving ring formation in the same manner as a
bar would. The oval is a bar-like feature sitting in the same place where a bar
and an inner ring would be. When a bar or an oval is viewed at an angle
intermediate between end-on and broadside on, the isovelocity contours bend
towards the bar as in NGC\,6300 (Buta 1987). Similar bending can be used to
identify an oval in an inclined galaxy (KK04).

\begin{figure}
\centerline{\includegraphics[width=\textwidth]{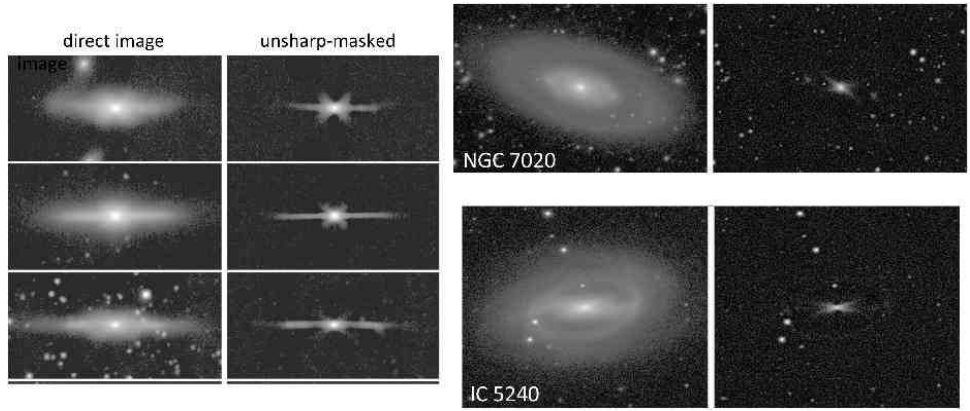}}
\caption{Examples of edge-on and non-edge-on X-galaxies.}
\label{x-gals} 
\end{figure}

Boxy or X-morphologies (Fig.~\ref{x-gals}) are commonly seen in edge-on disk
galaxies. Considerable evidence supports the idea that these features are simply
the projections of the vertical structure of bars (Bureau \& Freeman 1999;
Bureau \& Athanassoula 1999; Athanassoula \& Bureau 1999). However,
X-morphologies can also be seen in galaxies inclined less than 50$^{\rm o}$,
e.g., IC\,5240 (Buta \textit{et al}. 2007, see Fig.~\ref{x-gals}, lower right)
and IC\,4290 (Buta \& Crocker 1991). In such cases, the boxy/X-morphology
affects only the broad inner section and not the narrower ends, another argument
in favour of the ends of bars being flatter than their middle sections.

\begin{figure}
\centerline{\includegraphics[width=\textwidth]{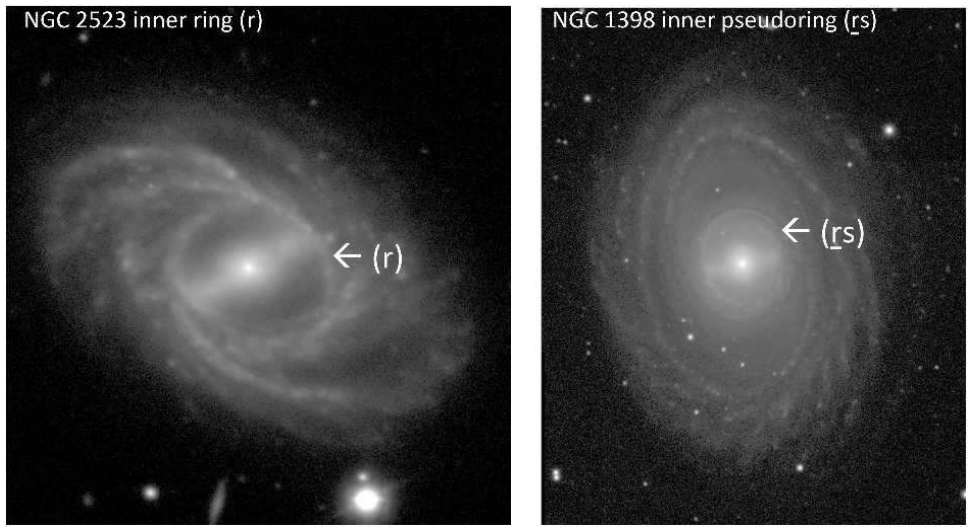}}
\caption{NGC\,2523 and NGC\,1398, two galaxies with conspicuous inner rings.}
\label{inner-rings} 
\end{figure}

Barred galaxies are well-known for their ring morphologies. The bar usually
fills one dimension of an inner ring, as in NGC\,2523 and NGC\,1398
(Fig.~\ref{inner-rings}). The inner ring of NGC\,1398 especially shows a tight
spiral\linebreak morphology that is mostly distinct from the rest of the galaxy's
extensive spiral structure.

\begin{figure}
\centerline{\includegraphics[width=\textwidth]{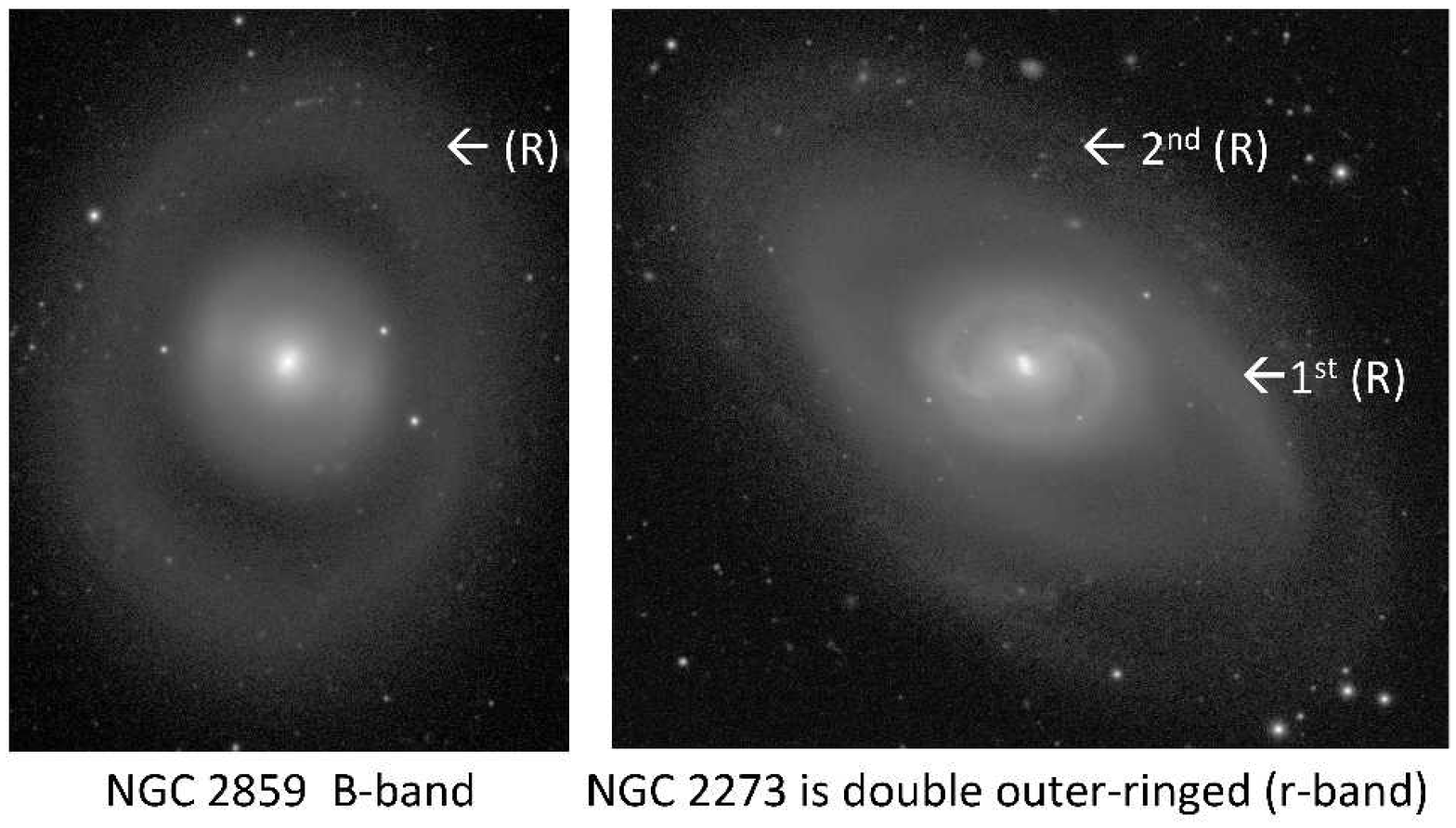}}
\caption{Two galaxies with conspicuous outer rings.}
\label{outer-rings} 
\end{figure}

\begin{figure}
\centerline{\includegraphics[width=\textwidth]{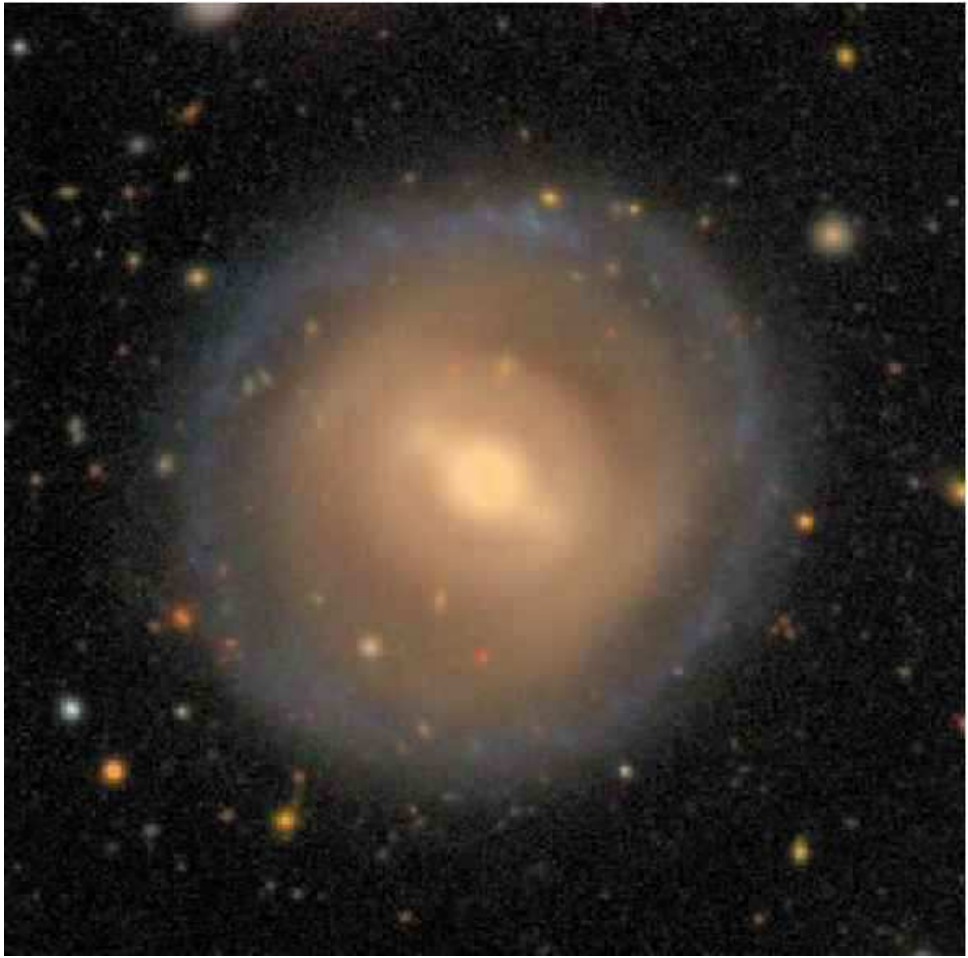}}
\caption{NGC\,1211, a double outer-ringed galaxy with an obvious stellar
population dichotomy.} 
\label{ngc1211-Rcolor} 
\end{figure}

Outer rings (R) are usually about 2--2.5 times the size of an inner ring/bar.
Usually there is only one (R), but very rare cases have two
(Fig.~\ref{outer-rings}). The stellar populations in outer rings are potentially
very interesting. In NGC\,2273, the two outer rings are smooth and both may be
characterised mostly by an older stellar population. However, the double outer
ring in NGC\,1211 has a population dichotomy that is well shown in an SDSS
colour image (Fig.~\ref{ngc1211-Rcolor}). The outer outer ring is blue while the
inner outer ring, the inner ring, the bar, and the bulge are defined by a much
older and redder stellar population. The suggestion is that the outer outer ring
is a recent acquisition, possibly an accretion feature where the material from a
disrupted companion has settled into the disk plane of a host (R)SB(rl)0$^+$
galaxy.  Alternatively, NGC\,1211's outer structure may be related to what Buta
(1995) referred to as an `R$_1$R$_2^{\prime}$' morphology, or double outer
ring/pseudoring (Buta \textit{et al.} 2007).  These also can show a population
dichotomy with the R$_1$ component being redder and made of older stars than the
R$_2^{\prime}$ component.  This kind of morphology has been linked to the outer
Lindblad resonance, and is described further later in this lecture.

The outer spirals in barred galaxies often have variable pitch angle and close
into what de Vaucouleurs referred to as an outer pseudoring, (R$^{\prime}$).
These features are most common among early- to intermediate-type spirals (Buta \&
Combes 1996). The morphologies of these features are distinctive enough to merit
special attention, as shown by the R$_1$R$_2^{\prime}$ morphology. This is also
discussed later in this lecture.

Rings are likely products of secular evolution in galaxies, in the following
sense. As reviewed by Buta \& Combes (1996), the best interpretation of barred
galaxy rings is that they form by gas accumulation at resonances, under the
continuous action of gravity torques due to the bar. In the presence of a bar
potential, gas clouds try and settle into the parent orbits of the potential,
but cannot do so without crossing other orbits. This leads to a spiral that
slowly can be torqued into rings near orbital resonances (Schwarz 1981, 1984a).

The resonance idea is not the only interpretation of barred galaxy rings that
has been put forward (see summary in B13 and also Athanassoula, this volume),
but it makes a number of predictions with respect to morphology that are easily
testable. To do so, we need to find a reliable way to judge the intrinsic shapes
and orientations of inner and outer rings. The way this can be done is to obtain
{\it distributions of apparent axis ratios and relative bar-ring position
angles}, and then to model these distributions under the assumptions of random
orientations of the disk planes and zero vertical thickness.  The latter
assumption is not unreasonable given that rings in spirals are often zones of
active star formation.

\begin{figure}
\centerline{\includegraphics[width=\textwidth]{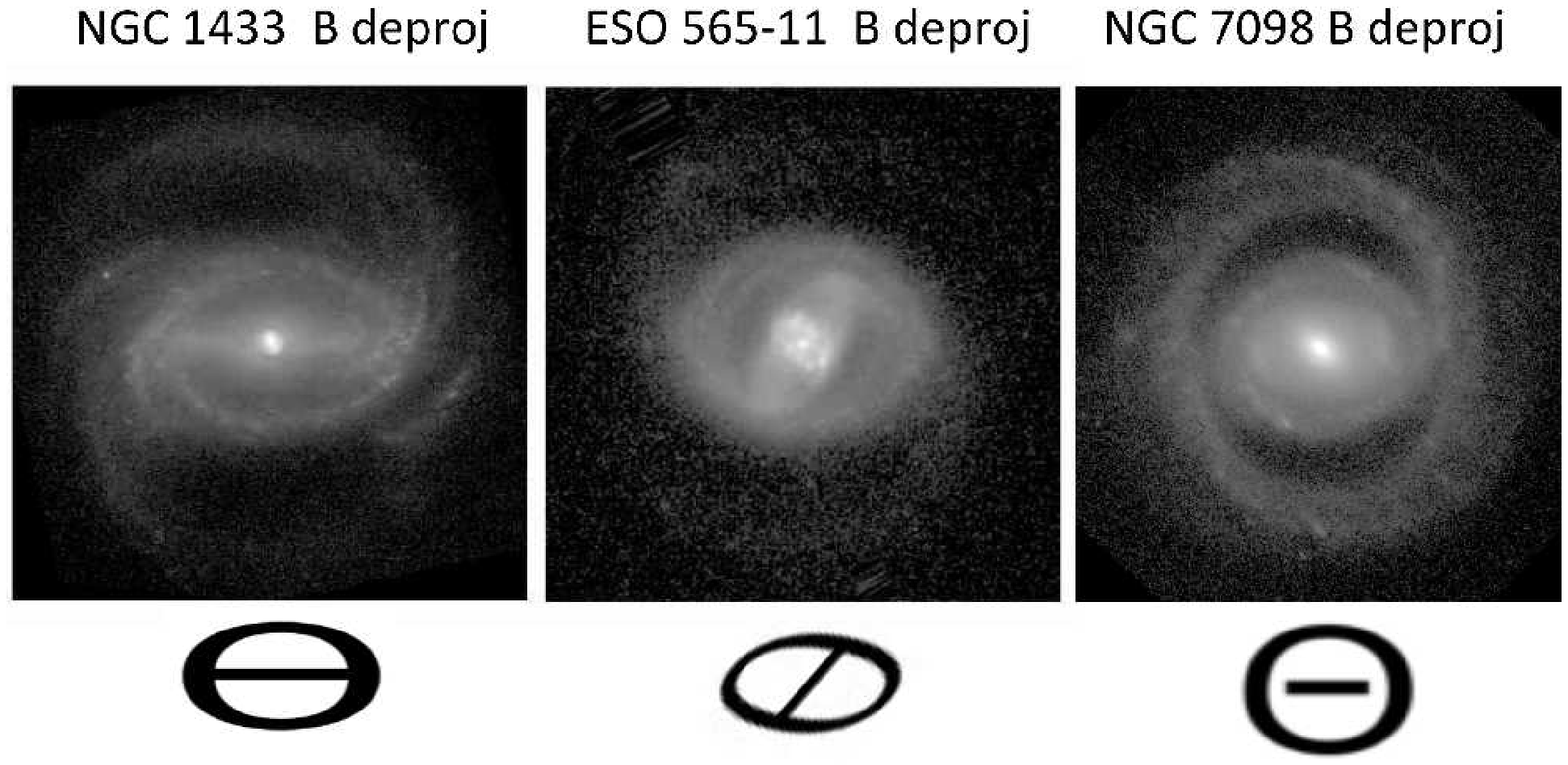}}
\caption{Intrinsic inner ring shapes and orientations in barred galaxies.} 
\label{ring-shapes} 
\end{figure}

The Catalogue of Southern Ringed Galaxies (CSRG, Buta 1995) was\linebreak designed to
evaluate the intrinsic shapes and orientations of inner and outer rings in this
manner. Inner SB rings are oval with intrinsic axis ratio 
$\langle q_{\rm o}\rangle$\,=\,0.81$\pm$0.06 and are aligned parallel to the bar
(Fig.~\ref{ring-shapes}, left panel).  The velocity field of the nearby galaxy
NGC\,1433 (Buta \textit{et al.} 2001) beautifully demonstrates this
characteristic alignment. The large inner ring in this galaxy has $q_{\rm o}$\,=\,0.63,
much more elongated than average. The velocity field betrays this shape by a
kinematic line of nodes that is nearly along the minor axis of the ring. Bars
may also underfill an inner ring or pseudoring, as in NGC\,7098
(Fig.~\ref{ring-shapes}, right panel). Bars generally do not overfill inner
rings, making H. Curtis's `$\phi$-type' characterisation not quite appropriate.

Although parallel alignment is clearly the `rule' for SB inner rings,
misalignments are sometimes seen. These are recognised in nearly face-on
galaxies as clear cases of inner pseudorings crossed at a large angle by a bar.
A kinematically confirmed example is ESO\,565-11 (Fig.~\ref{ring-shapes},
middle panel; see Buta \textit{et al.} 1995a). Other likely examples are
NGC\,309 (Sandage 1961) and CSRG\,1052 (Buta 1995).

ESO\,565-11 is such an extreme case that it suggests a possible evolutionary
effect. The galaxy is not only a misaligned bar-inner ring system, but also a
misaligned bar-oval system, because the inner pseudoring lies around the rim of
a massive oval, from which stellar outer arms emerge. Since long-term, stable
misalignment of massive nonaxisymmetric components is unlikely, the suggestion
is that the bar of ESO\,565-11 is relatively new, and that it formed within a
pre-existing oval that itself is a remnant of a past bar episode. Evidence in
support of this idea is the presence of the large and very unusual nuclear ring
(Buta \textit{et al.} 1999). The extreme elongated shape and size of this
feature (more than 3\,kpc in radius) are likely indicators of its youth.

\begin{figure}
\centerline{\includegraphics[width=\textwidth]{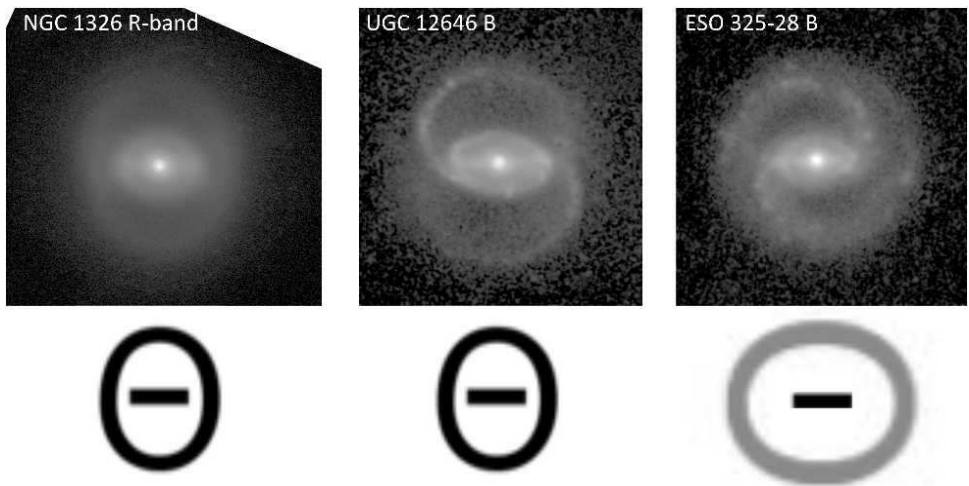}}
\caption{Intrinsic outer ring and pseudoring shapes and orientations in barred 
galaxies, as compared with deprojected images.} 
\label{SB-outer-rings} 
\end{figure}

For SB outer rings/pseudorings, statistics favor $\langle q_{\rm o}\rangle$\,=0.82$\pm$0.07 
and both parallel and perpendicular alignments (Buta 1995). The different
alignments are shown with deprojected inner rings in Fig.~\ref{SB-outer-rings}.
In each, note the exclusive parallel alignment of the inner rings in the same
galaxies.

\begin{figure}
\centerline{\includegraphics[width=\textwidth]{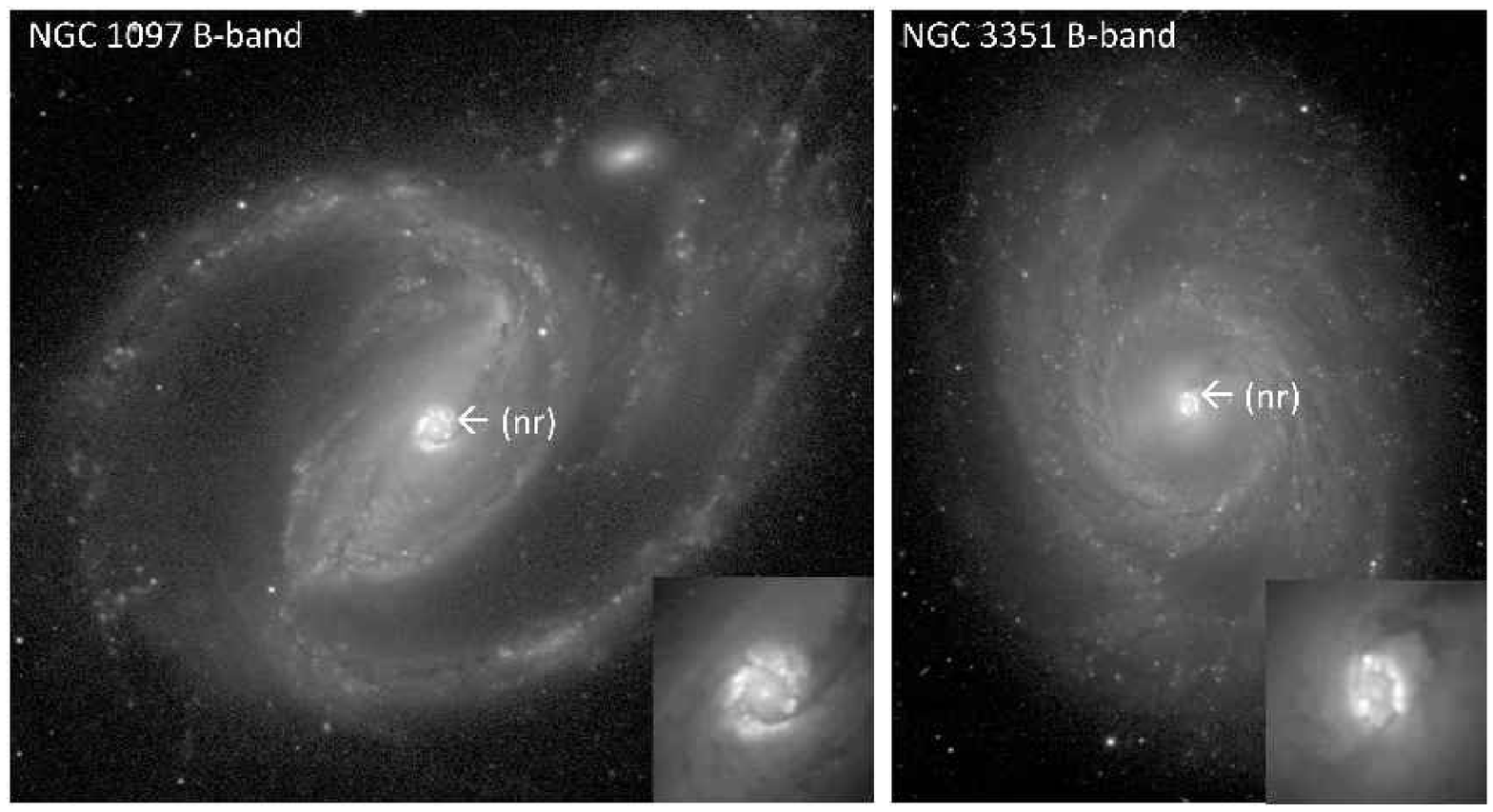}}
\caption{Two barred galaxies with nuclear rings.}
\label{SB-nuclear-rings} 
\end{figure}

\begin{figure}
\vspace{0.15cm}
\centerline{
\includegraphics[width=0.485\textwidth]{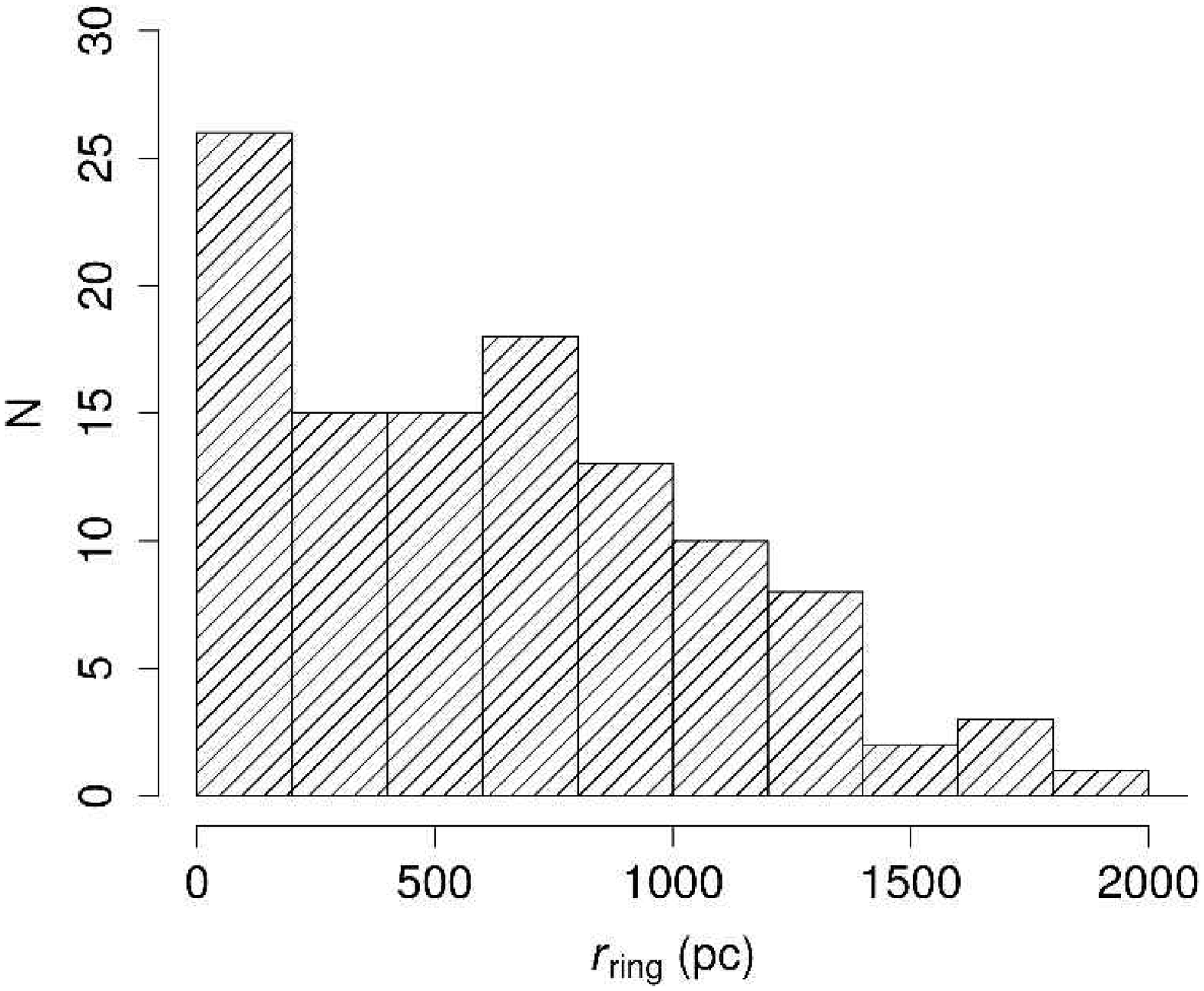}
\includegraphics[width=0.485\textwidth]{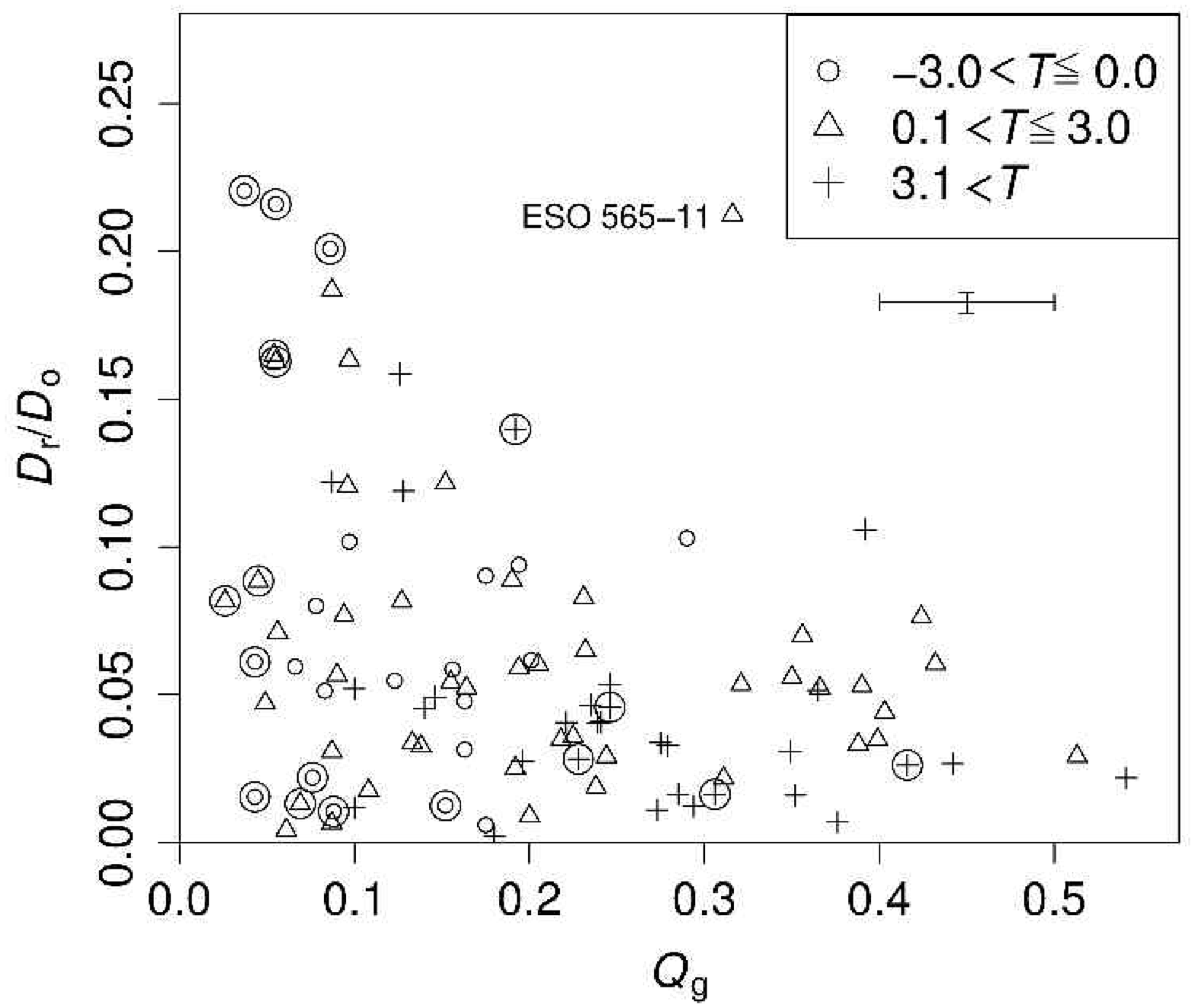}}
\caption{Two graphs from Comer\'on \textit{et al.} (2010, reproduced with
permission) showing the wide range of linear sizes of nuclear rings and the
sensitivity of relative ring size to relative bar torque strength $Q_{\rm g}$.} 
\label{nrings} 
\end{figure}

Nuclear rings (nr) are found well inside bars (Fig.~\ref{SB-nuclear-rings}).
Nuclear rings average 1.5\,kpc in diameter, and have roughly circular intrinsic
shapes and a morphology that can be shaped by dust. According to Knapen (2005)
and Comer\'on \textit{et al}. (2010) nuclear rings are found in 20\% of galaxies
in the type range S0$^-$ to Sd. Figure~\ref{nrings} shows two results from
Comer\'on \textit{et al.} (2010): that nuclear rings have a wide range of linear
diameters, from 200\,pc to several kpc, and that stronger bars tend to host
smaller nuclear rings, while weaker bars can host small and large rings. The
connections between bars and the properties of nuclear rings are discussed
further in Knapen (2010).

\begin{figure}
\centerline{\includegraphics[width=\textwidth]{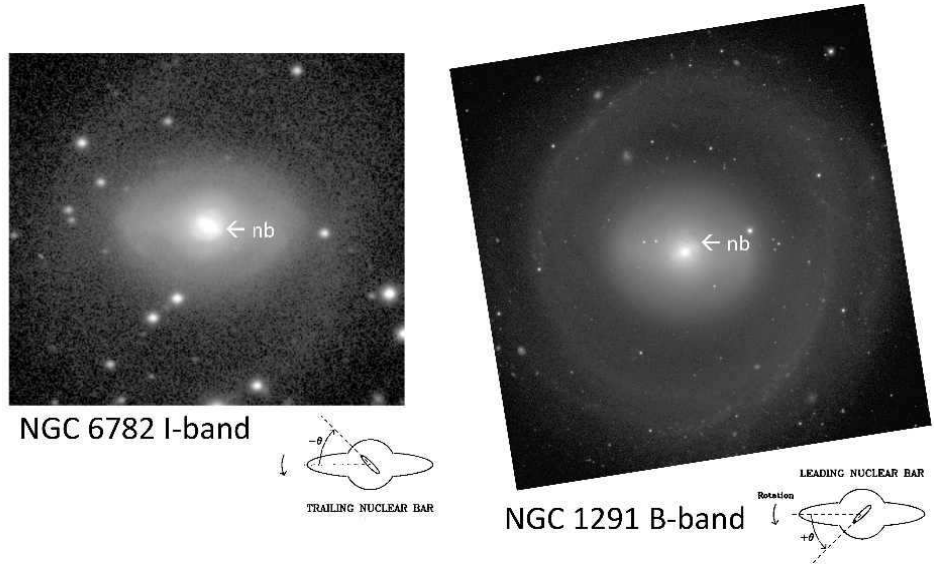}}
\caption{Two barred galaxies with secondary or nuclear bars. The one in
NGC\,6782 is tipped behind the primary bar and is called trailing, while
the one in NGC\,1291 is tipped ahead and called leading. Schematics are
from Buta \& Crocker (1993a).} 
\label{SB-nuclear-bars} 
\end{figure}

\begin{figure}
\centerline{\includegraphics[width=\textwidth]{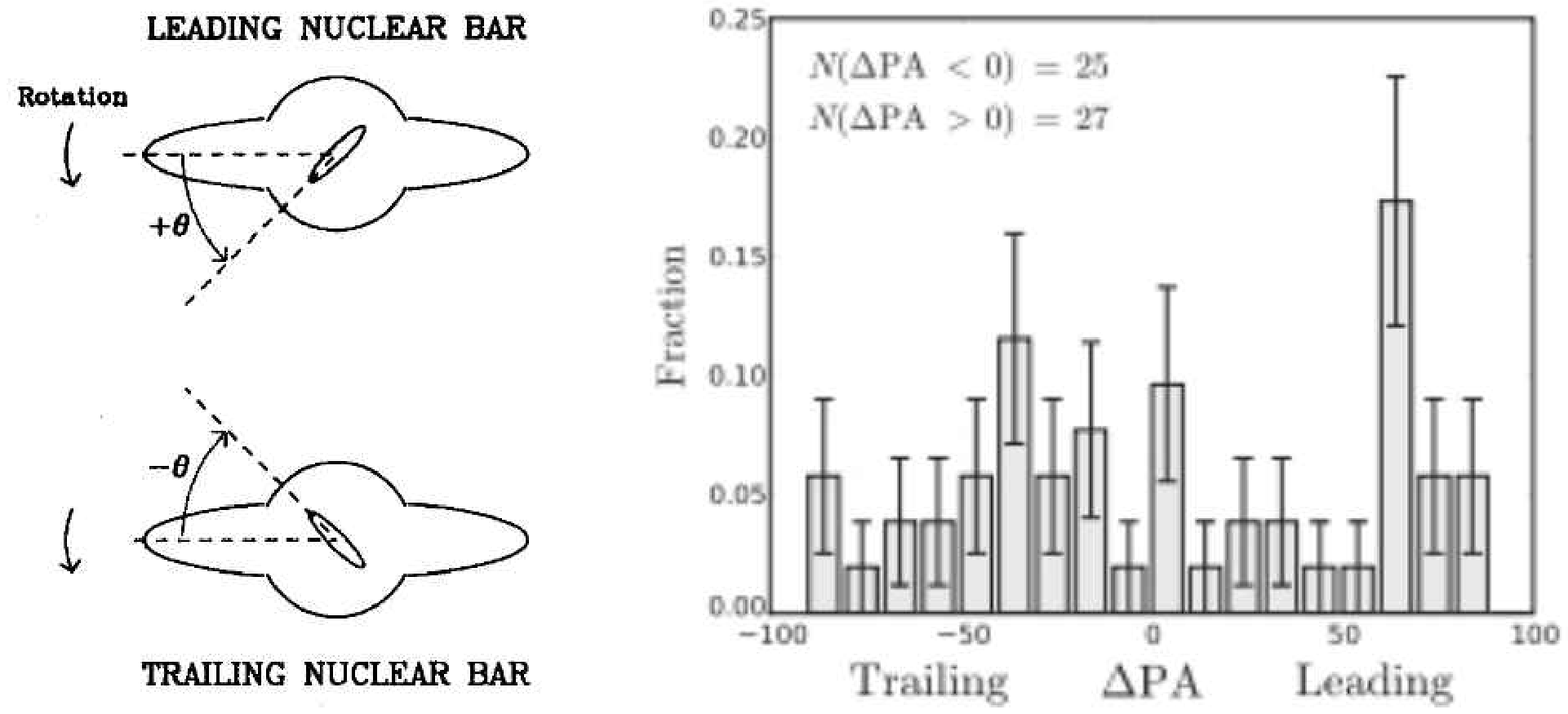}}
\caption{The schematic at left shows the definitions of the terms
`leading' and `trailing' as applied to nuclear bars in terms of the
sign of the angle $\theta$ (Buta \& Crocker 1993a). The graph at right
shows the distribution of $\theta$ values for a large sample of
secondary bars from Erwin (2004, reproduced with permission).} 
\label{bc93-erwin2004} 
\end{figure}

Found in the same area as nuclear rings are nuclear (secondary) bars (nb)
(Fig.~\ref{SB-nuclear-bars}). According to Erwin (2004, 2011), double bars are
found in $\sim$20\% of S0-Sb galaxies. The sense of a nuclear bar, tipped ahead
of (leading) or behind (trailing) a primary bar can be judged from the sense of
winding of spiral arms (assumed to be trailing). The existence of both types of
nuclear bars argue that these features have a different pattern speed from the
primary bar. Erwin (2004, 2011) shows that alignments between nuclear and
primary bars are random (Fig.~\ref{bc93-erwin2004}), and Corsini \textit{et al.}
(2003) find direct kinematic evidence for it in NGC\,2950.

In terms of secular evolution, nuclear rings are in a different domain from
inner and outer rings. Outer rings are thought to have a long timescale
(requiring more than a few Gyr to form; Schwarz 1981; Rautiainen \& Salo 2000),
while nuclear rings, being much closer to the centre, have a timescale of only
a few times $10^8$ years (Combes 1991). Only nuclear rings are prone to extreme
bursts of star formation. The presence of a nuclear bar inside a star-forming
nuclear ring (as, e.g., in NGC\,6782) should produce gravity torques that could
evolve the central regions (Shlosman \textit{et al.} 1989). Also, the
coexistence of nuclear, inner, and outer rings, each with a very different time
scale, in the same galaxy, suggests a persistent means of funnelling gas towards
the central regions.

\subsection{Quantifying bar strength}

One of the major goals de Vaucouleurs had during his long career was to find
ways of quantifying the three dimensions of the VRHS. He imagined setting the
stage $T$\,=\,$f$ (colour, surface brightness, H{\sc i} content, etc.), where $f$
is a numerical function of measured parameters, each of which correlate
individually with stage but with a large scatter. However, no effective way of
doing this was ever really found. $T$ is still best estimated visually. In
contrast, there is more than one way of reliably quantifying bars. Doing this
would be useful for several reasons: (1) to judge the actual significance of a
bar within its disk; (2) to examine the connection between bars and spirals
(e.g., we can ask: do the former `drive' the latter?); and (3) to investigate
how measureable properties of bar-associated features like rings correlate with
bar strength.

\begin{figure}
\centerline{
\includegraphics[width=0.55\textwidth]{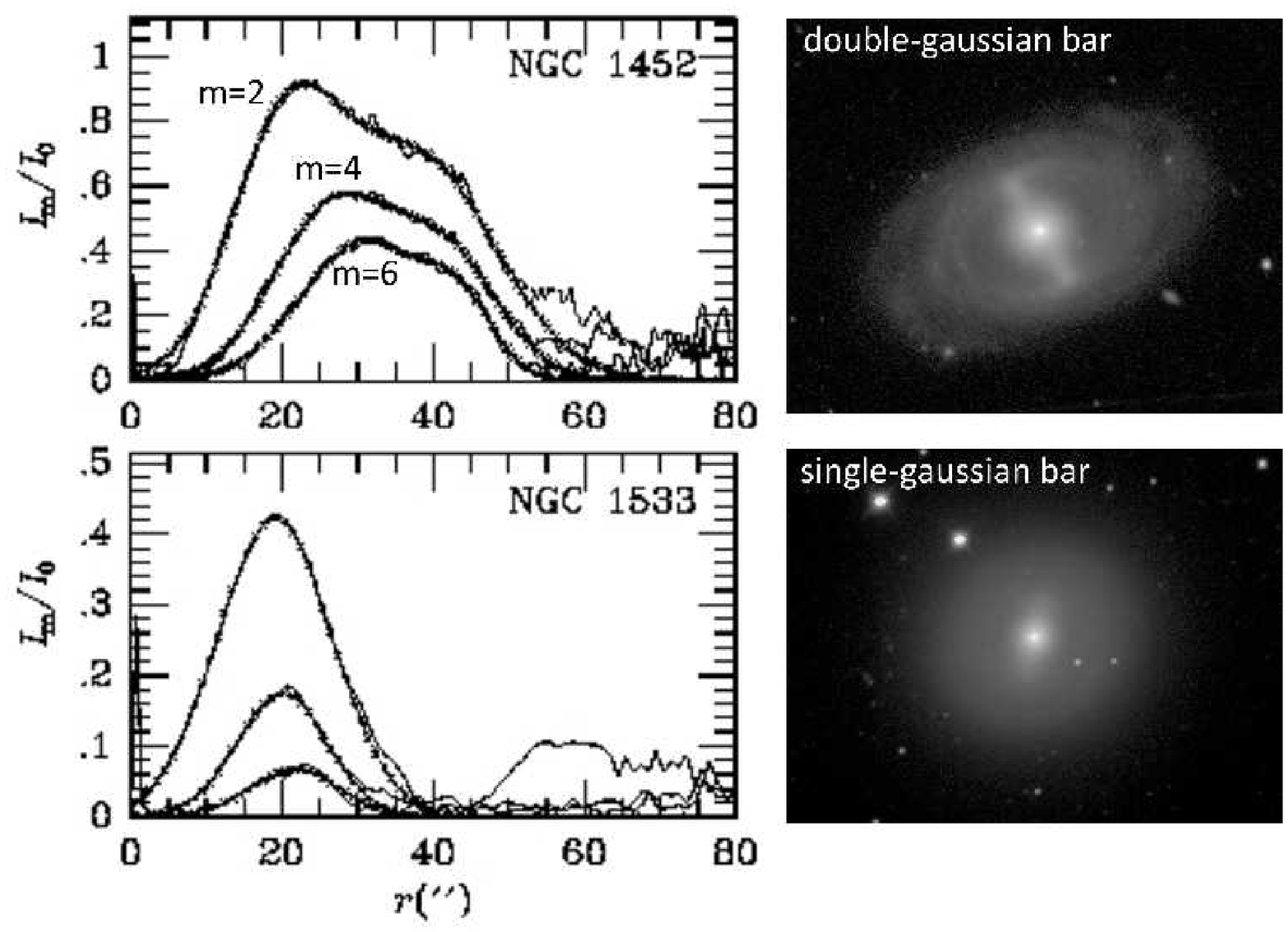}
\includegraphics[width=0.45\textwidth]{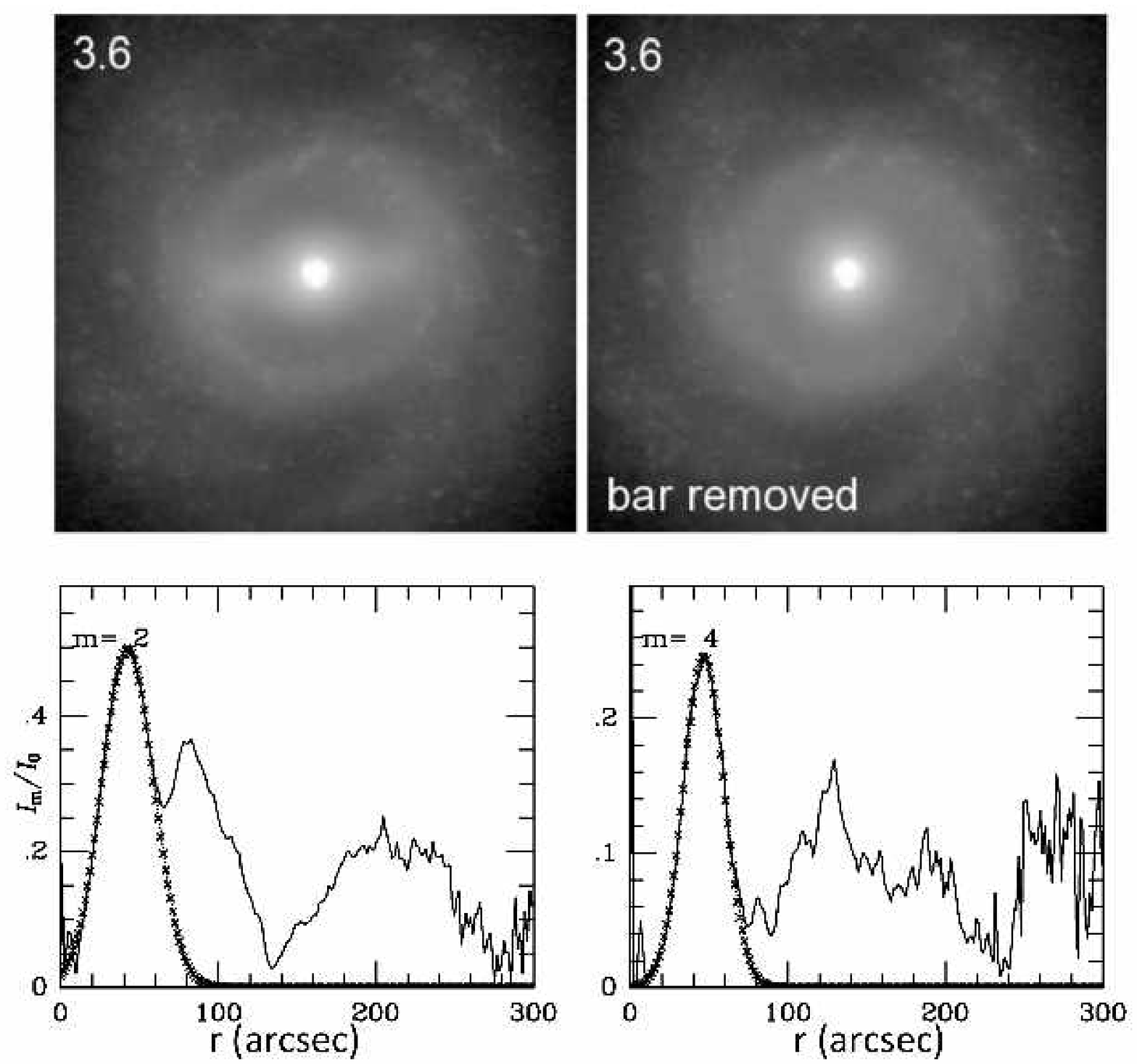}}
\caption{Examples of single and double-Gaussian Fourier bar profiles. (Left
frames): $m=2$, 4, and 6 $K_{\rm s}$-band relative Fourier intensity profiles and
$B$-band images of early-type galaxies NGC\,1452 and NGC\,1533 (Buta \textit{et
al.} 2006). (Right frames): 3.6\,$\mu$m Fourier profiles and images of
NGC\,3351, showing that a single Gaussian fits the $m=2$ and 4 Fourier
profiles well.}
\label{gaussians} 
\end{figure}

Relative Fourier intensity amplitudes can quantify the light perturbation
effectively. Two examples involving early-type barred galaxies are shown in
Fig.~\ref{gaussians}. $A_2$ is the maximum of the relative $m=2$ Fourier
amplitude $I_2/I_0$.  The amplitudes are derived from a Fourier expansion of the
brightness distribution:
\begin{equation}
I(R,\phi) = I_0(R) + \Sigma_{m=1}^{\infty} I_m(R){\rm cos}[m(\phi - \phi_m(R))].
\end{equation}
With such an expansion, the sine and cosine amplitudes are derived as
\begin{eqnarray}
I_{m{\rm c}}(R) &=& 2<I(R,\phi){\rm cos}(m\phi)>\nonumber\\
I_{m{\rm s}}(R) &=& 2<I(R,\phi){\rm sin}(m\phi)>\nonumber\\
I_m(R) &=& \sqrt{I_{m{\rm c}}(R)^2+I_{m{\rm s}}(R)^2}\nonumber\\
\phi_m(R) &=& {1\over m}{\rm tan}^{-1}({I_{m{\rm s}}(R)\over I_{m{\rm c}}(R)}).
\end{eqnarray}
An important finding from Buta \textit{et al.} (2006) is the single and double
Gaussian forms of the radial variations of $I_m/I_0$ for some galaxies,
especially early-types. The solid curves fitted to the amplitudes of the two
early-type examples (Fig.~\ref{gaussians}) shows how well these representations
work. The physical significance of these forms is unclear; however, similar
forms have been seen in numerical simulations.  For example, the relative
Fourier profile for NGC\,1452 is similar to a `massive halo' bar model from
Athanassoula \& Misiriotis (2002).  Note that more complex Fourier profiles are
seen that can be described as multi-Gaussian in nature, often occurring when the
bar is embedded in a massive oval.

Interestingly, single Gaussian bars are not restricted to early types like
NGC\,1533. The same kind of bar is seen in the SBb spiral NGC\,3351.
Figure~\ref{gaussians}, right, shows $m$\,=\,2 and 4 Fourier profiles
and their Gaussian representations for NGC\,3351. Also shown is how
effectively this representation removes the bar from an infrared image of the
galaxy. Single Gaussian bars appear to be the simplest type, but remain to be
explained.

\subsection{Bar strength from maximum relative gravitational torques}

Another way of quantifying bar strength is to use an infrared image to trace
stellar mass and infer the gravitational potential $\Phi$. Then the bar strength
can be estimated in terms of the maximum relative tangential forcing. Using Fast
Fourier Transform techniques (Binney \& Tremaine 1987, Section 2.9), a
two-dimensional image may be converted to a two-dimensional potential on a
Cartesian grid (Quillen {\it et al.} 1994). From this potential we derive the
parameter
\begin{equation}
Q_{\rm T}(R)=\left|{F_T(R)\over F_{0{\rm R}}}\right|_{\rm max},
\end{equation}
where $F_{\rm T}$\,=\,${1\over R}{\partial\Phi\over\partial\phi}$ is the tangential
force and $F_{0{\rm R}}$\,=\,${\partial\Phi_0\over \partial R}$ is the mean radial force.
This approach follows Sanders \& Tubbs (1980) and Combes \& Sanders (1981).

\begin{figure}
\centerline{\includegraphics[width=\textwidth]{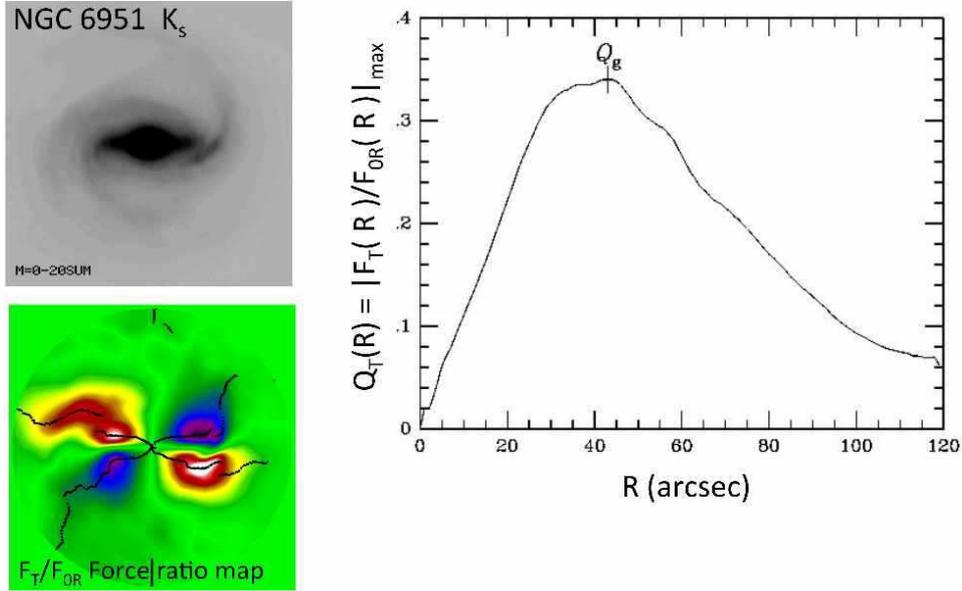}}
\caption{Derivation of $Q_{\rm g}$ for NGC\,6951, from Buta \textit{et al.} 
(2003).} 
\label{qtplot} 
\end{figure}

The actual application of this method is described by Buta \textit{et al.}
(2007) and highlighted in Fig.~\ref{qtplot}. Force ratio maps yield a
`butterfly' pattern defined by four `islands' of high tangential forcing.
Following the maximum of $F_T/F_{0{\rm R}}$ through each quadrant (dotted
curves, lower left of Fig.~\ref{qtplot}), and then averaging over the four
quadrants, gives a curve like that in the right panel of Fig.~\ref{qtplot}. For
the actual bar strength, we take the maximum of the average maxima, called
$Q_{\rm g}$. Strictly speaking, $Q_{\rm g}$ is the bar strength only if there are
no other perturbing features besides a bar, such as spiral arms. If arms are
present, then $Q_{\rm g}$ can include contributions from these as well. Buta
{\it et al.} (2003) describe a Fourier-based method of separating a bar from the
surrounding spiral. This allows the true bar strength $Q_{\rm b}$ to be
estimated, as well as an estimate of the spiral strength $Q_{\rm s}$. In
general, if the bar is strong compared to the spiral, then $Q_{\rm b}\approx
Q_{\rm g}$, while if the spiral is stronger than the bar, then $Q_{\rm s}\approx
Q_{\rm g}$.

Although the $Q_{\rm g}$/$Q_{\rm b}$ method sounds like a physically reasonable
approach to quantifying bar strength in galaxies, it has a number of problems
that limit its accuracy. First, deriving potentials requires making assumptions
about vertical thickness, and the thickness of bars may not be the same as that
of the disk. The broad inner section may be more 3D than the bar ends. Second,
since the method uses force ratios rather than light ratios, the impact of the
dark matter halo becomes an issue that really only can be evaluated
statistically. Finally, the method fails to distinguish different kinds of bars.
Two bars can have the same $Q_{\rm g}$ value and look very different.

\begin{figure}
\centerline{\includegraphics[width=0.45\textwidth]{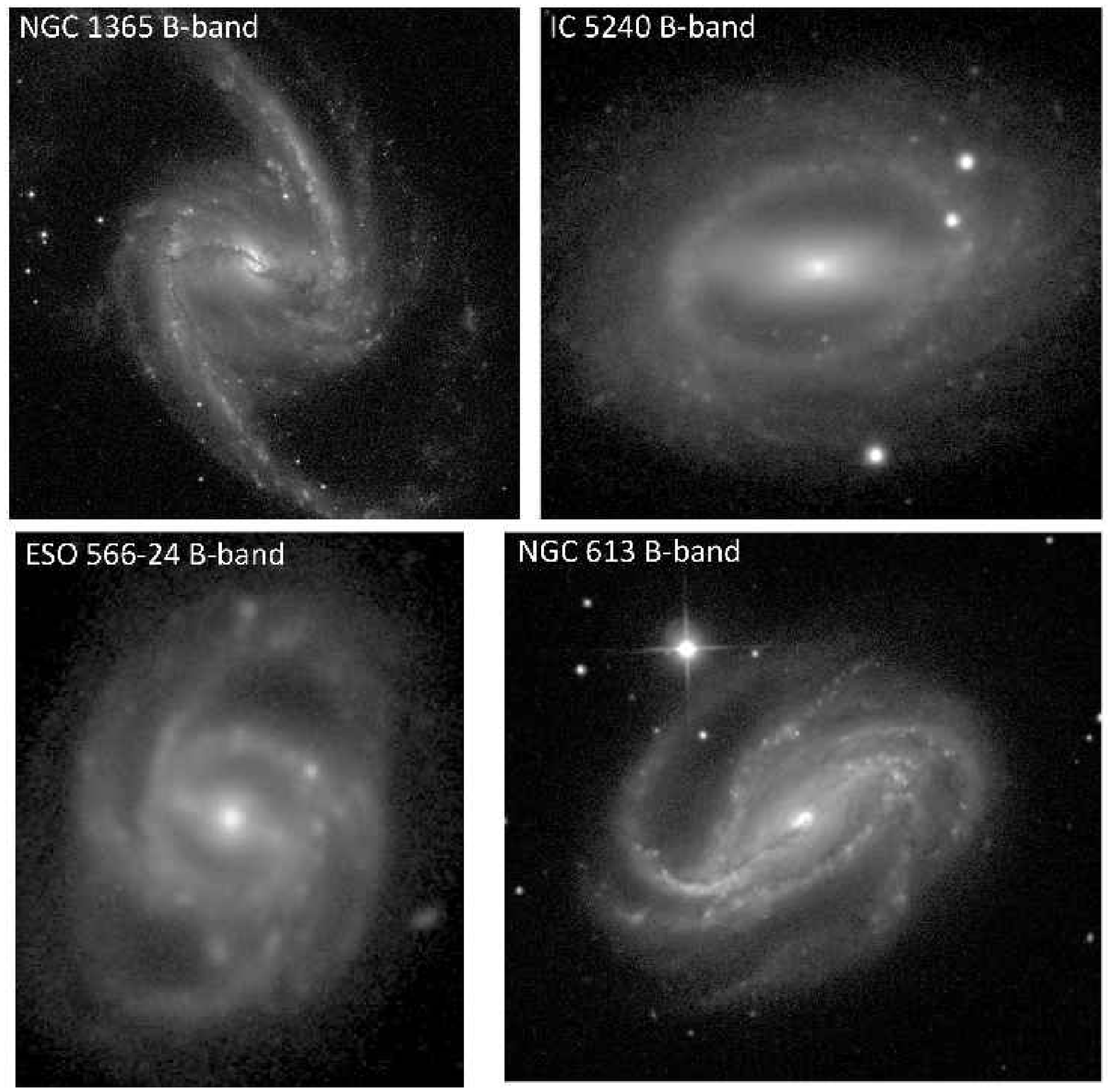}
\includegraphics[width=0.55\textwidth]{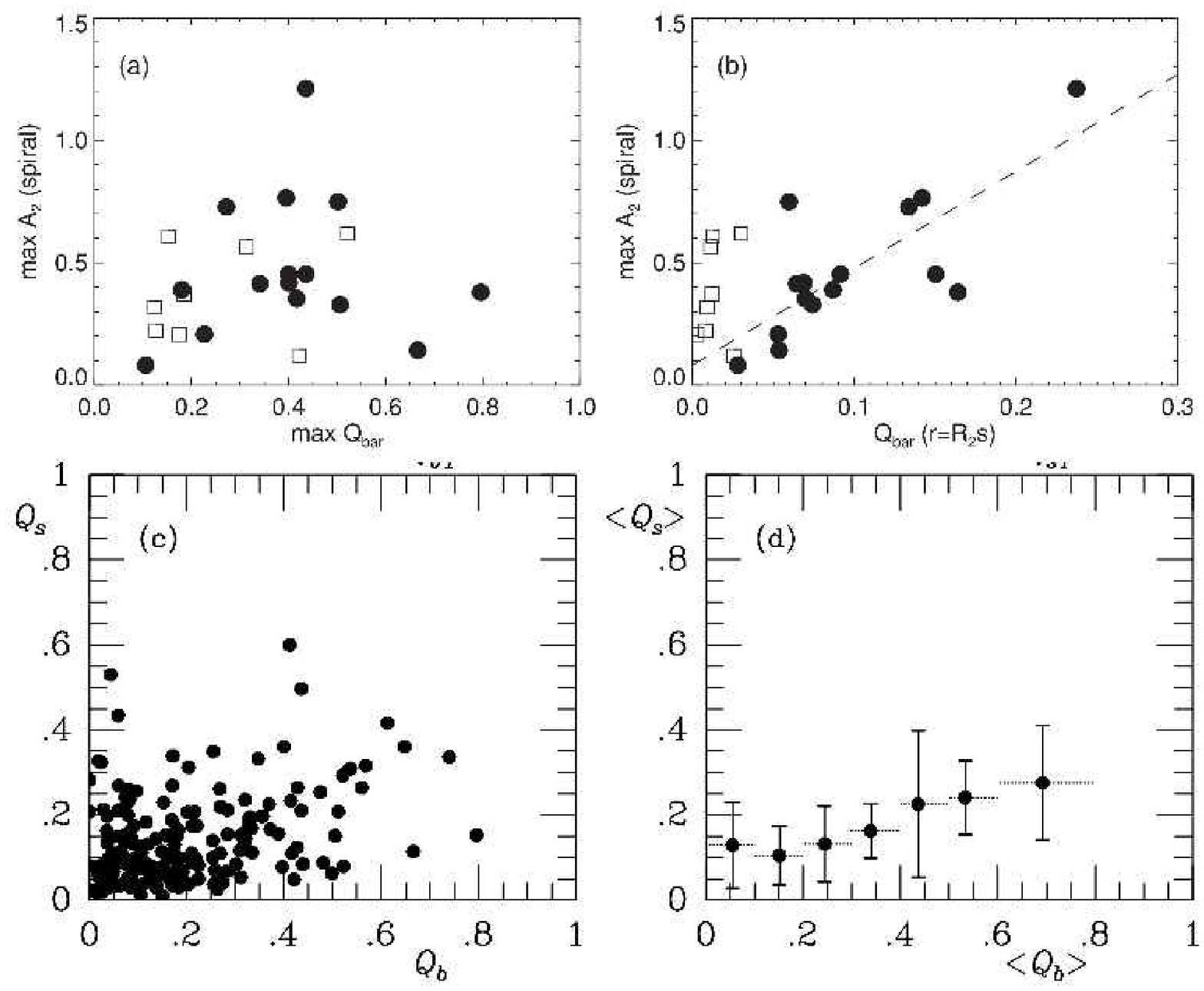}}
\caption{Examples of global SB spirals (left) and correlations of bar
and spiral strength from Buta \textit{et al.} (2009, bottom-right panels) and
Salo \textit{et al.} (2010, upper-right panels).} 
\label{spirals} 
\end{figure}

With these caveats in mind, let us examine how the strengths of spiral patterns
correlate with bar strength.  It is well-known that the spiral patterns in
barred galaxies are generally global in nature (Kormendy \& Norman 1979). Rare
examples have flocculent spirals instead. Figure~\ref{spirals} shows several
examples. An important question is, do the bars themselves `drive' these
patterns, or are the spirals mostly independent instabilities? In an article
titled `Do bars drive spiral density waves?', Buta \textit{et al.} (2009)
compared maximum relative spiral and bar torque strengths and detected a weak
positive correlation. In a follow-up paper `Bars do drive spiral density
waves', Salo \textit{et al.} (2010) compared local relative spiral and bar
torque strengths and found a much better correlation, confirming the conclusion
of Kormendy \& Norman (1979). This does not mean that all the spirals in barred
galaxies are necessarily driven by the bar. As will be described in Lecture 3
(see Section~\ref{sec:lecture3}), potential-density phase shift studies suggest 
that some spirals are decoupled from their bars.

\subsection{Inner ring shapes and bar strength}
As described previously, inner SB rings have a mean intrinsic axis ratio of
$q_{\rm o}$\,=\,0.81$\pm$0.06. However, individual cases suggest that inner rings can have a
shape over the much broader range $q_{\rm o}$\,=\,0.6--1.0 (Buta 1986). Another
question we can ask is: do inner ring shapes correlate with bar strength? 
Theoretically, such a correlation should exist because bar strength can
determine the shapes of periodic orbits, and rings are thought to take on the
shapes of specific orbits (e.g., Schwarz 1984b; Rautiainen \& Salo 2000).

\begin{figure}
\centerline{\includegraphics[width=0.5\textwidth]{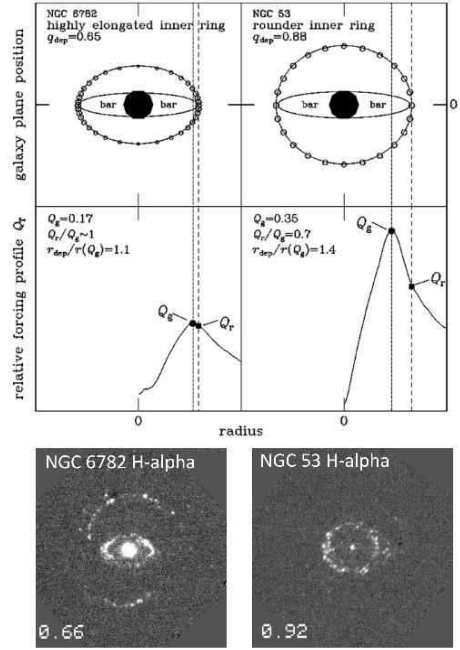}}
\caption{How inner ring shape connects to bar torque strength. The H$\alpha$
images are from Crocker \textit{et al.} (1996) while the schematics are from 
Grouchy \textit{et al.} (2010).}
\label{inner-rings-torque}
\end{figure}

It is an interesting aspect of SB inner rings that the way H{\sc ii} regions are
distributed around the rings is sensitive to intrinsic ring shape but not to
maximum relative bar torque strength $Q_{\rm g}$ (Buta 2002; Buta \textit{et al.}
2007). Nearly circular inner rings may be found around strong bars just as
highly elongated rings are found. However, if you compare the ring radius with
the radius of the bar torque maximum, then a correlation is found.
Figure~\ref{inner-rings-torque} shows that a highly elongated inner ring is
found when the semi-major axis radius of the ring $a_{\rm r}$\,$\approx$\,$R(Q_{\rm g})$, the
radius of the relative bar torque maximum. In such a case, the ring essentially
lies on the bar and can be considered a part of the bar. H{\sc ii} regions are
seen to `bunch up' around the ends of the bar, as in NGC\,6782. In contrast,
when $a_{\rm r}$\,$>$\,$R(Q_{\rm g})$, the bar maximum is far enough inside the ring
radius to allow the ring to be more circular, as in NGC\,53. In this case, the
H{\sc ii} regions are distributed more uniformly around the ring.

None of this really explains why some galaxies select one possible ring orbit
over another. According to Schwarz (1984b), an inner ring takes on the shape of
the largest 4:1 resonant orbit that is not cusped and which does not cross
another orbit. However, simulated inner rings could have more diamond-like
shapes if higher-order Fourier harmonics in the bar were important. Such a shape
seems present in the inner ring of NGC\,6782, and indeed Lin \textit{et al.}
(2008) interpreted the shape of this galaxy's inner ring in terms of an
interaction between waves excited at the inner 4:1 resonance and waves excited
at the inner Lindblad resonance, based on hydrodynamical simulations. These
authors also attributed the distribution of H{\sc ii} regions around the bar
ends as due to an interaction between the gas and curved shocks. 

It is unclear how the shapes of inner rings might evolve over time. The likely
evolution is from a pseudoring to a ring, while a change in ring shape might
link more to changes in the bar pattern speed, which affects the actual location
of resonances as I now describe.  In a model galaxy having a significant halo
and a moderately rising rotation curve, Rautiainen \& Salo (2000) found a time
sequence where a highly elongated inner ring becomes more circular.

\subsection{Resonances in barred galaxies}

The presence of even a weak bar sets up a pattern speed $\Omega_{\rm p}$ and
resonances in a differentially rotating disk. A perturbed orbit in a
weak bar potential has this shape (Binney \& Tremaine 2008):
\begin{equation}
R = R_{\rm o} + C_1{\rm cos}[\kappa\phi/(\Omega - \Omega_{\rm p})+\alpha] + 
C_2{\rm cos}(m\phi)/[\kappa^2 - m^2(\Omega - \Omega_{\rm p})^2].
\end{equation}
The resonances are in the $C_2$ terms and are as follows:

\begin{enumerate}[(a)]\listsize
\renewcommand{\theenumi}{(\alph{enumi})}

 \item Corotation resonance (CR): $\Omega = \Omega_{\rm p}$
 \item Inner Lindblad resonance (ILR): $\Omega - \Omega_{\rm p} = \kappa/2$
 \item Outer Lindblad resonance (OLR): $\Omega - \Omega_{\rm p} = -\kappa/2$
 \item Inner 4:1 resonance (I4R): $\Omega - \Omega_{\rm p} = \kappa/4$
 \item Outer 4:1 resonance (O4R): $\Omega - \Omega_{\rm p} = -\kappa/4$

\end{enumerate}

\noindent
where $\kappa = 2\Omega[1+{1\over 2}{R\over\Omega}{d\Omega\over dR}]^{1\over 2}$ 
is the epicyclic frequency and $\Omega$ is the circular angular velocity.

Are these resonances detectable in real galaxies? The answer is yes, and a
variety of methods is available to detect them in both barred and spiral
galaxies. Here we are mainly interested in how these resonances might be
manifested morphologically, and what their long-term impact on morphology\linebreak might
be.

\begin{figure}
\centerline{\includegraphics[width=\textwidth]{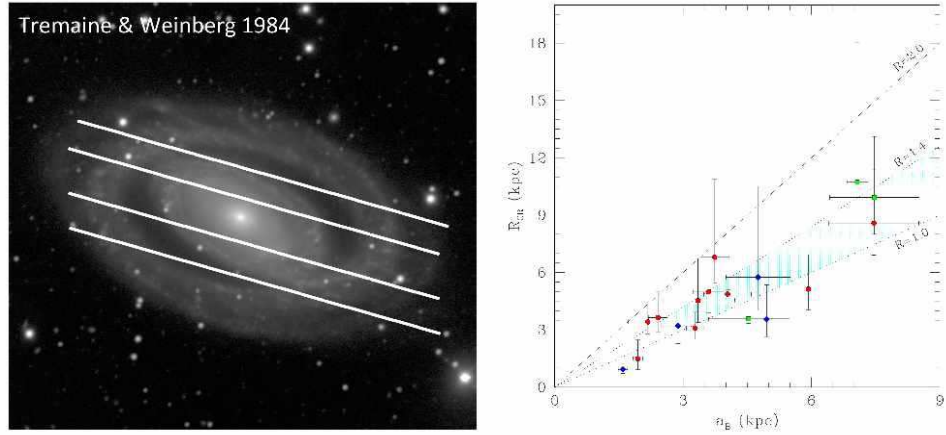}}
\caption{(Left): NGC\,7098 with a set of parallel slits along the major axis set
up to estimate the pattern speed using the Tremaine-Weinberg method. (Right):
graph from Corsini (2011) showing the preponderance of fast bars in a sample of
mostly SB0 galaxies. Reproduced with permission.}
\label{twm}
\end{figure}

The most direct method of locating resonances is to apply the Tremaine-Weinberg
(1984) approach, which uses the equation of continuity to estimate the pattern
speed $\Omega_{\rm p}$. Along a set of slits oriented parallel to the major axis
(Fig.~\ref{twm}, left), one derives the luminosity-weighted averages of the line
of sight velocity $\langle$V$\rangle$ and the position $\langle$X$\rangle$. A
graph of $\langle$V$\rangle$ versus $\langle$X$\rangle$ has the slope
$\Omega_{\rm p}\sin i$, where $i$ is the inclination. Then the CR is located by
using a rotation curve to calculate how $\Omega$ varies with galactocentric
radius. The method has been mostly applied to SB0 galaxies to derive the
parameter $\cal{R}$, where $\cal{R}$ = $R_{\rm CR}/R_{\rm bar}$. The results have suggested that
most bars in SB0 galaxies are fast with $\cal{R}\approx$1--1.4 (Corsini 2011;
Fig.~\ref{twm}, right).

\subsection{Rings and pseudorings as tracers of the bar pattern speed}

Rings were tied early on to resonances in galaxies harbouring density waves.
Many rings are zones of active star formation, and in colour index maps all ring
types, inner, outer, and nuclear, can show enhanced blue colours (e.g.,
NGC\,3081, Buta \& Purcell 1998).

\begin{figure} 
\centerline{
\includegraphics[width=0.4\textwidth]{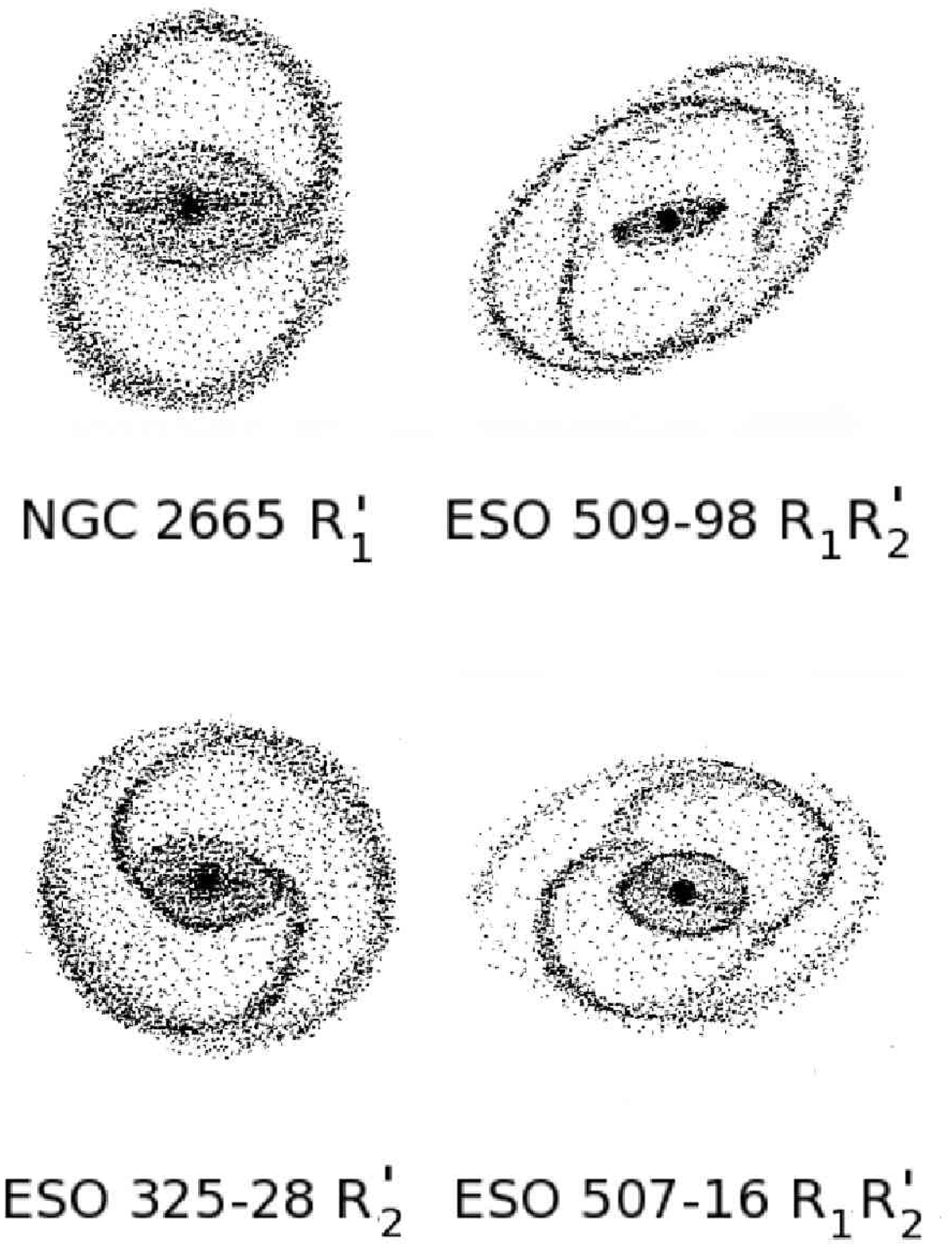}
\includegraphics[width=0.6\textwidth]{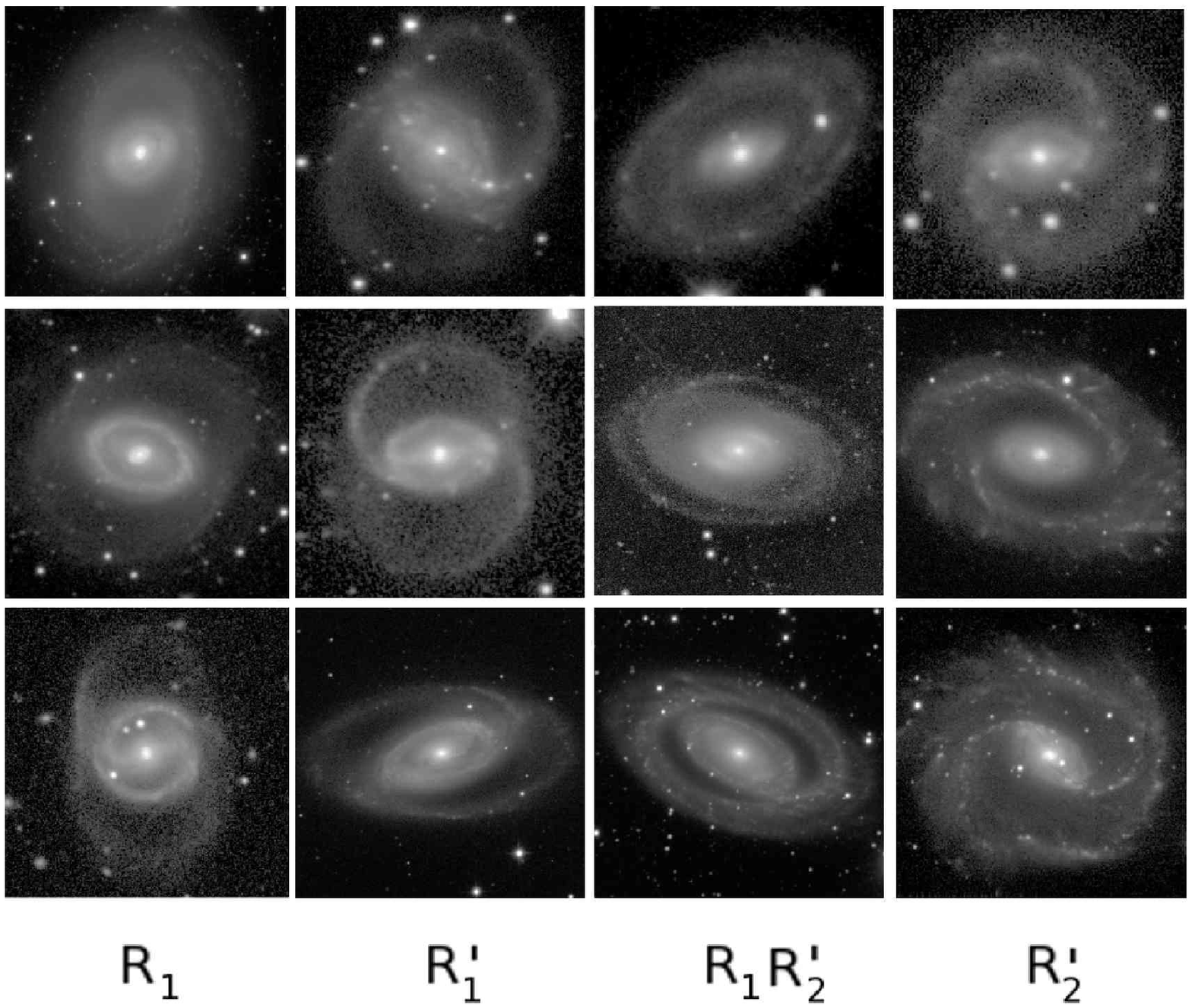}}
\caption{Schematics of OLR morphologies and examples (from B13).}
\label{olrclasses}
\end{figure}

Schwarz (1981) made test-particle simulations of barred galaxies that led to the
recognition of two morphologies that should be observed near the OLR. The first
type involves outer arms that wind 180$^{\rm o}$ and close into a pseudoring
aligned perpendicular to the bar, while the second type has arms winding
270$^{\rm o}$ and is aligned parallel to the bar. Within the limitations of
Schwarz's models, the first type was favoured if there was little gas outside
the OLR, while the second was favoured if there was more gas outside the OLR.
These types were searched for and found by Buta (1985, 1995), who called the
first type R$_1^{\prime}$ and the second type R$_2^{\prime}$, implying these
should be interpreted as outer pseudorings.  The argument that these features
trace the OLR was strengthened by the identification of galaxies showing a
combination of these two types, called R$_1$R$_2^{\prime}$ by Buta (1995). In
these types, you can almost see the two families of OLR periodic orbits that
Schwarz linked to the two main types originally. The Catalogue of Southern
Ringed Galaxies (Buta 1995) includes many examples of all three categories, and
many others can be found in the GalaxyZoo database ({\tt www.galaxyzoo.org} forums).
Figure~\ref{olrclasses} shows schematics of what are known as the `OLR
subclasses', while many examples are shown in Fig.~\ref{olrclasses}, right, and
the dVA.

\begin{figure}
\centerline{\includegraphics[width=\textwidth]{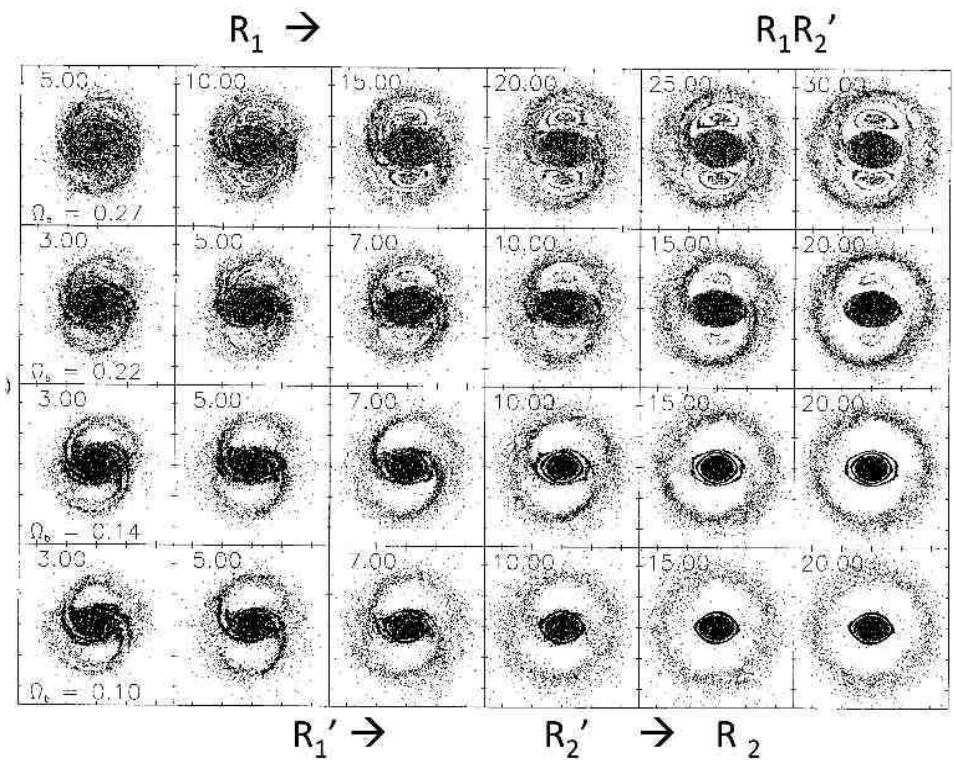}}
\caption{High pattern speed simulations of barred galaxies from Byrd 
\textit{et al.} (1994), showing the evolution of ring and pseudoring patterns at
the OLR.}
\label{byrdetal} 
\end{figure}

Schwarz (1981) did not predict the existence of the R$_1$R$_2^{\prime}$
morphology. How it might develop was shown by Byrd \textit{et al}. (1994) in
modified Schwarz-type simulations. These showed an evolution from R$_1$ to
R$_2^{\prime}$, and that an R$_1$R$_2^{\prime}$ morphology develops mainly in
cases where the pattern speed is relatively high and no inner resonances occur
(Fig.~\ref{byrdetal}). The more sophisticated $N$-body models of Rautiainen \&
Salo (2000) found an almost cyclic change between R$_1$ and R$_2$ morphologies,
due to the presence of a slower spiral mode in the outer disk. These authors
also explained the absence of rings as being due either to the non-existence of
a required resonance (like ILR) or, even when a resonance (like OLR) existed, 
to the greatly differing timescales of the different ring types.

The intrinsic shapes and orientations of rings also favour identification of
rings with resonances. In the simulations of Simkin {\it et al.} (1980), a
highly elongated inner ring aligned parallel to the bar forms at the inner 4:1
ultraharmonic resonance (I4R). This and the two alignments of outer pseudorings
are consistent with CSRG statistics.

Up to four resonant ring features may be present in the same galaxy. Example: 
the remarkable `resonance ring galaxy' NGC\,3081 (Buta \textit{et al.} 2004). 
Recognising all of its rings and a nuclear bar as well, here is a revised type 
for NGC\,3081: (R$_1$R$_2^{\prime}$)SAB(r,nr,nb)0/a.

\begin{figure}
\centerline{
\includegraphics[width=\textwidth]{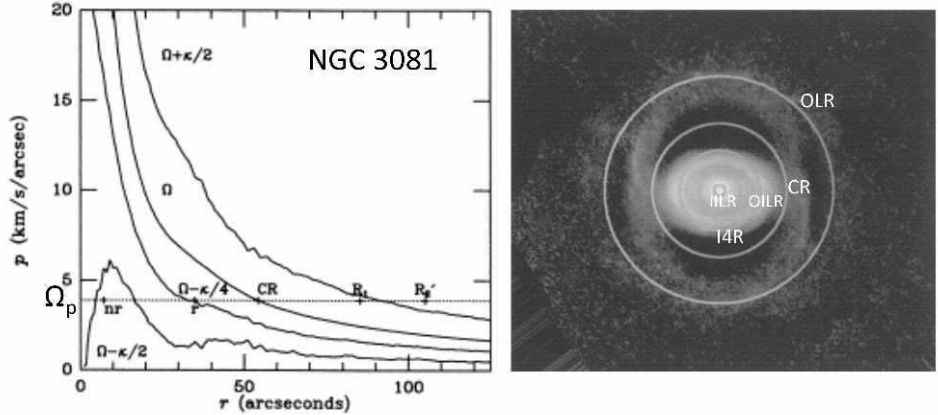}}
\caption{(Left): Precession frequency curves for NGC\,3081. (Right):
Resonance locations from rotation curve analysis (Buta \& Purcell 1998).}
\label{ngc3081-resonances}
\end{figure}

Can we use the rings in NGC\,3081 to estimate the pattern speed? Using a
Fabry-Perot velocity field and a representation of the rotation curve, Buta \&
Purcell (1998) derived $\Omega$ and the precession frequencies
$\Omega\pm\kappa/2$ and $\Omega-\kappa/4$ and set the OLR between the R$_1$ and
R$_2^{\prime}$ features. The resulting value of $\Omega_{\rm p}$ placed the
inner ring near the I4R and the nuclear ring between two ILRs, consistent with
the findings of Schwarz (1984a) and Simkin {\it et al.} (1980).
Figure~\ref{ngc3081-resonances} shows how well NGC\,3081 fits into barred spiral
theory. The images show the deprojected galaxy after subtraction of an
exponential disk model. If the R$_1$R$_2^{\prime}$ feature locates the OLR, then
CR lies in the `gap' between the inner ring and the R$_1$ outer ring.

A particularly interesting case with possible resonant features is the galaxy
NGC\,1433, which has an exceptionally well-defined and unusually highly
elongated inner ring (dVA). In addition, the galaxy has two secondary spiral
arcs off the leading sides of the bar called `plumes' by Buta (1984).
Treuthardt \textit{et al.} (2008) were able to simulate both the inner ring and
the plumes, as well as the outer R$_1^{\prime}$ pseudoring using a near-IR image
to define the potential. These simulations demonstrated that the outer
pseudoring in this case is likely related to the outer 4:1 resonance (O4R), not
the OLR as would have been expected from Schwarz's simulations. A similar result
was deduced for the outer R$_1$ ring of NGC\,6782 by Lin \textit{et al.}
(2008). 

As important as the `OLR subclasses' are for understanding galaxy structure,
there has been no systematic study of the global properties of these galaxies
and how their structure (e.g., disk scalelengths, bulge-to-total luminosity
ratios, etc.) compare with galaxies that lack these features. Interestingly,
P\'erez \textit{et al.} (2012) have used R$_2^{\prime}$ pseudorings to examine
the evolution of bar pattern speeds.

\begin{figure}
\centerline{
\includegraphics[width=0.5\textwidth]{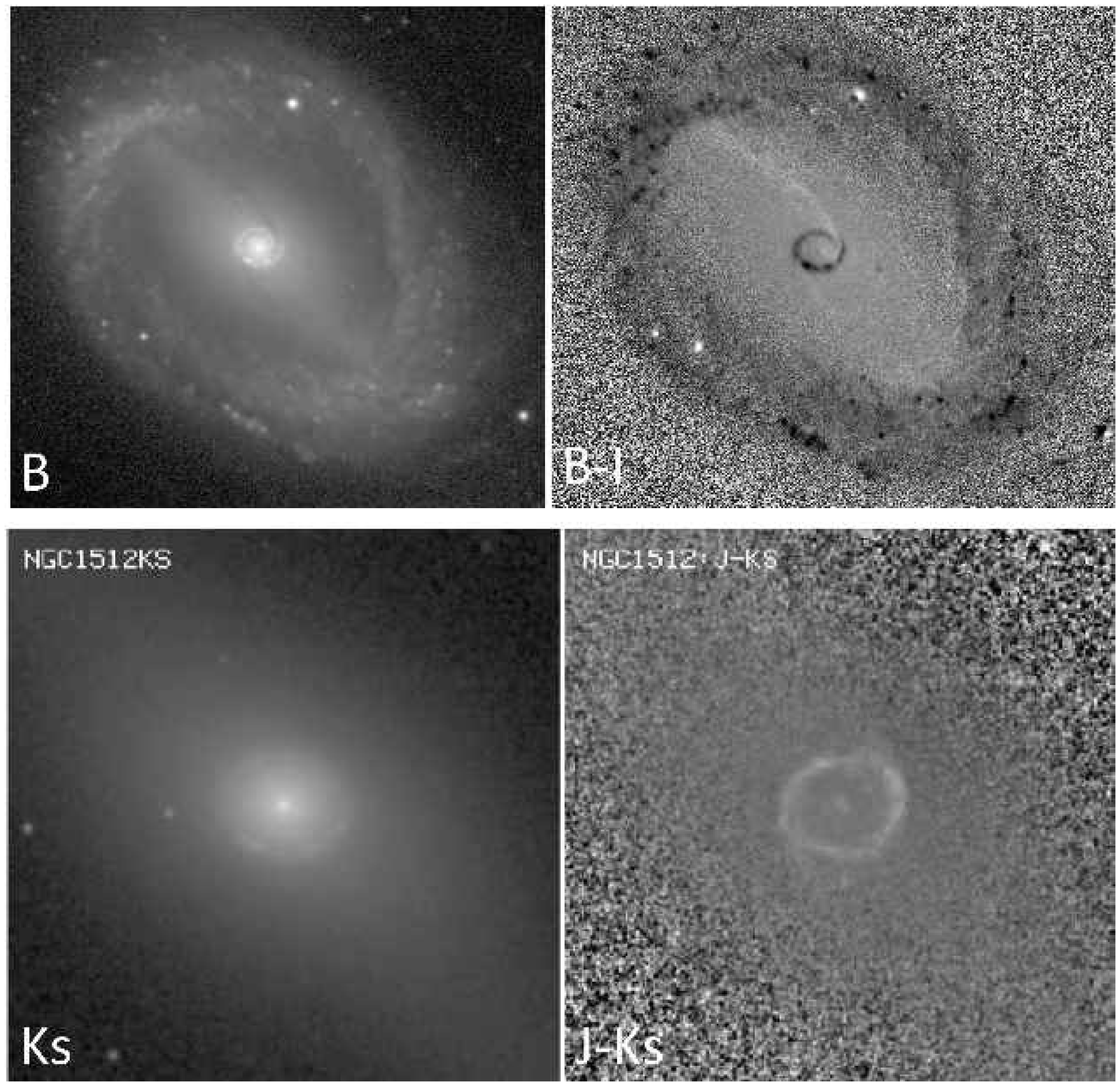}
\includegraphics[width=0.5\textwidth]{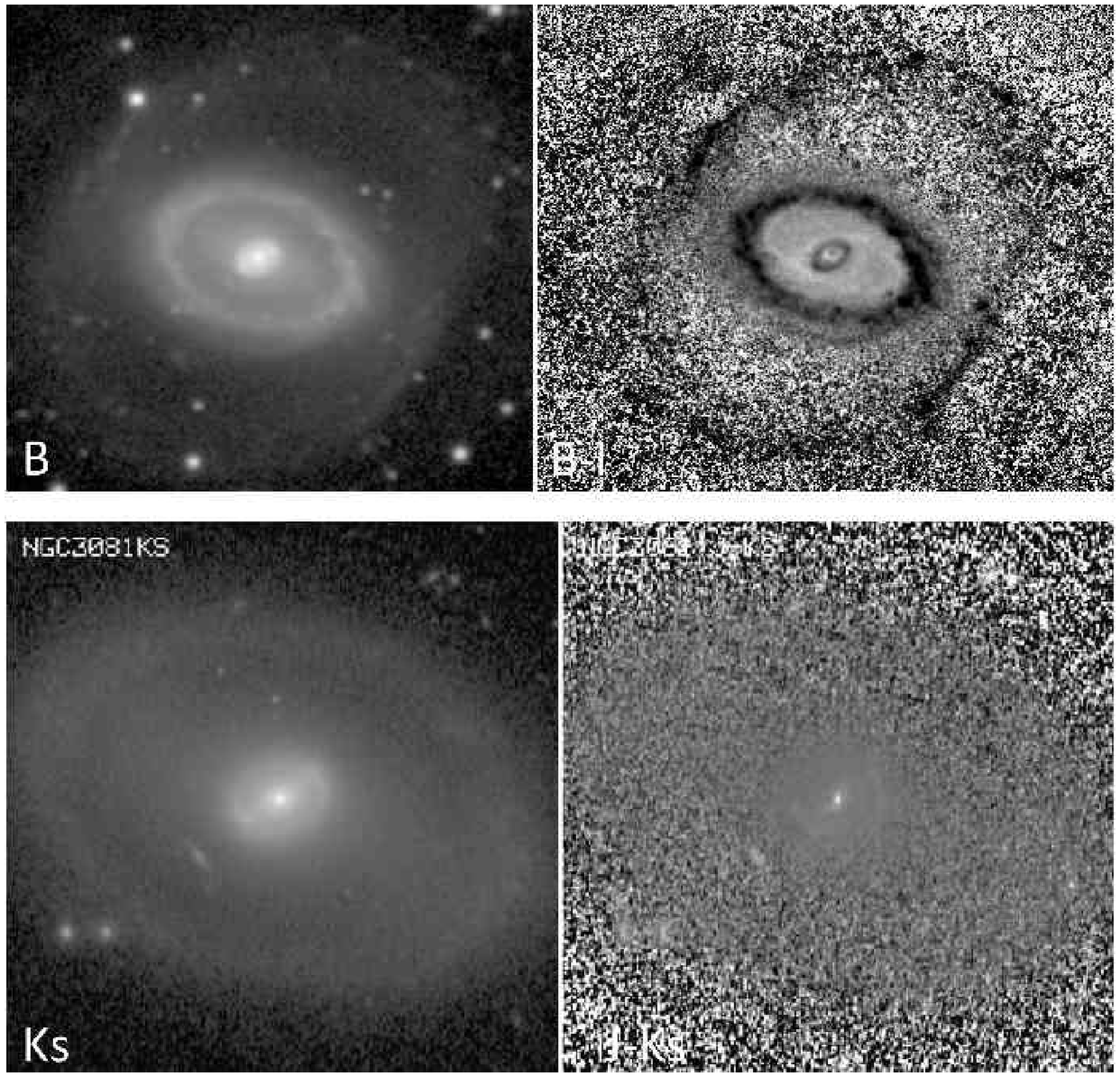}}
\caption{Optical and near-IR colours of nuclear rings.} 
\label{nrcolors}
\end{figure}

The possible evolution of nuclear rings might be evident in their observed
colours. Figure~\ref{nrcolors}, left, shows optical and near-IR images and
colour index maps of NGC\,1512, an early-type barred galaxy that harbours both
inner and nuclear rings. The nuclear ring in NGC\,1512 is blue in a $B-I$ colour
index map because of large numbers of young blue supergiants. However, the ring
is `red' in a $J-K_{\rm s}$ colour index map, implying the likely presence of
a large number of young red supergiants as well. In contrast, the nuclear ring
of NGC\,3081 is also blue in a $B-I$ colour index map (Buta \& Purcell 1998) but
is largely invisible in a $J-K_{\rm s}$ map (Fig.~\ref{nrcolors}, right),
implying far fewer red supergiants (and probably a more evolved ring than the
one in NGC\,1512).

\subsection{Are bars generally slightly skewed?}
\begin{figure}
\centerline{
\includegraphics[width=0.5\textwidth]{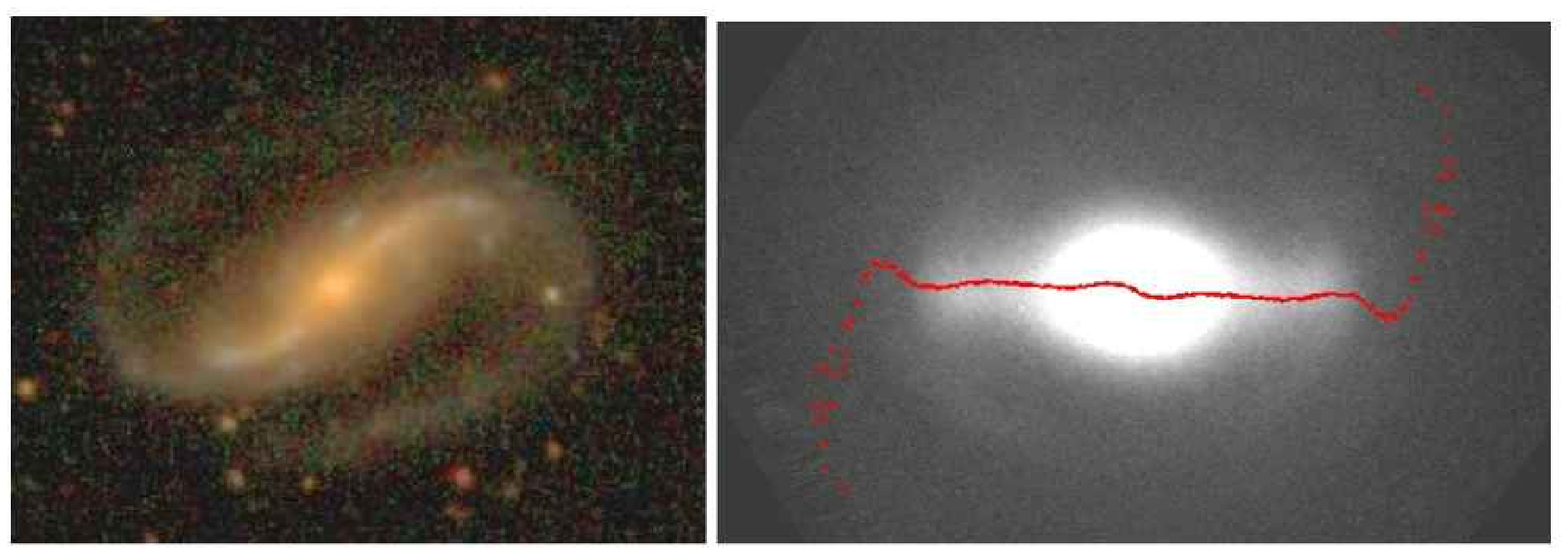}
\includegraphics[width=0.5\textwidth]{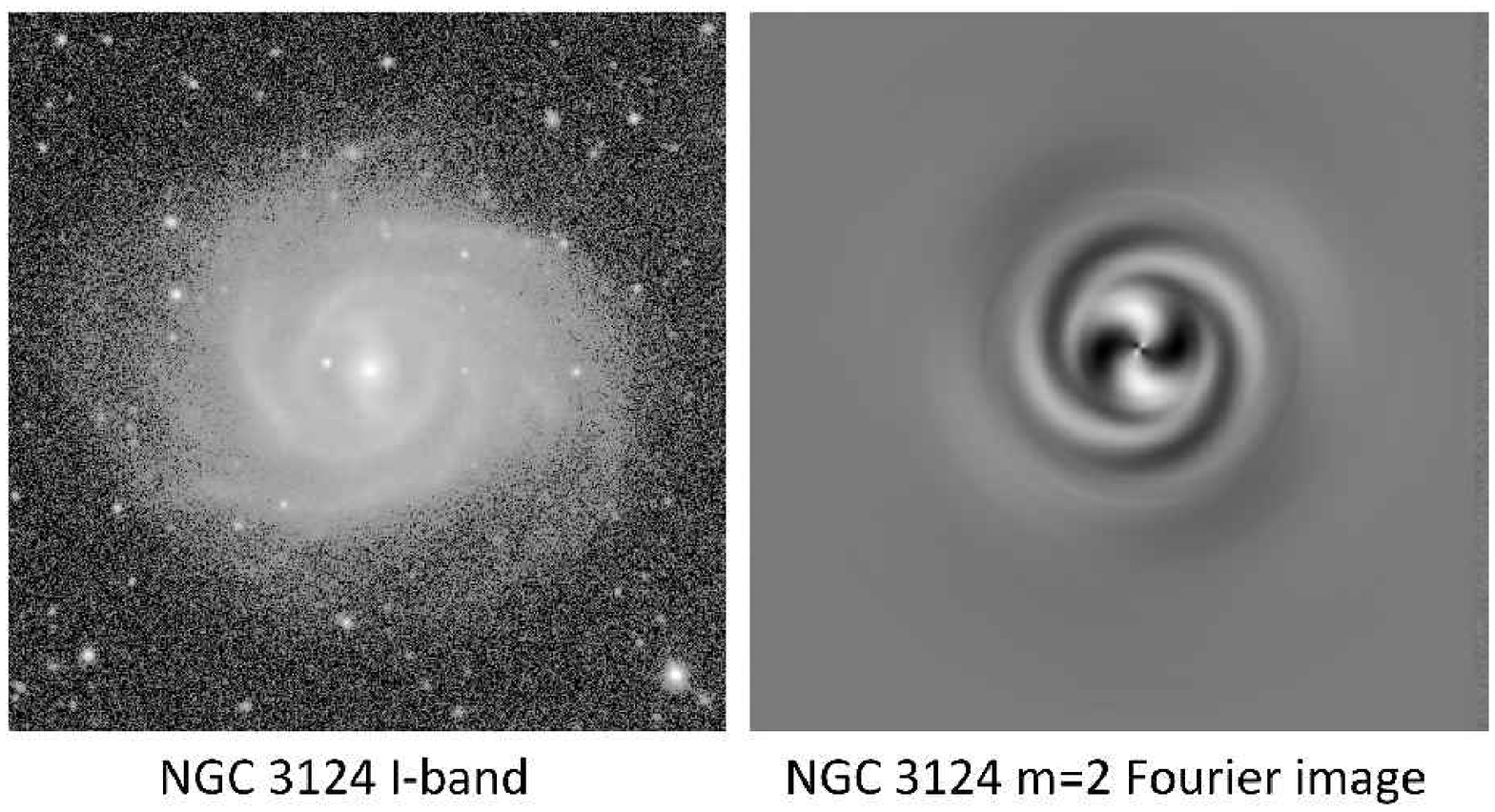}}
\caption{Examples of skewed bars. The SDSS galaxy at left was selected
from the GalaxyZoo Forum but its identity was not posted. The image of
NGC\,4596 is deprojected to face-on and the superposed curve shows the
phase of the $m$=2 component. At right is NGC\,3124, as it appears in an
$I$-band image and an $m$=2 Fourier image.}
\label{skewed-bars}
\end{figure}

The final morphological question I want to ask about bars is, how straight are
they? There is some indication that bars can be skewed in the form of very open
spirals. In the SDSS galaxy shown in Fig.~\ref{skewed-bars}, the bar is skewed
trailing, while in NGC\,4596 in the same figure, the bar is slightly skewed
leading, both after assuming trailing outer arms. The incidence and amplitude of
bar skewness has not been determined, but some interesting cases have been
identified. As shown in Fig.~\ref{skewed-bars}, the bar of NGC\,3124 is
distinctly skewed in a leading sense with little ambiguity. Bar skewness can
lead to secular evolution by the potential-density phase shift that would result
(Zhang 1996). This is described further in Lecture 3 (see
Section~\ref{sec:lecture3}).

\subsection{Secular evolution in barred galaxies}
\begin{figure}
\centerline{\includegraphics[width=0.6\textwidth]{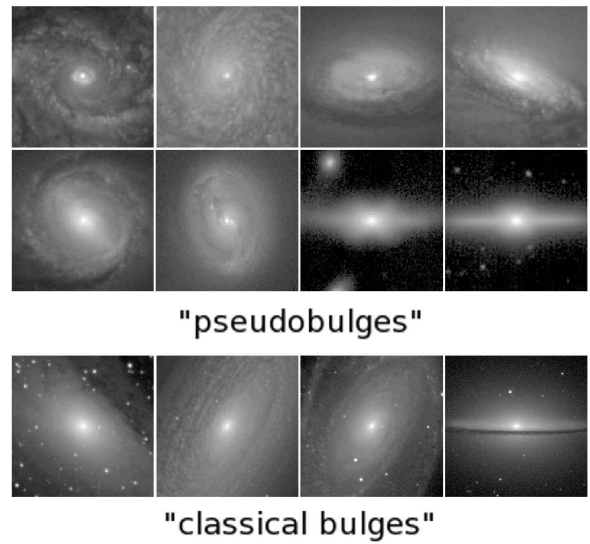}}
\caption{Examples of pseudobulges and classical bulges. From B13 based on KK04.}
\label{pseudobulges}
\end{figure}

The general view of secular evolution in barred galaxies is that the main effect
is the ability of a bar to drive interstellar gas into the centres of galaxies,
fuelling star formation and building up a pseudobulge (KK04).
Figure~\ref{pseudobulges} shows all of the types of features considered to be
pseudobulges by KK04 as compared to `classical' (merger-built) bulges.
Pseudobulges include not only `disky bulges' (as they are called by
Athanassoula 2005; see top row of examples), but also nuclear bars and
boxy-bulge galaxies. Pseudobulges generally have a low S\'ersic index (see
Lecture 3, Section~\ref{sec:lecture3}), are flatter, and have a higher ratio of
rotation to random motions compared to classical bulges.

Barred galaxies provide other possibilities for secular evolution:

\begin{enumerate}[(a)]\listsize
\renewcommand{\theenumi}{(\alph{enumi})}

\item Can a `normal' bar separate its middle section from its ends? Why would
it do this?  This is more a question of how ansae form and what they mean to
galactic dynamics. If S0 galaxies are stripped or otherwise gas-depleted former
spirals, and at the same time have a very different distribution of bar
strengths from spirals, then bar evolution is a possible avenue to explore for
the origin of ansae.

\item What causes intrinsic bar-ring misalignment? Is this a phase in bar
evolution?  This is such a rare phenomenon that it is likely to be relatively
short-lived.  Nevertheless, bar-ring misalignment (which can also mean bar-oval
misalignment) could point to an evolutionary mechanism: the destruction and
rejuvenation of bars. If destruction of a bar can lead to formation of a
lens-oval, and then external gas accretion cools the disk enough for bar
rejuvenation, could the new bar have a different pattern speed from the old one?
The misalignment\linebreak between primary and secondary bars has also been attributed to
different pattern speeds (e.g., Buta \& Crocker 1993a).

\item Why would a bar be skewed in a leading sense? I have shown that there are
suggestive examples where a bar is curved like a very open spiral. When the mass
associated with such spirals is converted into a gravitational potential, a
small phase difference can result that has relevance to secular evolution
(Lecture 3, Section~\ref{sec:lecture3}). But what is determining the sense of
spirality of the bar? Bar skewness supports the idea that a bar is a density
wave like an ordinary spiral.

\item What do double outer rings mean? Why should one ring be red and the other
blue?  Double outer rings seem much less common than double outer
rings/pseudorings.  The latter is usually found as the R$_1$R$_2^{\prime}$
morphology. But cases classified as `(RR)' as in the dVA (example: NGC\,2273)
are very rare. Could a double outer ring signify two episodes of ring formation
at different values of the pattern speed?

\item Is there an evolutionary connection between R$_1$ and R$_2$ outer rings? 
Did detached outer rings begin as pseudorings?  The simulations of Byrd
\textit{et al.} (1994) showed a clear trend of evolution from an R$_1^{\prime}$
to an R$_2^{\prime}$ outer pseudoring, followed by a detached outer ring. At
high pattern speeds, an R$_1$R$_2^{\prime}$ morphology developed. Rautiainen \&
Salo (2000) suggested a different evolutionary connection, in that the presence
of a slower spiral mode in the outer disk would cause cyclic changes from R$_1$
to R$_2$.

\end{enumerate}

%
%

\section{Lecture 3: The infrared experience}
\label{sec:lecture3}

I am titling this lecture as `the infrared experience' because moving to the
infrared (IR) opens such a big door to galactic studies. This is the wavelength
domain that allows us to see what is often referred to as the `backbone' of
the stellar mass distribution. It is also the wavelength domain that exposes the
interstellar medium (ISM) in a manner that no other domain can.

\begin{figure}
\centerline{\includegraphics[width=\textwidth]{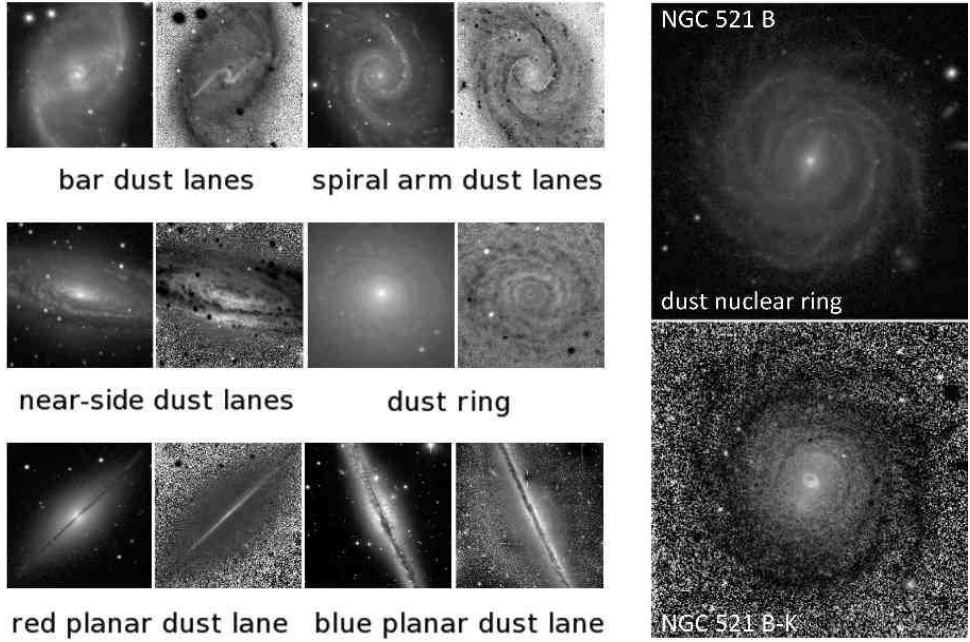}}
\caption{Dust morphologies in normal galaxies based on colour index maps. The
left panels are from B13 and show from top-left to lower-right: NGC\,1530,
NGC\,1566, NGC\,7331, NGC\,7217, NGC\,7814, and NGC\,891. The images of NGC\,521
are from Buta \textit{et al.} (2009).} 
\label{dust1} 
\end{figure}

\begin{figure}
\centerline{\includegraphics[width=\textwidth]{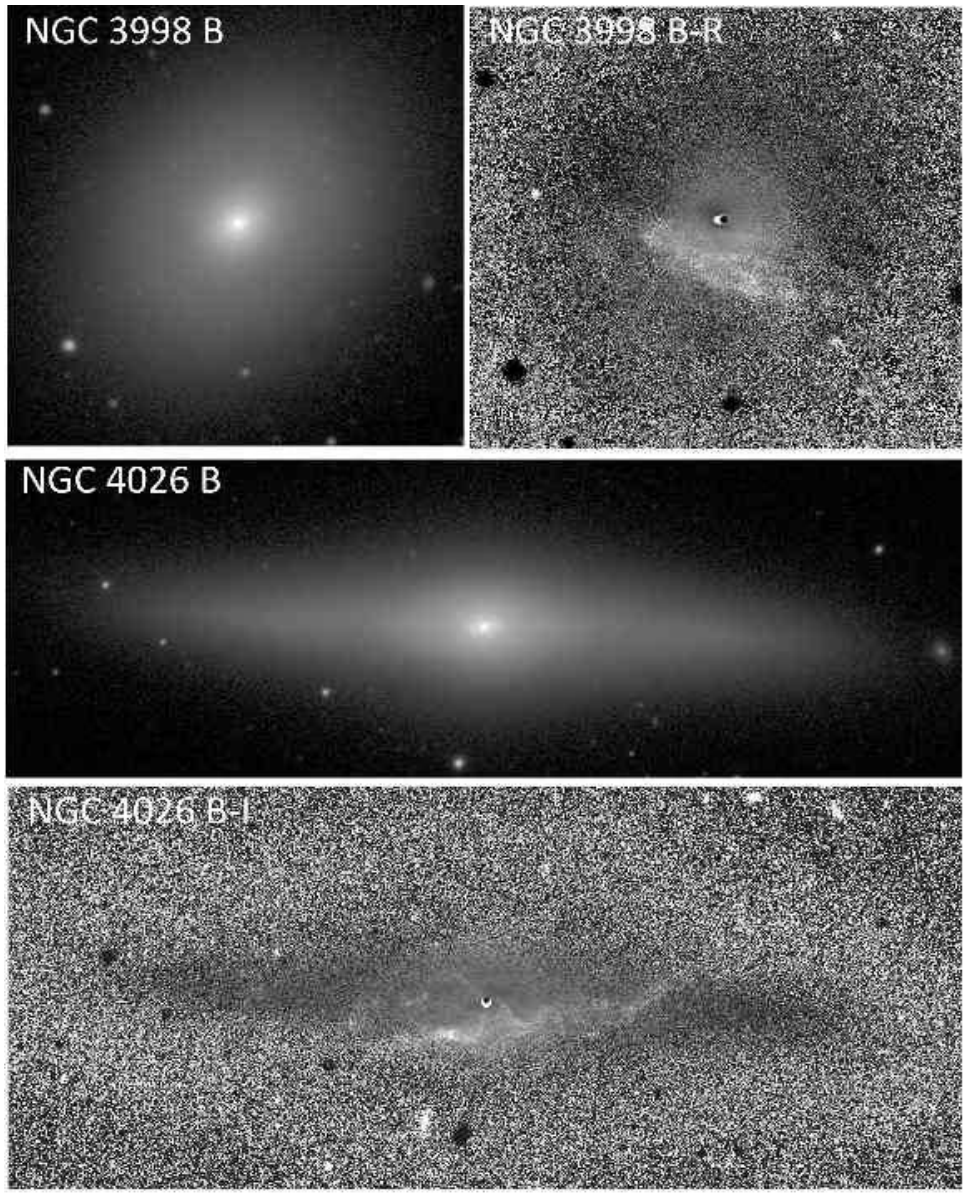}}
\caption{Examples of dusty S0s.} 
\label{s0dust} 
\end{figure}

\begin{figure}
\centerline{\includegraphics[width=\textwidth]{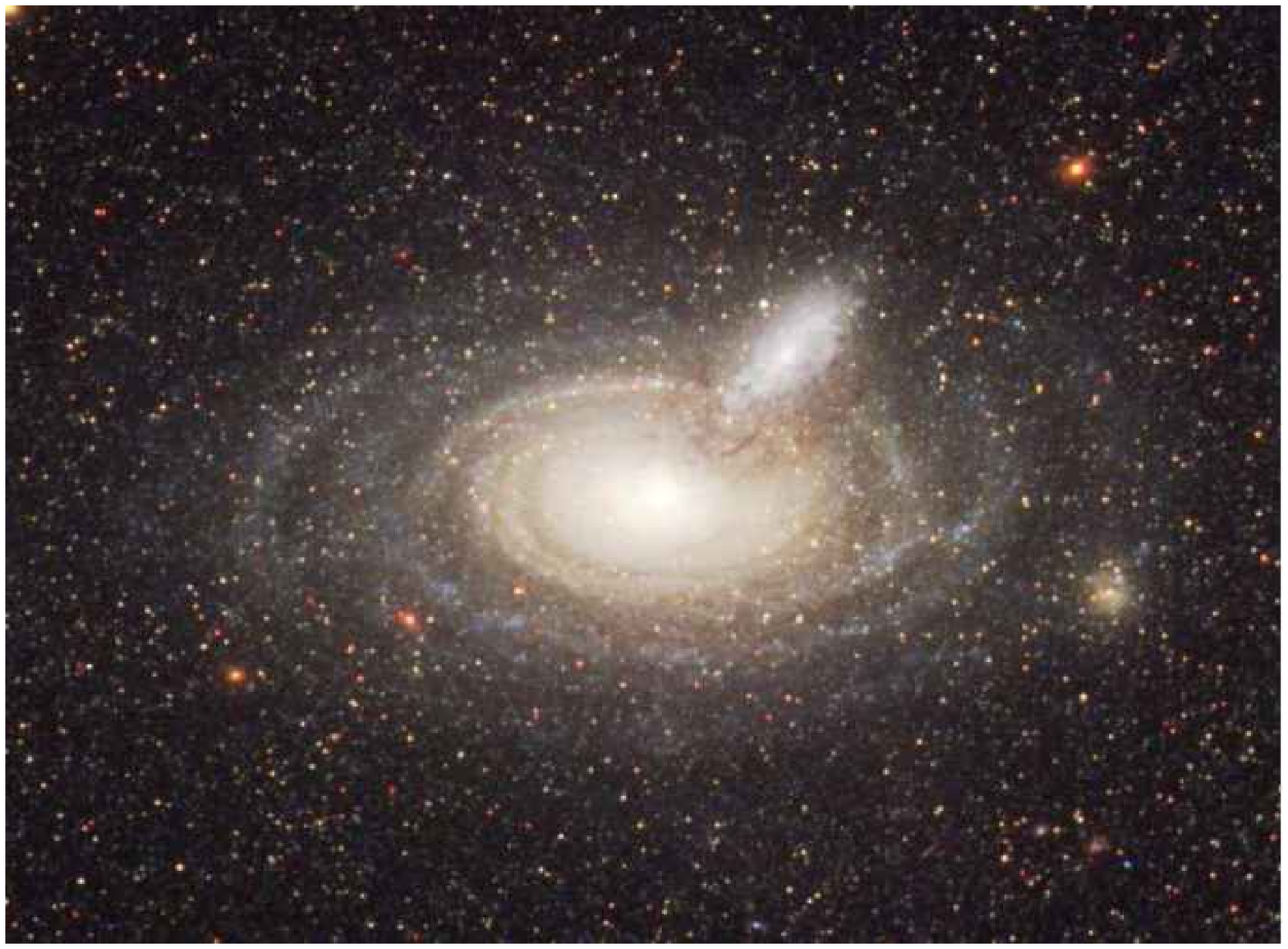}}
\caption{Example of an overlapping pair that reveals the significant
extent of the dust disk of the foreground galaxy (from Holwerda \textit{et al.}
2009, reproduced with permission).} 
\label{holwerda} 
\end{figure}

Observing at such wavelengths shows {\it dust-penetrated morphology}. At 2.2
microns, the extinction $A_{K_{\rm s}}\approx0.1A_V$, where $A_V$ is the visual
extinction, at 3.6 microns, the extinction is  $A_{3.6}\approx0.05A_V$. When we
observe in the near- and mid-IR, we minimise the effects of these features on
spiral galaxy morphology: bar dust lanes, spiral arm dust lanes, near-side dust
lanes, dust rings, red planar dust lanes, and blue planar dust lanes
(Fig.~\ref{dust1}, which includes colour index maps that show these features).
It is even advantageous to observe S0 galaxies in the IR, since extraplanar dust
is often seen in such galaxies (Fig.~\ref{s0dust}). Colour index maps are not
the only way to see the dust. In overlapping galaxy pairs, the dust is sometimes
detected well beyond the standard isophotal radius, $r_{25}$
(Fig.~\ref{holwerda}, Holwerda \textit{et al.} 2009).\looseness-2

\begin{figure}
\centerline{\includegraphics[width=\textwidth]{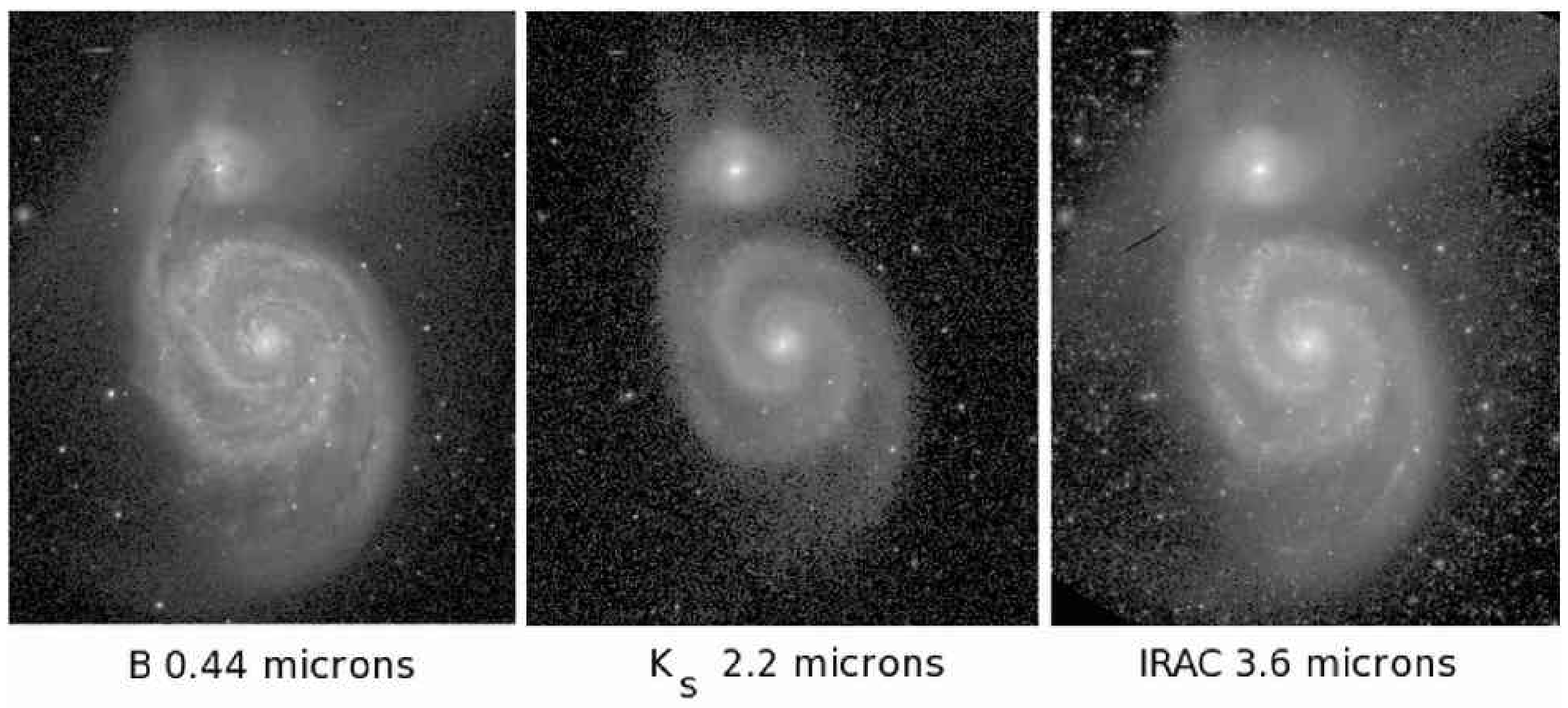}}
\caption{M\,51 from the $B$-band to the mid-IR.}
\label{m51-filters}
\end{figure}

The near- and mid-IR may be mostly free of the effects of extinction but they
are NOT completely free of the effects of star formation. Dust warmed by hot
young stars becomes increasingly visible over 2--5\,$\mu$m as shown in
Fig.~\ref{m51-filters}. In fact, the IR is an excellent domain for detailed
studies of star-forming regions and especially star-forming galaxies. Spectral
emission lines due to polycyclic aromatic hydrocarbons appear in three of the
four mid-IR bands of the Infrared Array Camera (IRAC) on the {\it Spitzer Space
Telescope} (e.g., Churchwell \textit{et al.} 2004).

The IR can also provide a great deal of information on the properties of bulges
and disks in galaxies. Because of the reduced effects of extinction, standard
parameters of bulges and disks can be considered more characteristic of the
stellar mass distribution than they are in the optical. Before describing IR
galaxy morphology, let me first briefly describe the methods people are
currently using to derive these parameters.

Most photometric decomposition today is 2D, meaning the whole image is fitted
rather just the major or the minor axis as was once the standard method. The
most popular parametric function to use is the Sersic Law.

Let $R_{\rm e}$ be the radius of the isophote enclosing half the total 
luminosity (called the effective radius) and let $I_{\rm e}$ be the surface
brightness of this isophote. Define relative parameters $\rho = R/R_{\rm e}$ 
and $J_n=I_n(R)/I_{\rm e}$. The Sersic Law connects these parameters
through
\begin{equation}
{\rm log}(J_n) = -0.434(2n-0.324)(\rho^{1\over n}-1),
\end{equation}
\noindent
where $n$ is a characteristic index that determines the shape of the radial
profile. The Sersic law is a `jack of all trades':

\begin{enumerate}[(a)]\listsize
\renewcommand{\theenumi}{(\alph{enumi})}

\item $n$=1: exponential law,
\item $n$=4: de Vaucouleurs $r^{1\over 4}$ law.

\end{enumerate}

Two-dimensional decomposition today has gone well beyond merely bulges and
exponential disks. Barred galaxies are no longer ignored because bars can be
fitted and their contribution allowed for. For example, one can use a Ferrers
bar model (e.g., Binney \& Tremaine 2008):
\begin{eqnarray}
I(m^2) &=& I_0 (1-m^2/a^2)^n, \quad (m\leq a)\nonumber\\
I(m^2) &=& 0, \quad\quad\quad\quad\quad\quad\quad(m>a),
\end{eqnarray}
where $m$ is the radius along the bar axis. The Sersic law can also be used for
bars. Values of $n$\,$<$\,1 can fit some bars well. Sersic fits have also been used
for secondary bars and lenses. If these features are not fitted, then the
bulge-to-total luminosity ratio $B/T$ will tend to be overestimated (Laurikainen
\textit{et al.} 2006). With initial guesses of all parameters, the best fit uses
$\chi^2$ minimisation, usually with an adopted weighting scheme. Initial
guesses can be based on 1D profile analysis.

\subsection{Significant IR morphological surveys}
I would like to describe now three recent extensive surveys that shed a great
deal of light on IR galaxy morphology.

\subsubsection{The Ohio State University Bright Spiral Galaxy Survey}
The Ohio State University Bright Spiral Galaxy Survey (OSUBSGS, Eskridge
\textit{et al.} 2002) was a comprehensive imaging survey of 205 bright nearby
spiral galaxies. The main IR imaging filter used was the $H$-band at
1.65\,$\mu$m.  The galaxies in the survey were classified in the de Vaucouleurs
(1959) system and special attention was paid to how near-IR types differ from
the optical classifications of galaxies. The authors found several galaxies
classified as SA in the Third Reference Catalogue of Bright Galaxies (RC3) for
which they assigned a classification of SB. They also estimated the frequency of
bars in their sample (Lecture 2, Section~\ref{sec:lecture2}). In general, it was found
that intermediate-type spirals were classified approximately one stage interval
earlier in the $H$-band than in the $B$-band. The authors also challenged the
idea, implied by earlier IR studies based on lower-quality imaging, that the
Hubble tuning fork breaks down in the near-IR.

\subsubsection{The Near-Infrared S0 Survey}

The Near-Infrared S0 Survey (NIRS0S, Laurikainen \textit{et al.} 2011) provides
a near-IR atlas (in the $K_{\rm s}$ or 2.15\,$\mu$m bands) of 174 mostly S0 and
some S0/a and Sa galaxies. The analysis described by Laurikainen \textit{et al.}
(2011) and in earlier papers in the series includes sophisticated 2D
multi-component decompositions and classifications in a modified dVA system
(with special recognition of lenses). A summary of the main findings of the
NIRS0S is provided by Laurikainen \textit{et al.} (2012).

\begin{figure}
\centerline{\includegraphics[width=\textwidth]{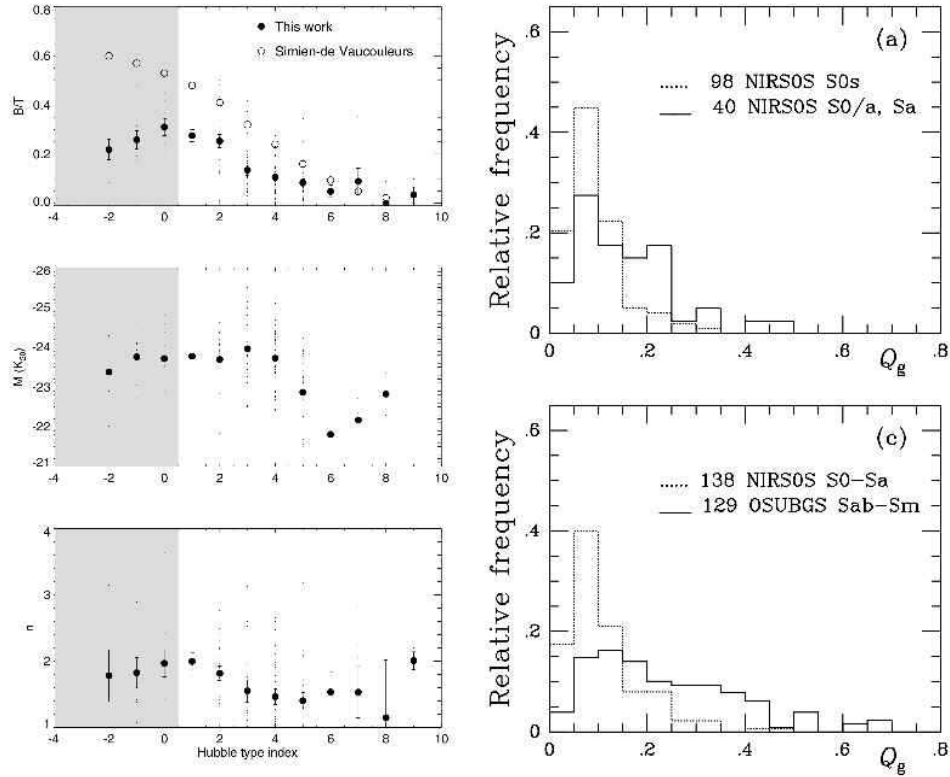}}
\caption{NIRS0S results from Laurikainen \textit{et al.} (2007, left frames) and 
Buta \textit{et al.} (2010a, right frames).}
\label{nirs0s1}
\end{figure}

NIRS0S 2D decompositions have modified our views of bulges along the Hubble
sequence. Laurikainen \textit{et al.} (2007) found that most bulges are
pseudobulges and that $B/T$ ratios of S0--Sa galaxies are much less than was
determined in past studies (Fig.~\ref{nirs0s1}, left). Across the Hubble
sequence, the average value of the Sersic index $n$ is generally less than 2,
consistent with one of the pseudobulge criteria adopted by KK04.

Two-dimensional decompositions allow a more reliable deprojection of galaxy 
images, because the bulge is modelled and can be removed from the image. Then
the remaining disk light can be deprojected and the bulge added back. With such
images, the bar strength parameter $Q_{\rm g}$ can be derived and the
distribution of bar strengths for early-type galaxies can be compared with that
for late-type galaxies. This was done by Buta \textit{et al.} (2010a) who used
NIRS0S deprojected images for S0 to Sa galaxies and OSUBSGS images for spirals
(Buta \textit{et al.} 2004). The histograms in Fig.~\ref{nirs0s1}, right, show
that S0 bars tend to be weak. This had been evident previously based on other
bar strength indicators (e.g., Aguerri \textit{et al.} 2009; Laurikainen
\textit{et al.} 2009). Part of the difference will be due to the presence of
significant bulges in early-type galaxies, which add forcing to the axisymmetric
background and thereby dilute the bar torques. However, all of the difference
cannot be explained this way, and it is possible that there is a process that
allows bar evolution to continue even after gas depletion in S0 galaxies.
Laurikainen \textit{et al.} (2006) showed that radial disk scalelengths of S0s
correlate with bulge effective radii, which favours secular evolution of S0
bulges from disk material.

How could the bar of an S0 evolve? Block \textit{et al.} (2002) argued that an
S0 bar should be a fossil relic, capable of no further change. But if S0s are
former spiral galaxies, why is it that S0 bars are weaker on average than those
in spirals and why are ansae bars so prevalent among S0 galaxies? The answer
could be secular evolution of the {\it stellar mass distribution}.

\begin{figure}
\centerline{\includegraphics[width=\textwidth]{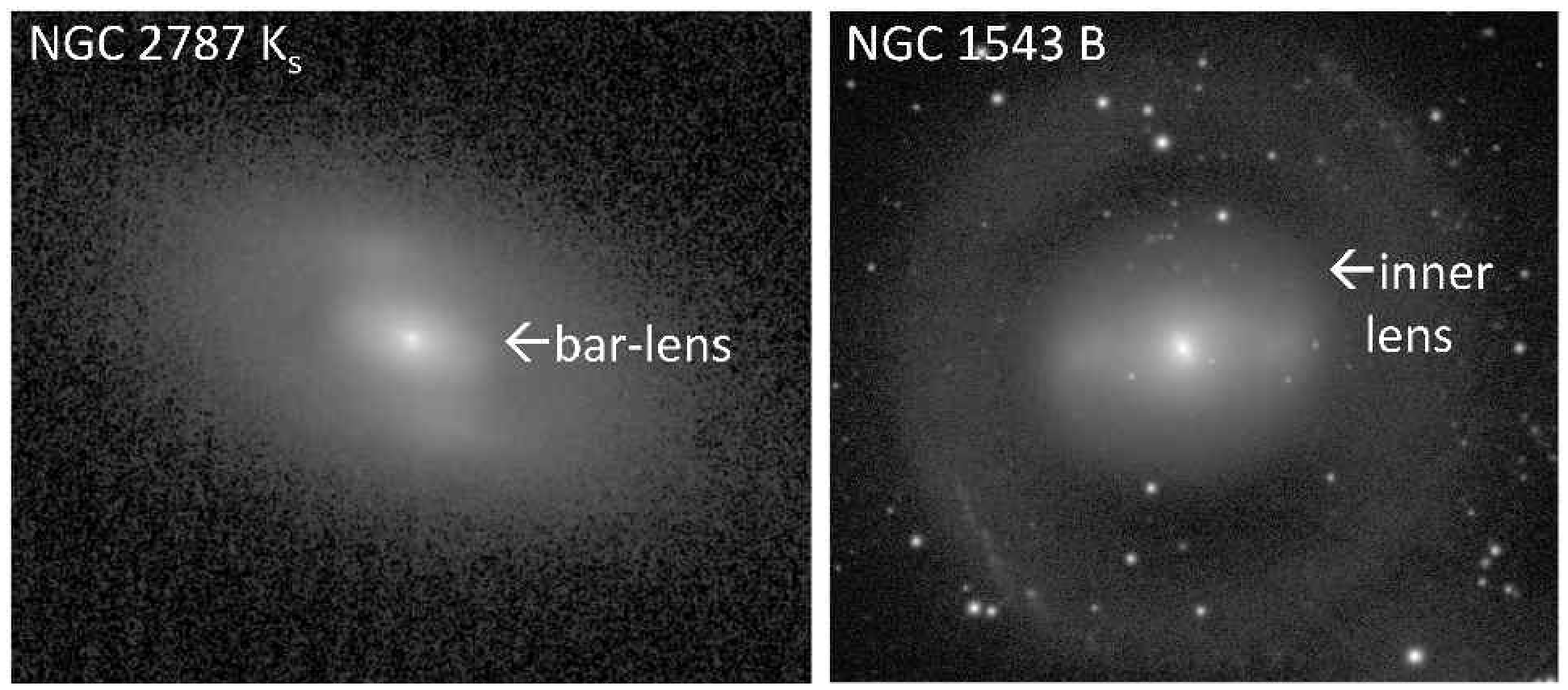}}
\centerline{\includegraphics[width=\textwidth]{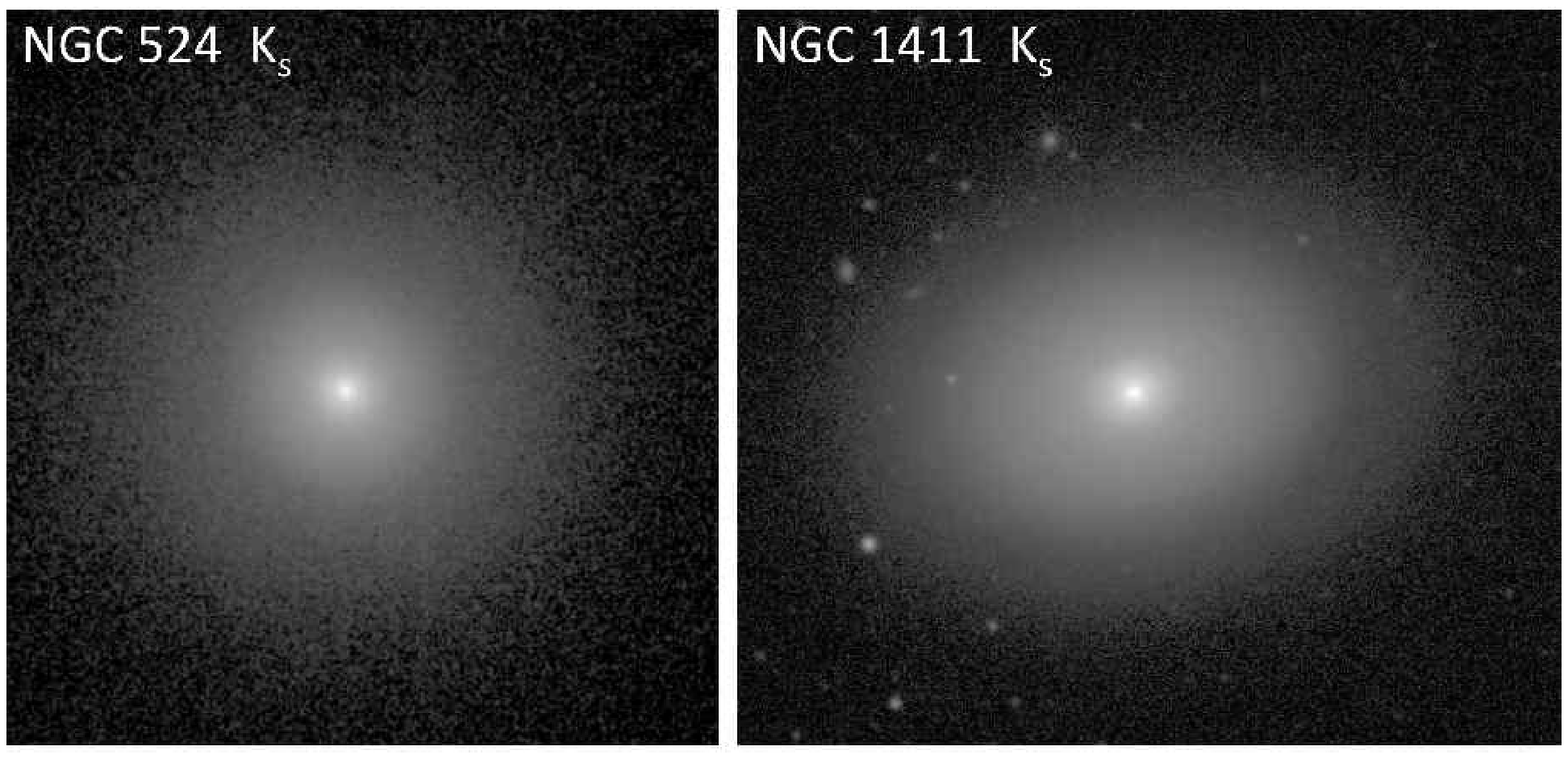}}
\caption{NIRS0S results from Laurikainen \textit{et al.} (2011). (Top): the
`barlens' as compared to a typical SB galaxy `lens'. (Bottom): two
examples of SA galaxies having multiple lenses.} 
\label{nirs0s2}
\end{figure}

Another important finding from the NIRS0S is the recognition of a new type of
lens: the `barlens' or inner component of the bar (Laurikainen \textit{et al.}
2011, 2012).  In NGC\,2787 (Fig.~\ref{nirs0s2}, top), the barlens is separated
from the bar ends, leaving the familiar `ansae' previously discussed (Lecture
2, Section~\ref{sec:lecture2}). At first sight, the barlens looks like nothing more
than a part of the bulge, but 2D decompositions do not support this
interpretation. To show that a barlens is a distinct feature, the right panel of
Fig.~\ref{nirs0s2} shows a normal Kormendy-type lens in the face-on ringed
galaxy NGC\,1543. A normal lens envelops and is filled by the bar,
while a barlens is {\it part of the bar} although it may not contribute much to
the actual torquing action of the bar. Bar dissolution, if it occurs, would
mainly mean the loss of the bar ends.

The nature of lenses also gains greater immediacy when we find them in multiple
form in nonbarred early-type galaxies. Two important and very similar examples
are shown in Fig.~\ref{nirs0s2}, bottom. In both NGC\,524 and NGC\,1411, three
distinct lens features are detectable, and each galaxy is classified as
(L)SA(l,nl)0$^{\rm o}$, where (L) refers to an {\it outer lens}, (l) to an {\it
inner lens}, and (nl) to a {\it nuclear lens}. The (L) and (l) notations are due
to Kormendy (1979), who also suggested that inner lenses (l) represent the
dissolution of bars, so that in principle a lens in a nonbarred galaxy could
imply that the galaxy was once barred. However, bar evolution cannot necessarily
explain all types of lenses.

\begin{figure}
\centerline{\includegraphics[width=\textwidth]{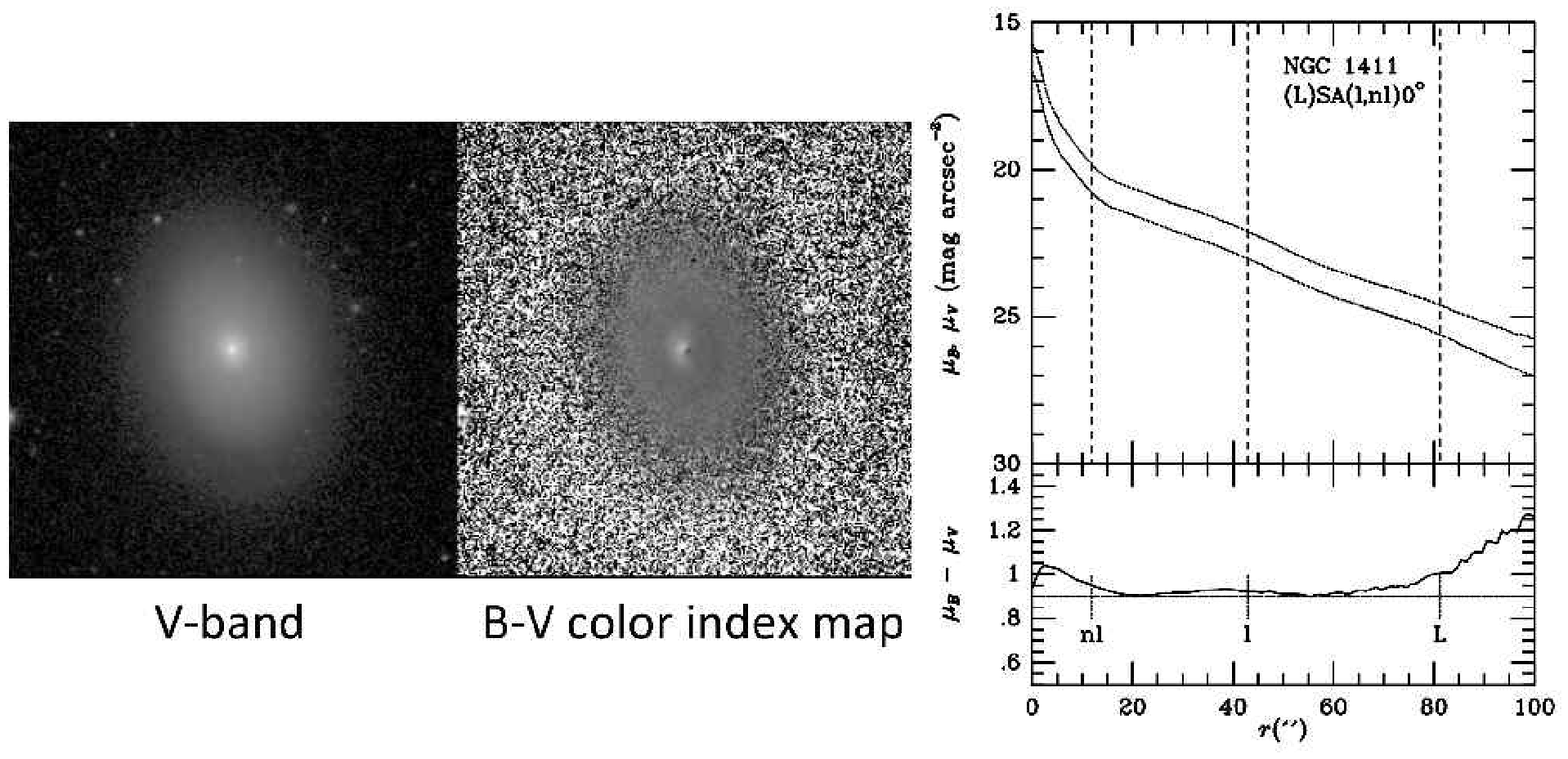}}
\caption{This shows NGC\,1411 and a colour index map that highlights the
enhanced blue colours that lie between the apparent `edges' of the
lenses.} 
\label{ngc1411} 
\end{figure}

One way to possibly shed light on the origin of lenses is to measure their
colours. Figure~\ref{ngc1411} shows a $V$-band image, a $B-V$ colour index map,
and azimuthally averaged $B$, $V$, and $B-V$ surface brightness profiles for
NGC\,1411. On the graph, the radial locations of the `edges' of the three
lenses of the galaxy are indicated. The $B-V$ colour index profile shows two
zones of slightly bluer colours, one between the (nl) and (l), and one between
the (l) and (L). The apparent sharp edges appear to delineate zones with
slight stellar population differences. The $B-V$ map also shows some reddening
on the east side of the centre, which could signify a minor merger.

\subsubsection{The {\it Spitzer} Survey of Stellar Structure in Galaxies}
The {\it Spitzer} Survey of Stellar Structure in Galaxies (S$^4$G, Sheth
\textit{et al.} 2010) is a comprehensive survey of 2331 nearby galaxies in the
3.6 and 4.5\,$\mu$m IRAC bands, the only filters that were usable for the {\it
Spitzer} `warm mission'. The images sample the Rayleigh-Jeans limit for all
stars with $T>2000$\,K, and the [3.6]$-$[4.5] colour is roughly constant with
radius, independent of stellar population (but see also the Peletier lectures,
this volume). The goal of S$^4$G is to examine stellar structure in galaxies
free of the effects of extinction and to a depth greater than groundbased
near-IR imaging could achieve.

The selection criteria for S$^4$G are as follows: galactic latitude $b\ge30^{\rm
o}$, isophotal diameter $D_{25}>1.0$\,arcmin, H{\sc i} radial velocity
$V>3000$\,km s$^{-1}$, and apparent photographic magnitude $m_B<15.5$. The
sample includes 1734 newly observed galaxies and 597 archival objects.
Histograms presented in Sheth \textit{et al}. (2010) show that the bulk of
S$^4$G galaxies are of types Sb and later, a consequence of the use of an H{\sc
i} velocity as a selection criterion. The most statistically complete subset of
S$^4$G galaxies is for masses ${\rm log}(M/M_{\odot})<9.2$.

\begin{figure}
\centerline{\includegraphics[width=\textwidth]{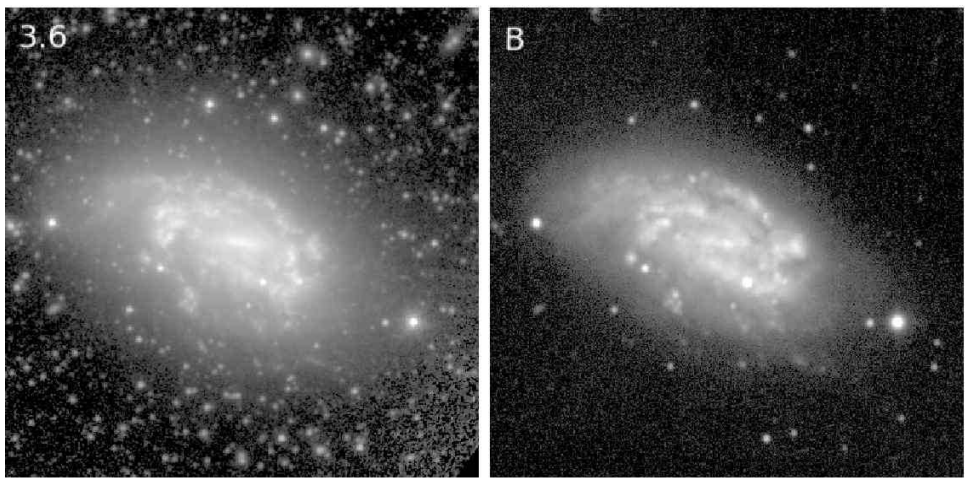}}
\caption{The late-type spiral NGC\,1559 has a very similar morphology in
blue and 3.6\,$\mu$m light.}
\label{ngc1559}
\end{figure}

\begin{figure}
\centerline{\includegraphics[width=\textwidth]{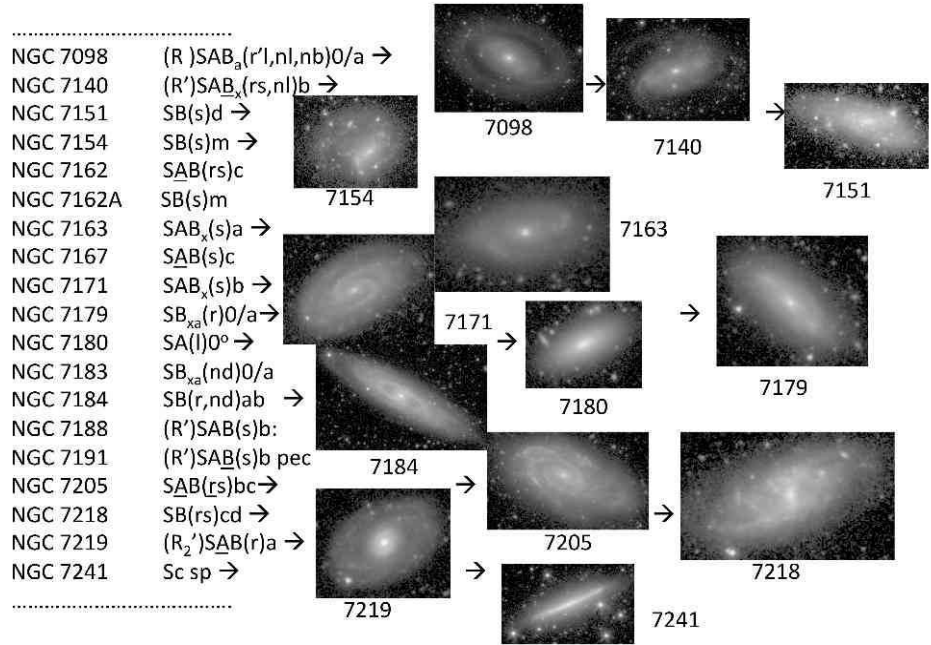}}
\caption{A set of classifications of S$^4$G galaxies using a modified
version of dVA types.}
\label{midIR}
\end{figure}

In a preliminary analysis of 200 mostly archival S$^4$G galaxies, Buta
\textit{et al.} (2010b) showed that galaxies can be classified in 3.6\,$\mu$m
images in much the same manner as for the historically standard $B$-band images
(essentially the same as was found by Eskridge \textit{et al.} 2002 for the
$H$-band). In fact, for many galaxies, 3.6\,$\mu$m morphology is very similar to
blue light morphology, absent the effects of obscuring dust
(Fig.~\ref{ngc1559}). Because mid-IR galaxy morphology is so similar to optical
morphology, we can apply the old classification systems to S$^4$G data as an
exploratory step. The system being used is a further modified version of that
outlined in the dVA.  Figure~\ref{midIR} shows some classifications of S$^4$G
galaxies in Buta \textit{et al.} (2013).

\begin{figure}
\centerline{\includegraphics[width=\textwidth]{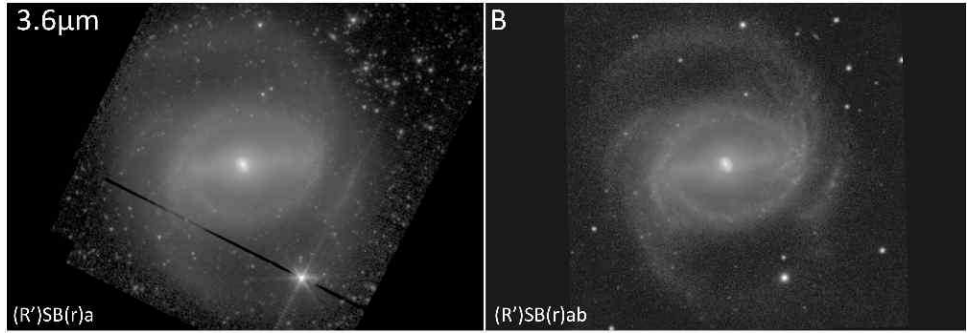}}
\caption{NGC\,1433 at 3.6\,$\mu$m (left), showing the `earlier' effect as
compared to the $B$-band (right).} 
\label{ngc1433-midIR} 
\end{figure}

Buta \textit{et al.} (2010b) also showed that galaxies of types S0/a to Sbc can
appear about one stage interval `earlier' at 3.6\,$\mu$m compared to $B$. Types
S0$^+$ and earlier and Sc and later can look almost the same. This is similar to
what Eskridge \textit{et al.} (2002) concluded for near-IR $H$-band images. The
effect is shown for NGC\,1433 in Fig.~\ref{ngc1433-midIR}. Some $B$-band
flocculent spirals look more globally armed at 3.6\,$\mu$m; many do not change
(Elmegreen \textit{et al.} 2011). S$^4$G images also show many galaxies where a
late-type spiral or irregular is embedded in an early-type disk. One example is
NGC\,5713, which has a clear SB(rs)m galaxy embedded in a background S0/a disk.
Other examples shown by Buta \textit{et al.} (2010b) are IC\,750, IC\,3392, and
NGC\,3769. Buta \textit{et al.} (2010b) show two $B$-band Magellanic irregulars
(Im) that appear more regular at 3.6\,$\mu$m.  If placed in a cluster
environment, these could become progenitors of Kormendy spheroidal (Sph) or
nucleated (Sph,N) galaxies. Also possible progenitors of spheroidals are
Magellanic barred spirals like NGC\,3906 and NGC\,4618, shown also in Fig.~10 of
Buta \textit{et al.} (2010b).

\begin{figure}
\centerline{\includegraphics[width=\textwidth]{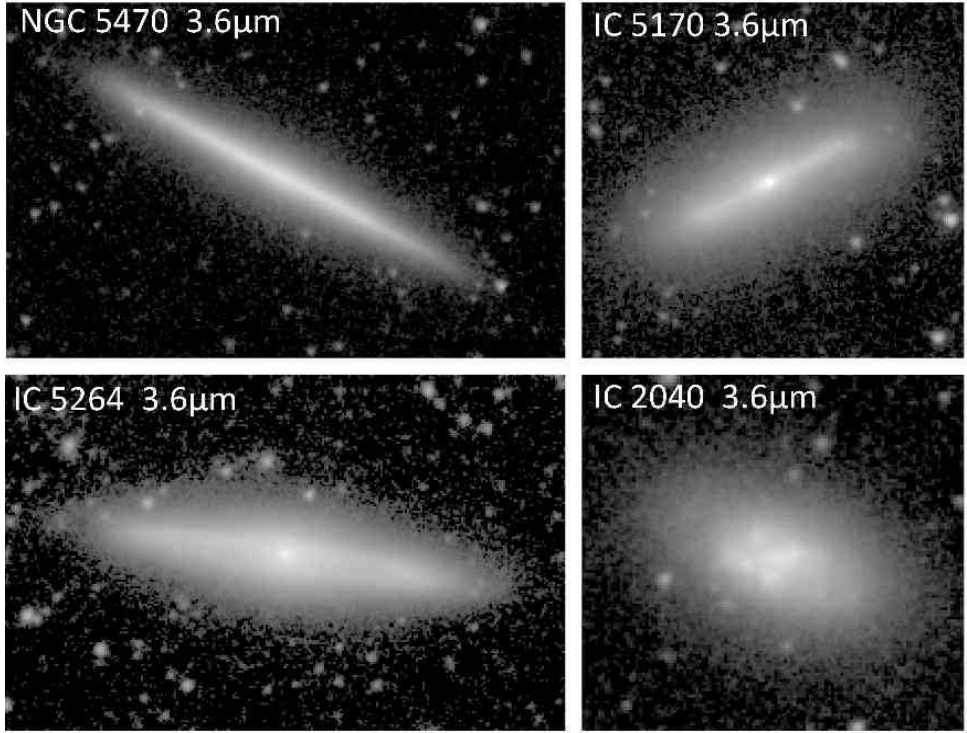}}
\caption{Four S$^4$G galaxies showing dust-free views of edge-on disks
and a Magellanic irregular.} 
\label{disks-midIR} 
\end{figure}

\begin{figure}
\centerline{\includegraphics[width=\textwidth]{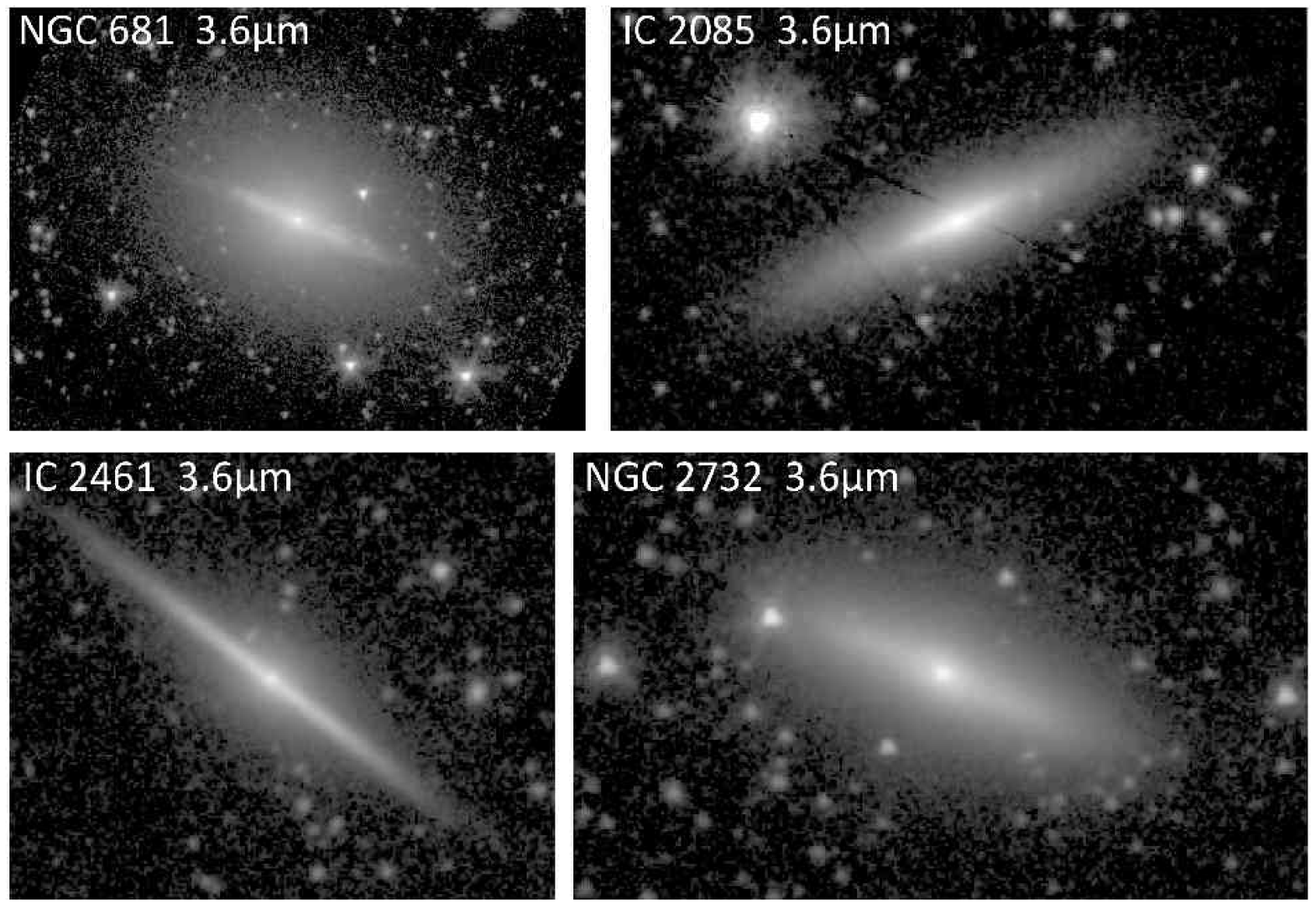}}
\caption{Four S$^4$G galaxies showing `embedded disks' in what appear
to be spheroidal systems.}
\label{special1}
\end{figure}

The dust-penetrated nature of S$^4$G images also can reveal unusual optical
misclassifications. NGC\,5470 is classified as type Sb, but at 3.6\,$\mu$m we
see an edge-on S0 with virtually no bulge and both thick and thin disks
(Fig.~\ref{disks-midIR}). In the mid-IR the galaxy should be classified as type
S0d, which alludes to the van den Bergh (1976) parallel sequence classification
of S0s. Other interesting edge-on disks, some of which look like they are
embedded in Kormendy spheroidals, are shown in Fig.~\ref{special1}.

\begin{figure}
\centerline{\includegraphics[width=\textwidth]{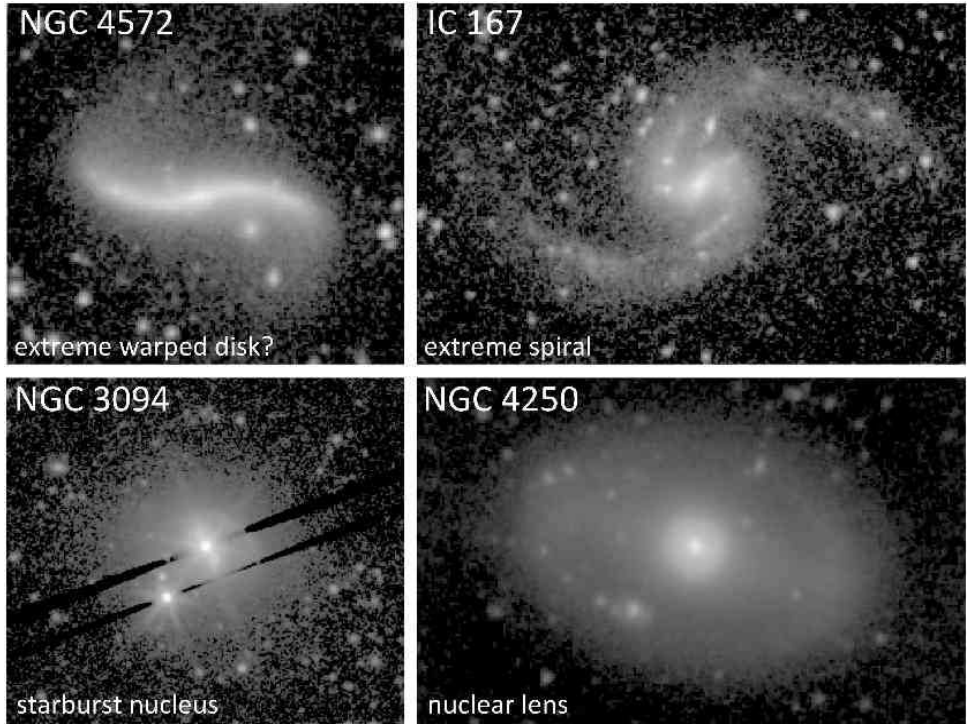}}
\caption{Four extreme S$^4$G cases.}
\label{special2}
\end{figure}

Several special S$^4$G cases are shown in Fig.~\ref{special2}. NGC\,3094 looks
like a barred spiral with a dusty bar in optical images, but at 3.6\,$\mu$m it
shows an extremely bright starburst nucleus that was measured by Narayanan
\textit{et al.} (2005) to have ${\rm log}(L_{\rm IR})=10.73$. NGC\,4250 looks
like an early-type galaxy with an oval ring/lens, a very faint outer ring, and
an exceptionally well-defined nearly circular nuclear lens. The type of the
galaxy at 3.6\,$\mu$m is (R)SAB(rl,nl)0$^+$. NGC\,4572 is a very unusual case
which at first sight appears to be a very open spiral, but which is more likely
to be an extremely warped edge-on disk. IC\,167 is an extreme case of an open
spiral in the S$^4$G sample.

\begin{figure}
\centerline{\includegraphics[width=\textwidth]{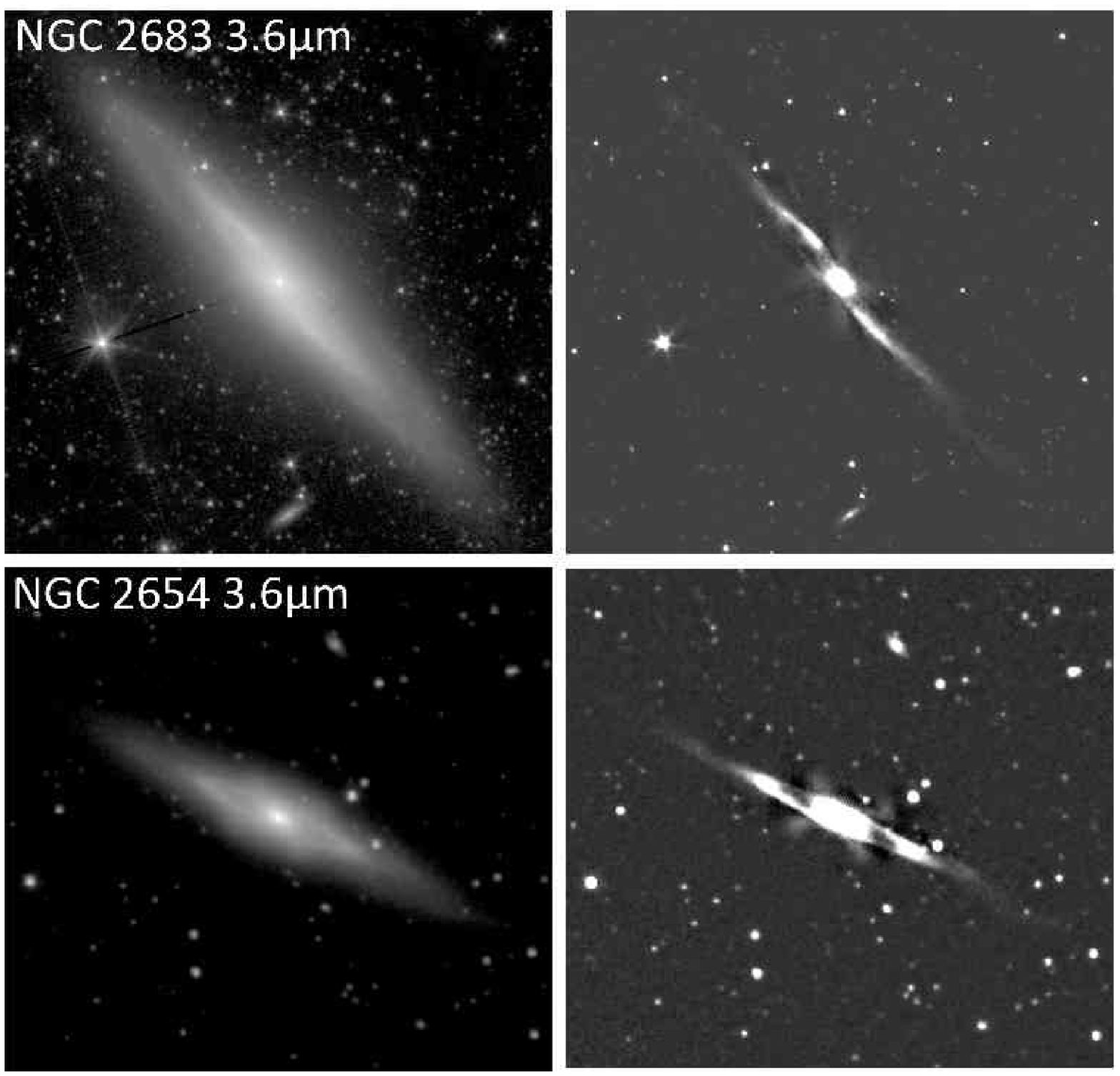}}
\caption{S$^4$G X-galaxies NGC\,2654 and NGC\,2683.}
\label{special3}
\end{figure}

Figure~\ref{special3}, left, shows S$^4$G 3.6\,$\mu$m images of two nearly
edge-on galaxies showing a clear X-pattern in the central regions.
Figure~\ref{special3}, right, shows unsharp-masked versions of the same images
of NGC\,2654 and NGC\,2683, both optical Sb spirals. The X-pattern in NGC\,2654
is more distinct than that in NGC\,2683 and is detectable even in an optical
SDSS image. These features point to the presence of a bar in each galaxy (see
also Kuzio de Naray \textit{et al.} 2009).

\subsection{The S$^4$G bar fraction}
The differing characteristics of bars along the de Vaucouleurs revised
Hubble-Sandage sequence of types complicate the interpretation of the
bar fraction. Excluding highly inclined galaxies (spindles), here are
some preliminary results from my classification of 2184 S$^4$G galaxies
(and a few of their companions):

\begin{enumerate}[(a)]\listsize
\renewcommand{\theenumi}{(\alph{enumi})}
\item Types S0$^-$ to Scd: bar fraction (SAB, SA$\underline{\rm B}$, and SB) = 57.2\%,
\item Types Sd and later: bar fraction (SAB, SA$\underline{\rm B}$, and SB) = 83.8\%.
\end{enumerate}

Thus, bars are very abundant in extreme late-type galaxies! This result is
consistent with the recent SDSS study by Barazza \textit{et al.} (2008), who
found an optical bar fraction of $\approx$87\% for what they called `class 3'
(bulgeless) galaxies.

\subsection{Inferring stellar mass from S$^4$G images}
The ultimate value of IR imaging is to infer stellar mass and the
gravitational potential in galaxies. With such potentials you can, among other
things:

\begin{enumerate}[(a)]\listsize
\renewcommand{\theenumi}{(\alph{enumi})}

\item Estimate maximum relative bar torques $Q_{\rm b}$ (or $Q_{\rm g}$) and 
mass-flow rates;

\item Derive mass models and run simulations to estimate pattern speeds (e.g., 
Aguerri \textit{et al.} 2001; Treuthardt \textit{et al.} 2008; 
Kalapthorakis \textit{et al.} 2010; Rautiainen \textit{et al.} 2008; 
Lin \textit{et al.} 2008);

\item Measure potential-density phase shifts and infer locations of resonances 
(Zhang 1996, 1998, 1999, hereafter Z96, Z98, Z99).

\end{enumerate}

Stellar surface mass density maps can be made from 2D images using surface
colours as indicators of stellar mass-to-light ratio (Bell \& de Jong 2001).
Calibrated surface brightness maps can be converted to units of $L_{\odot}$
pc$^{-2}$, and then multiplied by colour-inferred $M/L$ values in Solar units to
give the surface mass density $\Sigma$ $(i,j)$ in units of $M_{\odot}$ pc$^{-2}$
at pixel coordinate $(i,j)$. 

Bell \& de Jong (2001) give linear relationships between the logarithm of the
$M/L$ ratio in a given passband and a variety of colour indices in the
Johnson-Cousins systems. Bell \textit{et al.} (2003) give the same kinds of
relations for SDSS filters, as well as an improved calibration for $B-V$ and
$B-R$ colours that allows for the scatter in metallicities that affects near-IR
mass-to-light ratios. For 3.6\,$\mu$m images, colours such as $B-V$, $B-R$,
$V-I$, or even $g-i$ can be used to give the stellar mass-to-light ratio in the
$K$-band, depending on which of these colours is available. Then a simple
relationship between $M/L_K$ and $M/L_{3.6}$ (Oh \textit{et al.} 2008) can be
used to estimate the surface mass densities:
\begin{equation}
{M\over L_{3.6}} = 0.92{M\over L_K} - 0.05.
\end{equation}
In practice, it is useful to use two colours and two base IR images to check for
consistency of results. For example, X. Zhang and R.J. Buta (2013, in
preparation) have used SINGS (Kennicutt \textit{et al.} 2003) 3.6\,$\mu$m images
with $B-V$ or $B-R$ surface colours, and SDSS $i$-band images with $g-i$ surface
colours, to estimate mass flow rates (see also Zibetti \textit{et al.} 2009;
Foyle \textit{et al.} 2010).

As already noted, 3.6\,$\mu$m images are affected by hot dust connected with
star-forming regions and by a prominent 3.3\,$\mu$m emission feature due to a
polycyclic aromatic hydrocarbon compound that also is associated with
star-forming regions. These star-forming regions appear as conspicuous `knots'
lining spiral arms in 3.6\,$\mu$m images.

In practice, the star-forming region problems in the 3.6\,$\mu$m band can be
reduced using an 8.0\,$\mu$m image if available. IRAC 8.0\,$\mu$m images show
mainly the ISM with a small contribution from starlight. Kendall \textit{et al.}
(2008) describe how to correct the 3.6\,$\mu$m image. The first step is to match
the coordinate systems of the 3.6 and 8.0\,$\mu$m images and then subtract a
fraction (0.232) of the 3.6\,$\mu$m flux from the 8.0\,$\mu$m image to correct
the latter for continuum emission (Helou \textit{et al.} 2004). Then, a fraction
(0.059--0.095; Flagey \textit{et al.} 2006) of the net dust map is subtracted
from the 3.6\,$\mu$m map to give an image corrected for the hot dust emission.
These `contaminants' can also be eliminated using Independent Component
Analysis and [3.6]$-$[4.5] colours (Meidt \textit{et al.} 2012) if no
8.0\,$\mu$m image is available. This is the case for most of the S$^4$G sample.

\subsection{Secular evolution and the potential-density phase shift}
The availability of large numbers of IR images that can be converted into
stellar mass maps provides us with an opportunity to examine a promising new
approach to understanding the actual {\it mechanism} of secular evolution in
many galaxies. In any galaxy having a skewed bar, oval, or spiral perturbation,
there will be a radius-dependent phase shift between the density perturbation
and the potential perturbation (Zhang 1996). This phase shift distribution can
drive secular stellar and gaseous mass redistribution because it leads to an
interaction between the perturbation and the {\it basic state} of a galactic
disk.\footnote{The basic state is defined by the radial distributions of mass
density, velocity dispersion, and rotation speed of the bulge and disk
components (Bertin \textit{et al.} 1989).} The approach interprets bars, ovals, and
spirals as quasi-steady modes that arise spontaneously from an originally
featureless disk. In the presence of such modes, the interaction resulting from
the phase shift secularly evolves the basic state in the sense that a late-type
spiral galaxy will evolve to an earlier-type galaxy. The actual interaction
between the perturbation and the basic state involves collective effects that
scatter and change stellar orbits without actual collisions (called
`collisionless shocks').

An important result from Zhang's studies is that the phase shift changes from
positive (density spiral leads potential spiral) inside CR to negative (density
spiral lags potential spiral) outside CR. Given a surface mass density map of a
disk galaxy, one can convert the densities into a potential and measure the
phase shift distribution directly. This provides a direct way of locating the CR
resonance {\it without kinematic data}. Thus, in addition to providing
information on secular mass flow in galactic disks, the potential-density phase
shift can be used to interpret galaxy morphology, since CR is tied to a
fundamental property of a spiral or bar/oval: its pattern speed $\Omega_{\rm
p}$.

The method of locating CR using the potential-density phase shift (PDPS) method is
outlined by Zhang \& Buta (2007=ZB07) and Buta \& Zhang (2011=BZ11). The result
of this is that, if spirals and bars are quasi-steady modes, a single mode will
be characterised by one positive phase shift hump followed by one negative hump.
The negative part has to be there to take the angular momentum transferred
outward by the positive part. Thus CR will be located at a positive-to-negative
(P/N) crossing. The mode may end at a negative-to-positive (N/P) crossing, which
then leads into the next mode if present. To measure phase shifts:

\begin{enumerate}[(a)]\listsize
\renewcommand{\theenumi}{(\alph{enumi})}

\item First approx.: use IR image assuming constant $M/L$;

\item Second approx.: use IR image and 2D colour-dependent variable $M/L$
distribution;

\item Third approx.: add in atomic and molecular gas.

\end{enumerate}

\begin{figure} 
\begin{center}
\includegraphics[width=0.5\textwidth]{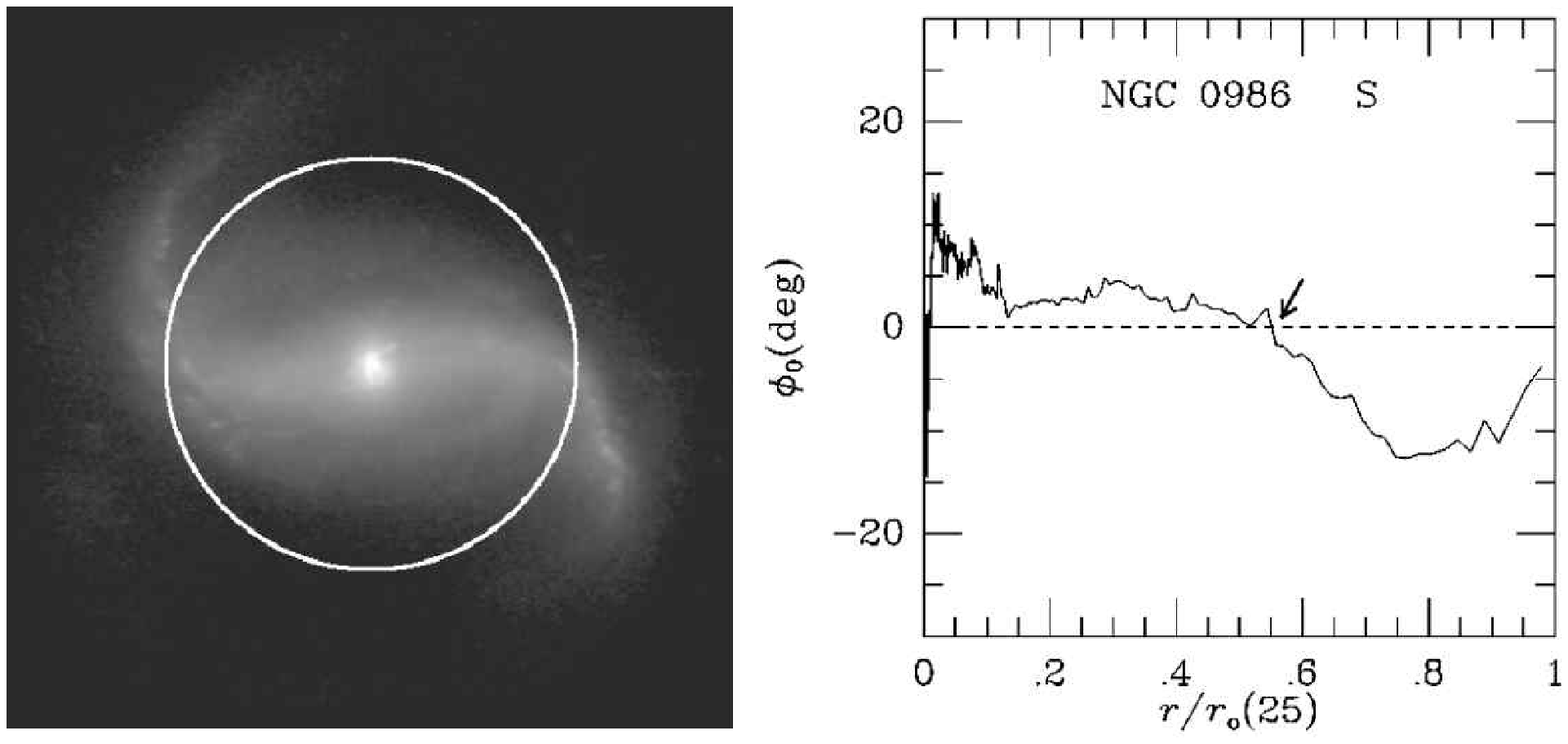}
\end{center} 
\begin{center}
\includegraphics[width=0.5\textwidth]{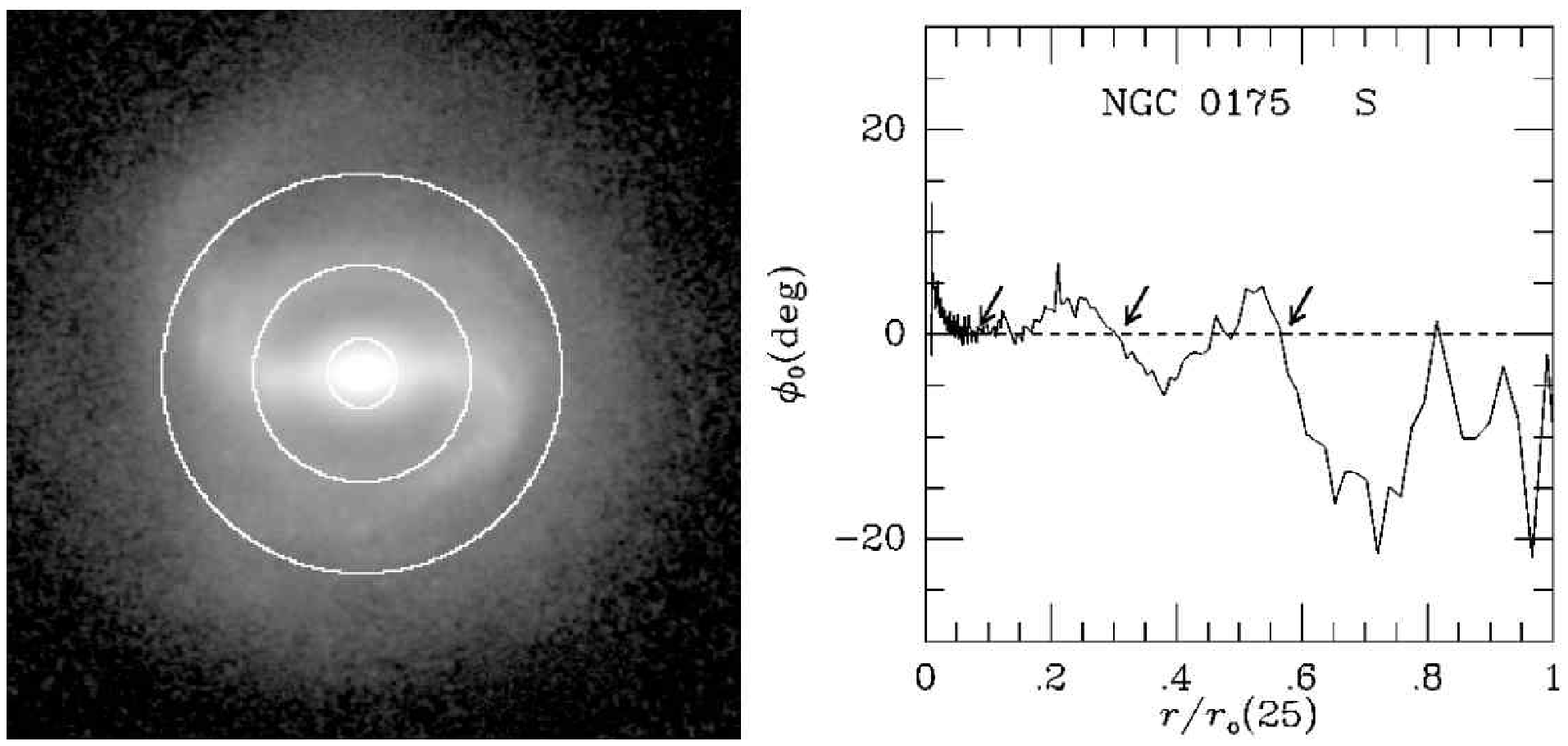}
\end{center} 
\begin{center}
\includegraphics[width=0.5\textwidth]{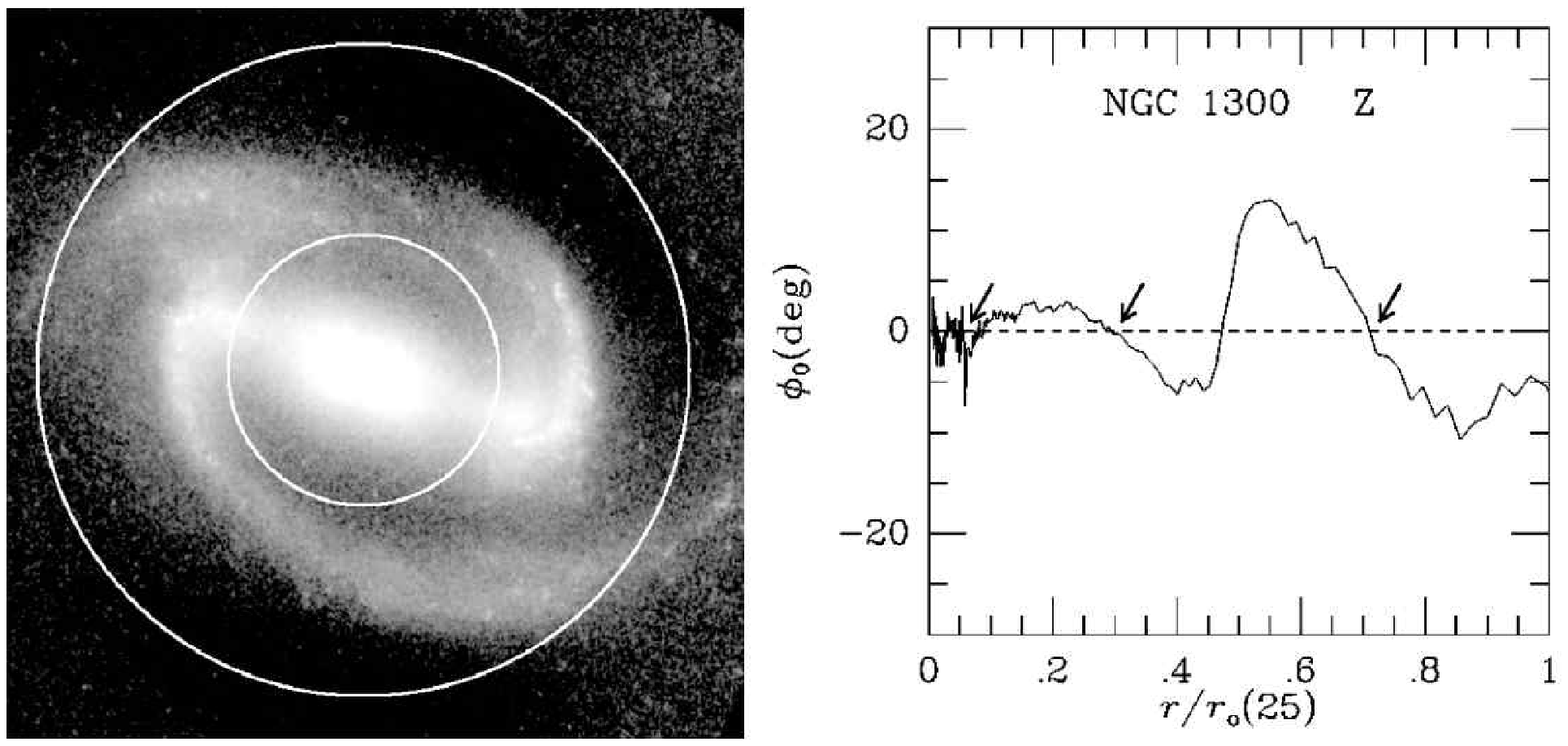}
\end{center} 
\caption{ \footnotesize $K_{\rm s}$-band images and phase shift
distributions of three galaxies: (top) NGC\,986; (middle) NGC\,175;
(bottom) NGC\,1300. The arrows on the phase shift plots show the main
P/N crossings, and hence the implied CR radii. The circles superposed
on the images show these radii.} 
\label{phaseshifts} 
\end{figure}

BZ11 describe first approximation phase shift analyses of several well-known
galaxies (Fig.~\ref{phaseshifts}). NGC\,986, a barred galaxy with a strong
spiral curving off the bar ends, shows a distribution with only a single major
P/N crossing that places CR exactly around the bar ends.  In this case, the bar
and the spiral would have the same pattern speed, and BZ11 interpret NGC\,986 as
a genuine example of a `bar-driven spiral'. In contrast, NGC\,175 is a barred
spiral with two clear P/N crossings.  The first CR circles the ends of the bar
and the mode includes the inner part of a bright spiral pseudoring, which is
interpreted as being bar-driven. The second CR refers to an independent outer
spiral pattern. In both of these cases, the ratio $\cal{R}$\,=\,$R_{\rm CR}({\rm
bar})/R_{\rm bar}$\,$\approx$\,1. Buta \& Zhang (2009=BZ09) highlighted two
cases, NGC\,1493 and NGC\,3686, which may be genuine examples having a `slow
bar', where $\cal{R}$\,$>$\,1.4 (Debattista \& Sellwood 2000).

\begin{figure} 
\begin{center}
\includegraphics[width=\textwidth]{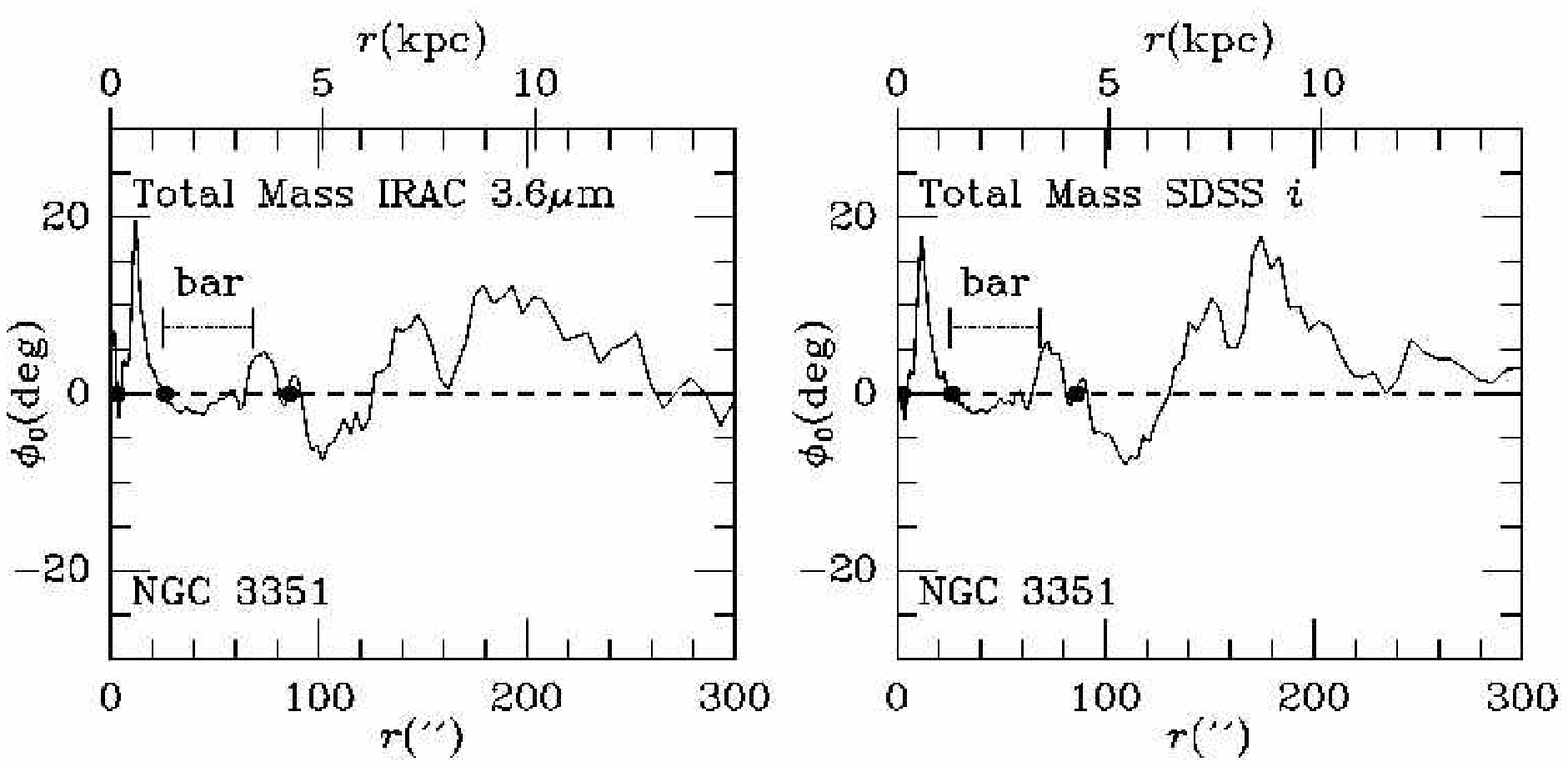}
\includegraphics[width=0.5\textwidth]{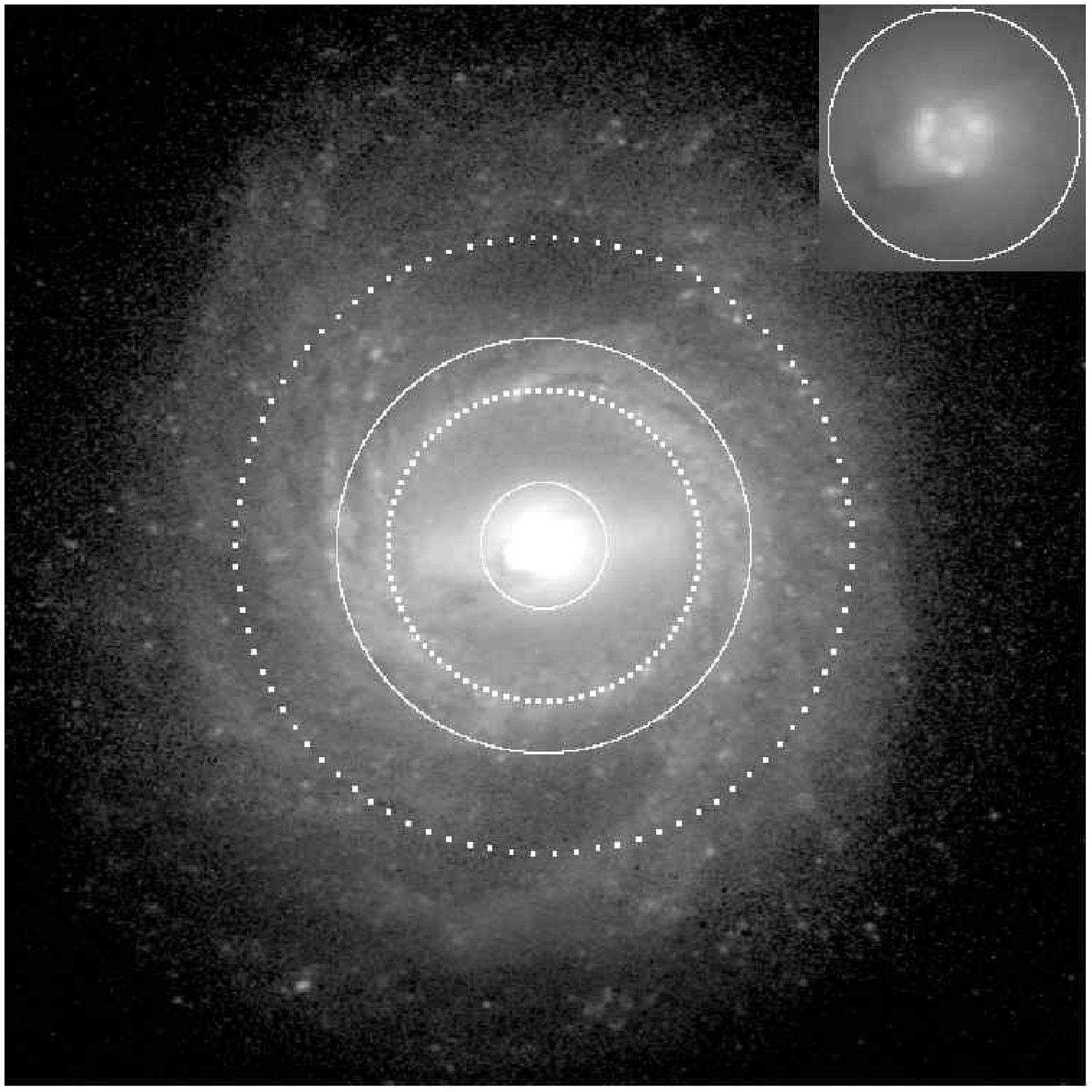}
\end{center} \caption{(Top): Phase shift analysis of NGC\,3351 based on
IRAC 3.6$\mu$m and SDSS $i$-band total (star plus gas) mass maps. The
bar extent is indicated. (Bottom): P/N (solid circles) and N/P (dotted
circles) phase shift crossings superposed on a deprojected SDSS $g$-band
image of NGC\,3351, based on an average of the values derived from the
IRAC and SDSS images. Since the bar extends to an N/P crossing, it is
of the `superfast' type (from ZB12).} 
\label{ps3351mass} 
\end{figure}

The most enigmatic (and by default the most controversial) result from
the phase shift method of locating CR is the unusual `superfast bars',
where the bars extend {\it beyond the bar CR}. This appears to violate
an almost cherished rule that bars cannot extend beyond their CR,
because orbits outside CR cannot support the bar (Contopoulos 1980).
This was a conclusion based on what ZB07 called `passive orbit
analysis', an analytic approach where the Jacobi integral is conserved
along a stellar orbit and secular change in the\linebreak average radius of the
orbit cannot occur.  BZ11 describe the case of NGC\,1300 as an example.
In NGC\,1300, there is a clear skewness to the apparently strong bar,
meaning it is a very open spiral. In such a case, it is possible that
the bar extends to its OLR and not to its CR.  BZ09 show two other
examples of superfast bars (NGC\,4902 and NGC\,5643).  BZ11 also show the
phase shift distribution for the strong grand-design spiral NGC\,5247.
In this case, CR lies at an intermediate position within the arms and
the arms likely extend to their OLR.

Zhang \& Buta (2012=ZB12) have carried out a more sophisticated (third
approximation) analysis of the well-known ringed, barred spiral NGC\,3351. An
IRAC 3.6$\mu$m SINGS (Kennicutt \textit{et al.} 2003) image and an SDSS $i$-band
image were converted into surface mass density maps using colour-dependent $M/L$
ratios as described above.  The appearance of NGC\,3351 can garner much
preconception.  The bar is surrounded by a bright inner spiral pseudoring that
could easily be interpreted in terms of the inner 4:1 resonance of the bar, as
suggested by Schwarz (1984a,b). This would imply that the bar CR is outside the
inner pseudoring. The phase shift distributions, however, suggest a very
different interpretation.  Figure~\ref{ps3351mass} shows the phase shift
distributions for NGC\,3351 based on total mass maps that include both H{\sc i}
and H$_2$ gas. The figure shows that the 3.6$\mu$m and SDSS $i$-band maps give
very similar results: the phase shift is negative throughout the main part of
the bar. This means that of the two CR radii indicated, CR$_1$ is the bar CR,
and thus NGC\,3351 is another example of a `superfast' bar.  In this
circumstance, the spiral pseudoring around the bar is a {\it separate mode}, not
a bar-driven spiral.

Why is the phase shift relevant to secular evolution? The phase shift brings
attention to the role of collective effects on galactic dynamics, and may be the
principal driver of secular evolution in disk galaxies. The idea is that if the
basic state is changed by the presence of a spontaneous, self-sustaining density
wave, then it means orbits of stars will migrate, slowly building up a
`pseudobulge' and spreading out the stellar disk.

\subsection{Summary}

\begin{enumerate}[(a)]\listsize
\renewcommand{\theenumi}{(\alph{enumi})}

\item Near- and mid-IR images are goldmines for studies of galaxy morphology.
\item The essential features of optical galaxy morphology are preserved in the 
IR, but are unambiguously detected because of the minimal effects of extinction.
\item Morphological analysis of S$^4$G images confirms a high fraction of bars 
in extreme late-type galaxies, more than 80\%.
\item IR images offer a way to locate CRs and to probe secular evolution of the 
stellar mass distribution using the PDPS method.

\end{enumerate}

%
%

\section{Lecture 4: Environmental effects, exotic morphologies, and\\ morphological databases}
\label{sec:lecture4}

In my final lecture in this series, I want to cover a \textit{potpourri} of
topics to finish out the richness of galaxy morphology and to highlight some
thoughts of what future studies of galaxy morphology and secular evolution might
be like.  Examining environmental effects may seem a little off the topic of
secular evolution, but such effects do not have to be rapid or violent.
Environmental effects can also produce exotic and very rare morphologies that
are worthy of discussion.  A few such morphologies (double detached outer rings,
intrinsic bar-ring misalignment) have already been discussed in Lecture 2 (see
Section~\ref{sec:lecture2}).

\subsection{Environmental effects in clusters}

Gravitational interactions, ranging from minor, distant or fast encounters, to
violent collisions and major/minor mergers, certainly play a role in galaxy
morphology. Up to 4\% of bright galaxies are currently involved in a major
interaction (Knapen \& James 2009).

The cluster environment offers a number of processes that can modify or even
transform galaxy morphology. The cluster environment also brings attention to
the idea of {\it environmental secular evolution} (see Kormendy's lectures, this
volume), that adds a new dimension to morphological studies. The main 
processes or effects are:

\begin{enumerate}[(a)]\listsize
\renewcommand{\theenumi}{(\alph{enumi})}

\item Gas stripping
\item Star-forming disk truncation
\item H{\sc i} deficiency and truncated H{\sc i} disks
\item Harassment
\item Starvation
\item Transformation (morphology-density relation).

\end{enumerate}

Gas stripping has long been thought to be one of the dominant mechanisms of
transforming a spiral galaxy into an S0 galaxy.  Spitzer \& Baade (1951)
suggested that repeated collisions with other galaxies can strip a galaxy clean
of its ISM.  An alternative idea is that, as a galaxy moves in a cluster, it
interacts with the intracluster medium in the form of X-ray emitting gas and
can be stripped clean of its own interstellar gas, causing star formation to
cease and spiral arms to disappear (Gunn \& Gott 1972).  Van den Bergh (1998)
has argued that perhaps even some of the stellar mass is lost during the
transformation, which might explain why S0s tend to have a lower luminosity than
spirals.

\begin{figure}
\centering{\includegraphics[width=\textwidth]{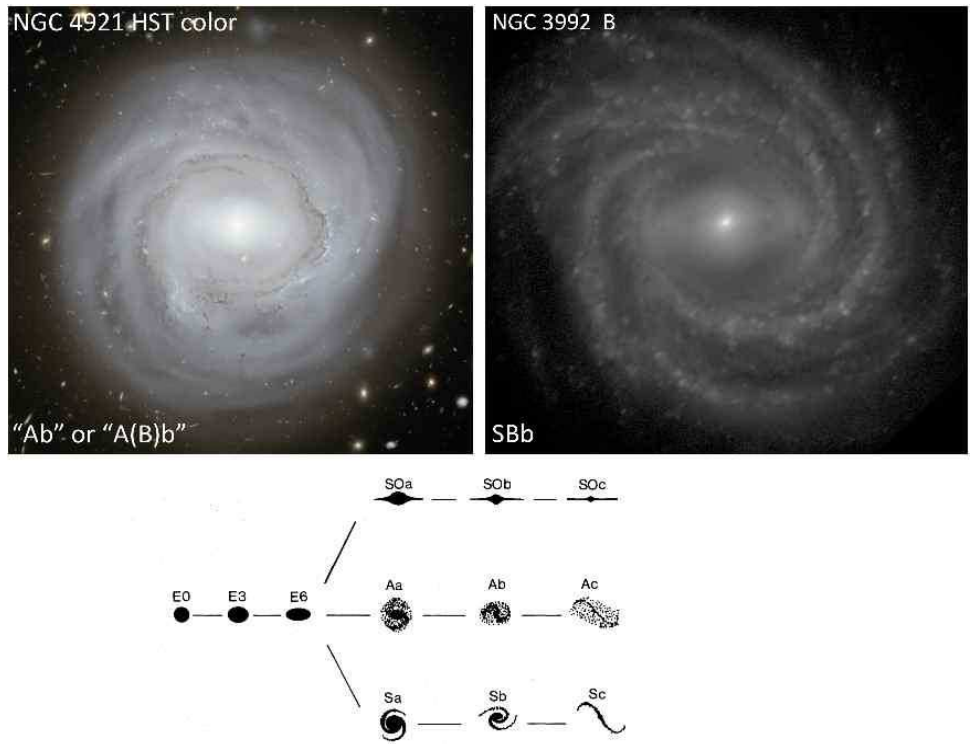}}
\caption{Comparison between the `anemic' barred spiral NGC\,4921 and the more 
normal barred spiral NGC\,3992. The van den Bergh (1976) parallel sequence 
classification is also shown.} 
\label{anemics}
\end{figure}

If this process occurs and takes time, then partially stripped spiral galaxies
ought to be observed. These are the galaxies referred to as `anemic' by van
den Bergh (1976), and morphologically, these seem to be a legitimate class of
objects. Figure~\ref{anemics} shows an example of an anemic spiral identified by
van den Bergh (1976). Anemic spirals are cluster spirals deficient in H{\sc i}
gas and as a result have a lower amount of dust and star formation.  NGC\,4921
is a Coma cluster spiral that strongly resembles a normal SBb spiral like
NGC\,3992, but has much smoother spiral arms. The smoothness indicates gas
deficiency probably caused by stripping due to the intracluster medium.
According to the hydrogen index parameter given in the RC3 (de Vaucouleurs
\textit{et al.} 1991), NGC\,4921 is H{\sc i} deficient compared to NGC\,3992 by
a factor of more than six. Bothun \& Sullivan (1980) argue that the galactic
anemia `look' can be caused by other factors besides H{\sc i} deficiency.

\begin{figure}
\centering{\includegraphics[width=\textwidth]{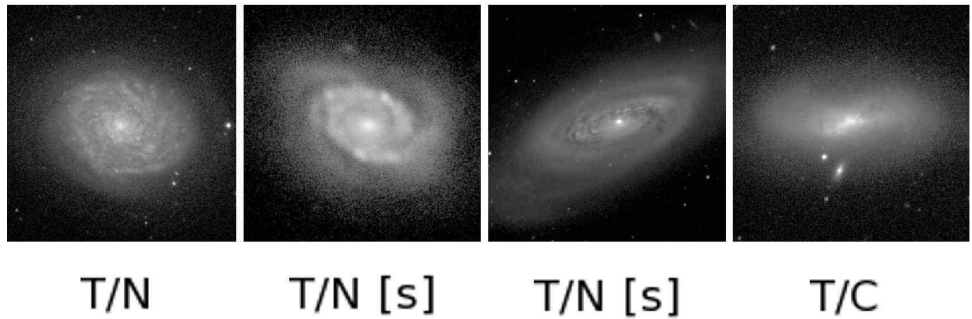}}
\caption{Koopmann \& Kenney (2004) environmental categories of Virgo cluster galaxies 
having truncated star-forming disks, illustrated using blue light images (from B13).} 
\label{koopmann}
\end{figure}

Environmental effects on the star-forming disks of Virgo cluster
galaxies were examined by Koopmann \& Kenney (2004) using H$\alpha$
images.  They recognised seven environmental categories based on star
formation rates and the H$\alpha$ disk as compared to a field sample.
Class N had normal star formation rates for the galaxy type. Class E
had star formation enhanced by a factor of three or more over normal,
while class A (anemic) had star formation a factor of three or more
below normal. The remaining categories are the truncated star-forming
disk morphologies, several examples of which are shown in
Fig.~\ref{koopmann}. Class T/N is `truncated/normal', meaning the
star-forming disk is confined to the inner regions, but in those
regions the star formation rates are normal. Class T/N[s] is similar
but the truncation of the disk is severe.  Class T/A is similar except
that the star-forming part of the disk is anemic. Finally, class T/C is
a truncated star-forming disk where the star formation is confined to a
compact central area.  Koopmann \& Kenney (2004) illustrate the
H$\alpha$ images of all of these categories, while Fig.~\ref{koopmann}
(from B13) illustrates the truncated categories in blue light.

\begin{figure}
\centering{\includegraphics[width=\textwidth]{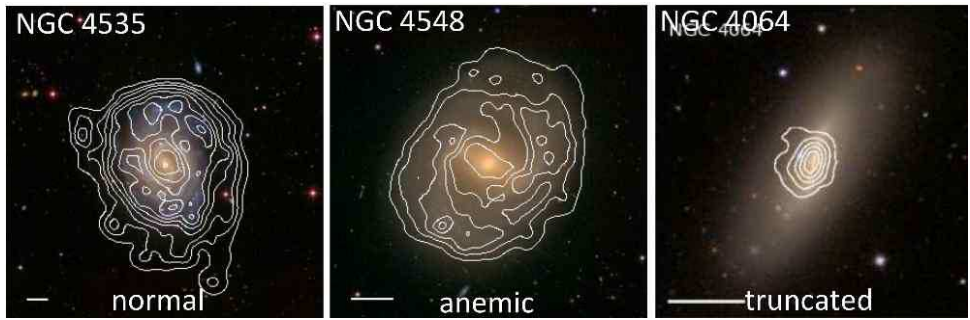}}
\caption{VIVA images of H{\sc i} disks of Virgo cluster galaxies (from Chung 
\textit{et al.} 2009, reproduced with permission).}
\label{viva}
\end{figure}

These optical categories translate into similar categories of H{\sc i}
distributions. The VLA (Very Large Array) imaging of Virgo spirals in Atomic gas
(VIVA) survey (Chung \textit{et al.} 2009) shows the interesting truncated
morphologies that are found near the centre of the cluster (Fig.~\ref{viva}).
Cases like NGC\,4064, NGC\,4405, and NGC\,4457 are among the most extreme
truncations.  The survey also found evidence of ongoing stripping and fallback,
the possibility that some galaxies have already fallen through the core and
reached farther distances from the cluster centre as stripped spirals, as well
as signs of gravitational interactions.

Yagi \textit{et al.} (2010) identified a dozen galaxies in Coma that show
ionised gas clouds mostly outside their disks. They suggest that these galaxies
are likely to be `new arrivals' to the core region of Coma in different phases
of stripping. These authors identified three morphological gas stripping
categories:

\begin{enumerate}[(a)]\listsize
\renewcommand{\theenumi}{(\alph{enumi})}

\item connected H$\alpha$ clouds with disk star formation, thought to be in an 
early phase of stripping,
\item connected H$\alpha$ clouds without disk star formation,
\item detached H$\alpha$ clouds, thought to be a later phase of stripping.

\end{enumerate}

A summary illustration of these categories can be found in B13.

\subsection{Gravitational encounter phenomena}

\begin{figure}
\centering{\includegraphics[width=\textwidth]{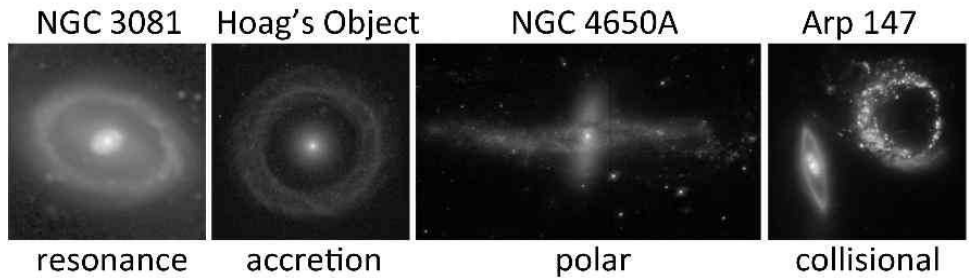}}
\caption{Examples of different galactic ring categories.}
\label{rings} 
\end{figure}

Among gravitational encounter morphologies, ring phenomena have captured a great
deal of attention. Here I discuss what I will refer to as `catastrophic
rings', or rings that have formed because of a collision between two galaxies.
The main classes of catastrophic rings: accretion, polar, and collisional, must
first be distinguished from the background of normal rings, usually referred to
as `resonance rings' which were discussed in Lecture~2 (see 
Section~\ref{sec:lecture2}). Figure~\ref{rings} shows an example of each ring 
type.

`Resonance rings' constitute the {\it vast majority} of all rings observed.
They are interpreted as features which form by gas accumulation at resonances,
owing to continuous action of gravity torques from a bar (Buta \& Combes 1996). 
The main resonances theoretically linked to rings are (Schwarz 1981, 1984a,\,b):

\begin{enumerate}[(a)]\listsize
\renewcommand{\theenumi}{(\alph{enumi})}
\item nuclear rings: ILR,
\item inner rings: I4R, and
\item outer rings: OLR.
\end{enumerate}

These theoretical links found strong support in statistics of intrinsic ring
shapes and orientations, as well as in morphology (Lecture 2, see
Section~\ref{sec:lecture2}).  However, there have been alternative
interpretations suggested over the years. Regan \& Teuben (2004) noted that the
concept of a `resonance radius' (i.e., a single radius where a resonance
occurs) is not really valid for strong bars because the epicyclic approximation
breaks down.  In the inner regions of a strongly barred galaxy, one instead has
a broad `ILR region'. Because rings in models seem to be
connected more to orbit transitions (i.e., orbit family changes, which are
usually accompanied by orientation changes), Regan \& Teuben (2004) preferred
the term `orbit transition rings' for inner and nuclear rings at least, but
still leaving outer rings and pseudorings in the `resonance' category.

A second alternative interpretation of rings is that they are connected to
`invariant manifolds', or collections of orbits which emanate from the $L_1$
and $L_2$ Lagrangian points of a bar potential (Athanassoula \textit{et al.}
2009a,\,b). This interpretation can successfully predict broad aspects of the
morphologies of inner and outer rings in barred galaxies.  Please see the
lectures by Athanassoula (this volume) for more details.

The PDPS provides additional interpretations of normal rings. As
described in Lecture 3 (Section~\ref{sec:lecture3}), the nearby barred
galaxy NGC\,3351 has a bright inner ring whose morphology fits well
into the resonance idea, yet the PDPS method suggests that the ring is
a spiral mode with a different pattern speed from the bar. ZB12
interpret the ring as a spiral located at its own inner 4:1 resonance
(not the bar's 4:1 resonance, because the bar in NGC\,3351 is
`superfast').

Accretion rings are thought to be made of material from an accreted satellite
(Schweizer \textit{et al.} 1987). Evidence for this is sometimes {\it
counter-rotation}, where the material in the ring rotates in the opposite sense
from the material in the main galaxy (Schweizer {\it et al.} 1989; de Zeeuw
\textit{et al.} 2002). The host galaxy can be an elliptical or disk galaxy.
Examples of the former are Hoag's object (Schweizer \textit{et al.} 1987) and
IC\,2006 (Schweizer \textit{et al.} 1989), while an example of the latter is
NGC\,7742 (de Zeeuw \textit{et al.} 2002). In the case of a disk galaxy like
NGC\,7742, the ring material mostly lies in the same plane as the receiving
disk. It is also possible, as shown by the interesting case of NGC\,1211
(Lecture 2, see Section~\ref{sec:lecture2}), that a `dead' resonance ring
galaxy could accrete a satellite in its outer regions and have a blue second
outer ring.

Polar rings are also accretion features, except that the accreting object is
usually a disk system, like an S0 (Schweizer \textit{et al.} 1983). The most
stable configuration is an accretion angle close to 90$^{\rm o}$. This limits
the ability of differential precession to cause the ring to settle into the main
disk (Schweizer \textit{et al.} 1983). S0s are preferred probably because the
disk is generally clean of interstellar gas and dust, allowing the
high-inclination orbital material to pass unimpeded through the plane.

The most easily recognisable polar ring galaxies are those where the two disks
are seen nearly edge-on (Whitmore \textit{et al.} 1990).  Cases where only one
disk is edge-on are less obvious unless the main disk is a spiral rather than an
S0. Cases like NGC\,660 (Whitmore \textit{et al.} 1990) and ESO\,235$-$58 (Buta
\& Crocker 1993b) both have been classified as barred spirals, but both show an
aligned dust lane along their apparent `bar' that demonstrates convincingly
that their bars are actually edge-on disks. Thus, these galaxies show what the
accreted high-inclination ring material looks like in a more face-on view. 
Cases where neither disk is edge-on are also possible but even harder to
recognise. Some possibilities include NGC\,1808, NGC\,4772, and NGC\,6870.

Collisional ring galaxies (also simply called `ring galaxies') are cases where
a small galaxy has crashed down the rotation axis of a larger disk galaxy,
triggering a radially expanding density wave (Lynds \& Toomre 1976). Multiple
rings are possible. Different morphologies represent different encounter
parameters and different time frames. To be classified as collisional, there
must be a viable intruder (Madore \textit{et al.} 2009). Struck (2010) provides
a useful overview of ring galaxy theoretical studies as compared with the
limited observational material available.

The three catastrophic ring types are all much rarer than resonance rings. For
example, according to Madore \textit{et al.} (2009), only one in a thousand
galaxies is a ring galaxy.

\begin{figure}
\centering{\includegraphics[width=\textwidth]{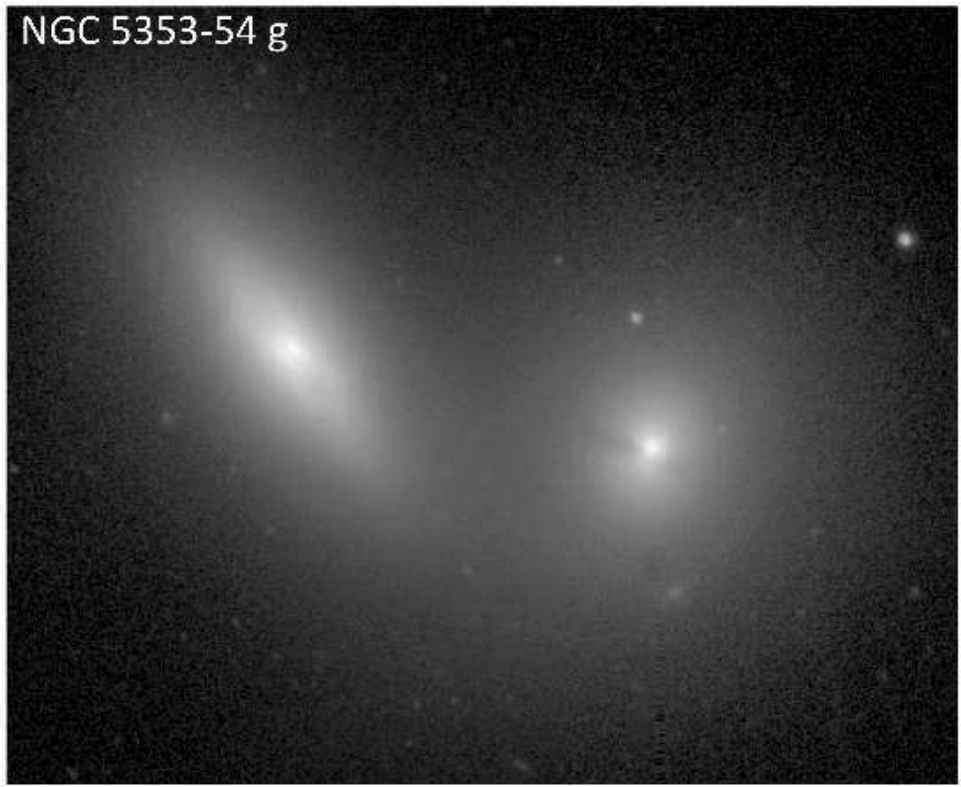}}
\caption{NGC\,5353 (right) is an example of a minor axis dustlane ETG. Its
companion, NGC\,5354, also has an inner dustlane.} 
\label{dust-lane-Es} 
\end{figure}

\subsection{Interaction and merger morphologies}

There are several categories of interaction and merger-driven morphologies. A
{\it dustlane} elliptical is an E or E-like galaxy showing lanes of obscuring
dust (Fig.~\ref{dust-lane-Es}). Minor axis, major axis, and misaligned lanes are
found.  Whether these should be classified as `ellipticals' or not was
controversial, as de Vaucouleurs had once quipped: `If an elliptical shows
dust, then it's not an elliptical'. Bertola (1987) established dustlane Es as
a class of interacting galaxies where a small gas-rich companion undergoes a
minor merger with a more massive E galaxy (Oosterloo \textit{et al.} 2002). The
current general view of these objects is to call them `dustlane early-type
galaxies' (or dustlane ETGs; Kaviraj \textit{et al.} 2011).

\begin{figure}
\centering{\includegraphics[width=\textwidth]{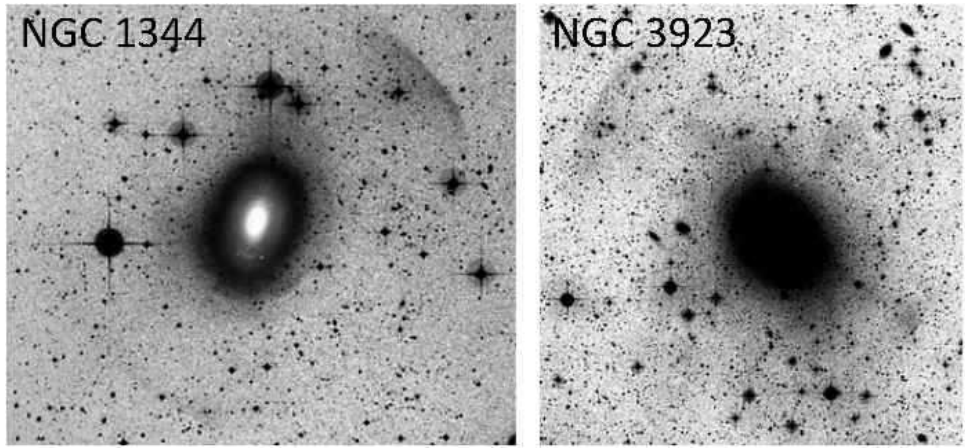}}
\caption{Two `shell' elliptical galaxies (copyright David Malin, reproduced with
permission).} 
\label{shells}
\end{figure}
\begin{figure}
\centering{\includegraphics[width=\textwidth]{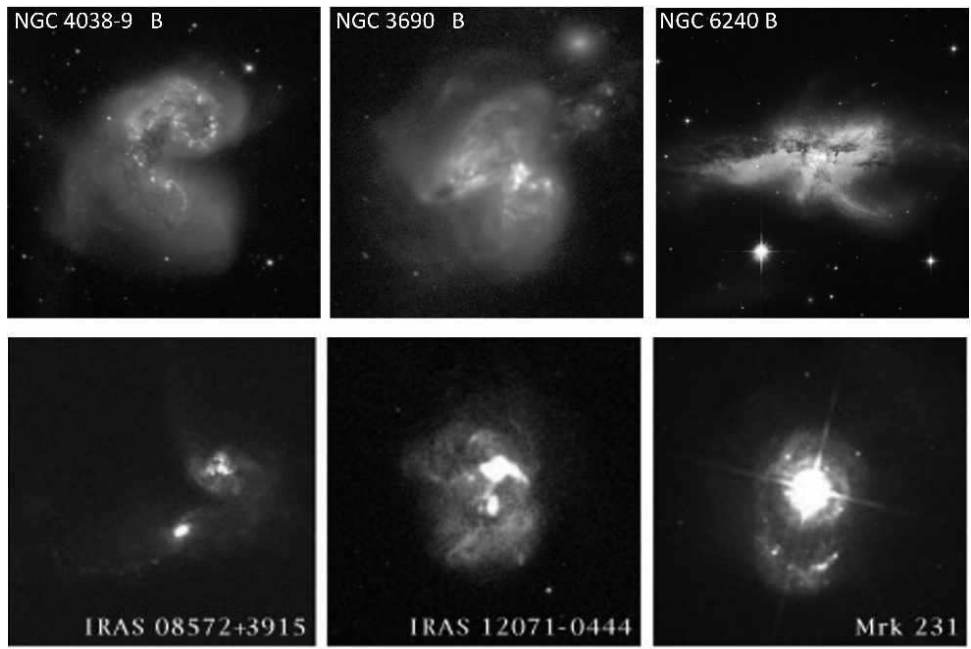}}
\caption{Three cases of advanced and ongoing mergers (top row, from B13) and 
three ULIRGs (bottom row, from Surace \textit{et al.} 1988, reproduced with
permission).} 
\label{mergers} 
\end{figure}

A shell galaxy is a normal elliptical or S0 galaxy showing faint ripples or
ellipsoidal, sharp-edged features in its outer regions. Well-known examples are
NGC\,1344 and NGC\,3923 (Fig.~\ref{shells}). Shells are likely to be 3D in\linebreak
geometry, not parts of a disk.  Schweizer \& Seitzer (1988) considered the term
`shell' as imposing an unjustified interpretation on the shapes of the
features, and suggested the term `ripples' instead.  The extent of shells is
huge, $>$\,100\,kpc in some cases.  Typically, the features are interleaved in
radius, although some may have an `all around' pattern (Prieur 1988). The best
theory suggests that shells/ripples are remnants of a minor merger between a
massive E galaxy and a small, cold, disk-shaped galaxy (e.g., Quinn 1984). The
shells/ripples are thought to represent the maximum excursions of the disrupted
orbits of the small companion's stars.

Tidal tails and bridges are common features of closely interacting but not
necessarily merging disk galaxies. Tidal tails are a consequence of the tidal
field due to the interaction and the shearing off of stars from the rotating
disks.  In M\,51-type interacting pairs, a small companion is seen near the end
of a bright spiral arm, as in M\,51.

Figure~\ref{mergers} shows three examples of ongoing and advanced mergers. In
these cases, the identities of the two separate galaxies are becoming less
distinct, although the two separate nuclei may still be seen. The most extreme
of such interactions may lead to ultra-luminous infrared galaxies (ULIRGs)
(Fig.~\ref{mergers}, lower frames). These objects have 
$L_{\rm IR}$\,$>$10$^{12}$\,$L_{\odot}$\,$\approx$\,$L_{\rm bol}$ of QSOs. As 
shown by Surace \textit{et al.} (1988), high-resolution \textit{HST} images of 
ULIRGs reveal clear evidence of likely strong interactions. These authors suggest 
that warm ULIRGs are transition cases to optical QSOs.

\begin{figure}
\centering{\includegraphics[width=\textwidth]{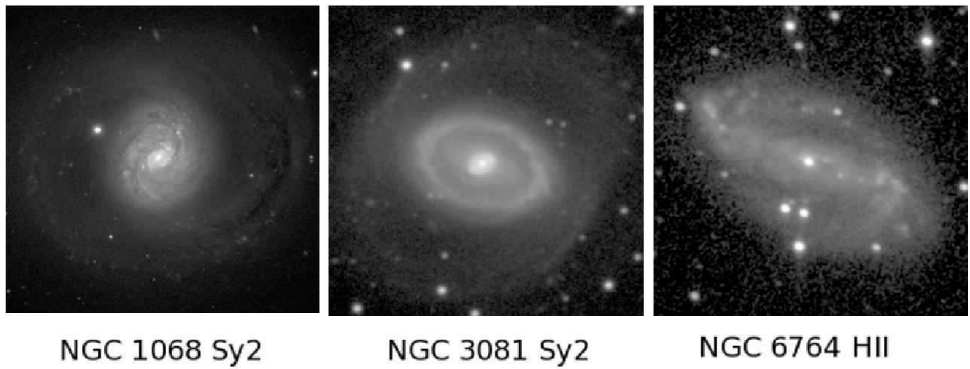}}
\caption{Three active galaxies (B13).}
\label{actives}
\end{figure}

\subsection{The morphology of active galaxies}

The global morphology of active galaxies has always been of interest as people
seek to understand the triggering mechanism of active galactic nuclei (AGN).
Figure~\ref{actives} shows three examples of active galaxies that highlight some
of the features that have been implicated: rings, bars, and interactions.\linebreak  Hunt
\& Malkan (1999) found that the frequencies of outer rings and inner/outer ring
combinations are three to four times higher in Seyferts than in normal spirals. 
Knapen \textit{et al.} (2000) also showed that bars are more frequent in active
galaxies than in non-active galaxies.

With \textit{HST} resolution, the details of the host galaxies of quasars have
been imaged. Bahcall \textit{et al.} (1997) obtained \textit{HST} images of
quasars that show a variety of host morphologies, including Es, interacting
pairs, systems with obvious tidal disturbances; and normal-looking spirals.
These authors concluded that interactions may trigger the quasar phenomenon.

\subsection{The morphology of brightest cluster members}

\begin{figure}
\centering{\includegraphics[width=\textwidth]{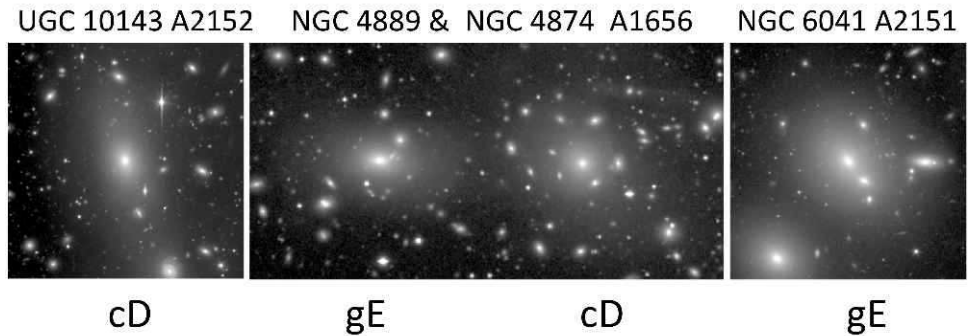}}
\caption{Four brightest cluster members as seen in optical images (B13).
Classifications are from Schombert (1986).} 
\label{bcms} 
\end{figure}

These are the extremely luminous galaxies often found in the centres of rich
galaxy clusters. Two of these were shown in Lecture 1 (see
Section~\ref{sec:lecture1_historical}, Fig.~\ref{morgan-cDs}). Figure~\ref{bcms} shows
several more examples. Schombert (1986, 1987, 1988) classified the brightest 
cluster members according to luminosity profile shape:

\begin{enumerate}[(a)]\listsize
\renewcommand{\theenumi}{(\alph{enumi})}

\item gE, giant ellipticals, 
\item D, larger and more diffuse, with shallower profiles, than gEs, and
\item cD, same as D but with a larger extended envelope. These are the Morgan supergiant types.

\end{enumerate}

The properties of the brightest cluster members fit well with merger 
simulations, including accretion and cannibalism (Schombert 1988).

\subsection{Warped disks}

\begin{figure}
\centering{\includegraphics[width=\textwidth]{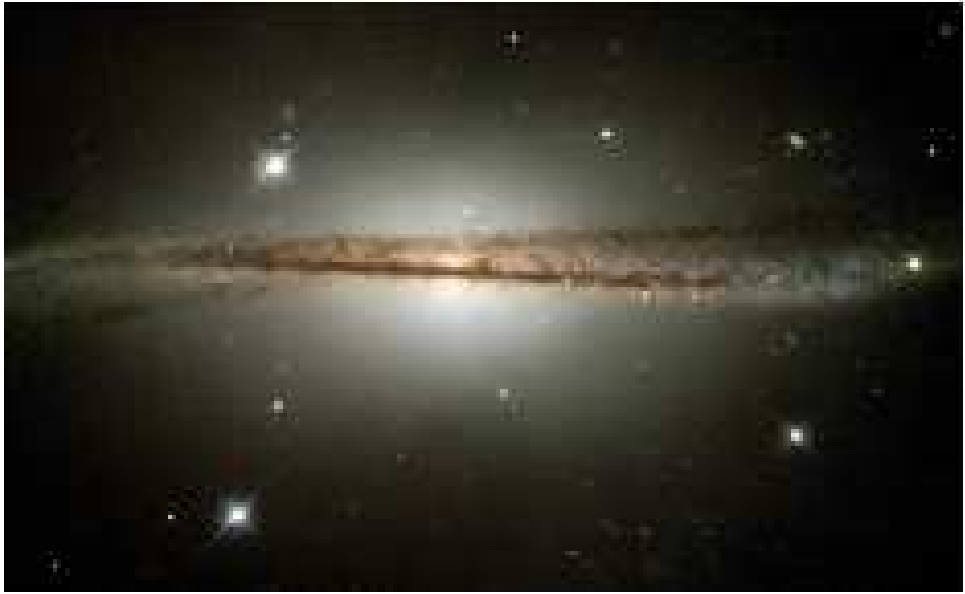}}
\caption{The warped disk galaxy ESO\,510-13 (Hubble Heritage).}
\label{eso510-13}
\end{figure}

Edge-on views of many disk galaxies reveal a clear warping of the outer disk
light. An excellent example is found in ESO\,510-13 (Fig.~\ref{eso510-13}). B13
shows three additional examples: NGC\,4762, NGC\,4452, and UGC\,3697. These are
all extreme cases detectable in optical light. Many more cases are found when
the H{\sc i} layer is considered. The most promising interpretation of warps is
that they are connected to the properties of dark matter haloes. For example,
the stellar disk may be slightly misaligned with the equatorial symmetry axis of
the dark halo. In this circumstance, the inner, more tightly bound part of the
disk remains perfectly flat, while the outer parts of the disk are bent towards
the equatorial plane of the halo, leading to warping of the disk at large radii
(Binney \& Tremaine 2008).

Do warps secularly evolve? Debattista \& Sellwood (1999) argued that a warp
would have a `winding problem' if the effects of a misaligned halo are not
taken into account. Dynamical friction with a misaligned halo can drive a
long-lived warp. Still, warps so driven are expected to be transient, because
the friction force decays over time.

\begin{figure}
\centering{\includegraphics[width=\textwidth]{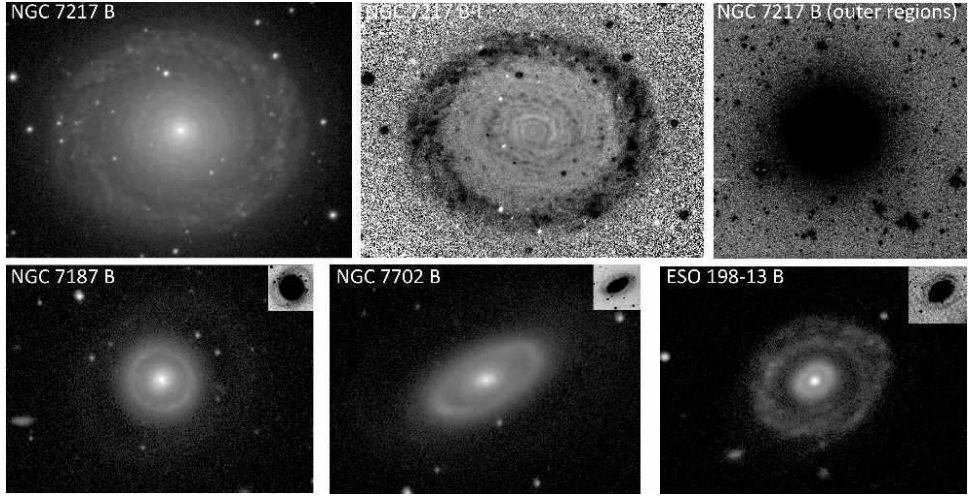}}
\caption{Examples of non-barred ringed galaxies.}
\label{non-barred-rings}
\end{figure}

\subsection{Non-barred ringed galaxies}

This is an important but mostly neglected class of galaxies. While the majority
of ringed galaxies are barred, a non-negligible fraction of non-barred ringed
galaxies exists. Unlike barred galaxy rings, which are fairly homogeneous in
metric and morphological properties, non-barred galaxy rings have a larger
dispersion in properties that seems to point to a variety of mechanisms of ring
formation.

NGC\,7217 is a well-studied nearby example showing a spectacular blue outer ring
in colour index maps (Buta \textit{et al.} 1995b; Fig.~\ref{non-barred-rings},
top row). At low light levels, the isophotes in NGC\,7217 become almost exactly
round, suggesting an extreme bulge-dominated system like the Sombrero Galaxy, 
NGC\,4594. The strong bulge could therefore be the reason NGC\,7217 is
non-barred. Buta \textit{et al.} (1995b) suggested that the strong outer ring
formed in response to a subtle broad oval in the mass distribution.

Other examples appear to be cases of former barred galaxies that in the course
of secular evolution lost their primary bars. A good possible example of this is
NGC\,7702, a double-ringed late S0 that looks very much like an early-type
barred galaxy with inner and outer rings and a nuclear bar (Buta 1991;
Fig.~\ref{non-barred-rings}, bottom row). However, no clear primary bar crosses
the bright inner ring, although this ring is undoubtedly oval in intrinsic shape
and therefore bar-like. Bar dissolution is a real possibility that may be tied
to the gradual build-up of a strong central mass concentration (e.g., Hasan \&
Norman 1990).

Other non-barred ringed galaxies could involve galaxies that accreted a small
companion, as described in Lecture 3 (see Section~\ref{sec:lecture3}). Hoag's
Object, IC\,2006, and NGC\,7742 would all be described as non-barred, and it is
clear that a mixture of such objects, ex-barred galaxies, and possibly
tidally driven rings like in NGC\,4622 could account for the larger dispersion
in the properties of non-barred galaxy rings.

\subsection{Counter-winding spirals}

This is a very rare, possibly interaction- or minor merger-driven morphology
where a disk-shaped galaxy has two non-overlapping spiral patterns that wind
outward in opposite senses.  While the bulk of spirals have been demonstrated to
be trailing the direction of rotation (de Vaucouleurs 1958), and the Toomre
(1981) swing amplification mechanism (as well as the Lynden-Bell \& Kalnajs 1972
mechanism) seems to explain why (Binney \& Tremaine 2008), counter-winding
spirals appear to present genuine examples of {\it leading} spiral arms.

\begin{figure}
\centering{\includegraphics[width=\textwidth]{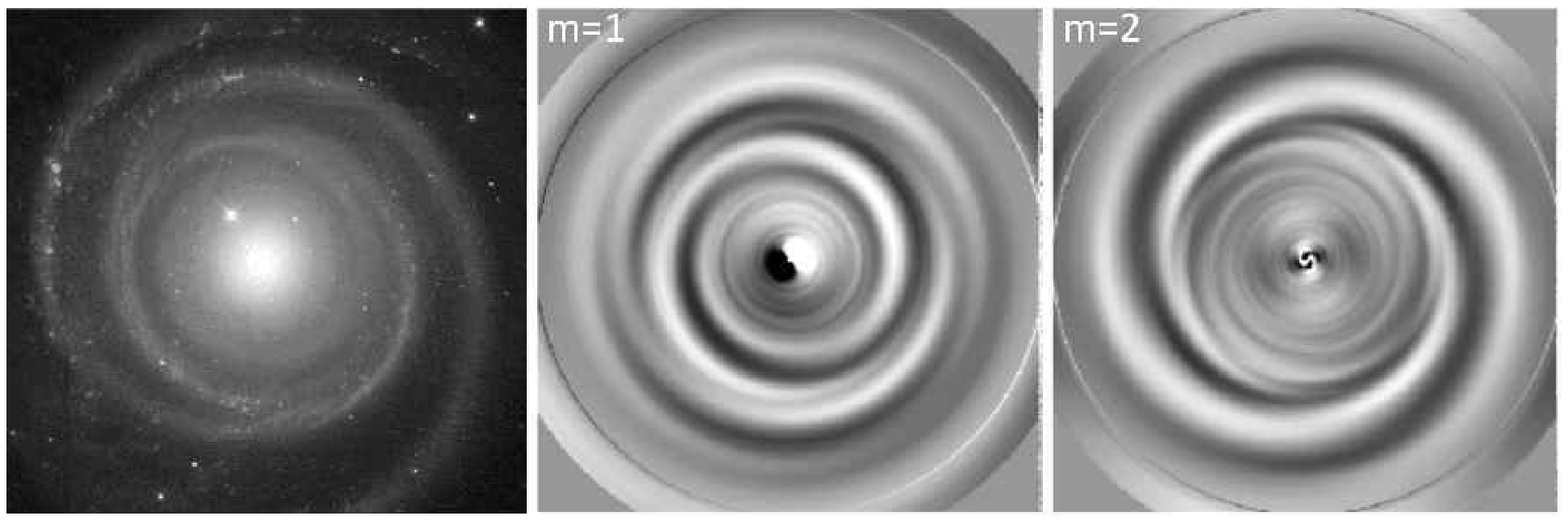}}
\centering{\includegraphics[width=\textwidth]{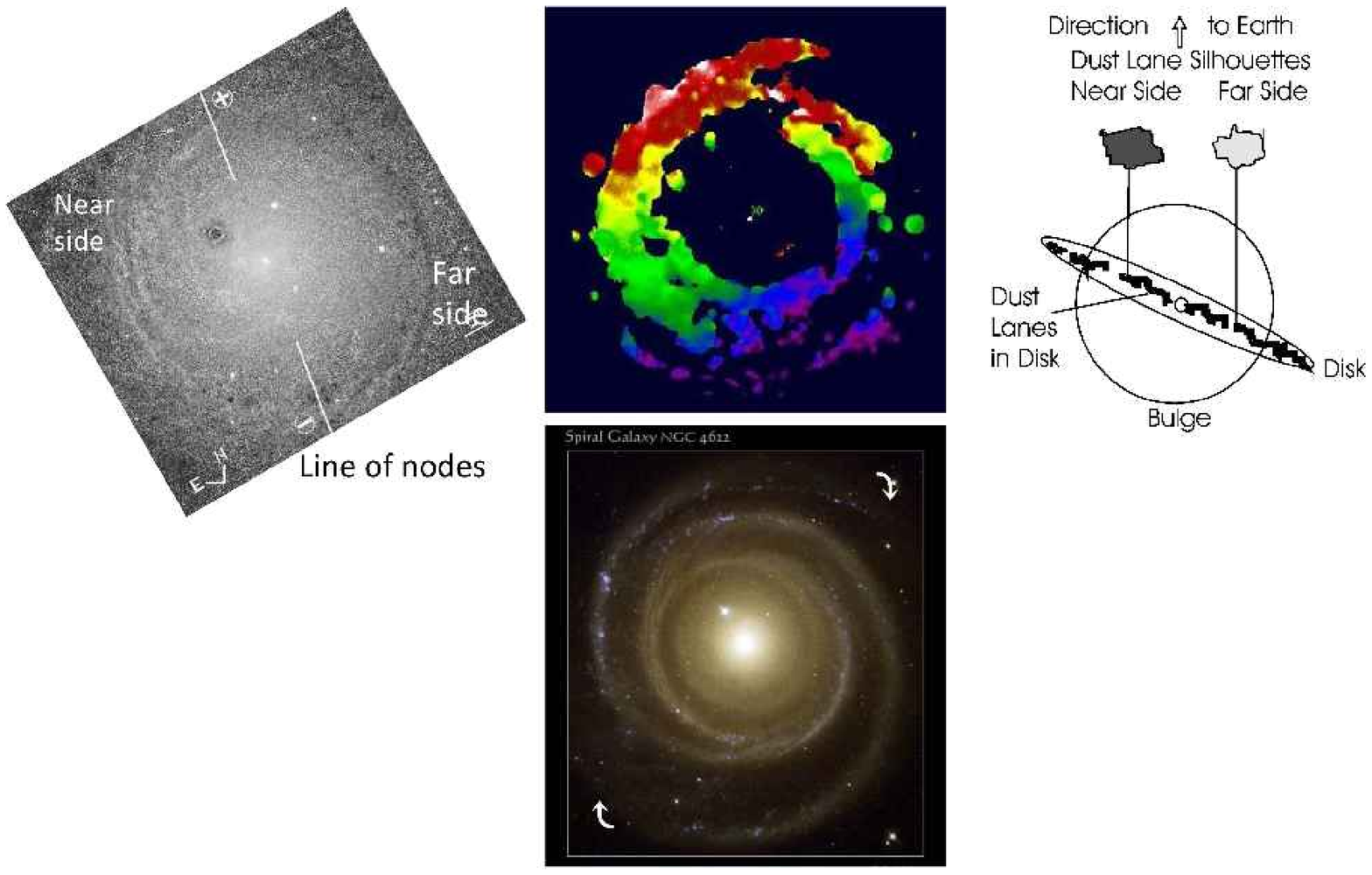}}
\caption{(Top row): $V$-band image of NGC\,4622 with $m$\,=\,1 and 2
Fourier decompositions (Buta \textit{et al.} 2003). (Bottom frames): $V$--$I$ colour
index map (left), Fabry-Perot velocity field (upper, middle), colour
image (lower, middle), all from Buta \textit{et al.} (2003). Also shown at
right is a schematic illustrating how tilt leads to an extinction and
reddening asymmetry across the galaxy minor axis. The arrows on the colour
image indicate the implied sense of rotation.}
\label{ngc4622} 
\end{figure}

Determining the sense of winding of spiral arms has typically depended on two
things: knowing which half of the galaxy is receding from us, and which side of
the galaxy is nearer to us. From the schematic in Fig.~\ref{ngc4622}, right, it
can be seen that on the near side, the bulge is viewed through the dust layer,
while on the far side, the dust layer is viewed through the bulge. This leads to
a reddening and extinction asymmetry that can be seen in the colour index map of
NGC\,7331 in Fig.~\ref{dust1} of Lecture 3 (see Section~\ref{sec:lecture3}).
Once the near-side is established, the velocity field tells us which
way the galaxy must be turning. With these two pieces of information
together, the sense of winding of the spiral arms can be reliably
judged. Of course, the less inclined a galaxy is, the less evident is
the nearside extinction and reddening effect.

The first counter-winding spiral identified, and the one that is best studied,
is the nearly face-on SA(r)a spiral NGC\,4622 located in the Centaurus cluster
(Byrd \textit{et al.} 1989). The galaxy has two high-contrast outer arms and a
single lower-contrast inner arm, with no overlap between the patterns
(Fig.~\ref{ngc4622}, top row). Instead, the single inner arm and the two outer
arms blend at the position of a conspicuous, offset inner ring. Buta {\it et
al.} (2003) obtained \textit{HST} imaging and a ground-based H$\alpha$ velocity
field, and used the dust distribution to judge the near-side and the velocity
field to judge the sense of rotation (Fig.~\ref{ngc4622}, bottom frames). A
$V$--$I$ \textit{HST} colour index map showed thin curved dust lanes all lying on
one side of the line of nodes, which can be identified as the near side. With
the velocity field telling us that the north side is receding relative to the
nucleus, this implied a\linebreak clockwise rotation of the disk. This gave the surprising
and unexpected result that the two strong outer arms, not the weaker inner arm,
have the leading sense, a result that is difficult to accept. In the centre a
small edge-on dust lane is found that suggests the galaxy has suffered a recent
minor merger that could be responsible for the peculiar morphology.

A second example of a counter-winding spiral was identified by Buta (1995). This
was the Sb-Sbc spiral ESO\,297-27. Although not in a cluster\linebreak environment,
ESO\,297-27 has similarities to NGC\,4622. In this case, the inner arm leads and
the outer arms trail (Grouchy \textit{et al.} 2008). Other galaxies with leading
spiral structure:

\begin{enumerate}[(a)]\listsize
\renewcommand{\theenumi}{(\alph{enumi})}

\item NGC\,3124 (bar in ring; Purcell 1998). See Fig.~\ref{skewed-bars} (Section~\ref{sec:lecture2}),
\item NGC\,6902 (bar in ring), and
\item IRAS\,182933413 (two leading arms, not counter-winding; Vaisanen 
\textit{et al.} 2008).

\end{enumerate}

What causes counter-winding spiral structure? The phenomenon is very rare, so
perhaps an unusual circumstance is at work.  If the two outer arms of NGC\,4622
are really leading, they would be difficult to explain in current theories of
spiral structure. Swing amplification depends on the swing of a leading density
wave. Also, the comprehensive study of Lynden-Bell \& Kalnajs (1972) showed that
only a trailing spiral pattern can transfer angular momentum outwards, which
allows the wave itself to be maintained.

\begin{figure}
\centering{\includegraphics[width=\textwidth]{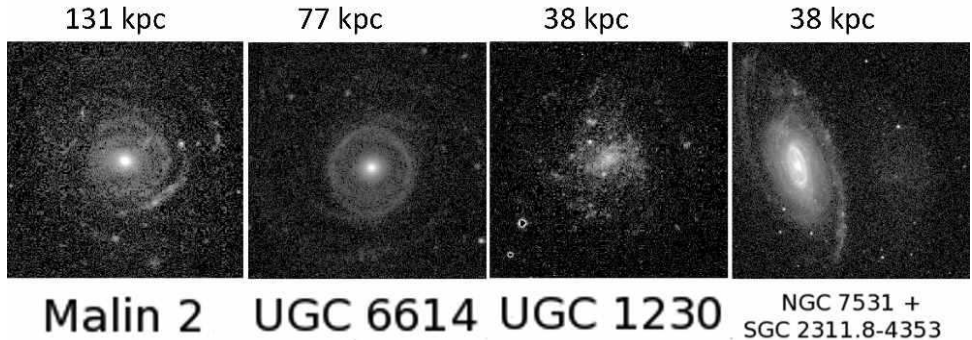}}
\caption{Giant low surface brightness galaxies, from B13 and references therein. 
The number above each box is the length of the top side at the distance of
the galaxy.} 
\label{glsbs} 
\end{figure}

\subsection{Giant low-brightness galaxies and stellar streams}

Giant low surface brightness galaxies were first identified by Bothun \textit{et
al.} (1987). They are galaxies having a relatively normal bulge and an extremely
low surface brightness, very large disk. The first example found was Malin~1
(Bothun \textit{et al.} 1987). Figure~\ref{glsbs} shows other examples: Malin~2,
UGC\,6614, UGC\,1230, and in addition a strange possibly unrelated case called
SGC\,2311.8$-$4353, which was first studied in detail by Buta (1987). 
SGC\,2311.8$-$4353 is a very low surface brightness companion to the giant
spiral NGC\,7531, and is notable for having an isophotal diameter 2/3 the size
of NGC\,7531. If at the same distance as NGC\,7531, SGC\,2311.8$-$4353 would be
a dwarf in luminosity but not in size. That is, it would be a large low surface
brightness galaxy, but not a giant. In a recent study of tidal streams in
late-type spirals, Mart\'inez-Delgado \textit{et al.} (2010) have interpreted
SGC\,2311.8$-$4353 as a companion in the act of tidal disruption by NGC\,7531.
These authors show how commonly `normal' late-type galaxies can show extremely
low surface brightness tidal features that are analogous to the streams found
around the Milky Way.

\begin{figure}
\centering{\includegraphics[width=\textwidth]{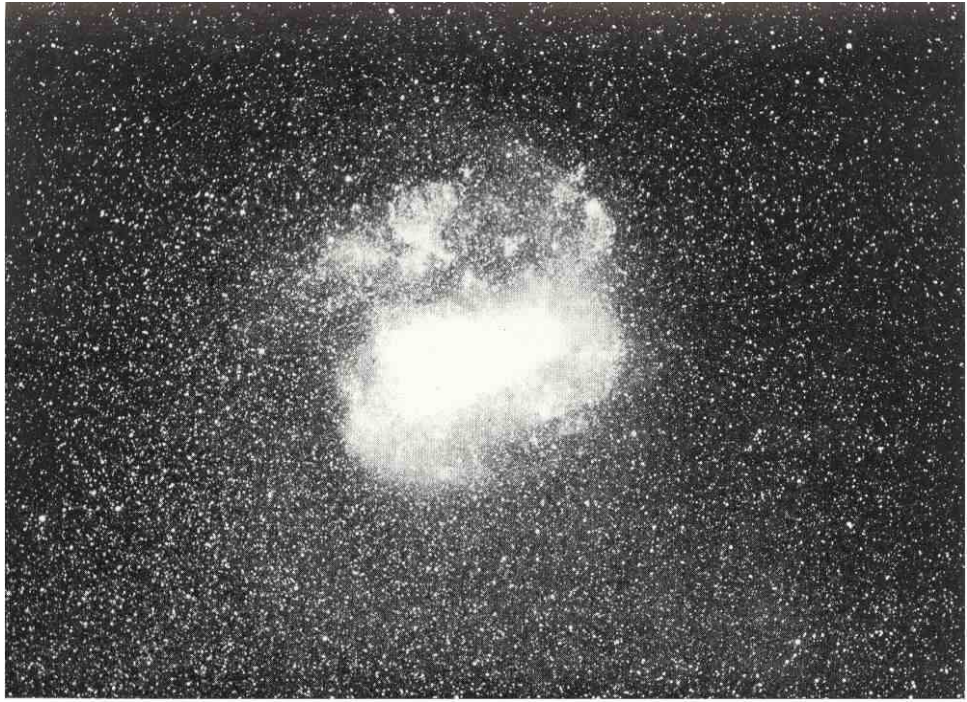}}
\caption{The LMC, the prototype of classification SB(s)m. Reproduced with
permission from de Vaucouleurs \& Freeman (1972).}
\label{lmc}
\end{figure}

\subsection{Magellanic barred spiral galaxies}

These were identified as a distinct morphological class by de Vaucouleurs \&
Freeman (1972; Fig.~\ref{lmc}; see also Lundmark 1927 and Lecture 1,
Section~\ref{sec:lecture1_historical}). The class is characterised by a bar with no
bulge, a single main spiral arm, shorter spiral features, and an offset of the
centre of the bar from the centre of outer isophotes. De Vaucouleurs \& Freeman
noted that SBm galaxies often come in pairs (e.g., the large Magellanic cloud
[LMC] and small Magellanic cloud [SMC], NGC\,4618 and NGC\,4625, NGC\,2537 and
NGC\,2537A), suggesting that the morphology may be interaction-driven. Although
the bar is usually offset in these galaxies, and in the LMC and SMC the bar is
also kinematically offset from the rotation centre, prominent cases like
NGC\,4027 were found to have a rotation centre coincident with the centre of the
bar (Pence \textit{et al.} 1988). Too few cases have been observed to
definitively prove that an offset rotation centre is characteristic of these
galaxies.

\begin{figure}
\centering{\includegraphics[width=\textwidth]{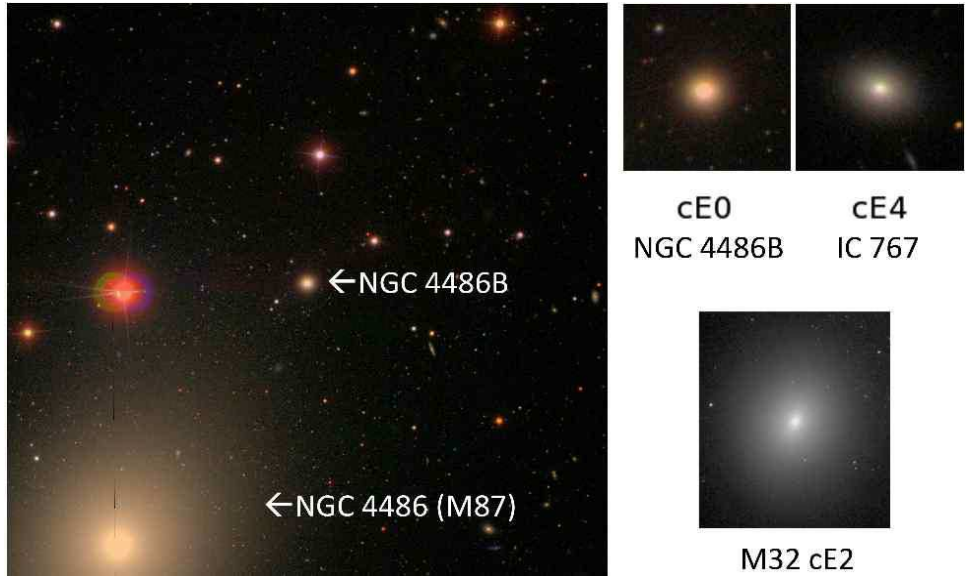}}
\caption{Examples of compact elliptical galaxies.}
\label{ces}
\end{figure}

\subsection{Compact ellipticals}

Compact galaxies (Fig.~\ref{ces}) have been known since Zwicky \& Zwicky (1971)
published a major catalogue. The departure of three compact elliptical (cE)
galaxies, M\,32, NGC\,4486B, and NGC\,5846A, from the norm in a graph of
$V$-band surface brightness versus absolute $V$-band magnitude for E galaxies
led Faber (1973) to suggest that these galaxies could be the stripped cores of
formerly larger elliptical galaxies. However, it is now known that cE galaxies
are simply the lower-luminosity tail of normal elliptical galaxies, based on
photometric parameter correlations (Kormendy 1985; Kormendy \textit{et al.}
2009). That is, cE galaxies are true `dwarf ellipticals'. These are to be
contrasted with a large number of Virgo cluster objects classified as type
`dE' by Bingelli {\it et al.} (1985) that have photometric properties more
akin to dwarf irregulars than to normal ellipticals. As I have already noted in
Lecture 1 (see Section~\ref{sec:lecture1_fork}), Kormendy \& Bender (2012) have
interpreted the Bingelli \textit{et al.} (1985) dE, dE,N, dS0, and dS0,N types
as environmentally driven morphologies, former irregular and very late-type
galaxies that were stripped, harassed, or otherwise windblown to lose their gas.

\begin{figure}
\centering{\includegraphics[width=\textwidth]{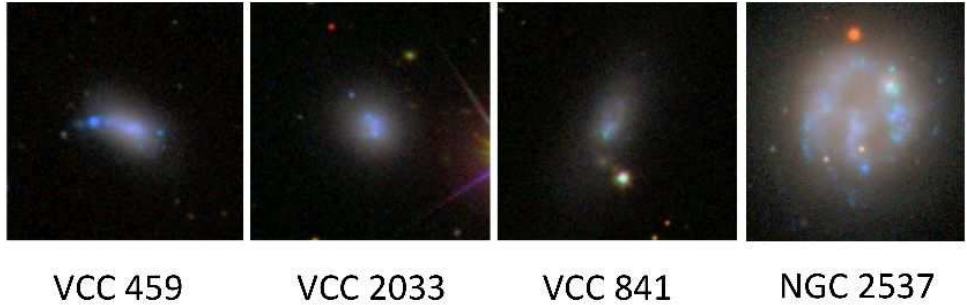}}
\caption{Examples of blue compact dwarf galaxies.}
\label{bcds}
\end{figure}

\subsection{Blue compact dwarf galaxies}

These are small, high surface brightness galaxies experiencing a strong
starburst (Fig.~\ref{bcds}). As defined by Gil de Paz \textit{et al.}
(2003), blue compact dwarf (BCD) galaxies have $(\mu_B-\mu_R)_{\rm
peak}\leq1$, $\mu_B<22$ mag arcsec$^{-2}$, and $M_K >-21$. On average,
BCDs have $B-R=0.7\pm0.3$, $M_B=-16.1\pm1.4$, and $\log L_{{\rm
H}\alpha}=40.0\pm0.6$. BCDs have been identified in the Virgo cluster
by Bingelli \textit{et al.} (1985; three examples are shown in
Fig.~\ref{bcds}), but are also known in field environments. In the
Local Group, IC\,10 has been identified as a BCD (Richer \textit{et
al.} 2001).  Figure~\ref{bcds} shows an SDSS colour image of BCD
NGC\,2537.

The BCD class has brought to light some of the most extreme cases of
star-forming galaxies. For example, the BCD I\,Zw\,18 was once thought to be a
genuine young galaxy with a high level of star formation and an optical
morphology greatly affected by stellar winds and supernovae from an earlier
starburst. However, deep imaging with \textit{HST} revealed a
definite older stellar population in I\,Zw\,18 (Contreras-Ramos \textit{et al.}
2011).

\subsection{Ultra-compact dwarf galaxies}

These are very compact, star-like galaxies with luminosities comparable to dE
galaxies (Drinkwater \textit{et al.} 2000, 2003). Hilker (2011) reviews the
properties of a significant population of ultra-compact dwarf galaxies (UCDs) in
the Fornax cluster. These objects have absolute magnitudes in the range
$-13.4$\,$<M_V$\,$<$\,$-11.4$, and are brighter than the brightest Milky Way globular
cluster but fainter than compact ellipticals. One interpretation of these
objects is that they are the threshed nuclei of dE,N galaxies (that is,
environmentally driven morphologies where a dE,N galaxy has lost all of its
stars, except for the `N').

\subsection{Isolated galaxies}
For investigations of the connection between morphology and secular evolution,
there can probably be no better sample than isolated galaxies. For such
galaxies, internal processes would have to be largely driving evolution.

A significant effort into establishing a reliable isolated galaxy
sample was made by Karachentseva (1973) and was further improved by the
Analysis of the Interstellar Medium in Isolated Galaxies (AMIGA)
project (Verdes-Montenegro \textit{et al.} 2005). Out of about 1000 galaxies in
a final sample, Sulentic \textit{et al.} (2006) found that the majority are
Sb-Sc spirals, a few of which were shown in Fig.~\ref{env-density}. The most 
striking thing about many of these galaxies is how
regular and well-defined their spiral patterns are. AMIGA galaxies are
selected to not have had a close encounter with a major galaxy in more
than 3\,Gyr. Do these galaxies support the idea that spiral structure can
arise spontaneously, independent of an interaction?

\subsection{Ultraviolet galaxy morphology}

The ultraviolet (UV) is an important window into star formation in galactic disks.
The recent launch of the \textit{Galaxy Evolution Explorer} (\textit{GALEX},
Martin \textit{et al.} 2005) has provided an extensive database of images at
0.15\,$\mu$m (far-UV, or FUV) and 0.22\,$\mu$m (near-UV, or NUV). Gil de Paz
\textit{et al.} (2007) outline what can be learned from such images:

\begin{enumerate}[(a)]\listsize
\renewcommand{\theenumi}{(\alph{enumi})}

\item young stars are mostly what you see at UV wavelengths in any star-forming 
galaxy,

\item massive early-type galaxies can show significant UV flux due
to the `UV-upturn', caused by hot low-mass horizontal branch stars,

\item UV flux is an excellent tracer of the star formation rate, and

\item UV flux absorbed by dust is re-emitted in the IR.

\end{enumerate}

\begin{figure}
\centering{\includegraphics[width=\textwidth]{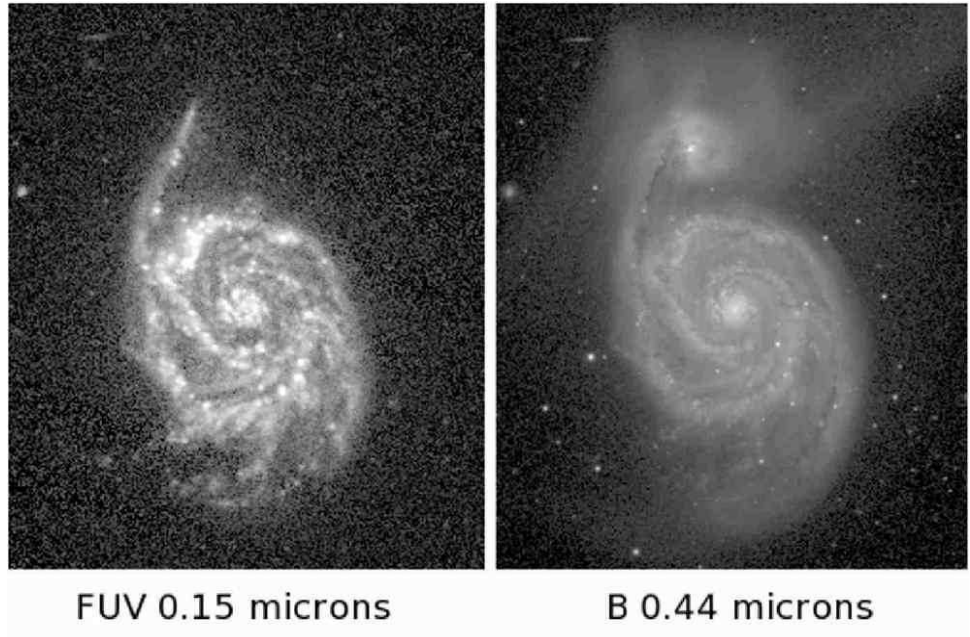}}
\caption{\textit{GALEX} UV image of M\,51 as compared to a typical $B$-band 
image.}
\label{uv-m51}
\end{figure}

Figure~\ref{uv-m51} shows the \textit{GALEX} FUV image of M\,51 compared to a
normal $B$-band image. The FUV image shows mainly hot stars younger than 10$^8$
years, while the $B$-band shows older stars in addition to dust and star-forming
regions. Most interesting is how the companion galaxy, NGC\,5195, and the
extensive tidal debris around it, are mostly invisible at FUV wavelengths. This
galaxy does not have a UV upturn that would make it more prominent in the FUV
image.

The NUV morphology of a spiral like M\,83 strongly resembles an H$\alpha$
image but with an older background disk.

B13 shows a montage of \textit{GALEX} NUV images of galaxies of different types.
In early-type galaxies, significant UV flux may occur if the galaxy has a
star-forming nuclear ring, as in NGC\,1317 and NGC\,4314. In an intermediate
type barred galaxy like NGC\,3351, the inner ring, nuclear ring, and outer arms
are conspicuous, while the bar is invisible.  In the SBc galaxy NGC\,7479, both
the bar and the arms are seen, while the late-type galaxies NGC\,628 and
NGC\,5474 have extensive UV disks. Most interesting was the discovery of an
extensive UV disk around the diminutive Sm galaxy NGC\,4625, which suggests that
the galaxy is currently forming most of its stars (Gil de Paz \textit{et al.}
2005). Finally, the intriguing star-forming galaxy NGC\,5253 shows bright UV
emission even from its extended diffuse disk.

\subsection{The morphology of the interstellar medium}

The ISM in galaxies has distinctive morphological qualities that are surely, in
some cases, tied to secular evolution. The H{\sc i} Survey of Nearby Galaxies
(THINGS, Walter \textit{et al.} 2008) provides some of the best information on
the H{\sc i} morphology of normal galaxies. In such galaxies we see:

\begin{enumerate}[(a)]\listsize
\renewcommand{\theenumi}{(\alph{enumi})}

\item extended H{\sc I} disks with spiral structure (NGC\,628), 
\item central holes (NGC\,2841) or central spots (M\,81),
\item large gaseous rings (NGC\,2841) or pseudorings (NGC\,2903),
\item small rings (NGC\,4736),
\item supernova-blown holes (WLM), and
\item sometimes huge H{\sc i}/optical sizes (DDO\,154).

\end{enumerate}

Other galaxies, possibly because of their environment, have H{\sc i} disks
comparable in extent to their optical disks, such as NGC\,1433 (Ryder \textit{et
al.} 1996) and NGC\,5850 (Higdon \textit{et al.} 1998). In these cases, the
H{\sc i} morphology follows the optical morphology closely. I have already
described previously the H{\sc i} morphology of galaxies in the central regions
of the Virgo cluster, where stripping has not only truncated the H{\sc i} disk
significantly, but also has erased morphological structures so thoroughly that
the diversity of H{\sc i} morphologies is greatly reduced.

The morphology of molecular hydrogen, H$_2$, is also of interest. The
Berkeley-Illinois-Maryland CO Survey of Nearby Galaxies (BIMA-SONG,
Helfer \textit{et al.} 2003) is one of the most extensive databases of molecular
galaxy morphology available. Using CO 2.6\,mm emission as a tracer of
H$_2$, BIMA-SONG maps reveal the following:

\begin{enumerate}[(a)]\listsize
\renewcommand{\theenumi}{(\alph{enumi})}

\item inner spiral arms (NGC\,628),
\item large rings (NGC\,2841, NGC\,7331), 
\item scattered giant molecular clouds (GMCs; NGC\,2403), 
\item primary bars (NGC\,7479), 
\item nuclear gas bars (NGC\,3351),
\item central spots (NGC\,4535), and
\item small pseudorings (NGC\,1068).

\end{enumerate}

CO morphology does not necessarily mimic H{\sc i} morphology and the CO disk may 
not extend as far as the H{\sc i} disk.

\begin{figure} 
\centering{\includegraphics[width=\textwidth]{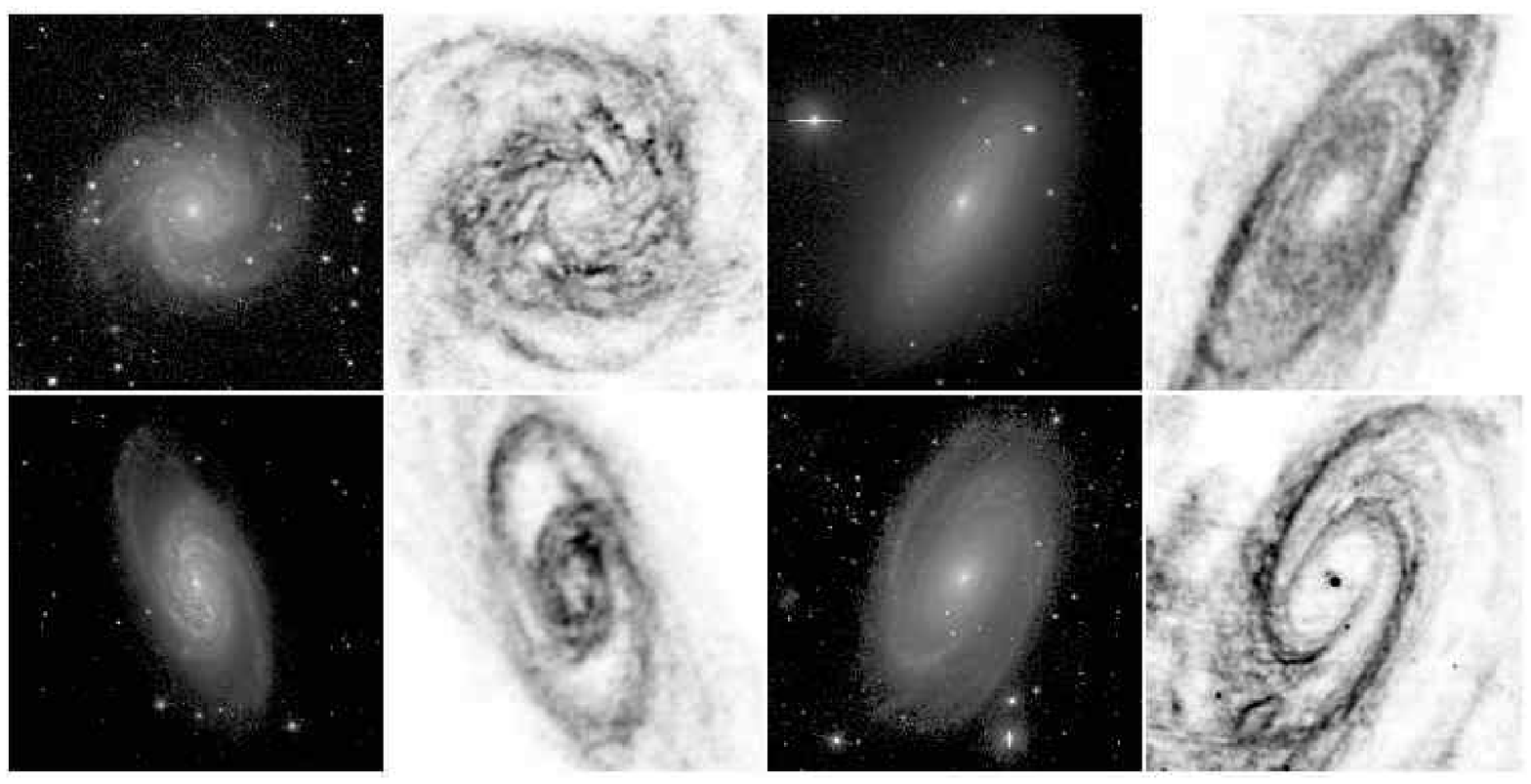}}
\centering{\includegraphics[width=\textwidth]{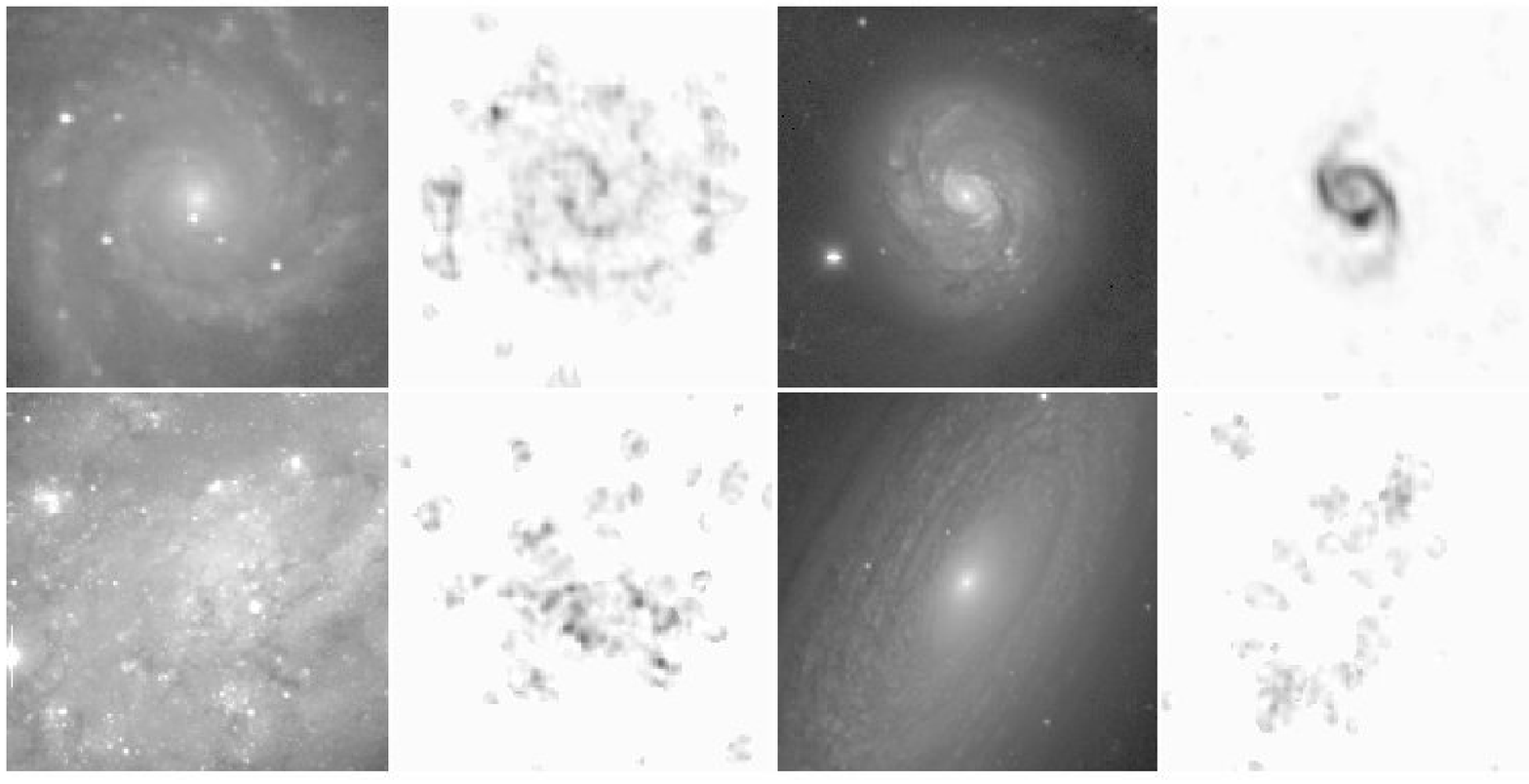}}
\caption{(Top rows): H{\sc i} maps of NGC\,628, NGC\,2841, NGC\,2903, and
NGC\,3031 from THINGS (Walter \textit{et al.} 2008); (bottom rows): CO maps of
NGC\,628, NGC\,1068, NGC\,2403, and NGC\,2841 from the BIMA-SONG (Helfer
\textit{et al.} 2003).}
\label{hi-co} 
\end{figure}

Figure~\ref{hi-co} shows several examples of H{\sc i} and CO maps of nearby
galaxies, with a corresponding optical image for comparison.

\begin{figure}
\centering{\includegraphics[width=\textwidth]{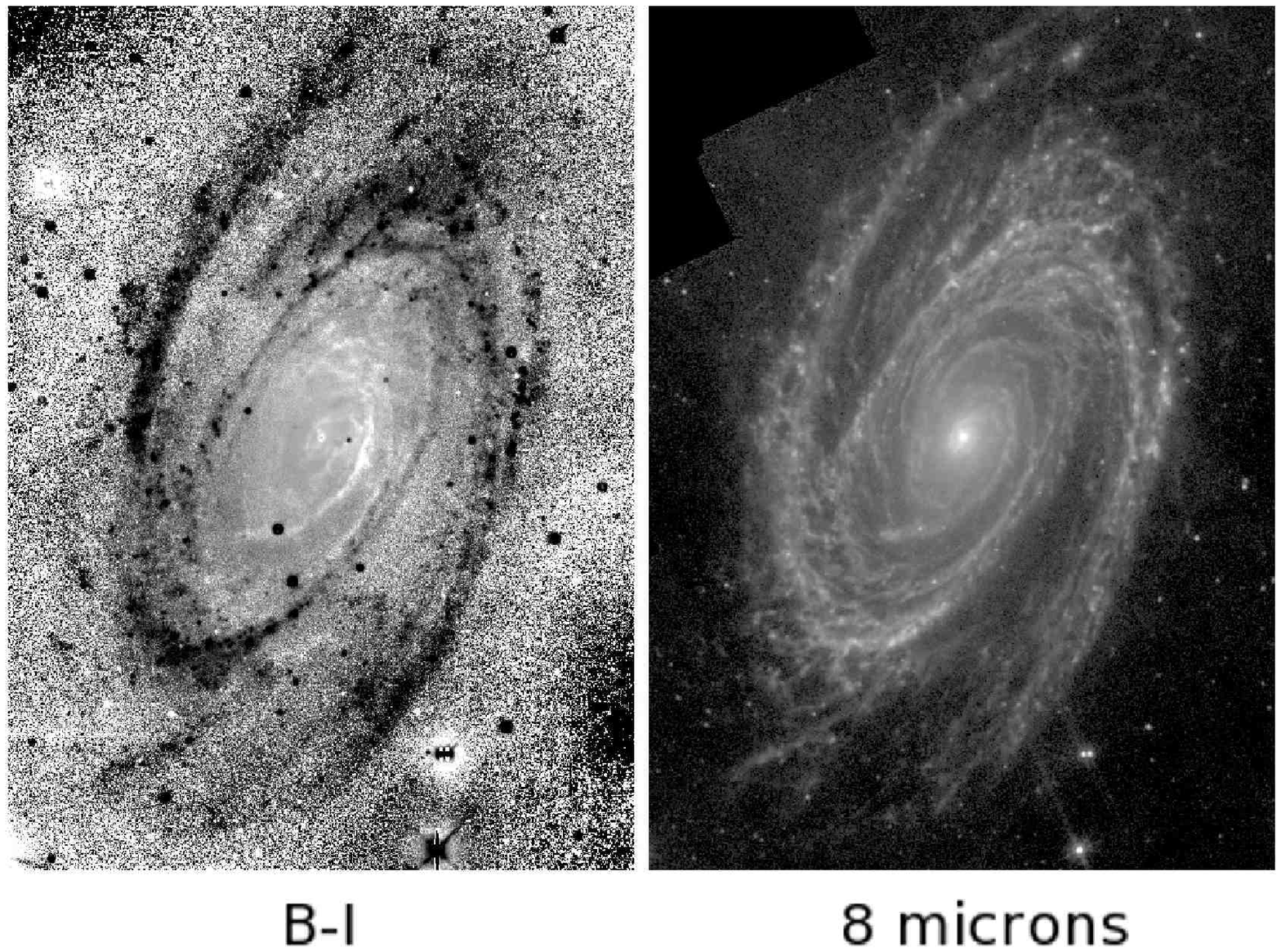}}
\caption{A $B-I$ colour index map of M\,81 as compared to an IRAC 8\,$\mu$m 
image (from B13).} 
\label{m81maps}
\end{figure}

Especially interesting morphology is revealed by 8\,$\mu$m emission maps
obtained with IRAC. This mid-IR wavelength is sensitive to
the warm dust associated with spiral arms and shows no near-side/far-side
asymmetry, as the comparison in Fig.~\ref{m81maps} shows. IRAC maps at 8\,$\mu$m
reveal the distribution of dust directly, instead of partly depending on how the
dust layer projects against the bulge. In optical images, the spiral structure
of M\,81 appears to nearly terminate in a large pseudoring in the outer parts of
the bright bulge. Spiral structure inside this pseudoring appears mainly in
near-side extinction arcs. In contrast, at 8\,$\mu$m there is complex spiral
structure inside the apparent pseudoring that continuously winds outward.

\subsection{High-redshift galaxy morphology}

There is no doubt that high-redshift galaxy morphology has come of age during
the past 10--15 years. The resolution provided by \textit{HST} with the Wide
Field Camera and the Advanced Camera for Surveys (ACS) has provided
morphological information on thousands of high-redshift galaxies. Using both
spectroscopy and photometric techniques, significant redshift ranges can be
isolated to examine galaxy evolution firsthand.

\begin{figure}
\centering{\includegraphics[width=\textwidth]{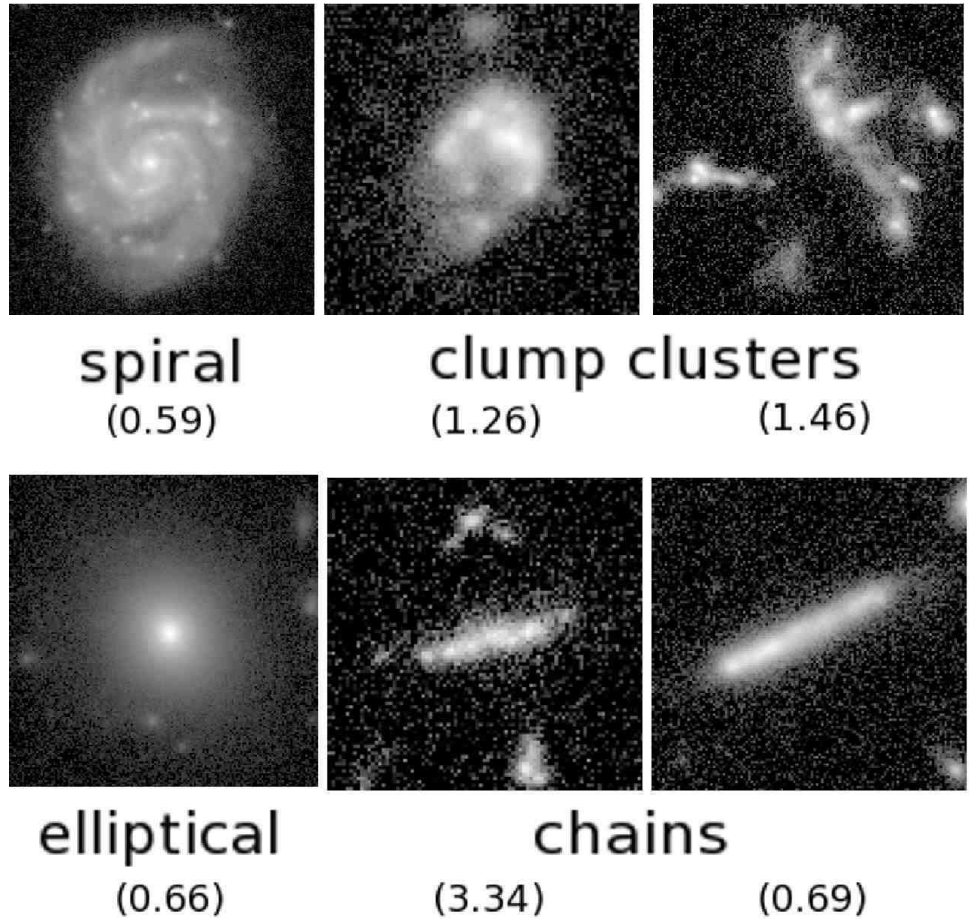}}
\caption{Several high-redshift morphological categories, from B13 and references 
therein. The number in parentheses is the redshift.} 
\label{highz}
\end{figure}

B13 reviews many papers on high-redshift galaxy morphology.  One of the
first results found from the early surveys is that high-redshift
galaxies reveal morphological categories that would fit poorly within
the various modern classification systems. Unusual irregular shapes
dominate the high-$z$ population. Normal spirals and ellipticals can be
recognised to $z\approx0.6$, but as $z\rightarrow1$, the number of
irregular-looking objects becomes more significant. Clump clusters,
linear and bent chains, `tadpoles', catastrophic rings, and mergers
are identified in papers by Elmegreen \textit{et al.} (2004, 2007),
van den Bergh \textit{et al.} (1996,2000), and Cowie \textit{et al.}
(1995). Clump clusters and chains are shown in Fig.~\ref{highz}.

\subsection{The Sloan Digital Sky Survey}

The SDSS (Gunn \textit{et al.} 1998; York \textit{et al.} 2000) is without a
doubt one of the most important assemblages of morphological information on
galaxies since the Palomar Sky Survey. The survey includes morphological,
photometric, and spectroscopic data for a million galaxies, and opened up the
new era of huge extragalactic digital databases of medium-high resolution
imagery. The SDSS also represents the advent of large-scale colour imagery for
galaxies of all types.

\begin{figure}
\centering{\includegraphics[width=\textwidth]{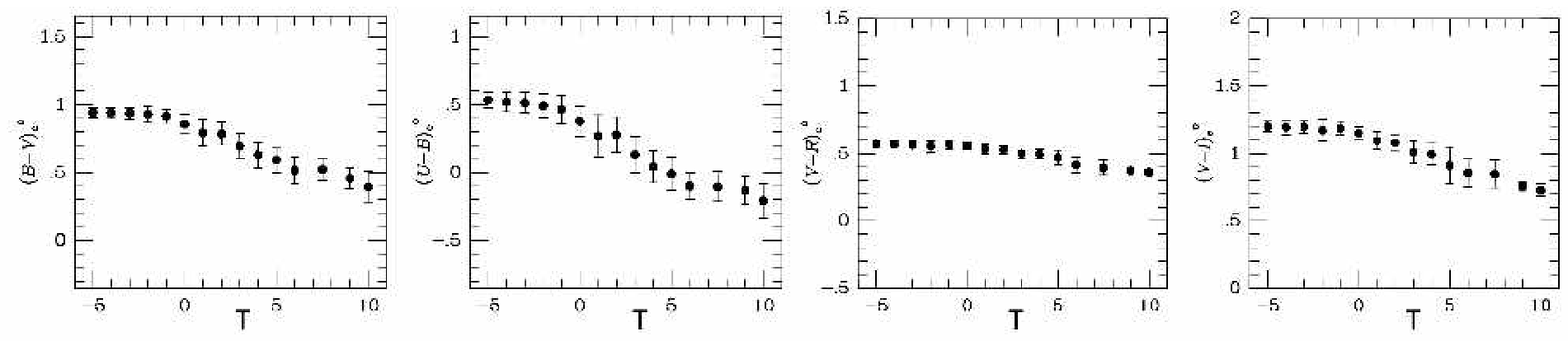}}
\centering{\includegraphics[width=\textwidth]{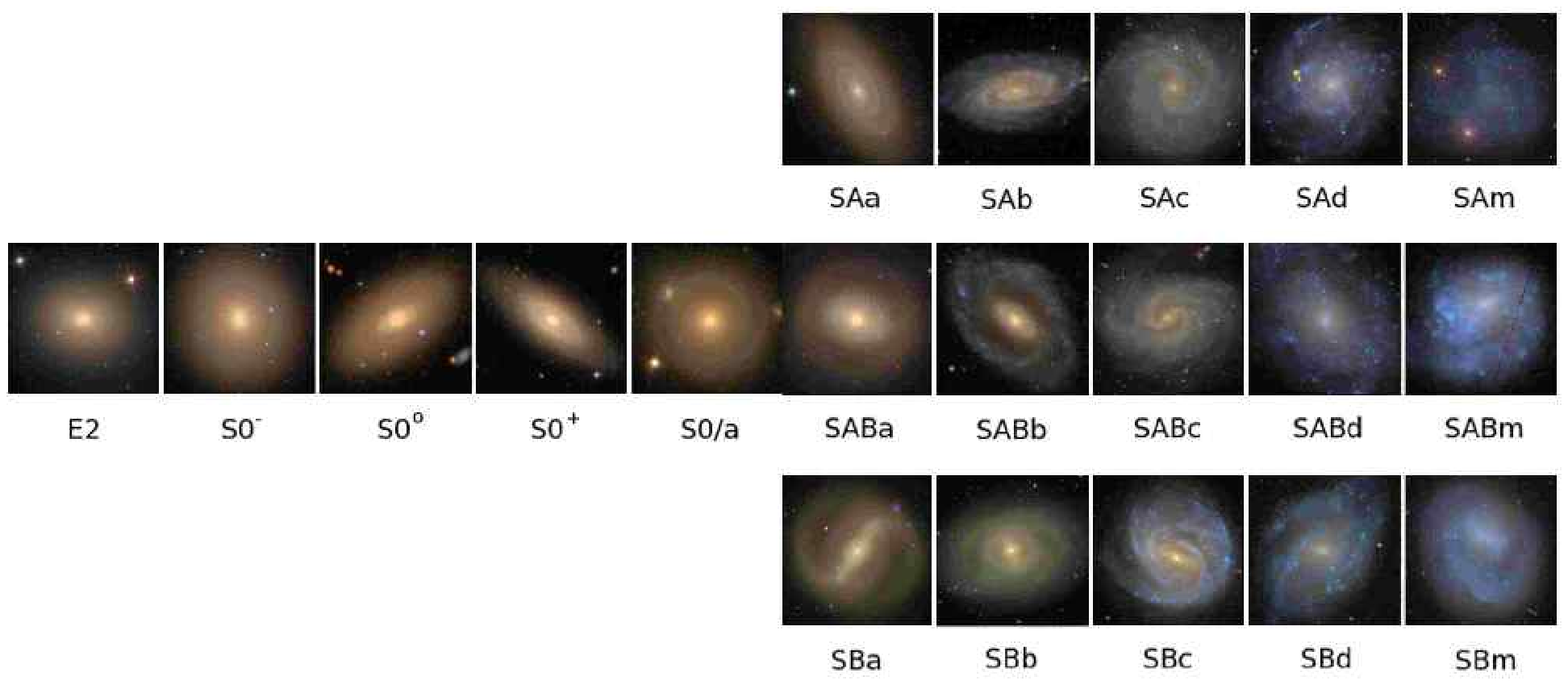}}
\caption{(Top row): quantitative colour versus type relations, from Buta 
\textit{et al.} (1994); from left to right, the colours are $(B-V)_{\rm eo}$,
$(U-B)_{\rm eo}$, $(V-R)_{\rm eo}$, and $(V-I)_{\rm eo}$, where the subscripts mean
the colours are within an effective (half-power) aperture and are
corrected for extinction. (Bottom): a colour tuning fork based on SDSS colour 
images.} 
\label{color-type-rels} 
\end{figure}

Figure~\ref{color-type-rels}, top, shows several colour-type relations based on
effective colour indices derived as outlined by Buta \textit{et al}. (1994) and
Buta \& Williams (1995). All show the same general trend: colours of E-S0
galaxies are the same within the scatter, but there is a smooth decrease in
colour indices from stage S0/a to Im. This morphology-colour relation is
beautifully illustrated with SDSS colour images in Fig.~\ref{color-type-rels},
bottom. The transition from redder to bluer colours begins with spiral arms,
leaving only bulges and bars remaining relatively red. However, as type
advances, bulges decrease in relative importance and bars become bluer. By type
Sm, the old stellar background is muted against the bright blue star-forming
disk.

\begin{figure}
\centering{\includegraphics[width=\textwidth]{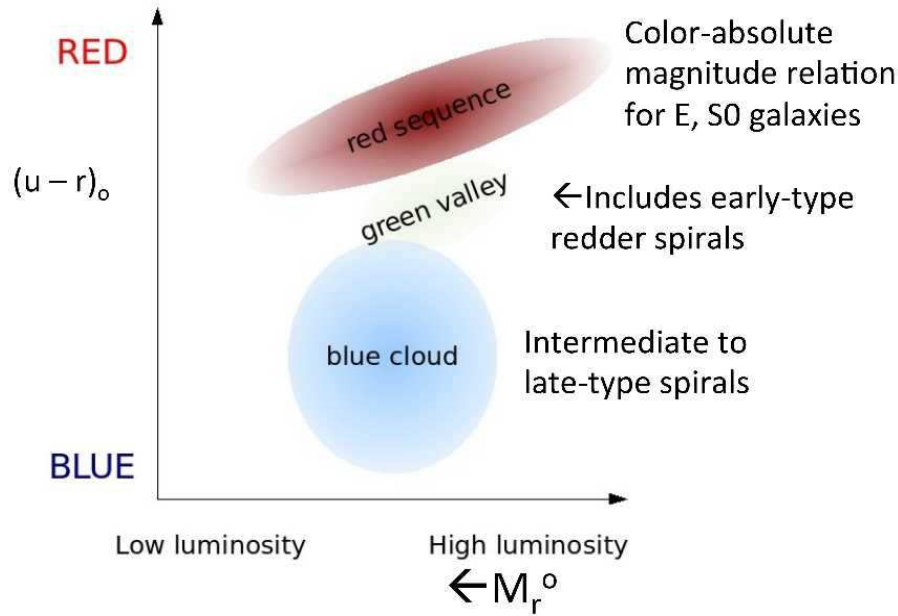}}
\caption{A `galactic Hertzsprung-Russell diagram' showing the three main groupings of 
galaxies, the `red sequence', the `blue cloud', and the `green valley'. 
(Adapted from Wikipedia).} 
\label{cmag} 
\end{figure}

The most famous result from SDSS multi-colour imaging is the `galactic Hertzsprung-Russell 
diagram', or colour-absolute magnitude diagram. This is shown in schematic form
in Fig.~\ref{cmag}. Galaxies that are made almost uniformly of old stars lie
along the `red sequence', the familiar colour-absolute magnitude relation for
early-type galaxies that was used extensively for distance scale studies in the
1980s. Star-forming galaxies, including spirals and irregulars, lie mainly in
the `blue cloud'. The `green valley' is the name given to the zone
intermediate between the red sequence and the blue cloud, and it is here where
redder, early-type spirals are usually found. Galaxies in the green valley are
thought to be evolving from the blue cloud to the red sequence, positioning
themselves according to their absolute luminosity.

\subsection{Galaxy Zoo and citizen science}

As I noted earlier, the SDSS is a goldmine for galaxy morphology. Not only are
the images of high quality for classification, but the sheer number of images,
on the order of a million, is beyond the capability of a small number of
experts.  An important question is, how to tap the information contained in the
survey in a reasonable amount of time? This is what Galaxy Zoo (GZ) was designed
to deal with: outsource galaxy morphology and classification to the Internet,
and allow non-professional volunteers to participate. The history of how the
project got started, and a summary of its results so far, is provided by Fortson
\textit{et al.} (2011).

\begin{figure}
\centering{\includegraphics[width=\textwidth]{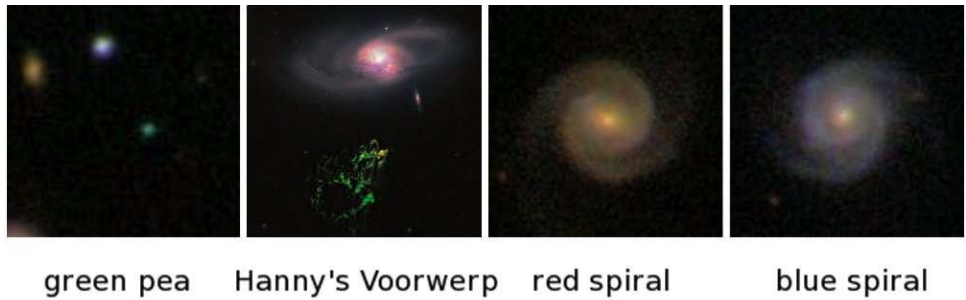}}
\caption{Highlights from GZ1 (B13). Hanny's Voorwerp reproduced with permission 
from Keel \textit{et al.} (2012).}
\label{zoo}
\end{figure}

Launched in 2006, GZ quickly  became a model example of scientists giving
something back to the general public: an opportunity to do science. The enlisted
volunteers became known as `citizen scientists', and although not a new
concept, the sheer number of such scientists who volunteered, more than 200\,000
forming a tight-knit community of galaxy morphologists, was surely a wonder to
behold. Eventually, two Zoo projects were activated:  GZ1, which asked very
basic questions about morphology, and GZ2 which was slightly more advanced. Some
results from GZ1 are (see also Fig.~\ref{zoo}):

\begin{enumerate}[(a)]\listsize
\renewcommand{\theenumi}{(\alph{enumi})}

\item decoupling of colour and morphology with high statistical significance, 
\item finding that 80\% of galaxies follow the usual colour-morphology 
correlation (meaning red early-types and blue late-types), 
\item attention brought to a significant number of red (passive) spirals and 
blue early-types,
\item studies showed that transformation from blue to red is faster than from 
spiral to early-type, 
\item no evidence for a preferred rotation direction in the Universe,
\item local fraction of mergers is 1--3\%, and
\item correlations between morphology and black hole growth.

\end{enumerate}

One thing that GZ offered to its volunteers was a real chance for discovery.
When examining so many images of galaxies that had not been studied in much
detail before, someone was bound to find something new and unusual. This was the
case with `Hanny's Voorwerp (Object)', a colourful cloud of ionised material
and some active star formation located near a faint spiral galaxy, IC\,2497.
Figure~\ref{zoo} shows an \textit{HST} image from Keel \textit{et al.} (2012).
Follow-up studies suggest that the Voorwerp is tidal debris from a past
encounter between IC\,2497 and another galaxy that drew out a tidal tail and
triggered infall into a supermassive black hole, producing a transient quasar
episode. The Voorwerp's light is an echo of the quasar phase acting on tidal
debris (Lintott \textit{et al.} 2009).

A second unusual type of object found in GZ is `green peas', which are compact
star-forming galaxies having a high equivalent width of [O{\sc iii}] emission
(Cardamone \textit{et al.} 2009). A recent study by Amor\'in \textit{et al.}
(2010) showed that green peas are a distinct class of metal-poor galaxies,
possibly affected by interaction-driven gaseous inflow over a short phase of
evolution.

Another result from GZ is that bars and bulge-dominated galaxies tend to be
found in denser environments than unbarred and disk-dominated counterparts
(Skibba \textit{et al.} 2012). Also, tidal dwarf galaxies, formed from ejected
debris during a merger, were studied by Kaviraj \textit{et al.} (2012) using the
GZ merger sample.

\subsection{Advanced galaxy morphology and classification}

While GZ took large-scale galaxy classification to the public, which led to
visual morphological information for hundreds of thousands of galaxies and
useful information for follow-up studies, the need for more detailed and
sophisticated types has led several professional groups to try and get more
information on fine structure details with the ultimate goal of facilitating
automated galaxy classification.  The key to success in such studies has been
the development of an interface that allows visual classification to be more
efficiently and more accurately carried out. This can be done either in
web-based fashion or off-web.

One of the first such studies was by Nair \& Abraham (2010), who used SDSS
$g$-band images to visually record morphological information on 14\,034 galaxies
brighter than magnitude $g^{\prime}=16$ and having $z<0.1$. The survey went
beyond RC3-style $T$-types to include recognition of rings, bars (regular and
ansae types), spiral arm morphology, dust, and tidal features.

Another advanced morphological project was the Extraction of the Idealised
Shapes of Galaxies from Imagery (EFIGI) survey (Baillard \textit{et al.} 2011),
which uses SDSS DR4 images of 4458 RC3 galaxies to get information on 16
morphological attributes, including features such as bulges, arms, bars, rings,
dust, flocculence, hotspots, inclination, and environment.  The procedure for
doing the `morphometry' was for 11 astronomers to first classify a common
subset of 100 galaxies to estimate relative biases and get each observer on a
common scale compared to the scale of morphological $T$-types in the RC3. Then
each observer individually classified 445 galaxies, or 0.1 of the final sample.
The final sets were then homogenised to give final classifications on the RC3
scale. The visual classification was done using a sophisticated interface called
`Manclass'. De Lapparent {\it et al.} (2011) analysed the statistical
properties of the final EFIGI catalogue. The goal of the EFIGI catalogue is to
set the stage for automated morphometry of the same attributes for a much larger
sample.

Another sophisticated morphology project was the Wide-field Nearby
Galaxy-clusters Survey (WINGS, Fasano \textit{et al.} 2012). This survey
provides morphological types of nearly 40\,000 galaxies in 76 nearby galaxy
clusters. The classification procedure began with 233 RC3 galaxies classified by
two astronomers to evaluate the reliability of visual $T$-types. Then a single
astronomer classified nearly a thousand randomly selected galaxies from the
cluster sample to train an automated tool called MORPHOT, which uses neural
networks and maximum likelihood techniques to extract types for the full cluster
samples.

All of these studies highlight much of the future of galaxy morphology and
classification, but most of all, they show that classical morphology still has
relevance to modern extragalactic research and still has much to offer as we
seek better understanding of galactic evolution.

\section{Summary}

Far from giving way to pure quantitative classification, galaxy
morphology today is a vibrant subject with a huge database of
material.

The Hubble tuning fork is being seriously modified. The placement of
S0s and the interpretation of dwarf early-types has led to the
resurrection of parallel sequence classification.

The high quality of digital images has allowed the old classification
systems to be modified to recognise the features of current interest in
galactic structure.

High-quality IR galaxy classification and morphology is now possible with
NIRS0S and the S$^4$G.

Early-type galaxies continue to be the focus of a great deal of
research. ATLAS$^{3{\rm D}}$ has been a major advance in understanding these
galaxies.

S$^4$G provides an opportunity to study the properties of extreme
late-type galaxies in great detail.

Various large imaging surveys, like SDSS, the \textit{Hubble} archives, deep
surveys like COSMOS, GOODS, HDF, HUDF, etc., continue to richly add to
morphological studies an evolutionary component.

The processes of secular evolution lie in the fine details of galaxy
morphology. Interpreting those details is the challenge of the coming
years.

\section*{Acknowledgments} 
I am deeply grateful to the organisers of this Winter School for giving me the
opportunity to participate as a lecturer in one of the most interesting topics
of astronomy today. It was an honour and a privilege to speak to the younger
astronomers and to be with such a great group of colleagues. I am also grateful
to Gerard de Vaucouleurs and Allan Sandage for inspiring my interest in galaxy
morphology almost 40 years ago, and for their encouragement and support over the
years that I knew them.  Finally, I am grateful to the US National Science
Foundation and to the National Aeronautics and Space Administration for past
financial support of my extragalactic studies.

This chapter uses images from the Sloan Digital Sky Survey (SDSS). Funding for
the creation and distribution of the SDSS Archive has been provided by the
Alfred P. Sloan Foundation, the Participating Institutions, NASA, NSF, the U.S.
Department of Energy, the Japanese Monbukagakusho, and Max Planck Society. This
chapter has also made use of THINGS, "The HI Nearby Galaxy Survey" (Walter
\textit{et al.} 2008), and BIMA-SONG, the Berkeley-Illinois-Maryland Survey of
Nearby Galaxies" (Helfer \textit{et al.} 2003). Other images are from the
archives of the {\it Hubble Space Telescope}, the {\it Spitzer Space Telescope},
and the {\it Galaxy Evolution Explorer} (GALEX). Observations with the NASA/ESA
{\it Hubble Space Telescope} were obtained at the Space Telescope Science
Institute, which is operated by the Association of Universities for Research in
Astronomy, Inc., under contract NAS 5-26555. The {\it Spitzer Space Telescope}
is operated by the Jet Propulsion Laboratory, California Institute of
Technology, under NASA contract 1407.  GALEX is a NASA mission operated by the
Jet Propulsion Laboratory. GALEX data is from the Multimission Archive at the
Space Telescope Science Institute (MAST). Support for MAST for non-HST data is
provided by the NASA Office of Space Science via grant NNX09AF08G and by other
grants and contracts.

%% file: buta.bbl
\begin{thebibliography}{}

\bibitem{}{Aguerri, J., Hunter, J., Prieto, M. {\it et al.} (2001), {\it A\&A}, {\bf 373}, 786}
\bibitem{}{Aguerri, J.~A.~L., M\'endez-Abreu, J., Corsini, E.~M. (2009), {\it A\&A}, {\bf 495}, 491}
\bibitem{}{Amor\'in, R.~O., P\'erez-Montero, E.,  V\'ilchez, J.~M. (2010), {\it ApJL}, {\bf 715}, 128}
\bibitem{}{Athanassoula, E. (2005), {\it MNRAS}, 358, 1477}
\bibitem{}{Athanassoula, E., Bureau, M. (1999), {\it ApJ}, {\bf 522}, 699}
\bibitem{}{Athanassoula, E., Misiriotis, A. (2002), {\it MNRAS}, {\bf 330}, 35}
\bibitem{}{Athanassoula, E., Romero-G\'omez, M.,  Masdemont, J.~J. (2009a), {\it MNRAS}, {\bf 394}, 67}
\bibitem{}{Athanassoula, E., Romero-G\'omez, M., Bosma, A.,  Masdemont, J.~J. (2009b), {\it MNRAS}, {\bf 400}, 1706}
\bibitem{}{Athanassoula, E., Morin, S., Wozniak, H. {\it et al.} (1990), {\it MNRAS}, {\bf 245}, 130}
\bibitem{}{Baade, W. (1963), {\it Evolution of Stars and Galaxies}, Cambridge: Harvard University Press}
\bibitem{}{Bahcall, J.~N., Kirhakos, S., Saxe, D.~H., Schneider, D.~P. (1997), {\it ApJ}, {\bf 479}, 642}
\bibitem{}{Baillard, A., Bertin, E., de Lapparent, V. {\it et al.} (2011), {\it A\&A}, {\bf 532}, 74}
\bibitem{}{Barazza, F., Jogee, S.,  Marinova, I. (2008), {\it ApJ}, {\bf 675}, 1194}
\bibitem{}{Bell, E.~F., de Jong, R.~S. (2001), {\it ApJ}, {\bf 550}, 212}
\bibitem{}{Bell, E.~F., McIntosh, D.~H., Katz, N., Weinberg, M.~D. (2003), {\it ApJS}, {\bf 149}, 289}
\bibitem{}{Bertin, G., Lin, C.~C., Lowe, S.~A., Thurstans, R.~P. (1989), {\it ApJ}, {\bf 338}, 78}
\bibitem{}{Bertola, F. (1987), in \textit{Structure and dynamics of elliptical galaxies}, IAU Symp., \textbf{127}, 135}
\bibitem{}{Binggeli, B., Sandage, A., Tammann, G.~A. (1985), {\it AJ}, {\bf 90}, 1681}
\bibitem{}{Binney, J., Tremaine, S. (1987), {\it Galactic Dynamics}, Princeton: Princeton University Press}
\bibitem{}{Binney, J., Tremaine, S. (2008), {\it Galactic Dynamics}, Second edition, Princeton: Princeton University Press}
\bibitem{}{Block, D.~L., Bournaud, F., Combes, F. {\it et al.} (2002), {\it A\&A}, {\bf 394}, 35}
\bibitem{}{Block, D.~L., Freeman, K.~C., Puerari, I. {\it et al.} (2004), in {\it Penetrating Bars Through Masks of Cosmic Dust}, D.~L. Block, I. Puerari, K.~C. Freeman, R. Groess, \& E. Block, eds., Astrophysics and space science library (ASSL) vol. 319. Dordrecht: Kluwer Academic Publishers, p. 15}
\bibitem{}{Bothun, G., Impey, C.~D., Malin, D.~F., Mould, J.~R. (1987), {\it AJ}, {\bf 94}, 23}
\bibitem{}{Bothun, G.~D., Sullivan, W.~T. (1980), {\it ApJ}, {\bf 242}, 903}
\bibitem{}{Bournaud, F., Combes, F. (2002), {\it A\&A}, {\bf 392}, 83}
\bibitem{}{Bureau, M., Aronica, G., Athanassoula, E. {\it et al.} (2006), {\it MNRAS}, {\bf 370}, 753}
\bibitem{}{Bureau, M., Athanassoula, E. (1999), {\it ApJ}, {\bf 522}, 686}
\bibitem{}{Bureau, M., Freeman, K.~C. (1999), {\it AJ}, {\bf 118}, 126}
\bibitem{}{Buta, R. (1984), {\it Proc. Astr. Soc. Australia}, {\bf 5}, 472}
\bibitem{}{Buta, R. (1985), {\it Proc. Astr. Soc. Australia}, {\bf 6}, 56}
\bibitem{}{Buta, R. (1986), {\it ApJS}, {\bf 61}, 609}
\bibitem{}{Buta, R. (1987), {\it ApJS}, {\bf 64}, 1}
\bibitem{}{Buta, R. (1991), {\it ApJ}, {\bf 370}, 130}
\bibitem{}{Buta, R. (1995), {\it ApJS}, {\bf 96}, 39}
\bibitem{}{Buta, R. (2002), in \textit{Disks of Galaxies: Kinematics, Dynamics and Peturbations}, E. Athanassoula, A. Bosma, and R. Mujica, eds., ASP Conf. Ser., \textbf{275}, 185}
\bibitem{}{Buta, R. (2013), in {\it Planets, Stars, and Stellar Systems. Volume 6: Extragalactic Astronomy and Cosmology}, T.~D. Oswalt, W.~C. Keels, eds., Springer, p. 1}
\bibitem{}{Buta, R., Block, D.~L.,  Knapen, J.~H. (2003), {\it AJ}, {\bf 126}, 1148}
\bibitem{}{Buta, R., Byrd, G.,  Freeman, T. (2003), \textit{AJ}, \textbf{125}, 634}
\bibitem{}{Buta, R., Byrd, G.,  Freeman, T. (2004), \textit{AJ}, \textbf{127}, 1982}
\bibitem{}{Buta, R., Combes, F. (1996), {\it Fund. Cosmic Phys.}, {\bf 17}, 95}
\bibitem{}{Buta, R., Corwin, H.~G.,  Odewahn, S.~C. (2007), {\it The de Vaucouleurs Atlas of Galaxies}, Cambridge: Cambridge University Press}
\bibitem{}{Buta, R., Crocker, D.~A. (1991), {\it AJ}, {\bf 102}, 1715}
\bibitem{}{Buta, R., Crocker D.~A. (1993a), {\it AJ}, {\bf 105}, 1344}
\bibitem{}{Buta, R., Crocker D.~A. (1993b), {\it AJ}, {\bf 106}, 939}
\bibitem{}{Buta, R., Crocker, D.~A., Byrd, G.~G. (1999), \textit{AJ}, \textbf{118}, 2071}
\bibitem{}{Buta, R. \textit{et al.} (2013), in preparation} 
\bibitem{}{Buta, R., Laurikainen, E., Salo, H. (2004), \textit{AJ}, \textbf{127}, 279}
\bibitem{}{Buta, R., Laurikainen, E., Salo, H., Block, D.~L., Knapen, J.~H. (2006), \textit{AJ}, \textbf{132}, 1859}
\bibitem{}{Buta, R., Laurikainen, E., Salo, H., Knapen, J.~H. (2010a), {\it ApJ}, \textbf{721}, 259}
\bibitem{}{Buta, R., Purcell, G.~B. (1998), {\it AJ}, {\bf 115}, 484}
\bibitem{}{Buta, R., Purcell, G.~B.,  Crocker, D.~A. (1995a), {\it AJ}, {\bf 110}, 1588}
\bibitem{}{Buta, R., Mitra, S., de Vaucouleurs, G., Corwin, H.~G. (1994), {\it AJ}, {\bf 107}, 118}
\bibitem{}{Buta, R., Knapen, J.~H., Elmegreen, B.~G. {\it et al.} (2009),  \textit{AJ}, \textbf{137}, 4487}
\bibitem{}{Buta, R., Ryder, S.~D., Madsen, G.~J. {\it et al.} (2001), {\it AJ}, {\bf 121}, 225}
\bibitem{}{Buta, R., Sheth, K., Regan, M. {\it et al.} (2010b), {\it ApJS}, \textbf{190}, 147}
\bibitem{}{Buta, R., van Driel, W., Braine, J. {\it et al.} (1995b), {\it ApJ}, {\bf 450}, 593}
\bibitem{}{Buta, R., Williams, K.~L. (1995), {\it AJ}, {\bf 109}, 543}
\bibitem{}{Buta, R., Zhang, X. (2009), {\it ApJS}, {\bf 182}, 559B}
\bibitem{}{Buta, R., Zhang, X. (2011), \textit{MSAIS}, \textbf{18}, 13}
\bibitem{}{Byrd, G.~G., Rautiainen, P., Salo, H., Buta, R., Crocker, D.~A. (1994), \textit{AJ}, \textbf{108}, 476}
\bibitem{}{Byrd, G.~G., Thomasson, M., Donner, K.~J., Sundelius, B., Huang, T.~Y., Valtonen, M. (1989), \textit{Celestial Mechanics}, \textbf{45}, 31}
\bibitem{}{Cappellari, M., Emsellem, E., Krajnovi\'c, D. {\it et al.} (2011), {\it MNRAS}, \textbf{416}, 1680}
\bibitem{}{Cardamone, C., Schawinski, K., Sarzi, M. {\it et al.} (2009), {\it MNRAS}, \textbf{399}, 1191}
\bibitem{}{Chung, A., van Gorkom, J.~H., Kenney, J.~D.~P., Crowl, H., Vollmer, B. (2009), \textit{AJ}, \textbf{138}, 1741}
\bibitem{}{Churchwell, E., Whitney, B.~A., Babler, B.~L. {\it et al.} (2004), \textit{ApJS}, \textbf{154}, 322}
\bibitem{}{Combes, F. (1991), in {\it Dynamics of Galaxies and Their Molecular Cloud Dstributions}, IAU Symp. 146, F. Combes \& F. Casoli, eds., Dordrecht, p. 255}
\bibitem{}{Combes, F., Sanders, R.~H. (1981), {\it A\&A}, \textbf{96}, 164}
\bibitem{}{Comer\'on, S., Knapen, J.~H., Beckman, J.~E. {\it et al.} (2010), {\it MNRAS}, \textbf{402}, 2462}
\bibitem{}{Contopoulos, G. (1980), {\it A\&A}, \textbf{81}, 198}
\bibitem{}{Contreras-Ramos, R., Annibali, F., Fiorentino, G. {\it et al.} (2011), {\it ApJ}, \textbf{739}, 74}
\bibitem{}{Corsini, E. (2011), \textit{MSAIS}, \textbf{18}, 23}
\bibitem{}{Corsini, E.~M., Debattista, V., Aguerri, J. (2003), \textit{ApJ}, \textbf{599}, 29}
\bibitem{}{Cowie, L., Hu, E., Songaila, A. (1995), \textit{AJ}, \textbf{110}, 1576}
\bibitem{}{Crocker, D.~A., Baugus, P.~D., Buta, R. (1996), {\it ApJS}, \textbf{190}, 147}
\bibitem{}{Curtis, H.~D. (1918), {\it Pub. Lick Obs. XIII}, Part I, 11}
\bibitem{}{Debattista, V., Sellwood, J.~A. (1999), {\it ApJL}, \textbf{513}, 107}
\bibitem{}{Debattista, V., Sellwood, J.~A. (2000), {\it ApJ}, \textbf{543}, 704}
\bibitem{}{de Lapparent, V., Baillard, A., Bertin, E. (2011), {\it A\&A}, \textbf{532}, 75}
\bibitem{}{de Vaucouleurs, G. (1958), {\it ApJ}, \textbf{127}, 487}
\bibitem{}{de Vaucouleurs, G. (1959), \textit{Handbuch der Physik}, \textbf{53}, 275}
\bibitem{}{de Vaucouleurs, G. (1963), {\it ApJS}, \textbf{8}, 31}
\bibitem{}{de Vaucouleurs, G., de Vaucouleurs, A., Corwin, H.~G., Buta, R., Paturel, G., Fouque\', P. (1991), \textit{Third Reference Catalogue of Bright Galaxies}, New York, Springer (RC3)}
\bibitem{}{de Vaucouleurs, G., Freeman, K.~C. (1972), \textit{Vistas in Astronomy}, \textbf{14}, 163}
\bibitem{}{Drinkwater, M., Gregg, M., Hilker, M. {\it et al.} (2003), \textit{Nature}, \textbf{423}, 519}
\bibitem{}{Drinkwater, M., Jones, J., Gregg, M, Phillips, S. (2000), \textit{Publ. Astr. Soc. Australia}, \textbf{17}, 227}
\bibitem{}{de Zeeuw, P.~T., Bureau, M., Emsellem, E. \textit{et al.} (2002), {\it MNRAS}, \textbf{329}, 513}
\bibitem{}{Dressler, A. (1980), {\it ApJ}, \textbf{236}, 351}
\bibitem{}{Elmegreen, D., Elmegreen, B.~G., Ferguson, T.~E., Mullan, B. (2007), {\it ApJ}, \textbf{663}, 734}
\bibitem{}{Elmegreen, D., Elmegreen, B.~G., Sheets, C. (2004), {\it ApJ}, \textbf{603}, 74}
\bibitem{}{Elmegreen, D., Elmegreen, B.~G., Yau, A. \textit{et al.} (2011), {\it ApJ}, \textbf{737}, 32}
\bibitem{}{Emsellem, E., Cappellari, M., Krajnovi\'c, D. \textit{et al.} (2007), {\it MNRAS}, \textbf{379}, 401}
\bibitem{}{Emsellem, E., Cappellari, M., Krajnovi\'c, D. \textit{et al.} (2011), {\it MNRAS}, \textbf{414}, 888}
\bibitem{}{Erwin, P. (2004), {\it A\&A}, \textbf{415}, 941}
\bibitem{}{Erwin, P. (2011), \textit{MSAIS}, \textbf{18}, 145}
\bibitem{}{Eskridge, P.~B., Frogel, J.~A., Pogge, R.~W. \textit{et al.} (2000), \textit{AJ}, \textbf{119}, 536}
\bibitem{}{Eskridge, P.~B., Frogel, J.~A., Pogge, R.~W. \textit{et al.} (2002), {\it ApJS}, \textbf{143}, 73}
\bibitem{}{Faber, S. (1973), {\it ApJ}, \textbf{179}, 423}
\bibitem{}{Fasano, G., Vanzella, E., Dressler, A. \textit{et al.} (2012), {\it MNRAS}, \textbf{420}, 926}
\bibitem{}{Firmani, C., Avila-Reese, V. (2003), \textit{RMxAC}, \textbf{17}, 107}
\bibitem{}{Flagey, N., Boulanger, F., Verstraete, L., Miville Deschenes, M.~A., Noriega Crespo, A., Reach, W.~T. (2006), {\it A\&A}, \textbf{453}, 969}
\bibitem{}{Fortson, L., Masters, K., Nichol, R. {\it et al.} (2011), arXiv/1104.5513}
\bibitem{}{Foyle, K., Rix, H.-W., Zibetti, S. (2010), {\it MNRAS}, \textbf{407}, 163}
\bibitem{}{Gil de Paz, A., Boissier, S., Madore, B.~F. \textit{et al.} (2007), {\it ApJS}, \textbf{173}, 185}
\bibitem{}{Gil de Paz, A., Madore, B.~F., Boissier, S. \textit{et al.} (2005), {\it ApJL}, \textbf{627}, 29}
\bibitem{}{Gil de Paz, A., Madore, B.~F., Pevunova, O. (2003), {\it ApJS}, \textbf{147}, 29}
\bibitem{}{Grouchy, R.~D., Buta, R.~J., Salo, H., Laurikainen, E. (2010), \textit{AJ}, \textbf{139}, 2465}
\bibitem{}{Grouchy, R.~D., Buta, R., Salo, H., Laurikainen, E., Speltincx, T. (2008), \textit{AJ}, \textbf{136}, 980}
\bibitem{}{Gunn, J.~E., Carr, M., Rockosi, C. \textit{et al.} (1998), \textit{AJ}, \textbf{116}, 3040}
\bibitem{}{Gunn, J.~E., Gott, J.~R. (1972), {\it ApJ}, \textbf{176}, 1}
\bibitem{}{Hasan, H., Norman, C. (1990), {\it ApJ}, \textbf{361}, 69}
\bibitem{}{Helfer, T.~T., Thornley, M.~D., Regan, M.~W. {\it et al.} (2003), {\it ApJS}, \textbf{145}, 259}
\bibitem{}{Helou, G., Roussel, H., Appleton, P. \textit{et al.} (2004), {\it ApJS}, \textbf{154}, 253}
\bibitem{}{Herschel, J. (1864), \textit{Catalogue of Nebulae and Clusters of Stars}, Philosophical Transactions of the Royal Society of London, Volume 154}
\bibitem{}{Higdon, J.~L., Buta, R.~J., Purcell, G.~B. (1998), \textit{AJ}, \textbf{115}, 80}
\bibitem{}{Hilker, M. (2011), \textit{European Astron. Society Publication Series}, \textbf{48}, 219}
\bibitem{}{Holwerda, B.~W., Keel, W.~C., Williams, B., Dalcanton, J.~J., de Jong, R.~S. (2009), \textit{AJ}, \textbf{137}, 3000}
\bibitem{}{Hoskin, M. (1982), \textit{Journal for the History of Astronomy}, \textbf{13}, 97}
\bibitem{}{Hubble, E. (1922), {\it ApJ}, \textbf{56}, 162}
\bibitem{}{Hubble, E. (1926), {\it ApJ}, \textbf{64}, 321}
\bibitem{}{Hubble, E. (1936), {\it The Realm of the Nebulae}, Yale: Yale University Press}
\bibitem{}{Hubble, E., Humason, M. (1943), {\it ApJ}, \textbf{74}, 43}
\bibitem{}{Hunt, L.~K., Malkan, M.~A. (1999), {\it ApJ}, \textbf{516}, 660}
\bibitem{}{Jeans, J. (1928), \textit{Astronomy and Cosmogony}, Cambridge: The University press}
\bibitem{}{Jedrzejewski, R. (1987), {\it MNRAS}, \textbf{226}, 747}
\bibitem{}{Jerjen, H., Kalnajs, A., Binggeli, B. (2000), {\it A\&A}, \textbf{358}, 845}
\bibitem{}{Kalapotharakos, C., Patisi, P., Grosbol, P. (2010), {\it MNRAS}, \textbf{408}, 9}
\bibitem{}{Karachentseva, V.~E. (1973), \textit{Astrofiz. Issled-Izv. Spets. Astrofiz. Obs.}, \textbf{8}, 3}
\bibitem{}{Kaviraj, S., Darg, D., Lintott, C., Schawinski, K., Silk, J. (2012), {\it MNRAS}, \textbf{419}, 70}
\bibitem{}{Kaviraj, S., Ting, Y.~S., Bureau, M. {\it et al.} (2012), \textit{MNRAS}, \textbf{423}, 49}
\bibitem{}{Keel, W.~C., Lintott, C., Schawinski, K. \textit{et al.} (2012), \textit{AJ}, \textbf{144}, 66}
\bibitem{}{Kendall, S., Kennicutt, R.~C., Clarke, C., Thornley, M.~D. (2008), {\it MNRAS}, \textbf{387}, 1007}
\bibitem{}{Kennicutt, R.~C., Armus, L., Bendo, G. \textit{et al.} (2003), {\it PASP}, \textbf{115}, 928}
\bibitem{}{Knapen, J.~H. (2005), {\it A\&A}, \textbf{429}, 141}
\bibitem{}{Knapen, J.~H. (2010), in \textit{Galaxies and their Masks}, D.~L. Block, K.~C. Freeman, \& I. Puerari, eds., Springer, p. 201}
\bibitem{}{Knapen, J.~H., James, P.~A. (2009), {\it ApJ}, \textbf{698}, 1437}
\bibitem{}{Knapen, J.~H., Shlosman, I., Peletier, R.~F. (2000), {\it ApJ}, \textbf{529}, 93}
\bibitem{}{Koopmann, R., Kenney, J.~D.~P. (2004), {\it ApJ}, \textbf{613}, 866}
\bibitem{}{Kormendy, J. (1979), {\it ApJ}, \textbf{227}, 714}
\bibitem{}{Kormendy, J. (1984), \textit{ApJ}, \textbf{286}, 116}
\bibitem{}{Kormendy, J. (1985), {\it ApJ}, \textbf{295}, 73}
\bibitem{}{Kormendy, J., Bender, R. (1996), {\it ApJL}, \textbf{464}, 119}
\bibitem{}{Kormendy, J., Bender, R. (2012), {\it ApJS}, \textbf{198}, 2}
\bibitem{}{Kormendy, J., Fisher, D.~B., Cornell, M.~E., Bender, R. (2009), {\it ApJS}, \textbf{182}, 216}
\bibitem{}{Kormendy, J., Kennicutt, R.~C. (2004), {\it ARA\&A}, \textbf{42}, 603}
\bibitem{}{Kormendy, J., Norman, C.~A. (1979), {\it ApJ}, \textbf{233}, 539}
\bibitem{}{Kuzio de Naray, R., Zagursky, M.~J., McGaugh, S.~S. (2009), \textit{AJ}, \textbf{138}, 1082}
\bibitem{}{Laurikainen, E., Salo, H., Buta, R., Knapen, J. (2007), {\it MNRAS}, \textbf{381}, 401}
\bibitem{}{Laurikainen, E., Salo, H., Buta, R., Knapen, J. (2009), {\it ApJL}, \textbf{692}, 34}
\bibitem{}{Laurikainen, E., Salo, H., Buta, R., Knapen, J. (2011), {\it MNRAS}, \textbf{418}, 1452}
\bibitem{}{Laurikainen, E., Salo, H., Buta, R., Knapen, J. (2011), \textit{Advances in Astronomy}, vol. 2011, id.516739}
\bibitem{}{Laurikainen, E., Salo, H., Buta, R., Knapen, J., Speltincx, T., Block, D.~L. (2006), {\it MNRAS}, \textbf{132}, 2634}
\bibitem{}{Lin, L.-H., Yuan, C., Buta, R. (2008), {\it ApJ}, \textbf{684}, 1048}
\bibitem{}{Lintott, C., Schawinski, K., Keel, W. \textit{et al.} (2009), {\it MNRAS}, \textbf{399}, 129}
\bibitem{}{Lundmark, K. (1927), \textit{Medd. Astr. Obs. Uppsala}, No. 30}
\bibitem{}{Lynden-Bell, D., Kalnajs, A.J. (1972), {\it MNRAS}, \textbf{157}, 1}
\bibitem{}{Lynds, C.~R., Toomre, A. (1976), {\it ApJ}, \textbf{209}, 382}
\bibitem{}{Madore, B.~F., Nelson, E., Petrillo, K. (2009), {\it ApJS}, \textbf{181}, 572}
\bibitem{}{Martin, D.~C., Fanson, J., Schiminovich, D. \textit{et al.} (2005), {\it ApJL}, \textbf{619}, 1}
\bibitem{}{Mart\'inez-Delgado, D., Gabany, R.~J., Crawford, K. \textit{et al.} (2010), \textit{AJ}, \textbf{140}, 962}
\bibitem{}{Mart\'inez-Valpuesta, I., Knapen, J.~H., Buta, R. (2007), \textit{AJ}, \textbf{134}, 1863}
\bibitem{}{Mart\'inez-Valpuesta, I., Shlosman, I., Heller, C. (2006), {\it ApJ}, \textbf{637}, 214}
\bibitem{}{Mateo, M. (1998), {\it ARA\&A}, \textbf{36}, 435}
\bibitem{}{Meidt, S., Schinnerer, E., Knapen, J.~H. \textit{et al.} (2012), {\it ApJ}, \textbf{744}, 17}
\bibitem{}{Morgan, W.~W. (1958), \textit{PASP}, \textbf{70}, 364}
\bibitem{}{Nair, P.~B., Abraham, R.~G. (2010), {\it ApJS}, \textbf{186}, 427}
\bibitem{}{Narayanan, D., Groppi, C.~E., Kulesa, C.~A., Walker, C.~A. (2005), {\it ApJ}, \textbf{630}, 269}
\bibitem{}{Oh, S., de Blok, W.~J.~G., Walter, F., Brinks, E., Kennicutt, R.~C. (2008), \textit{AJ}, \textbf{136}, 2761}
\bibitem{}{Oosterloo, T.~A., Morganti, R., Sadler, E.~M., Vergani, D., Caldwell, N. (2002), \textit{AJ}, \textbf{123}, 729}
\bibitem{}{Parsons, W. (1880), \textit{Observations of Nebulae and Clusters made with the 6 foot Reflector, from 1848 to about the year 1878}, Scientific Transactions of the Royal Dublin Society}
\bibitem{}{Pence, W.~D., Taylor, K., Freeman, K.~C. \textit{et al.} (1988), {\it ApJ}, \textbf{326}, 564}
\bibitem{}{Peng, C.-Y., Ho, L.~C., Impey, C.~D., Rix, H.-W. (2002), \textit{AJ}, \textbf{124}, 294}
\bibitem{}{P\'erez, I., Aguerri, J.~A.~L., M\'endez-Abreu, J. (2012), \textit{A\&A}, \textbf{540}, 103}
\bibitem{}{Purcell, G.~B. (1998), \textit{PhD Thesis}, Univ. of Alabama}
\bibitem{}{Quillen, A.~C., Frogel, J.~A., Gonz\'alez, R.~A. (1994), {\it ApJ}, \textbf{437}, 162}
\bibitem{}{Quinn, P.~J. (1984), {\it ApJ}, \textbf{279}, 596}
\bibitem{}{Rautiainen, P., Salo, H. (2000), {\it A\&A}, \textbf{362}, 465}
\bibitem{}{Rautiainen, P., Salo, H., Laurikainen, E. (2008), {\it MNRAS}, \textbf{388}, 1803}
\bibitem{}{Regan, M., Teuben, P. (2004), {\it ApJ}, \textbf{600}, 595}
\bibitem{}{Reynolds, J.~H. (1920), {\it MNRAS}, \textbf{80}, 746}
\bibitem{}{Richer, M., Bullejos, A., Borissova, J. {\it et al.} (2001), {\it A\&A}, \textbf{370}, 34}
\bibitem{}{Ryder, S.~D., Buta, R.~J., Toledo, H., Shukla, H., Staveley-Smith, L., Walsh, W. (1996), {\it ApJ}, \textbf{460}, 665}
\bibitem{}{Salo, H., Laurikainen, E., Buta, R., Knapen, J.~H. (2010), {\it ApJL}, \textbf{715}, 56}
\bibitem{}{Sandage, A. (1961), \textit{The Hubble Atlas of Galaxies}, Carnegie Inst. of Wash. Publ. No. 618}
\bibitem{}{Sandage, A. (2005), \textit{ARA\&A}, \textbf{43}, 581}
\bibitem{}{Sandage, A., Bedke, J. (1994), \textit{The Carnegie Atlas of Galaxies}, Carnegie Inst. of Wash. Pub. No. 638}
\bibitem{}{Sandage, A., Tammann, G.~A. (1981), {\it A Revised Shapley-Ames Catalog of Bright Galaxies}, Carnegie Institute of Washington Publ. No. 635 (first edition) (RSA)}
\bibitem{}{Sanders, R.~H., Tubbs, A.~D. (1980), {\it ApJ}, \textbf{235}, 803}
\bibitem{}{Schombert, J. (1986), {\it ApJS}, \textbf{60}, 603}
\bibitem{}{Schombert, J. (1987), {\it ApJS}, \textbf{64}, 643}
\bibitem{}{Schombert, J. (1988), {\it ApJ}, \textbf{328}, 475}
\bibitem{}{Schwarz, M.~P. (1981), {\it ApJ}, \textbf{247}, 77}
\bibitem{}{Schwarz, M.~P. (1984a), {\it MNRAS}, \textbf{209}, 93}
\bibitem{}{Schwarz, M.~P. (1984b), \textit{Proc. Astr. Soc. Aust.}, \textbf{5}, 464}
\bibitem{}{Schweizer, F., Ford, W.K., Jedrzejewski, R., Giovanelli, R. (1987), \textit{ApJ}, \textbf{320}, 454}
\bibitem{}{Schweizer, F., Seitzer, P. (1988), {\it ApJ}, \textbf{328}, 88}
\bibitem{}{Schweizer, F., van Gorkom, J., Seitzer, P. (1989), {\it ApJ}, \textbf{338}, 770}
\bibitem{}{Schweizer, F., Whitmore, B.~C., Rubin, V.~C. (1983), \textit{AJ}, \textbf{88}, 909}
\bibitem{}{Sheth, K., Regan, M., Hinz, J.~L. \textit{et al.} (2010), {\it PASP}, \textbf{122}, 1397}
\bibitem{}{Shlosman, I., Frank, J., Begelman, M.~C. (1989), \textit{Nature}, \textbf{338}, 45}
\bibitem{}{Simkin, S.~M., Su, H.~J., Schwarz, M.~P. (1980), {\it ApJ}, \textbf{237}, 404}
\bibitem{}{Skibba, R., Masters, K.~L., Nichol, R.~C. \textit{et al.} (2012), {\it MNRAS}, \textbf{423}, 1485}
\bibitem{}{Spitzer, L., Baade, W. (1951), {\it ApJ}, \textbf{113}, 413}
\bibitem{}{Struck, C. (2010), {\it MNRAS}, \textbf{403}, 1516}
\bibitem{}{Sulentic, J.~W., Verdes-Montenegro, L., Bergond, G. \textit{et al.} (2006), {\it A\&A}, \textbf{449}, 937}
\bibitem{}{Surace, J.~A., Sanders, D.~B., Vacca, W.~D., Veilleux, S., Mazzarella, J.~M. (1998), {\it ApJ}, \textbf{492}, 116}
\bibitem{}{Tremaine, S., Weinberg, M.~D. (1984), {\it ApJL}, \textbf{282}, 5}
\bibitem{}{Treuthardt, P., Salo, H., Rautinainen, P., Buta, R. (2008), \textit{AJ}, \textbf{136}, 300}
\bibitem{}{Vaisanen, P., Ryder, S., Mattila, S., Kotilainen, J. (2008), {\it ApJL}, \textbf{689}, 37}
\bibitem{}{van den Bergh, S. (1960a), {\it ApJ}, \textbf{131}, 215}
\bibitem{}{van den Bergh, S. (1960b), {\it ApJ}, \textbf{131}, 558}
\bibitem{}{van den Bergh, S. (1976), {\it ApJ}, \textbf{206}, 883}
\bibitem{}{van den Bergh, S. (1980), {\it PASP}, \textbf{92}, 409}
\bibitem{}{van den Bergh, S. (1998), \textit{Galaxy Morphology and Classification}, Cambridge, Cambridge University Press}
\bibitem{}{van den Bergh, S., Abraham, R.~G., Ellis, R.~S., Tanvir, N.~R., Santiago, B.~X., Glazebrook, K.~G. (1996), \textit{AJ}, \textbf{112}, 359}
\bibitem{}{van den Bergh, S., Cohen, J.~G., Hogg, D.~W., Blandford, R. (2000), \textit{AJ}, \textbf{120}, 2190}
\bibitem{}{Verdes-Montenegro, L., Sulentic, J., Lisenfeld, U. {\it et al.} (2005), {\it A\&A}, \textbf{436}, 443}
\bibitem{}{Walter, F., Brinks, E., de Blok, W.~J.~G. {\it et al.} (2008), \textit{AJ}, \textbf{136}, 2563}
\bibitem{}{White, S.~D.~M., Rees, M.~J. (1978), {\it MNRAS}, \textbf{183}, 341}
\bibitem{}{Whitmore, B.~C., Lucas, R.~A., McElroy, D.~B., Steiman-Cameron, T.~Y., Sackett, P.~D., Olling, R.~P. (1990), \textit{AJ}, \textbf{100}, 1489}
\bibitem{}{Wolf, M. (1908), \textit{Pub. Ap. Inst. K\"onig. Heidelberg}, Vol. 3, No. 5}
\bibitem{}{Yagi, M., Yoshida, M., Komiyama, Y. {\it et al.} (2010), \textit{AJ}, \textbf{140}, 1814}
\bibitem{}{York, D.~G., Adelman, J., Anderson, J.~E. \textit{et al.} (2000), \textit{AJ}, \textbf{120}, 1579}
\bibitem{}{Zhang, X. (1996), {\it ApJ}, \textbf{457}, 125}
\bibitem{}{Zhang, X. (1998), {\it ApJ}, \textbf{499}, 93}
\bibitem{}{Zhang, X. (1999), {\it ApJ}, \textbf{518}, 613}
\bibitem{}{Zhang, X., Buta, R. (2007), \textit{AJ}, \textbf{133}, 2584}
\bibitem{}{Zhang, X., Buta, R. (2012), \textit{MNRAS}, submitted (arxiv:1203.5334)}
\bibitem{}{Zibetti, S., Charlot, S., Rix, H.-W. (2009), {\it MNRAS}, \textbf{400}, 1181} 
\bibitem{}{Zwicky, F., Zwicky, M. (1971), \textit{Catalogue of Selected Compact Galaxies and Post-eruptive Galaxies}, Guemligen}

\end{thebibliography}
